\documentclass[runningheads,natbib]{svjour2}
\smartqed  
\usepackage{graphicx}
\usepackage{rotating}
\usepackage{amsmath}
\usepackage{amssymb}

\newcommand{\dd}{{\rm d}}
\newcommand{\biblab}[1]{}
\newcommand{\omegar}{\omega_{\rm r}}
\newcommand{\omegai}{\omega_{\rm i}}
\newcommand{\pt}{p_{\rm t}}
\newcommand{\rtr}{r_{\rm tr}}
\newcommand{\Eq}[1]{Eq.~(\ref{#1})}
\newcommand{\bolddelta}{\delta\kern-0.45em\delta\kern-0.45em\delta}
\newcommand{\boldeta}{\eta\kern-0.5em\eta\kern-0.5em\eta}
\newcommand{\boldOmega}{\Omega\kern-0.68em\Omega\kern-0.68em\Omega}
\newcommand{\bolddelr}{\bolddelta {\bf r}}
\newcommand{\boldF}{{\bf F}}
\newcommand{\Lrad}{L_{\rm rad}}
\newcommand{\CE}{{\cal E}}
\newcommand{\CF}{{\cal F}}
\newcommand{\CK}{{\cal K}}
\newcommand{\CP}{{\cal P}}

\newcommand{\muHz}{\,\mu {\rm Hz}}
\newcommand{\Msun}{\,{\rm M}_\odot}
\newcommand{\Gyr}{\,{\rm Gyr}}
\newcommand{\K}{\,{\rm K}}
\newcommand{\Teff}{T_{\rm eff}}
\newcommand{\ee}{{\rm e}}
\newcommand{\eye}{{\rm i}}
\journalname{Astronomy and Astrophysics Review}
{\catcode`\@=11 \gdef\SchlangeUnter#1#2{\lower2pt\vbox{\baselineskip 0pt
\lineskip0pt
  \ialign{$\m@th#1\hfil##\hfil$\crcr#2\crcr\sim\crcr}}}
}
\def\gtrsim{\mathrel{\mathpalette\SchlangeUnter>}}
\def\lessim{\mathrel{\mathpalette\SchlangeUnter<}}

\begin{document}

\title{Asteroseismology and interferometry
}


\author{ M.~S.~Cunha \and C.~Aerts \and  J.~Christensen-Dalsgaard \and A.~Baglin \and L.~Bigot \and T.~M.~Brown \and C.~Catala \and O.~L.~Creevey \and A.~Domiciano~de~Souza \and P.~Eggenberger \and P.~J.~V.~Garcia \and  F.~Grundahl \and P.~Kervella \and D.~W.~Kurtz \and P.~Mathias \and A.~Miglio \and M.~J.~P.~F.~G.~Monteiro \and G.~Perrin \and F.~P.~Pijpers \and D.~Pourbaix \and A.~Quirrenbach \and K.~Rousselet-Perraut \and T.~C.~Teixeira \and F.~Th\'evenin \and M.~J.~Thompson
}

\authorrunning{Cunha {\etal}} 
\institute{M.~S. Cunha and T.~C. Teixeira \at 
              Centro de Astrof\'\i sica da Universidade do Porto,   
              Rua das Estrelas, 4150-762, Porto, Portugal.
              \email{mcunha@astro.up.pt}
           \and
           C. Aerts \at
             Instituut voor Sterrenkunde, Katholieke Universiteit Leuven,
	     Celestijnenlaan 200 D, 3001 Leuven, Belgium; Afdeling Sterrenkunde, Radboud University Nijmegen, PO Box 9010, 6500 GL Nijmegen, The Netherlands.
           \and
            J.~Christensen-Dalsgaard and F. Grundahl \at
              Institut for Fysik og Astronomi, Aarhus Universitet, Aarhus, Denmark.
           \and
           A. Baglin and C. Catala and P. Kervella and G. Perrin \at
              LESIA, UMR CNRS 8109, Observatoire de Paris, France.
           \and
           L.~Bigot and F.~Th\'evenin \at
              Observatoire de la C\^ote d'Azur, UMR 6202, BP 4229, F-06304, Nice Cedex 4, France.
           \and
           T.~M. Brown \at
              Las Cumbres Observatory Inc., Goleta, CA 93117, USA.
           \and
           O.~L.~Creevey \at
              High Altitude Observatory, National Center for Atmospheric
           Research, Boulder, CO 80301, USA; Instituto de Astrofísica de Canarias, Tenerife, E-38200, Spain.
           \and 
           A.~Domiciano~de~Souza \at
              Max-Planck-Institut f\"ur Radioastronomie, Auf dem H\"ugel 69, 53121 Bonn, Germany. 
           \and 
           P.~Eggenberger \at 
              Observatoire de Gen\`eve, 51 chemin des Maillettes, 1290 Sauverny,
           Switzerland; Institut d'Astrophysique et de G\'eophysique de l'Universit\'e de Li\`ege All\'ee du 6 Ao\^ut, 17 B-4000 Li\`ege, Belgium.
           \and
           P.~J.~V.~Garcia \at
                  Centro de Astrof\'{\i}sica, Universidade do Porto,
      Rua das Estrelas, 4150-762 and
     Departamento de Engenharia F\'{\i}sica,
     Faculdade de Engenharia, Universidade do Porto,
     Rua Dr. Roberto Frias, 4200-465 Porto, Portugal.
          \and
          D.~W.~Kurtz \at
             Centre for Astrophysics, University of Central Lancashire, Preston PR1 2HE, UK.
          \and
	 P. Mathias \at
          Observatoire de la C\^ote d'Azur, UMR 6203, BP 4229, F-06304, Nice Cedex 4, France.  
          \and
           A. Miglio \at
	   Institut d'Astrophysique et de G\'eophysique de l'Universit\'e de Li\`ege All\'ee du 6 Ao\^ut, 17 B-4000 Li\`ege, Belgium.
          \and
          M.~J.~P.~F.~G.~Monteiro \at
            Centro de Astrof\'\i sica e Departamento de Matem\'atica Aplicada da Faculdade de Ci\^encias, Universidade do Porto, Porto, Portugal.
          \and
           F. P. Pijpers \at
             Space and Atmospheric Physics Group, Imperial College London, Exhibition Road, London, SW7 2AZ, UK.
           \and
           D. Pourbaix \at
             F.R.S-FNRS, Institut d'Astronomie et d'Astrophysique, Universit\'e Libre de Bruxelles, 1050 Brussels, Belgium
           \and
           A. Quirrenbach \at
             Zentrum f\"ur Astronomie der Universit\"at Heidelberg, Landessternwarte K\"onigstuhl, 69117 Heidelberg, Germany.
           \and
           K. Rousselet-Perraut \at
             Laboratoire d'Astrophysique de l'Observatoire de Grenoble, BP 53, 38041 Grenoble Cedex 9, France.
           \and
            M. J. Thompson \at
             School of Mathematics and Statistics, University of Sheffield, Hounsfield Road, Sheffield S3 7RH, UK.
}

\date{Received: date}

\maketitle

\begin{abstract}
Asteroseismology provides us with a unique opportunity to improve our understanding of stellar structure and evolution. Recent developments, including the first systematic studies of solar-like pulsators, have boosted the impact of this field of research within Astrophysics and have led to a significant increase in the size of the research community. In the present paper we start by reviewing the basic observational and theoretical properties of classical and solar-like pulsators and present results from some of the most recent and outstanding studies of these stars. We centre our review on those classes of pulsators for which interferometric studies are expected to provide a significant input. We discuss current limitations to asteroseismic studies, including difficulties in mode identification and in the accurate determination of global parameters of pulsating stars, and, after a brief review of those aspects of interferometry that are most relevant in this context, anticipate how interferometric observations may contribute to overcome these limitations. Moreover, we present results of recent pilot studies of pulsating stars involving both asteroseismic and interferometric constraints and look into the future, summarizing ongoing efforts concerning  the development of future instruments and satellite missions which are expected to have an impact in this field of research. 

\keywords{Stars: variables: general \and  Stars: interiors \and Techniques: interferometric}
\end{abstract}

\vspace{1cm}
\setcounter{tocdepth}{2}
\tableofcontents
\newpage

\section{Introduction}
\label{intr}

\subsection{Historical perspective}
Periodic variable stars have been known to exist since the $17{\rm th}$ century,
when Jan Fokkens Holwarda realized that the magnitude of Mira, a variable star
discovered almost a century earlier by David Fabricius, varied periodically with
a period of 11\,months \citep{hoffleit97}. The periodic variability observed by
Holwarda is intrinsic to Mira and results from consecutive contractions and
expansions of its surface, associated with waves that propagate within its
interior. Stars which owe their variability to this phenomenon are known
as {\it pulsating (variable) stars}.

Despite the early discovery by Holwarda, the understanding that some variable
stars owe their variability to intrinsic pulsations came almost three centuries
later, following a suggestion by \citet{shapley14} driven by the difficulties
with reconciling the observed variability with the hypothesis of binarity. At
about the same time, pulsating stars started being systematically used as tools
for astrophysics research following the discovery of the Period-Luminosity
relation by Henrietta Leavitt \citep{leavitt08,pickering12}. Classical pulsating
stars have since been used as ``standard candles" of the Universe and, still
today, are of fundamental importance for the determination of astrophysical
distances.

During the decades that followed the discovery by Henrietta Leavitt, the
theoretical studies of classical pulsating stars were directed towards the
mathematical description of the pulsations and the understanding of the
mechanism behind their excitation \cite[e.g.,][and references
therein]{zhevakin63}. In the early 60s, the discovery and interpretation of
solar oscillations \citep{leighton62,frazier68,ulrich70} opened a new era in the
study of pulsating stars. Solar oscillations have been shown to be amazing tools
to `look' inside the Sun and image its structure and dynamics. The study of
solar oscillations fostered the development of an entire new field of research
-- known as {\it Helioseismology}
\cite[e.g.,][]{Gough1977a,duvall82,duvall84,christ85} -- which has proved
extremely successful in probing the physics of the solar interior
\citep[see][for a recent review on helioseismology]{Christ2002}.

\begin{figure}
\centering
\includegraphics[width=0.85\textwidth]{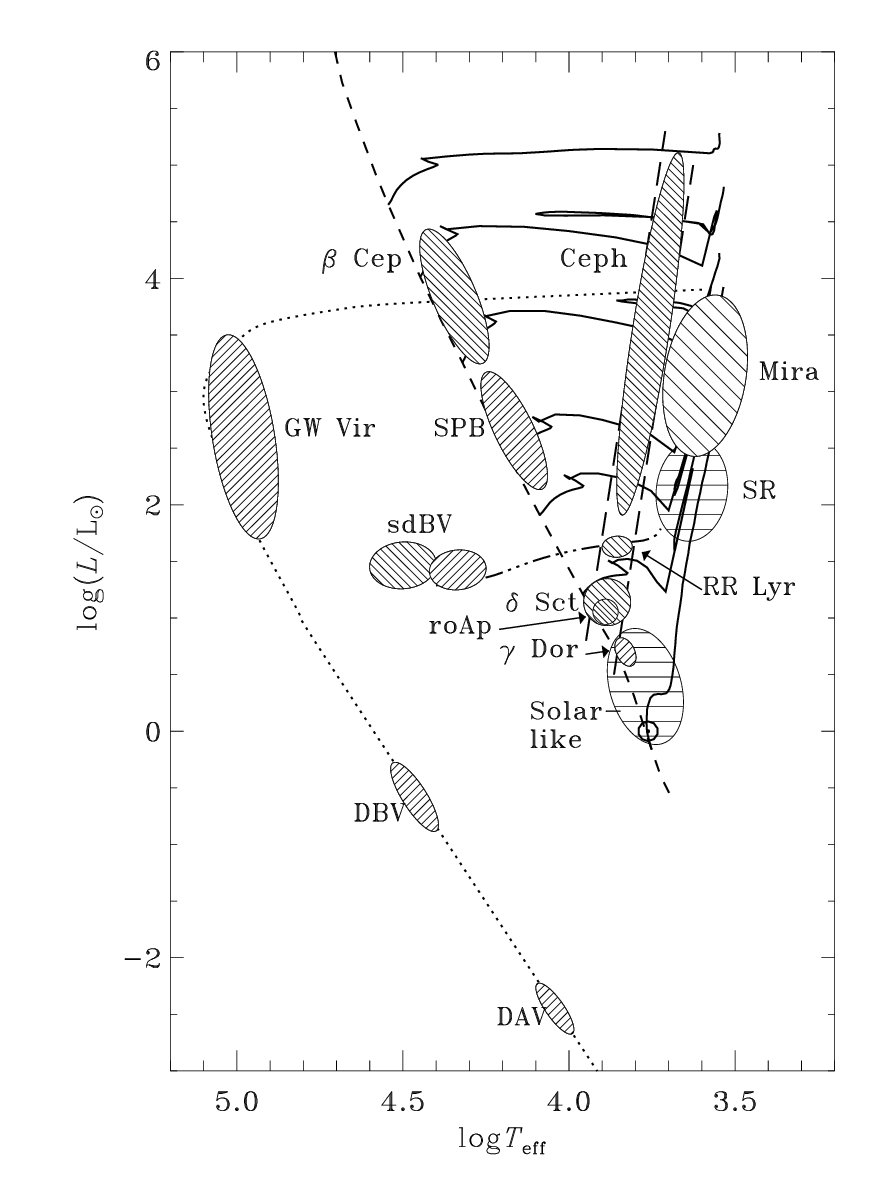}
\caption{Hertzsprung-Russell Diagram showing different classes of pulsating stars.
Some of these are named after a particular member of the class. Others are
acronyms, standing, respectively, for: rapidly oscillating Ap (roAp); Slowly
Pulsating B (SPB); subdwarf B variables (sdBV).  The group labelled GW Vir
includes what has formerly been known as the PNNV stars (for Planetary Nebulae
Nuclei Variables), and the variable hot DO white dwarfs (DOV); the DBV and DAV
stars are variable DB (helium-rich) and DA (hydrogen-rich) white dwarfs.
The parallel long-dashed lines indicate the Cepheid instability strip.
}
\label{HRD}       
\end{figure}

Following on the success of helioseismology, attempts to use intrinsic
oscillations of stars other than the Sun to learn about their internal structure
and dynamics have led to the development of {\it Asteroseismology}. However, the
road-map to success of asteroseismology proved much harder, as result of the
limitations brought about by our incapacity to resolve distant stars, and also
by the limited number of oscillation frequencies observed in most of them. In
fact, even though many different classes of pulsating stars are presently known
-- most of these are shown in Fig.~\ref{HRD} -- not all classes of pulsators are
suitable for asteroseismic studies (cf.\ Section~\ref{mspulsators}).

The $21^{\rm st}$ century has brought important new developments to
asteroseismology. With highly precise spectrometers such as CORALIE and HARPS
(at ESO-La Silla), UCLES (at AAT), and UVES (at ESO-Paranal) fully operational,
it was finally possible to make clear detections of solar-like oscillations in
stars other than the Sun \cite[see][for a recent review]{bedding07}, thus
confirming earlier evidence, found by different authors with different observing
techniques and instruments, of excess power in the oscillation spectra of some
of these stars \cite[e.g.,][]{brown91,schou01}. Our understanding of the Sun and
solar oscillations, as well as the tools developed in the context of
helioseismology, make us confident that the detection of solar-like pulsations
in stars other than the Sun will lead to major developments in our understanding
of stellar structure, dynamics and evolution.

\subsection{Stellar modelling and the ultimate goal of asteroseismology}
\label{sec:stelmod}
Broadly speaking, ``classical'' stellar observables, such as the effective temperature,
gravity, and metallicity, are insensitive to the
details of the internal structure of stars, and do
not provide sufficient constraints for the determination of the
basic stellar parameters, not to mention the calibration of
parameters commonly used in stellar modelling.

Pulsation frequencies, on the other hand, are very sensitive to the details of
the internal structure of stars. Each individual frequency probes the stellar
interior differently from all other frequencies. Thus, in principle, the
accurate measurement of a large number of oscillation frequencies in a given
star allows one to study the details of its internal structure. Moreover, such
frequencies may provide sufficient constraints to reduce the allowed space of stellar parameters to the
extent that one may be convinced that a star is indeed within a
particular error ellipse in the classical HR diagram.

There are many aspects about the physics and dynamics of
stellar interiors that are not fully understood and that may be
addressed with asteroseismology. To deal with some of the unknown
details of the physics of stellar interiors, stellar evolutionary
codes include parameters that can be tuned. Most of these relate to
the treatment of convection, diffusion and settling of heavy elements,
the equation of state, and opacities. Among these, the poor treatment of
convection and, to some extent, of diffusion, can have important
implications in the context of stellar modelling.

Global convective instabilities are present either in the core or
envelope of most stars. Nevertheless, the appropriate modelling of
convection remains one of the most difficult tasks in the context of
stellar astrophysics. The ``standard'' recipe for convection in the
context of stellar model fitting is the Mixing Length Theory \citep[MLT,][]{Bohm1958,Henyey1965}.  
An alternative treatment of
convection, developed by \citet{Canuto1991,CanMazz1992}, is also often used.
 Connected to
the treatment of convection is the problem of the extent of overshoot \citep[e.g.,][]{zahn91}.  
As observations improve, and, in particular, as more evolved stars are observed, details such as overshoot will become increasingly important in the
modelling. Strong uncertainties exist also in studies concerning the
interplay between convection and other physical phenomena, such as
rotation, magnetic fields, and radiation. All these limitations, in turn, limit
our ability 
to determine accurate
stellar ages and stellar global parameters.

Likewise, element diffusion occurs in most stars. Diffusion plays a
key role in a diversity of contexts in stellar astrophysics, including
studies of the Sun and of most main-sequence stars, and studies of
white dwarfs. As a result of diffusion, often the original chemical
composition of stars is hidden below the surface and cannot be reached
by direct observations. There are indeed
several physical mechanisms whose physical description is not yet very
well known but whose effect may imply a different mixture at the
surface than in the interior. These include gravitational settling,
radiative levitation, rotational mixing and mass loss through a stellar
wind \citep{theado05a,theado05b}. To test
deep diffusion models it is necessary to connect them to observations
which, in turn, can only be done if accurate models of the atmospheric
layers are available. Unfortunately, stars in which the effects of
diffusion are most obvious -- such as peculiar stars -- tend to have
complex atmospheres, making the task of directly testing diffusion
models hardly possible.  Asteroseismology allows us to perform
indirect tests to diffusion theories, and make inferences on the competition
between diffusion and mixing processes in the interior of pulsating
stars. Some attempts to perform such tests have already been carried out
using seismic data of roAp stars \citep{cunha04,vauclair04}.

Except for the case of the Sun, the current understanding of stellar interiors is very limited also concerning the presence of physical agents that may introduce deviations from spherical symmetry. Stellar evolutionary codes assume that stars are spherically symmetric and, thus, aspects such as rotation (let alone differential rotation) and magnetic fields, are commonly ignored when modelling the interior of stars. Until recently, models based on the assumption of spherical symmetry could reproduce reasonably well the observables. That has changed in the past few years, as a result of the development of new instrumentation with capability to resolve (even if, in most cases, indirectly) the surface of stars other than the Sun;
striking examples are cases where large departures from sphericity 
have been detected interferometrically \citep[e.g.,][]{Domici2003}.
This, together with constraints on departures from spherical symmetry also in stellar interiors
that may be revealed by stellar oscillations, will continue to motivate
the development of a new generation of stellar evolutionary codes 
that takes into account substantial departures from spherical symmetry,
particularly caused by rotation.
Encouraging progress has been made in recent years
\citep[e.g.,][]{Roxbur2004, roxburgh06, Jackso2005, rieutord06a,
MacGre2007} 
\citep[see also][for a review]{rieutord06b}.

\subsection{Origin and physical nature of the oscillations}
\label{origin}

The ability to fulfill the ultimate goal of asteroseismology depends critically
on the understanding of the physics underlying the observed phenomena. The
detailed dependence of the oscillations on the stellar interior, and hence their
asteroseismic diagnostic potential, obviously arises from their physical
nature. Consequently, in an asteroseismic study it is the nature and origin of
the observed pulsations that are considered. In fact, the same star may belong
to different pulsating classes (following the traditional classification), if
modes of different origin and/or different physical nature are excited.

Stars, including the Sun, display a broad range of pulsations.  The
large-amplitude pulsators detected initially can generally be understood in
terms of spherically symmetric, or {\it radial\/}, pulsations.  However, in many
stars, including the Sun, we observe oscillations with a variety of structure on
the stellar surface; in the solar case these extend to modes with a surface
wavelength of a few thousand kilometers. In distant stars, which are not
resolved, it is only the large scale structure that can be detected, since the
small-scale structure is averaged out in the observations. In practice this
means that the sensitivity of brightness variations is restricted to modes with
less than 3 to 4 node lines at the surface.

A rough measure of stellar pulsation periods is the {\it dynamical timescale\/}
\begin{equation}
t_{\rm dyn} = \left({R^3 \over G M} \right)^{1/2} \propto \bar \rho^{\,\,-1/2} \; ,
\label{eq:dyntime}
\end{equation}
where $R$ and $M$ are the surface radius and mass of the star, $G$
is the gravitational constant, and $\bar \rho$ is the mean density of
the star. Hence, even a simple measurement of a pulsation period gives
some indication of the overall properties of a star.

Concerning the physical nature of the oscillations, broadly speaking, the modes
are either of the nature of standing acoustic waves (commonly referred to as
pressure modes or {\it p modes\/}) or internal gravity waves ({\it g modes\/});
the latter involve departure from spherical symmetry and hence are nonradial
modes.  However, particularly in evolved stars, modes of a mixed nature are also
found.  In addition, the Sun shows modes that can be identified as surface
gravity waves, of relatively short surface wavelength.

Concerning their origin, the oscillations can be either intrinsically stable, or
intrinsically unstable. Intrinsically unstable oscillations result from the
amplification of small disturbances, through a heat-engine mechanism, acting in
an appropriate region of the star. The original perturbation grows until some
amplitude-limiting mechanism sets in, determining the final amplitude. This
mechanism may depend on subtle details of the mode, hence the resulting
amplitudes can vary strongly over a range of unstable modes.  Also, as discussed
below, it depends on the precise location of specific features in the opacity
within the star, and hence the instability tends to be confined to specific
regions in the HR diagram.  An example is the {\it Cepheid instability strip\/},
indicated in Fig.~\ref{HRD}.  
This type of
pulsations, originally observed in stars such as Cepheids, is commonly referred
to as {\it classical pulsations}.  In contrast with classical pulsations,
intrinsically stable oscillations are stochastically excited by an external
forcing -- typically near-surface convection. The resulting amplitudes are
determined by the balance between the energy input from the forcing and the
damping. Since the damping and forcing typically vary relatively slowly with
frequency this tends to lead to the excitation of modes in a substantial
frequency range. Pulsations of this type were first found in the Sun, and are
commonly referred to as {\it solar-like pulsations}.

\subsection{\label{mspulsators}Selected stellar pulsators}
Probably the most successful asteroseismic studies carried out thus far concern
white dwarf pulsators. Unfortunately, these stars are not suitable
interferometric targets, and, thus, we shall not discuss them in the present
review. For recent reviews on asteroseismology of white dwarf pulsators see,
e.g., \citet{Metcal2005,kepler07}.

From the point of view of asteroseismology a crucial requirement is to have a
substantial number of accurately determined frequencies.  This is satisfied both
for classical, heat-engine-driven pulsators, particularly in stars near the main
sequence, and for solar-like pulsators, and indeed both classes are very
interesting as asteroseismic targets.  However, there are also substantial
differences between them.  The classical pulsators tend to show much larger
amplitudes, and hence they are {\it a priori\/} easier to observe; after all,
this is precisely the origin of their `classical' status.  However, the
distribution of mode amplitudes amongst the modes that are expected to be
unstable is highly irregular, and so therefore is the selection of modes
observed to a given level of sensitivity. These and other properties of the
frequency spectra of classical and solar-like pulsators will be discussed
further in Section \ref{excitation}, where the mechanism driving the
oscillations in each of these types of pulsators will be looked at more closely.

Among the classes of pulsating stars along the main sequence that are likely to
be most promising when it comes to combined asteroseismic~-interferometric
studies are the {\it $\delta\,$Scuti}, {\it $\beta\,$Cephei} and the rapidly
oscillating Ap or {\it roAp} stars (shown in Fig.~\ref{HRD}).

The {\it $\delta\,$Scuti} stars are found at the crossing of the main sequence
and the Cepheid instability strip. They have masses between 1.5 and
2.5\,M$_\odot$, where M$_\odot$ is the mass of the Sun, and exhibit both radial
and non-radial pressure modes and gravity modes. Their observational properties
are analysed and reviewed by \citet{rodriguez01}. The $\delta\,$Scuti stars are
the main sequence classical pulsators for which the largest number of
oscillation modes have been observed. The record holder is FG\,Vir for which 79
frequencies were detected by \citet{Breger2005} and whose amplitude spectrum is
reproduced in Fig.\,\ref{fig:fgvir}.  The stars 4\,CVn \citep[34
frequencies,][]{breger00} and XX\,Pyx \citep[30 frequencies,][]{handler00}
44\,Tau \citep[29 frequencies,][]{antoci07} are competing for the second place.
The observed frequency spectra of these three stars show that the
$\delta\,$Scuti stars have complex oscillation patterns, possibly including
variable amplitudes from season to season.  As discussed by \citet{Breger2006}
such variability is difficult to distinguish from beating between extremely
close pairs of modes; in the case of the star FG Vir they identify the observed
variability as arising from such beating, although the extremely small frequency
separations are difficult to account for on physical grounds.

\begin{figure}
\centering
\rotatebox{270}{\resizebox{10cm}{!}{\includegraphics{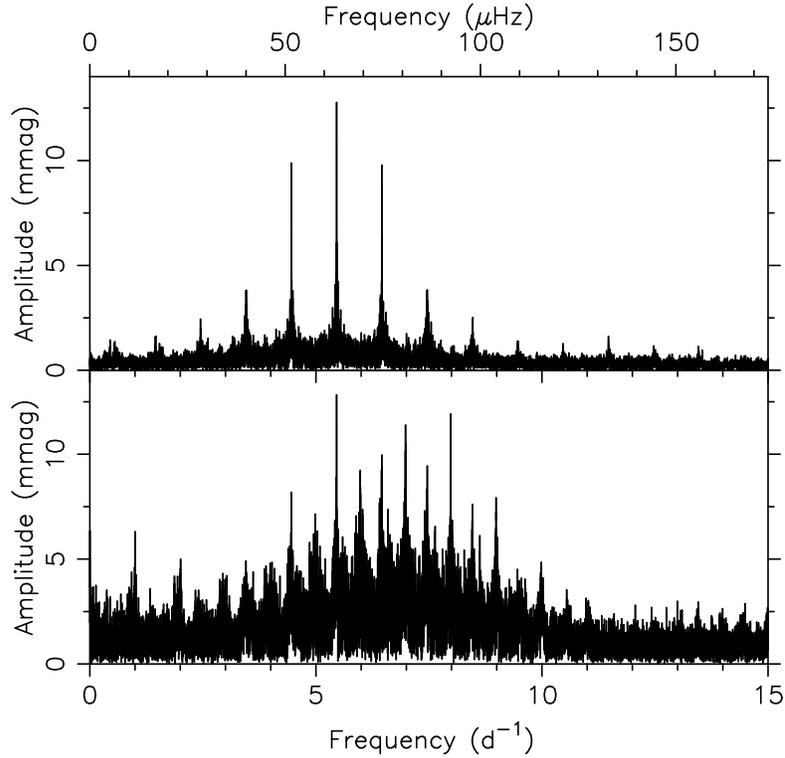}}}
\caption{Amplitude spectrum of the $\beta\,$Cephei star HD\,129929 derived from
single-site Geneva U data spanning 21 years (bottom panel). The top panel is the
spectral window, resulting from a noiseless sine generated with the observed
amplitude of the dominant mode and for the sampling of the data. (Data taken
from \citet{aerts03}).}
\label{hd129929}       
\end{figure}

{\it $\beta\,$Cephei} stars are the more massive analogues of the
$\delta\,$Scuti stars along the main sequence. They are stars with masses
between 8 and 18\,M$_\odot$ and also exhibit both radial and non-radial pressure
modes and gravity modes. Their observational characteristics have recently been
reviewed by \citet{Stanko2005}.  A typical amplitude spectrum resulting from 21
years of single-site photometry is shown in Fig.\,\ref{hd129929}.  Recent
progress in the seismic interpretation of selected $\beta\,$Cephei stars has
raised serious questions about the modelling of these stars, since current standard
stellar structure models of B stars are unable to explain the oscillation data
for the best-studied stars: HD\,129929 \citep{aerts03}, $\nu\,$Eri
\citep{Pamyat2004,Aussel2004}, and 12\,Lac \citep{Handle2006,ausseloos05}.
We return to this issue in Section~\ref{heatengine}.

Also at the crossing of the main sequence and the Cepheid
instability strip, we can find another interesting group of pulsators,
namely, the roAp stars \citep[see, e.g.,][for recent reviews]{cunha05,cunharev07,kurtzetal04}. While similar in mass and age to main sequence
$\delta$~Scuti stars, these stars differ from the latter in that they
have very peculiar atmospheres, particularly concerning their
chemical composition, and are permeated by strong magnetic
fields. Moreover, they rotate significantly slower than $\delta$~Scuti
stars. In principle, the chemical peculiarities observed at the
surface of Ap stars provide excellent environments to test diffusion
theories. However, the complexity of their atmospheres makes this task
rather hard. The pulsating properties of roAp stars are also
significantly different from those of $\delta$~Scuti stars. 
Oscillations of roAp stars
can be either mono-periodic or multi-periodic and result from acoustic modes
generally of significantly higher frequency than those found in
$\delta$~Scuti stars. The number of frequencies observed in the former
is also typically smaller than in the latter.

In addition to the classes mentioned above, other classical pulsators
exist along the main sequence such as, e.g., the $\gamma\,$Doradus stars
\citep[][and references therein]{kaye99} and the slowly pulsating B
or {\it SPB} stars \citep[][and references therein]{aerts99}. These two classes
consist of stars with multiperiodic gravity modes with periods of
order 0.5 to 3\,d and thus pose a very serious observational
challenge when it comes to detecting numerous modes.

\begin{figure}
\centering
  \includegraphics[width=0.85\textwidth]{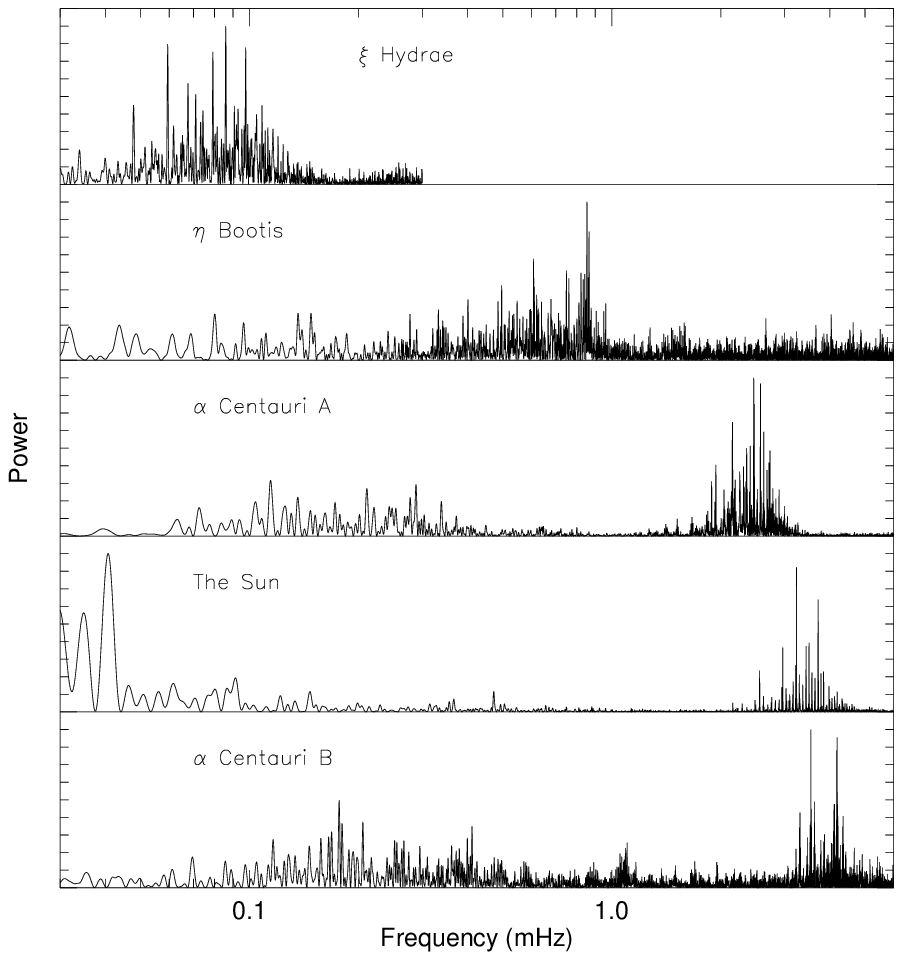}
\caption{Power spectra of solar-like oscillations of selected stars,
including the Sun, organized in the order of decreasing mean density
from bottom to top, and shown on the same frequency scale.
The ordinate is arbitrary.
The solar data were
obtained with GOLF instrument on the SOHO spacecraft, using Doppler observations
in light integrated over the disk of the Sun \citep[e.g.,][]{Garcia2001}.
The $\alpha$ Cen binary system brackets the Sun in mass and
has a slightly higher age;
data for the A component, with mass around $1.1 \Msun$, are from
\citet{Butler2004},
whereas data for the B component,
with a mass around $0.9 \Msun$, are from \citet{Kjelds2005}.
The spectrum for $\eta$~Boo, a subgiant in the hydrogen shell-burning
phase, was obtained by \citet{Kjelds1995a},
whereas the data for $\xi$ Hya, a G7 giant likely in the core helium-burning
phase, are from \citet{Frands2002}.
}
\label{fig:solarlikespect}       
\end{figure}

Last, but not least, solar-like pulsators are found among main-sequence core,
and post-main-sequence shell, hydrogen-burning stars, on the cool side of the
Cepheid instability strip.  Figure~\ref{fig:solarlikespect} shows examples of
frequency spectra of solar-like pulsators spanning a broad range in stellar
properties.  The spectra are much denser than those of a typical classical
pulsator (cf.\ Fig.\,\ref{hd129929}). Moreover, the amplitudes are much
smaller than those generally found in classical pulsators and they show an
``envelope" shape, which is typical of stochastically excited oscillations and
is remarkably similar for all the stars illustrated.

\subsection{The role of interferometry}
Even though the first astronomical applications of interferometry date back to the late 19th century and the first measurement of an angular diameter of a star other than the Sun using this technique dates back to the early 20s \citep{michelson21}, it was mostly during the last decade that optical interferometers became available to a wide astronomical community. During this period, a large window of research opportunities has been opened by a new generation of interferometers, such as CHARA (at Mount Wilson), VLTI (at ESO-Paranal), and Keck (at Mauna Kea). 

Interferometry is expected to have a strong impact on studies of pulsating stars and, in that way, on the progress of our understanding of stellar structure and evolution. As we shall see, the success of asteroseismology depends strongly on the availability of accurate global parameters of pulsating stars and on our ability to correctly identify the modes of oscillation observed in each star. The possibility of measuring stellar angular diameters, of determining dynamical parallaxes of binaries, and of mapping stellar surfaces through differential interferometry, is, thus, expected to provide an important contribution to the success of asteroseismic studies.
\vspace{1cm}

The review is organized in three main sections. In Section~\ref{sec1} we present an extensive review of asteroseismology, with emphasis on those classes of pulsators that are within reach of present interferometric instruments. In particular, in Section~\ref{obsintr} we discuss the observational techniques and instruments currently used to observed stellar pulsations, in Sections~\ref{properties} and \ref{excitation} we describe the basic properties of the oscillations and their origin, in Section~\ref{inference} we present the methods used to infer information about the star from the analysis of asteroseismic data, and, finally, in Section~\ref{limitations} we highlight the most common difficulties currently faced in asteroseismic studies. 

In Section~\ref{sec2} we provide a brief review of optical interferometry, focused on aspects that are of relevance to asteroseismic studies. In particular, in Sections~\ref{principles} and \ref{stphys} we describe some basic principles of optical interferometry, and emphasize particular aspects of stellar physics that can gain from the recourse to this technique. In Sections~\ref{presentinf} and \ref{limitinter} we discuss the capabilities of the interferometric instruments currently available to the community and emphasize the main limitations faced when using such interferometers to measure stellar angular diameters. Finally, in Section~\ref{futureinterf} we discuss the instrumental developments that may be expected in the near future.
 
Section~\ref{sec3} explores important synergies between asteroseismology and interferometry. Focusing on the main difficulties faced by current asteroseismic studies, described in Section~\ref{limitations}, we discuss ways in which interferometry may help overcome these difficulties, both through direct and indirect contributions to the determination of accurate global parameters of pulsating stars, discussed in Section~\ref{gp}, and through the contribution to the identification of the modes of oscillation, discussed in Section~\ref{mi}. Finally, in Sections~\ref{cs} and \ref{futurea} we briefly describe recent and on-going studies of particular stellar pulsators involving a combination of asteroseismic and interferometric data.  

We conclude the review with a brief look into the future, presented in Section~\ref{future}, where we describe relevant ground-based and space projects under study, as well as the impact they are expected to have on the development of this research field.

\section{Asteroseismology}
\label{sec1}

\subsection{Observations}
\label{obsintr}

\subsubsection{Photometry, spectroscopy and polarimetry}
\label{obstechniques}

The mainstay of stellar observations, and therefore also of
asteroseismology, are UV, visible and IR light. The flux of neutrinos, gravitational
waves, and other more exotic particles is essentially restricted to
more extreme states of matter, and can, therefore, be set aside for this
discussion.

To observe the oscillations of stars it is necessary to understand the
effect the oscillations have on the emergent flux of light. In principle, the
emergent flux of light involves an integral over the
source function along the line of sight, and hence is influenced by
the physical state of the plasma along the latter. In practice
this integral is dominated by a particular region, which is wavelength dependent, namely the intersection of
the line of sight with a spatially thin shell centred on a surface with
an optical depth $\tau=2/3$. Since stars
are normally unresolved, the observed flux is an average over
the $\tau=2/3$ surface, modulated by a function of the angle with the
line of sight. By their influence on the location of, and the physical
conditions at, the $\tau=2/3$ surface, stellar oscillations modulate
the emergent flux. 

There are three types of measurements that may be carried out on light. In
particular, one can measure the flux within bands of medium to low spectral
resolution, the relative flux at high resolution, and the polarisation state of
the light.  These correspond, respectively, to the observational techniques of
photometry, spectroscopy, and polarimetry. Polarimetry has found very little use
in asteroseismology, for a variety of reasons. Normal stars are to a very high
degree spherically symmetric and, therefore, the stellar disk on the sky is to a
high degree axially symmetric. Hence, in normal stars polarization signatures
generally cancel out, when the light is integrated over a stellar disk. Although
non-radial pulsations of stars can break the axial symmetry, in particular in
the presence of rotation, the integrated effect remains small, also because the
intrinsic degree of polarisation of most emission mechanisms is itself not
usually high. Combined with the difficulty of carrying out polarimetric
measurements, polarimetry will probably not make a major impact as an
observational technique except perhaps for strongly magnetic stars such as roAp
stars. An example of where polarimetry is used in tandem with asteroseismology
is \cite{Ryab+97}, but there too polarimetry is not used to carry out the
asteroseismic observations.

Both photometry and spectroscopy are widely used as asteroseismic
techniques. Intensity variations, resulting from the intrinsic pulsation of the
star, can be studied from photometric time series. Spectroscopic time series, on
the other hand, allow us to measure directly the changes of the surface
velocity.  It is important to note that observations in intensity and velocity
of pulsations in one and the same star sample the same pulsation properties, but
not in exactly the same way.  In the case of p modes, both the Doppler shift of
spectral lines and the intensity variations are weighted averages of the
pulsation amplitude over the $\tau=2/3$ surface \citep[e.g.,][]{Dziemb1977,
ChristGough1982}, severely reducing the sensitivity to modes with a high number $l$
of nodal lines on the surface.  However, for the Doppler observations the
projection of the velocity onto the line of sight gives rise to one more factor
of $\cos\theta$ in the weighting function, where $\theta$ is the angle with the
line of sight.  This difference in weighting generally increases the response of
velocity observations to modes of moderate $l$, compared with intensity
observations.  An important example, particularly for solar-like oscillations of
low intrinsic amplitudes, are the $l = 3$ modes which have been detected in
velocity observations but not in intensity observations.  Also, the difference
in response can in principle be used to help in identifying the value of $l$
associated with a given period of
oscillation, which is one of the major problems in asteroseismology. A
similar purpose is served by carrying out photometric observations in
various bands \citep[e.g.,][]{Stamfo1981, Watson1988}.
Section~\ref{sec3} provides further details on mode identification.

\subsubsection{Oscillation frequency spectra}
\label{spectra}
The photometric and spectroscopic time series acquired during observations of
pulsating stars can be used to generate frequency spectra. If the
signal-to-noise ratio is sufficiently high, the oscillations will generate peaks
above the noise, at the corresponding oscillating frequencies. In turn, these
frequencies are often used as starting points for asteroseismic studies.

If the time series were ``perfect'' -- i.e.,\ continuous, with no gaps, and
lasting forever -- one could in principle identify, without ambiguity, the
frequencies of all global oscillations of a star, as far as their signal were
above the noise. However, the time series acquired are usually far from perfect,
and the identification of which peaks correspond to true oscillation frequencies
is not always trivial, as described below.

For a known function of time $f(t)$ the amplitude in the frequency domain
$F(\omega)$ is obtained by carrying out a Fourier integral:
\begin{equation}
F(\omega) = \int_{-\infty}^{\infty} {\rm d}t f(t) \ee^{-\eye \omega t} \; .
\end{equation}
If, for example, the signal were a single-frequency harmonic function, $f(t) = A
\cos(\omega_0 t + \delta_0)$, with $A$ the amplitude, $\delta_0$ a phase and
$\omega_0$ the {\it angular frequency\/,}%
\footnote{In addition, particularly when discussing observations, 
one often also uses the {\it cyclic frequency\/} $\nu = \omega/2 \pi$.}
$F(\omega)$ would have delta-function
peaks at $\omega = \omega_0$ and \hbox{$\omega = - \omega_0$}.

In reality, the collection of data, whether it be intensity or velocity,
requires a finite amount of time of `integration' during which the detector,
normally a CCD, is exposed to radiation; after this there is a dead
time during which the CCD is read out and the digitized signal stored
on computer or transmitted to a receiving station. The integration
time itself is usually short. The time series is therefore not measured
continuously but sampled discretely. Normally the detection process is
automated to sample the time series equidistantly at a rate that is
sufficiently high to resolve the relevant time scales of variation.

In the ideal case the data would therefore be discretely sampled at a constant
rate, and one would apply a discrete Fourier transform (DFT) to obtain the
signal as a function of frequency within a band.  The highest useful frequency
to search for is termed the {\it Nyquist frequency\/} $\omega_{\rm Nyq}$. For
evenly spaced data, with a sampling interval $\Delta_t$, we have,
\begin{equation}
\omega_{\rm Nyq} = \frac{\pi}{\Delta_t} \; .
\end{equation}
In the case of unevenly spaced data, the Nyquist frequency can be quite
different from this value, particularly if numerous large gaps or serious
undersampling occur.

Since a real data stream has a finite length 
{\bf $T$}, in the Fourier domain there is
also a finite frequency resolution, $\Delta_{\omega}$, such that,
\begin{equation}
\Delta_{\omega} \propto \frac{1}{T} \; .
\end{equation}
Generally, if two or more signals present in a time series are more closely
spaced than the frequency resolution, these signals will not appear as multiple
peaks in the Fourier domain. Instead, single or deformed peaks, or even no peaks
at all will be seen. Moreover, signals with frequencies below this resolution
cannot be detected with any confidence.

Besides the limitations associated with the finite length of the time series and
with the discrete sampling of the data, in the analysis of real observations
there are often difficulties associated also with the problem of `missing
data'. Because of technical problems, or poor observing conditions, data can be
lost either pointwise or in blocks. These missing data are usually distributed
randomly within the time series. Moreover, regular gaps in the time series are
usually present in observations carried out from ground-based telescopes, due to
the incapacity of these telescopes to observe the stars during the day and
during the time that the stars are below the horizon. These regular gaps produce
additional peaks in the Fourier domain, which can hamper considerably the
interpretation of the data, particularly of multiperiodic pulsators. Together
these properties of the data sampling are known as the {\it window function} in
the time domain and as the {\it spectral window} in the Fourier domain.  The
spectral window has multiple peaks, even for a monochromatic underlying signal,
as can be seen in the upper panel of Fig.\,\ref{hd129929}.

A time series containing multiple signals sampled in the same way produces a
rather complex Fourier spectrum. In fact, even in the absence of measurement
noise, the task of identifying which of the peaks correspond to true oscillation
frequencies and which are sidelobes generated by the sampling can be extremely
hard. A number of methods have been developed to deal with this problem.  A
well-known technique is the Lomb-Scargle periodogram \citep{scargle82}.  A more
general discussion of several methods in use in astronomy can be found in
\cite{Ado95} and \cite{Vio+00}. A discussion directed more specifically to
asteroseismology can be found in \cite{KoeLom95}, \cite{Koe99} and
\cite{Pijper2006}.  \citet{Schwarzenberg-Czerny1997} showed that the Fourier
transform is optimal to deduce the frequencies, in the sense that it is
equivalent to all methods relying on phase binning and variance estimates, for a
given sampling, binning, and weighting of the data. While gap-filling techniques
and methods to hide aliases have been developed, these cannot extract more
frequency information than the Fourier transform \citep{kurtz02}.

\subsubsection{Ground-based facilities}
\label{groundbased}

For the classical pulsators, with relatively high amplitudes, photometry is most
commonly used. This is due to the fact that most small and medium-sized
telescopes have facilities for doing absolute or relative photometry in a
variety of photometric systems (the Johnson, Str\"o{}mgren, and Geneva systems
being most common). As mentioned before, asteroseismic studies require long time
series of observations. Thus, temporary networks of small and medium size
telescopes equipped with photometers are often organized by researchers working
in this field.

The Whole Earth Telescope (WET)\footnote{\tt http://wet.physics.iastate.edu},
Delta Scuti Network (DSN)\footnote{\tt
http://www.univie.ac.at/tops/dsn/intro.html} and the STEllar PHotometry
International network (STEPHI)\footnote{\tt
http://www.lesia.obspm.fr/$\sim$stephi/} are examples of consortia of
astronomers who create such temporarily networks of telescopes for continuous
photometric observations of asteroseismic target stars -- networks that for a
few weeks or even months mimic the dedicated helioseismic networks such as GONG\footnote{\tt
http://gong.nso.edu/},
BiSON\footnote{\tt
http://bison.ph.bham.ac.uk/}, IRIS, TON\footnote{\tt http://ton.phys.nthu.edu.tw/index.htm} and others. Besides these organised networks, also individual
scientists have taken the initiative to organise large multisite campaigns on
dedicated stars \cite[e.g.,][]{Handle2004,Handle2006}.

These photometric multisite campaigns have been a great success, with many
notable discoveries from seismic inferences.  PG\,1159--035, a DOV pulsating
white dwarf star \citep{Winget1991} and GD\,358, a DBV pulsating white
dwarf star \citep{Winget1994}, both observed during WET campaigns, are still
record holders for number of independent pulsation modes detected and identified
in compact stars. Significant understanding of white dwarf mass, envelope mass,
luminosity, differential rotation and other discoveries have come from these
studies. More recently the exciting question of crystallisation in the DAV white
dwarf BPM\,37093 \citep{2005A&A...432..219K,brassard04,fontaine05} has been
addressed. The ground-based limit for photometric precision has been pushed to
14\,$\mu$mag in a WET study of the roAp star HR\,1217
\citep{2005MNRAS.358..651K} and the amazing eclipsing subdwarf B pulsator
PG\,$1336 - 018$ has been studied \citep{2003MNRAS.345..834K}. The simultaneous
presence of p and g modes was discovered in the $\beta\,$Cephei stars $\nu\,$Eri
\citep{Handle2004} and 12\,Lac \citep{Handle2006} and differential rotation was
found in the interior of $\nu\,$Eri \citep{Pamyat2004}, after it had been
discovered already for the similar main-sequence B star HD\,129929
\citep{aerts03}.

Nevertheless, even with the extraordinary efforts of these large teams,
duty cycles for the asteroseismic targets do not come close
to the 100\% that is desired for run lengths of weeks.  For that either
satellite missions, an asteroseismic telescope in the polar regions, or a
permanently dedicated network are needed (see also Section~\ref{future}).

As mentioned in Section\,\ref{mspulsators}, the amplitudes of the variations in
solar-like pulsators are much smaller than in classical pulsators. To date, all
ground-based attempts have failed to detect solar-like intensity variations due
to the limitations imposed by the Earth's atmosphere \cite[for a very ambitious
attempt, involving most of the then-largest telescopes available,
see][]{Gillil1993}.  Thus, to study solar-like pulsators, one must either
acquire photometric data from space, or carry out spectroscopic measurements at
large or highly optimized ground-based telescope facilities.

During the past $\sim$5 years there has been a number of detections of
solar-like oscillations in solar-type stars. The technical background behind
these detections is the marked increase in the achievable precision of
radial-velocity measurements spurred by the detection of extrasolar planets.  In
order to achieve the high precision needed for the detection of solar-like
oscillations ($\sim$5$\,{\rm m\,s^{-1}}$) highly specialized spectrographs are
needed.  Consequently, to date only relatively few instruments are available for
this.

So far the following spectrographs have produced oscillation spectra with
unambiguous excess power and in several cases also detections of individual
oscillation frequencies in solar-like pulsators: {\it HARPS} at the ESO 3.6-m
telescope and {\it UVES} on the VLT\footnote{\tt http://www.eso.org/public/},
{CORALIE} at the 1.2-m Euler telescope\footnote{\tt
http://obswww.unige.ch/$\sim$naef/CORALIE/coralie.html} on La Silla, {SARG} at
the TNG\footnote{\tt http://www.pd.astro.it/sarg/} on La Palma and {\it UCLES}
at the AAT\footnote{\tt http://www.aao.gov.au/about/aat.html} . Other telescopes
have contributed, but the main results come from the five mentioned above. A summary of recent observations of solar-like pulsations among main-sequence and subgiant stars is provided by \cite{bedding06}.

To date, time at any of these telescopes (except {CORALIE}) has only been
allocated in blocks shorter than 11 nights -- and no facility exists for
coordinated observations. In the northern hemisphere {SARG} was, until
recently, the only instrument available, but recently SOPHIE at OHP\footnote{\tt
http://www.obs-hp.fr/www/guide/sophie/sophie-eng.html} and FIES at the
NOT\footnote{\tt http://www.not.iac.es/instruments/fies/} have become
operational. In the southern hemisphere, of the remaining four instruments,
three are located in Chile and one is located in Australia.  A few successful
campaigns have been run by coordinating observations from the instruments in
Chile and the AAT \citep[e.g.,][]{Butler2004, Beddin2006, Kjelds2005,
Beddin2007}.  
The lack of a dedicated network of telescopes with capabilities in this field is
a major limitation for further progress in ground-based asteroseismic efforts
for studying solar-like pulsators. Studies of such a dedicated network are under way. This, and other ground-based projects under study aimed at acquiring continuous spectroscopic time-series of solar-like pulsators are discussed in Section~\ref{future}.

\subsubsection{Space missions}
 
Space very early appeared as a preferred place for asteroseismology, using the
technique of ultra-high-precision photometry. Observations from space can
provide long and almost continuous datasets on the same objects.  Moreover, with
very moderate apertures, space photometry can be used to track oscillations with
amplitudes around 1\,ppm, and with very high sampling rates during extended runs
it is in principle possible to access characteristic time scales that are out of
reach from the ground.

The past two decades have seen a succession of proposals for national and
international space projects dedicated to asteroseismology.  Moreover, some
non-dedicated instruments, such as HST \citep[e.g.,][]{zwintz00,castanheira05}
and WIRE \citep{bruntt05} have been used (in the former case occasionally, in
the latter case regularly) to obtain data for asteroseimic studies.

The Canadian microsat MOST \citep{Walker2003, matthews05}, launched in June
2003, is able to measure light variations of bright stars for around 40\,d
continuously and demonstrated the efficiency of the technique.  Striking results
have been obtained on pulsations of different kinds of pulsators, including a
pulsating Oe star \citep{Walker2005a}, pulsating Be stars
\citep{Walker2005b,saio07}, a $\beta$~Cephei star \citep{Aerts2006a}, a slowly
pulsating B stars \citep{Aerts2006b}, a pulsating B supergiant \citep{saio06}, a
subdwarf B star \citep{Randal2005a} and Wolf-Rayet stars \citep{lefevre05}. In
all these examples, ground-based photometry was largely exceeded, both in
quality and in duty cycle.

Launched in December 2006, the satellite CoRoT \citep{baglin03} is expected to
produce a major step in the domain of space-based asteroseismology.  CoRoT is a
French CNES-led mission with contributions from seven other entities: Austria,
Brazil, Belgium, Germany, Spain, and the RSSD~/ESTEC division of ESA.  With its
27-cm diameter telescope, CoRoT is expected to detect periodic signals of
amplitude as small as a few ppm with lifetimes of a few days, in $6^{\rm th}$
magnitude stars. 
Early, so far unpublished, results provide very encouraging indications
that this goal will in fact be met.

This mission will observe at least 100 stars (half of them for 150\,d, the
others for 20 to 40\,d), and will produce a dense mapping of the seismic
properties (frequencies, amplitudes and life-times) of pulsators all across the
HR diagram.  Simulations show that, with the frequency resolution and the
signal-to-noise ratio that are expected to be reached during the long runs of
150\,d, it will be possible to measure the size of the convective cores, the
extent of the radiative zones, and the angular momentum distribution in
intermediate rotators.  In addition, the observation of 100 000 faint stars for
the CoRoT exoplanet search programme will produce a lot of complementary seismic
information.  A further advance in space-based asteroseismology will result
from the asteroseismic programme on the Kepler mission
\citep[e.g.,][]{Christetal2007}; this will carry out photometric observations, with
a precision sufficient to detect solar-like oscillations, for thousands of stars, with the
option to observe selected stars continuously or repeatedly over a period of at
least four years. This and other space-based missions including asteroseismic programmes are discussed further in Section~\ref{future}.

\subsection{Basic properties of the oscillations\label{properties}}

We restrict the discussion to small-amplitude oscillations of stars
that do not rotate very rapidly.  In this case the oscillations can be
treated as small perturbations about a static equilibrium structure,
and effects of rotation (and some other perturbations to the
spherically symmetric structure of the star, such as buried magnetic
fields) can also be treated using a perturbation analysis.  Thus the
basic description of the oscillations is obtained as linearized
perturbation equations around a spherical equilibrium structure.  It
follows that individual modes of oscillation depend on co-latitude
$\theta$ and longitude $\phi$, in spherical polar coordinates $(r,
\theta, \phi)$ where $r$ is the distance to the centre of the star, as
a spherical harmonic $Y_l^m(\theta, \phi)$.  As before, here the {\it degree\/}
$l$ indicates the number of nodal surface lines and hence
describes the overall complexity of the mode and the {\it
azimuthal order\/} $m$ determines the number of nodes around the
equator.
In addition, the modes are characterized by their {\it radial order\/}
$n$ which, roughly speaking, measures the number of nodes in the radial
direction
\citep[e.g.,][]{Scufla1974, Osaki1975, Takata2005}.

\subsubsection{The properties of the modes}
\label{modeprop}

\begin{figure}
\centering
  \includegraphics[width=0.75\textwidth]{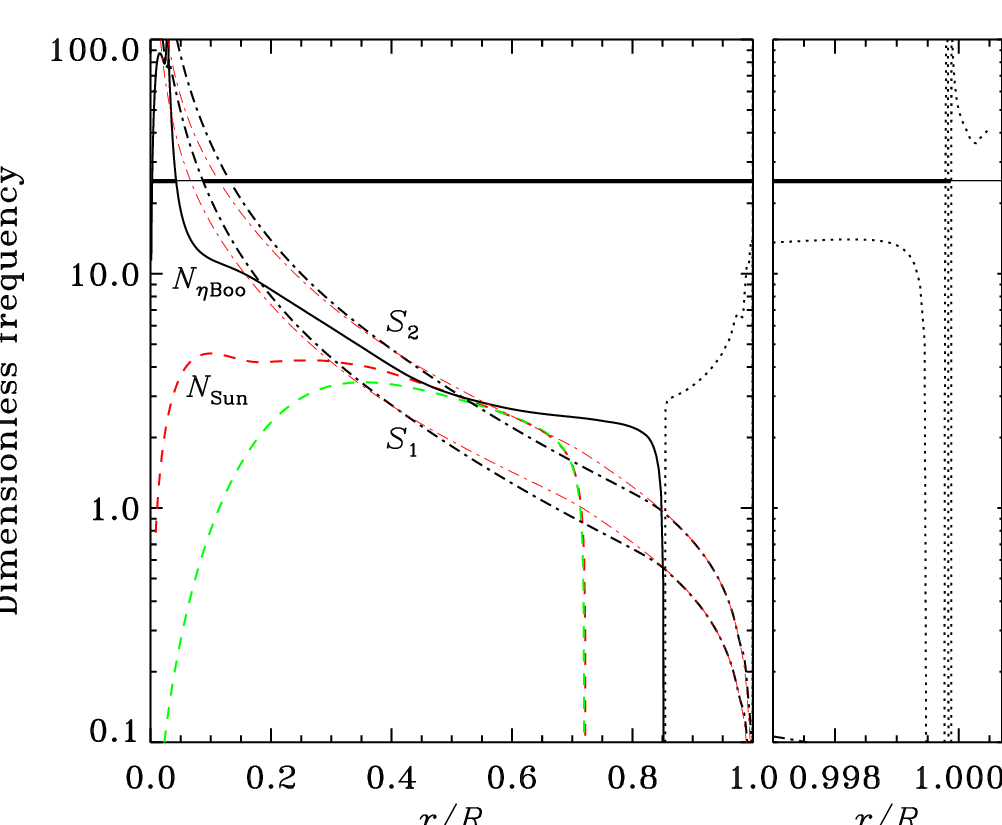}
\caption{Dimensionless characteristic frequencies,
in units of $(G M/R^3)^{1/2}$,
for a 1~M$_\odot$ ZAMS model, a model of the present Sun and of $\eta$~Boo,
a 1.63 M$_\odot$ subgiant.
The dot-dashed curves show $S_l$ for $l = 1$ and $2$, in the solar model (red)
and the model of $\eta$~Boo (black).
The dashed curves show $N$ in the ZAMS model (green) and the solar model (red),
whereas the solid curve shows $N$ in the model of $\eta$~Boo.
Finally, the dotted curve shows $\omega_{\rm c}$ in the model of $\eta$~Boo.
For clarity, $N$ is not shown in the atmosphere and $\omega_{\rm c}$
not below the base of the convective envelope.
The horizontal line marks a typical frequency of stochastically excited modes;
it is shown as thicker in regions where such a mode with $l = 1$ propagates
in the $\eta$~Boo model.
} 
\label{fig:charfreq}       
\end{figure}

The frequencies of oscillation of stellar models can be determined with
good precision in the adiabatic approximation, assuming that 
the perturbations $\delta p$ and $\delta \rho$ in pressure $p$ and
density $\rho$ in a mass element following the motion
(the so-called {\it Lagrangian perturbations\/}) are related by
\begin{equation}
{\delta p \over p} = \Gamma_1 {\delta \rho \over \rho} \; ,
\label{eq:adiab}
\end{equation}
where $\Gamma_1 = (\partial \ln p / \partial \ln \rho)_{\rm ad}$,
the derivative corresponding to an adiabatic change.
Given this approximation, it is straightforward to compute precise
frequencies of a stellar model,
and the frequencies therefore provide direct diagnostics of the
stellar interior; 
this is obviously the basis of asteroseismology.
However, to understand the diagnostic potential of the frequencies,
asymptotic analyses of the oscillation equations are extremely instructive.
A convenient approximate equation for this purpose has been derived by
Gough \citep[see][]{Deubne1984, Gough1986, Gough1993}, 
based on an analysis by \cite{Lamb1909}:
\begin{equation}
{\dd^2 X  \over \dd r^2} + K(r) X = 0 ,
\label{eq:gougheq}
\end{equation}
where
\begin{equation}
K(r) = {\omega^2 \over c^2}
\left[ 1 - {\omega_{\rm c}^2  \over \omega^2}
- {S_l^2  \over \omega^2} \left( 1 - {N^2  \over \omega^2} \right) \right] \; ;
\label{eq:goughasymp}
\end{equation}
here $X = c^2 \rho^{1/2} {\rm div} \bolddelr$, where 
$c$ is the adiabatic sound speed and $\bolddelr$ is the displacement
vector.
The behaviour of the mode is determined by three characteristic frequencies:
the acoustic (or Lamb) frequency $S_l$, given by
\begin{equation}
S_l^2 = {l(l+1) c^2 \over r^2} \; ,
\label{eq:sl}
\end{equation}
the buoyancy (or Brunt-V\"ais\"al\"a) frequency $N$, given by
\begin{equation}
N^2 = g \left( {1 \over \Gamma_1} {\dd  \ln p \over \dd r } -
{\dd  \ln \rho \over \dd r }\right) \; ,
\label{eq:buoy}
\end{equation}
and the acoustical cut-off frequency $\omega_{\rm c}$, given by
\begin{equation}
\omega_{\rm c}^2
= {c^2  \over 4 H^2} \left( 1 - 2 {\dd H \over \dd r }\right) \; ,
\label{eq:cutoff}
\end{equation}
where $H = - (\dd \ln \rho / \dd r)^{-1}$ is the density scale height.
Examples of these frequencies are plotted in Fig.~\ref{fig:charfreq}
for selected stellar models.

For a fully ionized ideal gas (a good approximation in much of most stars,
as long as radiation pressure or degeneracy can be neglected)
\begin{equation}
c^2 \simeq {5 \over 3} {k_{\rm B} T \over \mu m_{\rm u}} ,
\label{eq:csqideal}
\end{equation}
where $k_{\rm B}$ is Boltzmann's constant, $T$ is temperature,
$\mu$ is the mean molecular weight and  $m_{\rm u}$ is the atomic mass unit.
Also,
\begin{equation}
N^2 \simeq {g^2 \rho \over p} ( \nabla_{\rm ad} - \nabla + \nabla_\mu) \; ,
\label{eq:buoyideal}
\end{equation}
where, following the usual convention,
\begin{equation}
\nabla = {\dd \ln T \over \dd \ln p} \; , \qquad
\nabla_{\rm ad} = 
\left({\partial \ln T \over \partial \ln p} \right)_{\rm ad} \; , \qquad
\nabla_\mu = {\dd \ln \mu \over \dd \ln p} \; ,
\label{eq:nablas}
\end{equation}
and $g$ is the local gravitational acceleration.
In regions of nuclear burning, $\mu$ increases with increasing
depth and hence increasing pressure, 
and therefore the term in $\nabla_\mu$ makes a positive
contribution to $N^2$.

\begin{figure}
\centering
  \includegraphics[width=0.45\textwidth]{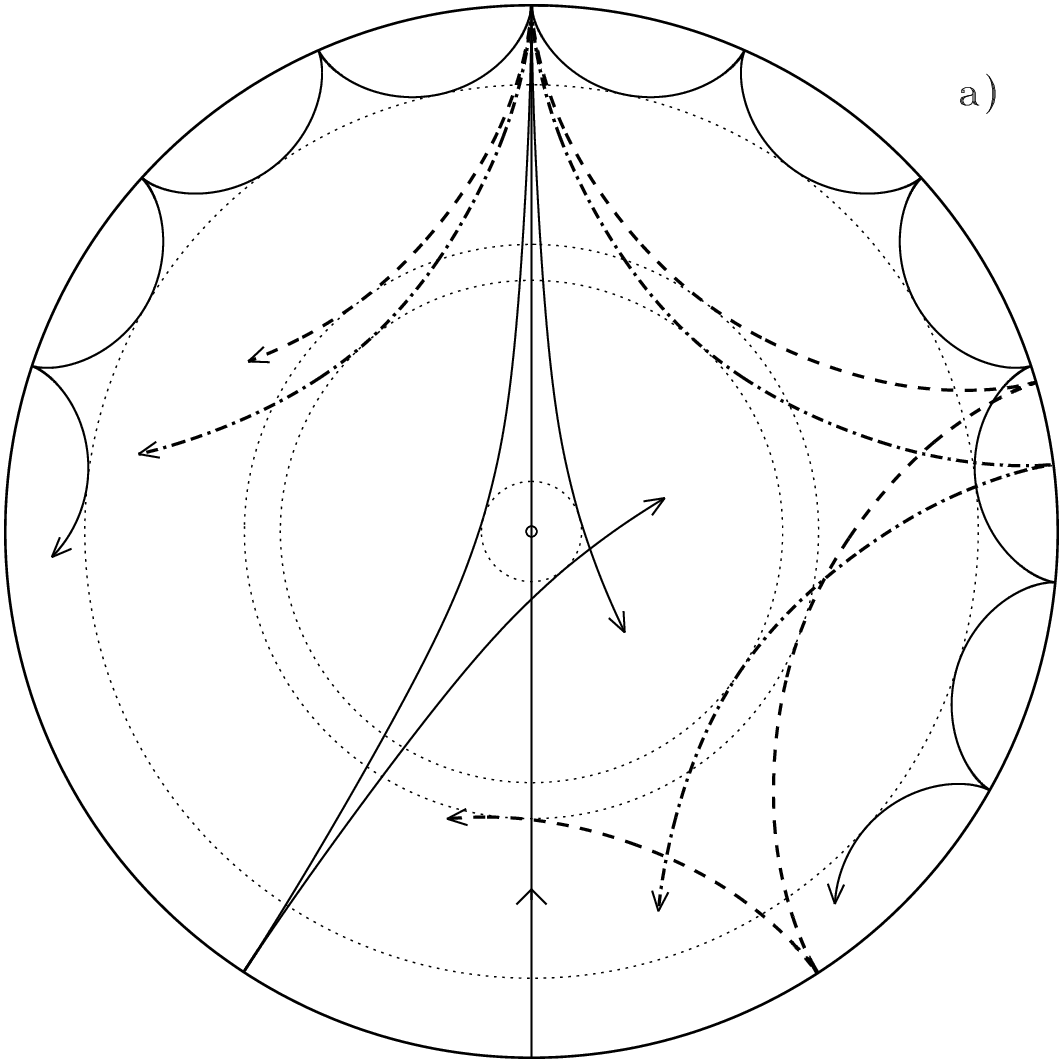}
  \includegraphics[width=0.45\textwidth]{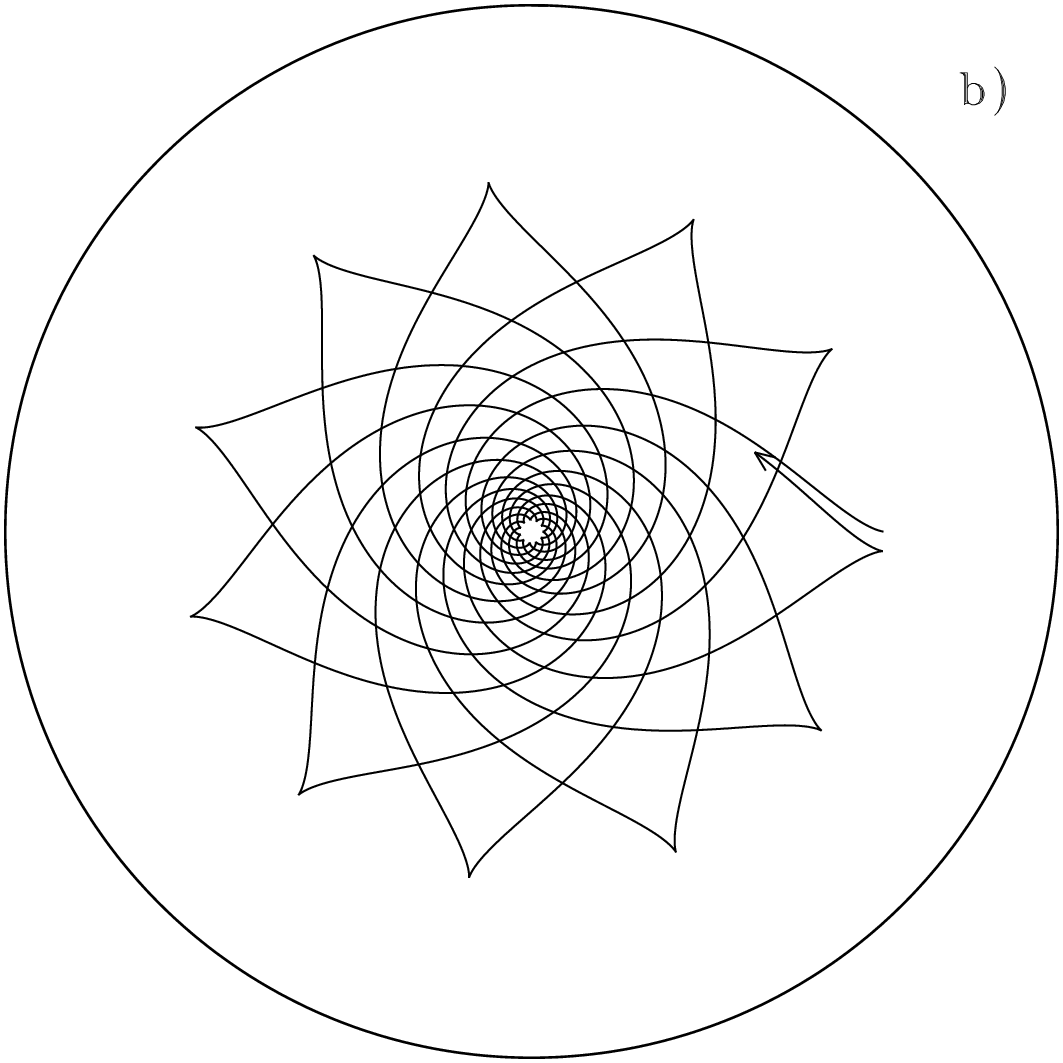}
\caption{
Propagation of rays of sound or gravity waves in a cross-section of the
solar interior. The acoustic ray paths (panel a)
are bend by the increase in sound speed
with depth until they reach {\em the inner turning point}
(indicated by the dotted circles) where they undergo total internal
refraction, at the distance $r_{\rm t}$ determined by \Eq{eq:rt}.
At the surface the acoustic waves are reflected by the rapid decrease
in density.
Rays are shown corresponding modes with frequency $3000 \muHz$ and degrees
(in order of increasing penetration depth) $l = 75$, $25$, $20$ and $2$;
the line passing through the centre schematically illustrates the behaviour
of a radial mode.
The gravity-mode ray path (panel b) corresponds to a mode of frequency
$190 \muHz$ and degree 5.
}
\label{fig:raypath}       
\end{figure}

A mode oscillates as a function of $r$ where $K(r) > 0$
and varies exponentially where $K(r) < 0$;
in the former regions the mode is said to be {\it propagating\/}
and in the latter the mode is {\it evanescent\/}.
Points where $K(r) = 0$ are {\it turning points\/} of the mode.\footnote{More precisely, a turning point is a  property of the equation that describes the eigenmode, rather than of the mode itself. Hence, the depth $r_{\rm t}$ at which is located the lower turning point of a given mode, for instance, can be slightly different depending on whether the radial displacement or the Lagrangian pressure perturbation is used to describe the mode \cite[e.g.,][]{Gough1993,schmitz98}.}
In most cases the mode has large amplitude in just one propagating region
which therefore mainly determines its frequency;
the mode is said to be {\it trapped\/} in such a region.
Near the surface typically $S_l \ll \omega$ and the behaviour of the
mode is controlled by $\omega_{\rm c}$.
Modes with frequency below the atmospheric value of $\omega_{\rm c}$
decay exponentially in the atmosphere and hence are trapped within the star.
In the rest of the star $\omega_{\rm c}$ plays a smaller role and
the properties of the modes are controlled by $S_l$ and $N$.
In unevolved stars $N$ is relatively low throughout the star.
In that case a high-frequency mode is predominantly controlled by
the behaviour of $S_l$;
the eigenfunction oscillates as a function of $r$ between the
near-surface reflection where $\omega = \omega_{\rm c}$ and
a {\it lower turning point\/} at $r = r_{\rm t}$, such that
$\omega = S_l(r_{\rm t})$ or
\begin{equation}
{c(r_{\rm t})^2 \over r_{\rm t}^2} = {\omega^2 \over l(l+1)} \; .
\label{eq:rt}
\end{equation}
This is typical of a p mode.
Obviously, at low degree $r_{\rm t}$ is small and the mode samples most
of the star, including parts of the core.
In particular, radial p modes extend essentially to the centre of the star.
The behaviour of these modes can be illustrated in a {\it ray plot\/},
such as illustrated in Fig.~\ref{fig:raypath}a;
as shown, the waves undergo total internal reflection at a surface of
radius $r_{\rm t}$, given by \Eq{eq:rt}.

For low-frequency modes $\omega \ll S_l$ in much of the star, and
the mode is oscillatory in a region approximately determined by $\omega < N$;
this characterizes g modes.
As shown by Fig.~\ref{fig:raypath}b such modes can similarly be illustrated
by the propagation of rays of internal gravity waves;
here, for a model of the present Sun, reflection takes place near the centre 
and at the base of the convective envelope.
However, as is obvious from Fig.~\ref{fig:charfreq} and \Eq{eq:buoyideal},
$N$ may reach very large values in the core of evolved stars.
Here $K$ may be positive at relatively high frequency both
in the outer parts where $\omega$ is greater than both $S_l$ and $N$,
the mode thus behaving as a p mode, and in the core where
$\omega < S_l, N$ and the mode behaves like a g mode.
This is illustrated in Fig.~\ref{fig:charfreq} in the case of $\eta$~Boo.
Such {\it mixed modes\/} have a particularly interesting 
diagnostic potential (see Section \ref{mixmodes}).

An important global property of the oscillation modes is their {\it inertia\/}
\begin{equation}
E = {\int_V |\bolddelr|^2 \rho \dd V \over M |\bolddelr|^2_{\rm ph}} 
\equiv {M_{\rm mode} \over M} \; ,
\label{eq:inertia}
\end{equation}
also defining the {\it mode mass\/} $M_{\rm mode}$,
where $M$ is the mass of the star, $|\bolddelr|_{\rm ph}$ is the norm
of the photospheric displacement and the integral is over the volume $V$
of the star.
Evidently, for modes trapped in the deep interior of the star with an
extensive evanescent region between the trapping region and the surface
the value of $E$ is expected to be large.
With this definition, the kinetic energy of the oscillation is given by
\begin{equation}
\CE_{\rm kin} = {1 \over 2} M_{\rm mode} V_{\rm rms}^2
= {1 \over 2} M E V_{\rm rms}^2 \; ,
\label{eq:ekin}
\end{equation}
where $V_{\rm rms}$ is the photospheric rms velocity.
On this basis one might expect that it is more difficult to excite
a mode to a given amplitude, the higher the value of $E$;
this is certainly the case for stochastically excited modes
which tend to be excited to roughly the same energy at a given frequency.

\subsubsection{Departures from spherical symmetry}
\label{nonsphere}

We have so far neglected the dependence of the oscillations
on the azimuthal order $m$.
In fact, for a spherically symmetric star the properties of the modes
are independent of $m$: the definition of $m$ depends on the orientation
of the coordinate system used to describe the star and this has
no physical meaning in the spherically symmetric case.
This frequency degeneracy is lifted by departures from spherical symmetry,
of which the most important example is rotation.
For a slowly rotating star, with an angular velocity $\Omega(r)$
that depends only on $r$, the effect of rotation on the frequencies,
as observed in an inertial frame, can be expressed as
\begin{equation}
\omega_{nlm} = \omega_{nl0} + m \beta_{nl} \langle \Omega \rangle_{nl} \; ,
\label{eq:rotsplit}
\end{equation}
where $\langle \Omega \rangle_{nl}$ is an average of $\Omega$ that depends
on the properties of the eigenfunctions in the non-rotating model
\citep[e.g.,][]{Ledoux1951, Hansen1977, Gough1981};
the constant $\beta_{nl}$ is sometimes expressed as $1 - C_{nl}$,
where $C_{nl}$ is the {\it Ledoux constant\/}.
For high-order or high-degree acoustic modes $\beta_{nl} \simeq 1$;
thus the rotational splitting is given by the average angular velocity.
On the other hand, for high-order g modes $\beta_{nl} \simeq 1 - 1/[l(l+1)]$
\cite[e.g.,][]{Wolff1974}.

The extent to which the splitting resulting from slow rotation can be detected
obviously depends on the rotation rate and the intrinsic frequency width of the
observed modes, which in turn depends on the time span of the observation or the
mode life time.  The inclination of the rotation axis relative to the line of
sight to the observer is an additional important factor, for stellar
observations where only light integrated over the disk of the star is observed.
\cite{Gizon2003} studied this effect in the case of stochastically-excited
modes.
In the special case where the rotation axis points towards the observer, only
modes with $m = 0$ are visible in integrated light and no rotational splitting
is observed.  The solar case corresponds to the opposite extreme: here an
Earth-bound observer is always close to the equatorial plane and disk-integrated
observations are sensitive essentially only to modes where $|l - m|$ is even,
including the modes with $m = \pm l$ which display the largest splitting.  For
intermediate inclinations the sensitivity to the different $m$ is essentially a
geometrical effect, affected by the distribution of intensity across the disk as
described by the limb-darkening function.%
\footnote{For radial-velocity observations an additional effect may arise as a
result of the rotational shift of the spectral lines across the stellar disk;
e.g., \citep{Brooke1978, Christ1989a}.}
\cite{Gizon2003} investigated the possibility of determining both the rotation
rate and the inclination of the rotation axis from the observed splittings and
ratios between the amplitudes of the split modes.  \cite{Ballot2006} carried out
a similar analysis, but addressing specifically the potential of the CoRoT
mission and carrying out Monte Carlo simulations to estimate the errors in the
inferred quantities.  They concluded that for a rotation rate twice solar it was
possible to determine both the inclination and the rotation rate, whereas for
rotation at the solar rate the inclination was poorly determined, whereas it was
still possible to infer the rotation rate.  \cite{Gizon2004} extended the
analysis to show that with observation of $l = 2$ modes with $m = \pm 1$ and
$\pm2$, which is possible at intermediate inclinations, some information can be
obtained about the latitudinal variation of rotation, at least for rotation
rates exceeding four times solar. It is worth reminding here that this
type of analysis cannot be applied to heat-driven modes, because their intrinsic
amplitudes do not follow a regular pattern nor do we understand their selection
mechanism.

Equation (\ref{eq:rotsplit}) is valid for rotation sufficiently slow
that terms of order $\Omega^2$ and higher can be neglected; 
in particular, this neglects the effect of rotation on the structure of
the star, through the centrifugal effects.
However, many stars rotate too rapidly for this to be valid.
Higher-order effects cause departures from the uniform splitting
given in \Eq{eq:rotsplit} and may lead to complex frequency spectra,
greatly complicating the analysis of the observed spectra.
Second-order effects were considered by, for example,
\cite{Chlebo1978}, \cite{Saio1981}, \cite{Gough1990a} and \cite{Dziemb1992},
whereas third-order effects were treated by \cite{Soufi1998}.
As a result of the higher-order effects, the surface behaviour of a mode
is no longer a pure spherical harmonic; this must be taken into account,
e.g., in the observational mode identification \citep{Daszyn2002}.
Also, \cite{Goupil2004}
showed {\it inter alia} that mode coupling caused by rotation 
can significantly perturb the small frequency separation 
which is used as a diagnostic of the star's core structure
(see Section \ref{basics}) and they demonstrated
how to correct for this at least for stars 
which are not latitudinally differentially rotating.
For even more rapid rotation solutions in terms of expansions 
in $\Omega$ are no longer applicable.
Initial results for fully two-dimensional, non-perturbative calculations
are becoming available for polytropic models 
\citep[e.g.,][]{Lignie2006, Reese2006a, Reese2006b};
interestingly, even at equatorial velocities less than 50\,${\rm km \, s^{-1}}$
(in a 1.9 M$_\odot$ model) the precision of the third-order expansion becomes 
insufficient, compared with the expected observational precision of the
CoRoT project \citep{Reese2006b}.
The modelling of the equilibrium structure of 
such rapidly rotating stars is evidently also problematic
in the physically realistic case, although some progress is being made
(see Section~\ref{sec:stelmod}).

Rotation is not the only physical agent that can influence the oscillations in a
non-spherically symmetric way. Other symmetry-breaking agents, such as magnetic
fields and structural differences associated with stellar spots, can affect both
the frequency of the oscillations and the geometry of the pulsations in the
surface layers of some pulsating stars.  An important difference between the
effect on the oscillations of the Lorentz force produced by an axisymmetric
magnetic field and the Coriolis force, is that only the latter can detect the
sense of the axis of symmetry. Thus, unlike rotation, an axisymmetric magnetic
field affects in the same way modes with the same absolute value of $m$.

Stellar magnetic fields often influence the oscillations in a way that cannot be described in terms of a standard perturbation analysis. If the magnetic field is force free, it will essentially affect the oscillations directly, via the Lorentz force. Since in the outer layers of a star the Lorentz force is comparable to or larger than the pressure forces, even a relatively small magnetic field will influence in a significant manner the dynamics of the oscillations for those modes which have their maximum amplitude in this region, such as, e.g., high frequency acoustic modes.

The direct effect of strong axisymmetric force-free magnetic fields on stellar oscillations has been studied in the context of roAp stars, which are permeated by magnetic fields with typical magnitudes of a few kG \citep{dziembowski96,bigot00,cunha00,saio04,saio05,cunha06}. From these studies it becomes clear that the influence of a magnetic field on acoustic pulsations has three important consequences. Firstly, the frequencies of the oscillations are shifted from their non-magnetic values. This fact is particularly well illustrated by the data and by the results of model calculations for the prototype HR~1217 \citep{cunha01,kurtzetal02}. Secondly, the oscillations near the surface are distorted by the magnetic field in such a way that each mode of oscillation is no longer well described by a single spherical harmonic function $Y_l^m$, but rather by a sum of spherical harmonics with different degrees, $l$. Thirdly, the coupling of the oscillations with the magnetic field near the surface of the star generates running waves that take away part of the pulsation energy; these include slow Alv\'enic waves in the interior that are expected to dissipate as they propagate inwardly, towards regions of higher density, and acoustic waves in the atmosphere which will propagate outwardly, along inclined magnetic field lines.  Some of these features depend mostly on the magnetic field configuration and on the degree of the modes, while others depend essentially on the magnetic field magnitude and on the structure of the surface layers of the star. A study of the combined effect of rotation and magnetic field in the context of roAp stars has also been carried out by \citet{bigot02}. They showed that for the cases of moderate
magnetic fields ($< 1$~kG), the quadratic effects of rotation
 in roAp stars lead to an inclination of the mode axis with respect to the
magnetic one. So far, there is no clear evidence of roAp stars for which the rotational and magnetic effects are comparable. Nevertheless, even if weak, the Coriolis effect is essential to interpret the observed multiplets in some roAp
stars. In fact, being the only  agent that can cause an asymmetry between the $m =\pm 1$ components of a mode, the Coriolis effect must be responsible for the amplitude inequalities found in multiplets observed in some of these stars.

\subsection{\label{excitation}The causes of stellar oscillations}

The characteristics of the frequency spectra, and our ability to interpret them, determine whether or not useful information about the pulsators can be inferred from the seismic data.  As mentioned in Section \ref{mspulsators}, the frequency spectra of classical pulsators and solar-like pulsators show significant differences, which are associated with the mechanisms that are responsible for driving the pulsations in each case. Understanding the origin of these differences, and acknowledging the limitations that are intrinsic to the frequency spectra for each of these types of pulsators, is an important step towards improving our capability to infer information from them.
Also, an understanding of mode excitation is needed for the planning of
future observations.
Finally, the observed amplitudes and lifetimes of the modes are
directly related to the excitation and damping mechanisms;
thus, they provide diagnostics of, for example, the properties of convection
in solar-like pulsators \cite[e.g.,][]{Christ1989b, Samadi2006}.

\subsubsection{Heat-engine-driven pulsators}
\label{heatengine}

To illustrate the damping and excitation properties we write the
time dependence of the oscillations as $\exp(-\eye \omega t)$
separating the angular frequency $\omega$ into a real and imaginary
part as $\omega = \omegar + \eye \omegai$.
Obviously the amplitude of the mode grows or decays with time depending
on whether $\omegai > 0$ or $\omegai < 0$.
{}From the equations of stellar oscillations follows the so-called work integral
for $\omegai$,
\begin{eqnarray}
\label{eq:workint}
\omegai & \simeq &- {1 \over 2 \omegar}
{\displaystyle {\rm Im} \left[ \int_V {\delta \rho^* \over \rho} \delta p \,
\dd V \right] \over
\int_V \rho |\bolddelr|^2 \dd V} \\
&=& - {1 \over 2 \omegar}
{\displaystyle {\rm Im} \left[ \int_V {\delta \rho^* \over \rho} \delta \pt
\dd V \right] \over
\int_V \rho |\bolddelr|^2 \dd V} 
+ {1 \over 2 \omegar^2}
{\displaystyle {\rm Re} \left[\int_V {\delta \rho^* \over \rho} (\Gamma_3 - 1)
\delta (\rho \epsilon - {\rm div}\,\boldF) \dd V \right] \over
\int_V \rho |\bolddelr|^2 \dd V} \; , \nonumber  
\end{eqnarray}
where Im and Re indicate the imaginary and real parts, respectively, and $*$ indicates the complex conjugate;
also, 
$\epsilon$ is the rate of energy generation per unit mass,
$\boldF$ is the energy flux,
and $\Gamma_3 - 1 = (\partial \ln T / \partial \ln \rho)_{\rm ad}$.
Finally, $\delta \pt$ is the Lagrangian perturbation to the turbulent pressure.
We write \Eq{eq:workint} as 
\begin{equation}
\eta \equiv {\omegai \over \omegar}
\simeq  \eta_{\rm t} + \eta_{\rm g} 
= {W_{\rm t} \over I} + {W_{\rm g} \over I} \; ,
\label{eq:workint1}
\end{equation}
where
\begin{eqnarray}
\label{eq:workint2}
W_{\rm t} &=&
 -  \omegar \, {\rm Im} \left[ \int_0^R {\delta \rho^* \over \rho} \delta \pt
r^2 \dd r \right] \; , \nonumber \\
W_{\rm g} &=&
{\rm Re} \left[\int_0^R {\delta \rho^* \over \rho} (\Gamma_3 - 1)
\delta (\rho \epsilon - {\rm div}\,\boldF) r^2 \dd r \right] \; , \nonumber \\
I &=& 2 \omegar^3 \int_0^R \rho |\bolddelr|^2 r^2 \dd r \; .
\end{eqnarray}
The expression for $W_{\rm g}$
highlights the heat-engine aspect of the excitation:
$\delta (\rho \epsilon - {\rm div}\,\boldF)$
is the perturbation to the heating;
regions where the heating is positive at compression, i.e., with positive
$\delta \rho$, therefore
contribute to the excitation of the mode, as expected for a heat engine.
Obviously, other regions may have the opposite phasing and 
excitation of the mode requires that the regions contributing to the
excitation dominate in the integral. 
In this expression the term in $\delta (\rho \epsilon)$ typically contributes
to the excitation, since compression leads to an increased temperature 
and hence increased energy generation rate.
However, this contribution to $W_{\rm g}$ is small in most cases.
Thus the excitation or damping is dominated by the perturbation to the
energy flux.

It should be noticed that it is often convenient to consider the
partial integrals for $W_{\rm t}$ and $W_{\rm g}$ as functions
of an upper limit $r$ of the integral.
In this way the regions of the star dominating the excitation and
damping can be identified.
Similarly, e.g., $\eta_{\rm g}(r) = W_{\rm g}(r)/I$ can be regarded
as a function of the upper limit.

In most cases of relevance here the processes take place near the stellar
surface, and the modes can for this purpose be approximated as being radial.
Neglecting also $\delta(\rho \epsilon)$ we obtain
\begin{equation}
W_{\rm g} = 
- {\rm Re} \left[\int_0^R {\delta \rho^* \over \rho} (\Gamma_3 - 1)
{\dd \delta L \over \dd r} \dd r \right] \; ,
\label{eq:lumdriv}
\end{equation}
where $\delta L$ is the Lagrangian perturbation of the luminosity $L$.
Assuming also the diffusion approximation for the radiative flux, the
contribution of radiation to the luminosity perturbation
in \Eq{eq:workint} can be written as
\begin{equation}
\delta \Lrad = \left[ 4 {\delta r \over r} 
+ (4 - \kappa_T) {\delta T \over T}
- \kappa_\rho {\delta \rho \over \rho} 
+ ({\dd \ln T / \dd r})^{-1} {\dd \over \dd r} \left({\delta T \over T} \right)
\right] \Lrad \; ,
\label{eq:radlum}
\end{equation}
where $\kappa_\rho = (\partial \ln \kappa / \partial \ln \rho)_T$,
$\kappa_T = (\partial \ln \kappa / \partial \ln T)_\rho$,
and $L_{\rm rad}$ is the radiative luminosity.
In the deep interior, the oscillations are essentially adiabatic and
hence the temperature perturbation is related to the density perturbation
by
\begin{equation}
{\delta T \over T} \simeq (\Gamma_3 - 1) {\delta \rho \over \rho} \; .
\label{eq:tpert}
\end{equation}
On the other hand, very near the surface the energy content in the stellar
matter is too small to affect significantly the perturbation to the flux.
The transition takes place approximately at a location $\rtr$ 
such that the thermal timescale $\tau_{\rm th}$
of the layer outside $\rtr$, i.e., the time required to radiate the
thermal energy of the layer, matches the pulsation period.
Thus
\begin{equation}
\Pi \simeq \tau_{\rm th} = {\langle c_V T \rangle \Delta m \over L} 
= L^{-1} \int_{\rtr}^R c_V T 4 \pi r^2 \rho \dd r \; ,
\label{eq:trans}
\end{equation}
where $\Pi$ is the pulsation period,
$c_V$ being the specific heat at constant volume, 
$\Delta m$ the mass outside $\rtr$, $L$ the luminosity
and $\langle \ldots \rangle$ indicates an average 
over the region outside $\rtr$.
Inside $\rtr$ the oscillations are nearly adiabatic, \Eq{eq:tpert}
being satisfied, whereas outside $\rtr$ the flux perturbation is
`frozen in' and the contribution to the work integral is negligible.

These considerations form the basis for a simple qualitative understanding
of the properties of unstable modes excited by the radiative flux,
developed by J.~P.\ Cox \cite[see][]{Cox1968, Cox1974}.%
\footnote{The analysis of stellar instability goes back to
\cite{Edding1926},
who pointed out the potentially crucial role of the phasing of the
heat leak and hence the opacity for driving the oscillations.
The detailed understanding of the Cepheid instability strip was developed
in a series of papers by S. A. Zhevakin \cite[e.g.,][]{Zhevak1953}
and J.~P.\ Cox and C. Whitney \cite[e.g.,][]{Cox1958}.
See also the review by \cite{zhevakin63}.}
In the quasi-adiabatic region we approximate $W_{\rm g}$ by
\begin{equation}
W_{\rm g} \simeq 
\int_0^R (\Gamma_3 - 1) {\delta \rho \over \rho} 
{\dd \over \dd r}\left(\psi_{\rm rad} {\delta \rho \over \rho} \right) 
\dd r \; ,
\label{eq:quasiad}
\end{equation}
where the eigenfunctions were taken to be real, energy transport was
assumed to be dominated by radiation, and
$\delta L_{\rm rad} \simeq - \psi_{\rm rad} \delta \rho/\rho$, with
\begin{equation}
\psi_{\rm rad} = 
(\kappa_T - 4) (\Gamma_3 - 1) + \kappa_\rho \; ;
\label{eq:psirad}
\end{equation}
thus we neglected the terms in $\delta r/r$ and the derivative
of $\delta T/T$ in \Eq{eq:radlum}.
In the outer parts of the star $\delta \rho/\rho$ typically increases with $r$;
also, except very near the surface, $\psi_{\rm rad}$ is generally negative.
Thus if $\psi_{\rm rad}$ does not vary substantially the contributions
to the integral in \Eq{eq:quasiad} are negative, giving a negative
contribution to $W_{\rm g}$ and hence contributing to damping the mode.
The exception are regions where the opacity derivatives (or $\Gamma_3 - 1$)
vary rapidly,
leading to rapid variations in $\psi_{\rm rad}$.
A region where $\psi_{\rm rad}$ increases towards the surface
contributes to the driving;
however, this would typically be followed by a region where
$\psi_{\rm rad}$ decreases and hence,
as long as \Eq{eq:quasiad} is valid, the integrated effect of
such regions largely cancels.
On the other hand, if the transition defined by $\rtr$
falls near the region where $\psi_{\rm rad}$ varies rapidly,
such that only the lower, driving, part of that region contributes,
the net effect may be to make the mode unstable.
Thus instability tends to be found in stars and for periods where the
transition region coincides with a region of rapid variation in the opacity,
e.g., associated with a region of partial ionization of an important opacity
source.
This is the case for the excitation of the classical Cepheids and other stars
in the Cepheid instability strip (cf. Fig.~\ref{HRD});
here the dominant effect is associated with the second ionization of helium, except for roAp stars, with much shorter periods, for which the dominant effect is associated with the region of hydrogen ionization \citep{balmforth01,cunha02}.
\begin{figure}
\centering
  \includegraphics[width=0.75\textwidth]{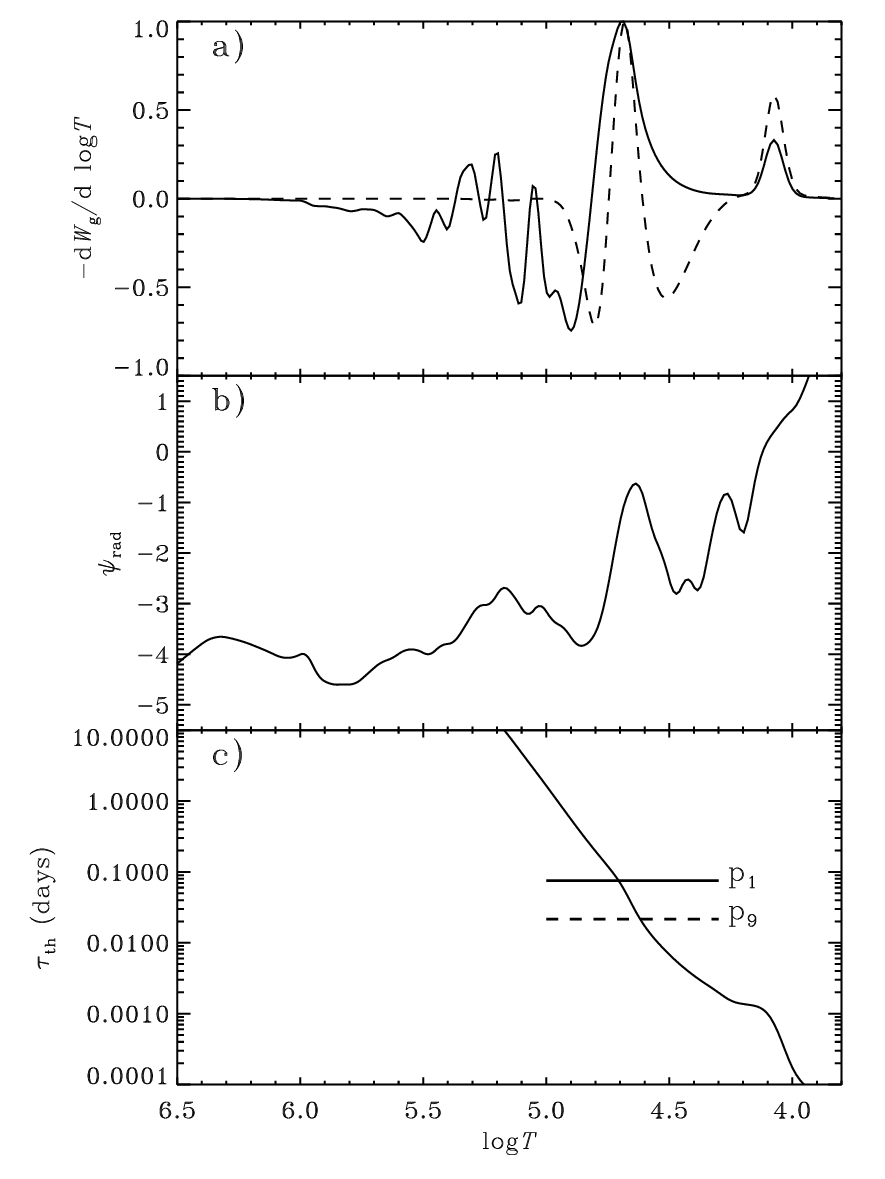}
\caption{
Differential work integral $- \dd W_{\rm g} /\dd \log T$
(cf.\ \Eq{eq:workint2}; the minus sign is included for consistency
with that equation), the function $\psi_{\rm rad}$ 
(cf.\ \Eq{eq:psirad}) determining the response of the opacity and
the thermal timescale $\tau_{\rm th}$ 
(cf.\ \Eq{eq:trans}), in a model of a $\delta$~Scuti star
of mass $1.8 \Msun$, age $0.8 \Gyr$ and effective temperature $7500 \K$.
The solid and dashed curves in panel a) and the solid and dashed
heavy horizontal lines in panel c) correspond to radial modes
of radial order $n = 1$ and 9, respectively.
\cite[See][]{Pamyat1999}.
}
\label{fig:dscutiex}       
\end{figure}

An extensive overview of the excitation of oscillations through
the opacity mechanism was given by \cite{Pamyat1999}.
Following him, Fig.~\ref{fig:dscutiex} shows the differential work integral for
two modes in a model of a $\delta$~Scuti star, as well as the quantity
$\psi_{\rm rad}$, which determines the luminosity perturbation in terms
of the density perturbation, and the thermal time scale.
Evidently, regions where $\dd W_{\rm g}/\dd \log T$ is positive contribute
to the excitation of the mode.
The peak in $\psi_{\rm rad}$ at $\log T \simeq 4.6$ corresponds to
the second ionization of helium.
As expected, the rising part of this peak causes excitation for both
the $n = 1$ and $n = 9$ modes.
For the $n = 1$ mode the transition to strongly nonadiabatic oscillations,
with $\tau_{\rm th}/\Pi \ll 1$, takes place so close to the
region of excitation that the corresponding damping contribution
is insignificant;
for the $n = 9$ mode, on the other hand,
there is strong damping in the outer parts
of the helium peak in $\psi_{\rm rad}$.
The net effect is that the $n = 1$ mode is unstable while the $n = 9$
mode is stable.
(The secondary excitation around $\log T \simeq 4.1$ arises from the
first ionization of helium and the ionization of hydrogen but makes a
relatively modest contribution to the overall excitation.)

With increasing effective temperature the helium peak in $\psi_{\rm rad}$
moves closer to the surface, in the direction of decreasing $\tau_{\rm th}$.
Consequently, higher-order modes, with smaller periods, have a tendency
to be excited. 
Thus we may expect to see excitation of higher-order modes on 
the blue (high-temperature) side of the instability strip.
This is confirmed by the detailed calculations presented by
\cite{Pamyat1999}.
As discussed by Pamyathykh, a similar effect operates amongst the B stars.
These are excited by an opacity feature (visible in Fig.~\ref{fig:dscutiex}b
around $\log T \simeq 5.2$) caused by contributions from iron-group elements.
This causes excitation of low-order acoustic and gravity modes with periods
of a few hours
in the $\beta$~Cephei stars, with masses of around $10 \Msun$ and
effective temperatures around $25\,000 \K$.
The same mechanism causes excitation of the high-order g-mode 
oscillations, with periods of a day or more,
in the Slowly Pulsating B stars with masses around $4 \Msun$ and
effective temperatures around $15\,000 \K$.%
\footnote{The apparently analogous situation regarding the 
long-period oscillations in the $\gamma$~Doradus stars on the cool side 
of the instability strip probably has a different physical origin, 
since the oscillations in these stars appear to be excited through
`convective blocking'; see below.}

Diffusive processes may play an important role in the excitation of modes in hot
stars.  In the case of the subdwarf B variables (sdBV stars), \citet{Fontai2003}
found that radiative levitation of iron was required to increase the iron
abundance in the driving region to the point where the observed modes were
unstable.  Interestingly, \citet{Aussel2004} and \citet{Pamyat2004} found that
in standard models of the $\beta$~Cephei star $\nu$~Eri, which fitted all the
observed frequencies, not all the observed modes were unstable.  They inferred
that a similar enhancement of the iron abundance in the driving zone might be
required to account for the excitation.  So far, no consistent evolution
calculations of diffusive processes have been carried out in $\beta$~Cephei
models, however.  Alternatively, \citet{miglio2007} found that the instability
of modes in B stars could be enhanced without considering diffusive processes,
but by using the revised solar mixture by \citet{Asplund2005} (in controversy with helioseismology) and OP opacities
\citep[e.g.,][]{Seaton2005}; this may contribute to the solution of the issue of
mode excitation.

\begin{figure}
\centering
  \includegraphics[width=0.75\textwidth]{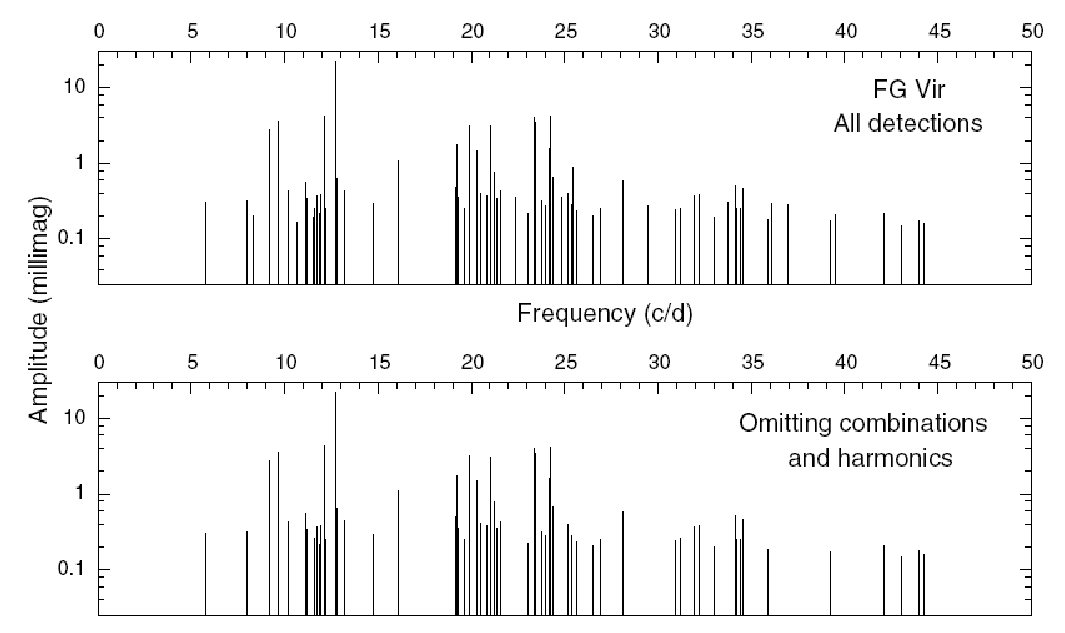}
\caption{
Observed modes in the $\delta$~Scuti star FG~Vir, from extensive
photometric observations spanning more than a decade.
In the lower panel combination frequencies, including harmonics, resulting
from the non-sinusoidal nature of the oscillations have been removed;
this therefore represents the true spectrum of independent oscillations in
the star.
\citep[From][with permission.]{Breger2005}.
}
\label{fig:fgvir}       
\end{figure}

This analysis determines which modes are linearly unstable and hence have a
tendency to grow, but it says nothing about the resulting amplitudes.  Formally,
the amplitude grows exponentially, as $\exp( \omegai t)$, when $\omegai > 0$;
clearly, other mechanisms must set in to limit the amplitudes to finite values.
For high-amplitude pulsators, such as Cepheids or RR Lyrae stars, a single mode
typically dominates and it is reasonable to assume that the driving mechanism is
saturated by nonlinear effects in this mode; in fact, early nonlinear
calculations of radial oscillations \citep[e.g.,][]{Christ1966} were quite
successful in reproducing the observed amplitudes and light- and velocity curves
in such stars.  For low-amplitude pulsators near the main sequence the situation
is far more complex.  Here typically a substantial number of modes are excited
to observable amplitudes; although the frequency interval covered by the
observed modes generally corresponds reasonably well with the interval in which
unstable modes are predicted, their amplitudes typically vary greatly and with
no obvious pattern.  An example is illustrated in Fig.~\ref{fig:fgvir}.  It was
suggested by \cite{Dziemb1985} that in this case the dominant amplitude limiting
mechanism could be resonant mode coupling between an unstable mode and a pair of
stable lower-frequency modes (typically g modes).  The resulting amplitudes would
therefore depend on the complex spectrum of g modes and the extent to which the
resonance conditions were satisfied.  The mechanism was examined in detail by
\cite{Nowako2005}.  He concluded that, even taking into account rotational
splitting, the resonant-coupling mechanism was inadequate to explain the
observed amplitudes and proposed that nonlinear saturation of the driving,
involving interactions between the unstable modes, must be invoked.  It is
clear, however, that we still do not have a general understanding of the
physical phenomenon that determines the amplitudes in low-amplitude
multi-periodic stars, let alone the ability to predict the observed amplitudes
in detail.  An impressive attempt to remedy the situation are the computations
for radial modes by \citet{smolec07}, who found that collective saturation of
the driving mechanism by several acoustic modes can predicted amplitudes to the
observed level.

We have so far assumed that each mode of oscillation is a simple harmonic
function of time.
This is true as long as linear theory is adequate;
however at larger amplitudes, the behaviour is typically distorted.
When only a single frequency is present, the outcome is obviously 
a set of harmonics of the basic frequency, which together defines
the actual light or velocity curve.
In multimode pulsators the effect of the non-linearities 
in the power spectrum is to
generate a set of linear combinations of the basic frequencies;
this is visible for the $\delta$~Scuti star FG Vir in the upper
panel of Fig.~\ref{fig:fgvir} where the harmonics and combination
frequencies are included.
This behaviour is particularly striking in observations of white dwarfs
\citep[e.g.,][]{Winget1994};
it was noted by \cite{Wu2001} and \cite{Montgo2005} that the amplitudes
and phases of the combinations can be used as diagnostics for the
convection zone in such stars.

So far we have neglected the convective contributions to the stability
of the mode.
In addition to the term in the turbulent pressure, shown explicitly in
\Eq{eq:workint}, these include the convective contribution to the flux.
The inclusion of these contributions is obviously hampered by the lack of
a reliable procedure to compute the effects of convection, even in the
static case of the equilibrium model.
Various time-dependent convection formulations have been developed to
deal with the convection-pulsation interactions, based on differing 
physical models.
The formulation by \cite{Unno1967}
was further developed by \cite{Gabrie1974, Gabrie1975, Gabrie1996, Gabrie2000},
as summarized by \cite{Grigah2005}.
\cite{Gough1977b} developed a somewhat different formulation based on
a detailed physical description of the dynamics of convective elements;
this was extended to a non-local formulation by \cite{Balmfo1992},
based on the non-local description of convection in a static
model developed by \citet{Gough1977c} and illustrated in a solar model
by \citet{Gough1985a}.
Alternative formulations were proposed by \cite{Stelli1982}
\cite[further developed by, e.g.,][]{Kuhfus1986, Gehmey1992, Feucht1999}
and \cite{ Xiong1997}.
A review of these different formulations was provided by \cite{Baker1987}.

Simple analyses show that convection may contribute to the excitation 
in the extreme cases of very long and very short convective time scales,
compared with the pulsation period.
In the former case it is plausible that the convective flux does not
react to the pulsations, leading to a negligible convective-flux
perturbation $\delta L_{\rm con}$.
It was noted by \cite{Cox1968} that this might cause excitation
if the radiative flux and hence the radiative-flux perturbation
become small in the convection zone:
since $\psi_{\rm rad}$ is typically negative, $\delta L$ is in phase
with $\delta \rho/\rho$ just beneath the convection zone;
thus the change to very small $\delta L$ in the convection zone
corresponds to a negative $\dd \delta L /\dd r$ in \Eq{eq:lumdriv}
and hence to a contribution to the driving.
Physically this effect arises from the effective blocking by
convection of the luminosity perturbation at the base of the
convection zone, leading to heating in phase with compression.
Thus the mechanism was dubbed `convective blocking' by \cite{Pesnel1987}.
It has been shown that this may account for the driving 
of the long-period g modes in the
$\gamma$~Doradus stars \citep[e.g.,][]{Guzik2000, Warner2003};
here the convection zones are sufficiently deep that the convective 
timescale is substantially longer than the pulsation period.
More detailed calculations by \cite{Dupret2004,Dupret2005a}, using the
convection formulation of \cite{Gabrie1996}, essentially confirmed that
convective blocking dominates the driving of these oscillations.

The opposite extreme, the convective timescale being much shorter than
the pulsation period, is relevant to pulsating white dwarfs with
significant outer convection zones. 
The g-mode pulsations in such stars have periods far exceeding
the convective timescales in the relatively thin convection zones.
It was argued by \cite{Brickh1983, Brickh1991} for DA white dwarfs
that in this case also the convection zone may act to excite the modes;
this was confirmed in a more detailed analysis by \cite{Goldre1999}.
In this case the energy input to the convection zone by the 
flux perturbation at its base is redistributed throughout the convection
zone as a result of the short convective timescale, causing heating of
the convection zone;
since the radiative flux perturbation is in phase with the density 
perturbation at the top of the radiative region, as argued above,
and the density perturbation varies little through the convection zone,
heating is in phase with the density perturbation throughout the
convection zone, which therefore contributes to the excitation of the mode.

\begin{figure}
\centering
  \includegraphics[width=0.75\textwidth]{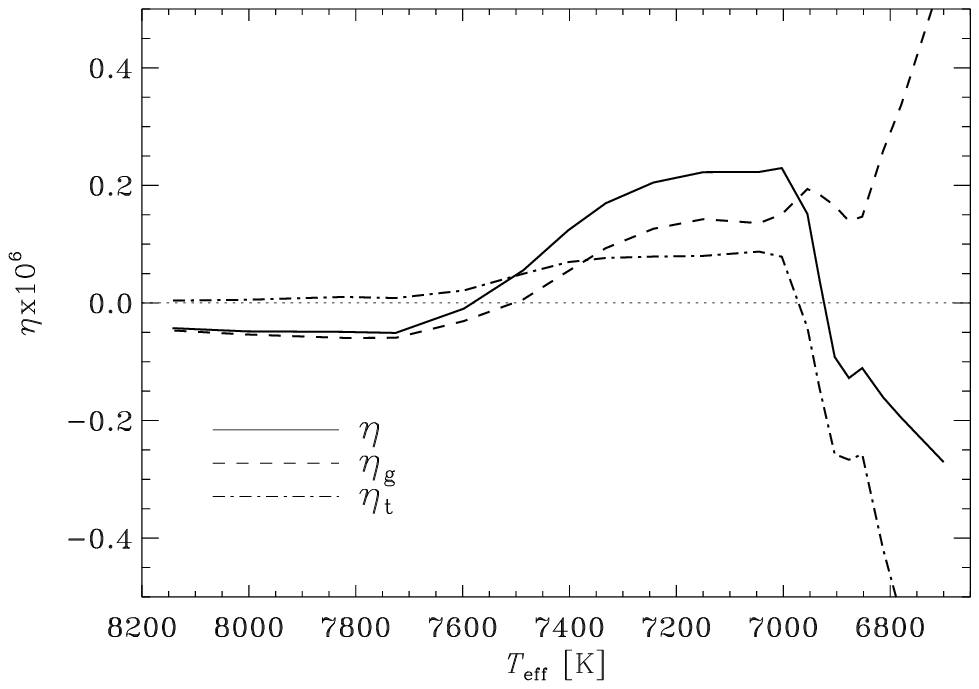}
\caption{
Stability coefficients $\eta = \omega_{\rm i}/\omega_{\rm r}$ 
for the fundamental radial mode as a 
function of effective temperature $\Teff$, for a 
$1.7 \Msun$ model of a $\delta$~Scuti star evolving during the core
hydrogen-burning phase.
The dashed and dot-dashed curves show the contributions from the 
thermal effect and the perturbation to the turbulent pressure,
respectively (see Eqs (\ref{eq:workint1}) and (\ref{eq:workint2})),
and the solid curve shows their sum.
\citep[Adapted from][]{Houdek2000}.
}
\label{fig:cephstab1}      
\end{figure}

\begin{figure}
\centering
  \includegraphics[width=0.75\textwidth]{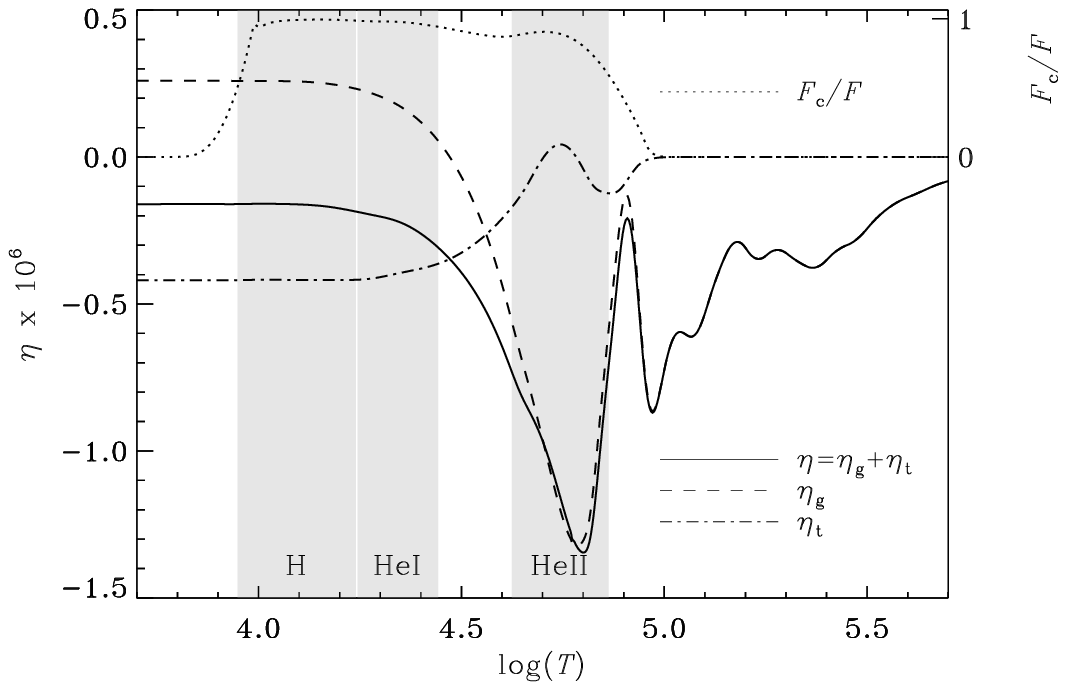}
\caption{
Relative work integrals, 
as functions of the upper limit of integration expressed in terms of $\log T$,
for the fundamental radial model in a model with $\Teff = 6813 \K$
in the $1.7 \Msun$ sequence illustrated in Fig.~\ref{fig:cephstab1}.
The dashed and dot-dashed curves show the contributions from the 
thermal effect and the perturbation to the turbulent pressure,
respectively (see Eqs (\ref{eq:workint1}) and (\ref{eq:workint2})),
and the solid curve shows their sum.
The shaded areas indicate the regions of hydrogen ionization (H) and
first and second helium ionization (HeI and HeII, respectively).
The dotted curve, using the right-hand ordinate scale, shows the
contribution $F_{\rm c}/F$ of convection to the total flux in
the equilibrium model.
\citep[Adapted from][]{Houdek2000}.
}
\label{fig:cephstab2}       
\end{figure}

Early detailed calculations of the excitation of modes in the 
Cepheid instability strip \citep[e.g.,][]{Baker1962, Baker1965}
found reasonable agreement between the upper limit in effective
temperature of instability and the observed blue edge of the instability strip.
However, the models remained unstable at much lower effective temperature
than observed.
In these calculations the interaction between convection and pulsations
was ignored.
An early demonstration that convective effects can in fact delimit the Cepheid
instability strip on the cool side was obtained by \cite{Baker1979},
using the formulation by \cite{Gough1977b}, for models of RR Lyrae stars.
\cite{Gonczi1981} similarly found stability at the red edge, for
Cepheid models, using the formulation of \cite{Unno1967}.
\cite{Stelli1984} also delimited the RR Lyrae instability strip using
the formulation by \cite{Stelli1982}. 
\cite{Gehmey1993} studied the suppression of instability by convection
at the red edge of the RR Lyrae region, in a fully nonlinear calculation,
while \cite{Xiong2001} used the non-local time-dependent theory
of \cite{Xiong1997} to locate the red edge of the $\delta$~Scuti
instability strip.
However, although these calculations generally find that convective effects
can account for the stabilization of the cool models, it is striking that
the physical description of the effect depends strongly on the detailed
convection formulation employed.

To illustrate the effects of convection on the location of the instability 
strip we consider calculations carried out by \cite{Houdek2000},
using a formulation developed by \cite{Gough1977b} and \cite{Balmfo1992}.
Figure~\ref{fig:cephstab1} shows the resulting relative growth rates,
for an evolution sequence of $1.7 \Msun$ models evolving through
the Cepheid instability strip during core hydrogen burning and hence
corresponding to $\delta$ Scuti stars.
The onset of instability, at the blue edge of the strip,
takes place at $\Teff \simeq 7600 \K$.
Interestingly, the thermal contribution $\eta_{\rm g}$ to excitation, 
which includes the effects of the perturbation to the convective flux,
grows strongly as the model $\Teff$ decreases;
however, this is more than balanced by the large negative values of
$\eta_{\rm t}$ arising from the perturbation to the turbulent pressure,
leading to overall return to stability at a red edge of the instability
strip near $\Teff = 6900 \K$.
Further details on the contributions to excitation and damping 
are illustrated in Fig.~\ref{fig:cephstab2}, for a model just on the
cool side of the instability strip. 
Obviously, the bulk of the HeII ionization zone contributes to the
excitation of the mode, essentially as in Fig.~\ref{fig:dscutiex};
but the contribution from the turbulent pressure is damping in most
of the convection zone, leading to overall stability.

Extreme cases of interaction between convection and pulsations are probably
found in the {\it Mira variables\/} on the extreme red-giant branch,
with very extensive outer convection zones.
Their oscillations are likely excited by the heat-engine mechanism, undoubtedly
involving a strong effect of convection, but the details are so far
highly uncertain
\citep[e.g.,][]{Xiong1998, Muntea2005, Olivie2005}.
Owing to their huge radii these stars are excellent targets for interferometry,
including early work by \citet{Quirre1992}
\citep[see also, for example][]{Ohnaka2005, Wittko2007},
yielding detailed information about their complex atmospheric structure.
Since they typically show only one, or at most a few, modes the stars are
less interesting from an asteroseismic point of view,
although further studies might help elucidate aspects of the 
convection-pulsation interactions.

\subsubsection{Stochastically excited pulsations\label{stochastic}}

It appears that modes in stars on the cool side of 
the Cepheid instability strip, including the Sun, are generally stable.
The excitation of the observed modes is attributed to stochastic forcing
by the near-surface convection, a process first discussed in detail in
the stellar case by \cite{Goldre1977}. 
Indeed, as pointed out by \cite{Lighth1952} turbulent flows with 
speeds approaching the speed of sound generate sound very efficiently,
and the near-surface convection is thus a source of strong sound generation
in stars with substantial outer convection,
exciting the normal modes of the star.
\cite{Stein1968} applied Lighthill's theory to the solar convection zone
and atmosphere, to estimate the acoustic energy flux generated by convection.

\begin{figure}
\centering
  \includegraphics[width=0.75\textwidth]{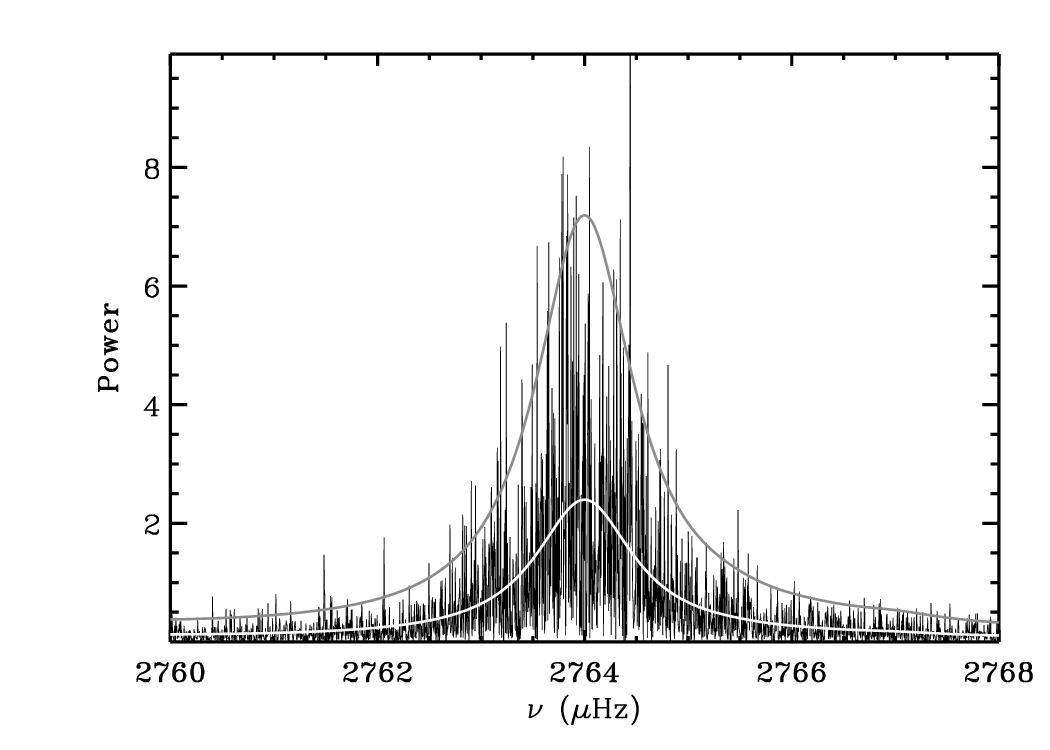}
\caption{Observed spectrum, from coherent analysis of eight years of
Doppler observations
with the BiSON network, of a single radial mode of solar oscillations
(of radial order $n = 19$).
The smooth white curve shows the fitted Lorentzian profile,
while the grey curve shows the same profile, multiplied by three.
\cite[See][]{Chapli2002a}.
}
\label{fig:bispeak}       
\end{figure}

\begin{figure}
\centering
  \includegraphics[width=0.75\textwidth]{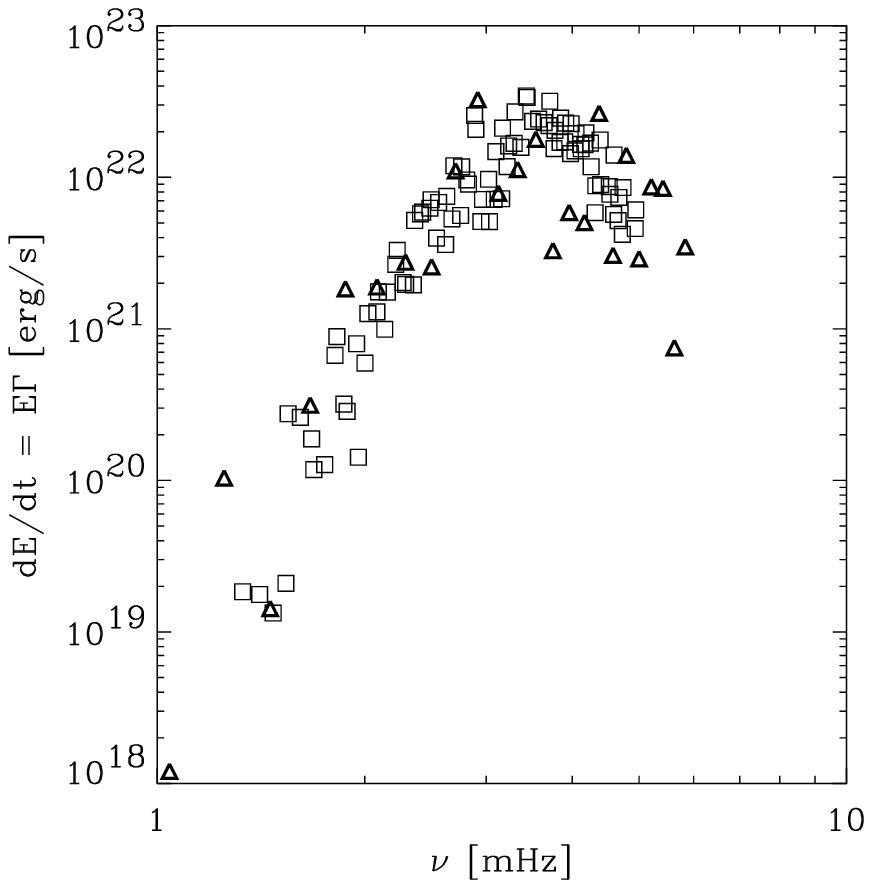}
\caption{Comparison of observed and computed energy input to low-degree
solar acoustic modes (cf.\ \Eq{eq:energyin}).
The squares are observed values, from Doppler-velocity observations of
modes of degree $l = 0$ to $3$, with the GOLF instrument on the SOHO spacecraft
\citep{Roca1999}.
The triangles result from hydrodynamical simulations of solar convection
\citep{Stein2001}.
}
\label{fig:energyin}
\end{figure}

\cite{Batche1956} considered the general problem of the stochastic
excitation of a damped oscillator.
It is straightforward to show that the resulting power spectrum, for
a mode of frequency $\omega_0$ and damping rate $\omegai$, is
given by
\begin{equation}
P(\omega) \simeq {1 \over 4 \omega_0^2} {P_f(\omega) \over 
(\omega - \omega_0)^2 + \omegai^2} \; ,
\label{eq:stochpow}
\end{equation}
where $P_f$ is the power of the forcing function
\citep[see also][]{Christ1989b}.
Thus if the forcing varies slowly with frequency the result is, on average,
a Lorentzian profile, with a width determined by $\omegai$.
An example of a single realization of such a spectrum, including the
fitted Lorentzian profile, is illustrated in Fig.~\ref{fig:bispeak},
based on 8 years of solar whole-disk observations with the
BiSON network.
Thus with well-resolved observations the damping rates can be determined
from the observed line widths.
{}From the observed surface amplitudes and damping rates in the solar case,
and the mode mass (cf.\ \Eq{eq:inertia}),
the required rate of energy input can be estimated as
\begin{equation}
{\dd \CE \over \dd t} = |\omega_{\rm i}| M_{\rm mode} A_V \; ,
\label{eq:energyin}
\end{equation}
where $A_V$ is the velocity power, corrected for the geometric effect of
averaging over the solar surface \citep{Roca1999}.
In Fig.~\ref{fig:energyin}
the result is compared with the stochastic energy
input calculated from a detailed hydrodynamical model of near-surface
convection \citep{Stein2001}.
The agreement between the observations and the computations is evidently
quite reasonable.
It should be noted, in particular, that in contrast to the heat-engine
excitation we obtain a definite theoretical prediction of the mode amplitudes
which furthermore tend to vary relatively slowly with frequency.
Thus in this case we expect to see the excitation of a fairly complete
set of modes over a substantial range of frequencies, obviously easing
the problem of mode identification and increasing the asteroseismic
information content.

\citet{Chapli2005} made a detailed analysis of the expected amplitudes
of stochastically excited oscillations, summarized by \citet{Houdek2006a}.
The result can be written as
\begin{equation}
\langle V^2 \rangle = {1 \over \omegai E} {\CP_{\rm f} \over E} \; ,
\label{eq:stochamp}
\end{equation}
where $\langle V^2 \rangle$ is the mean square oscillation velocity,
$E$ is the mode inertia (cf.\ \Eq{eq:inertia}) and $\CP_{\rm f}$ is a measure
of the forcing.
From Eqs (\ref{eq:workint}) and (\ref{eq:workint1}),
and the fact that the integrals in $W_{\rm t}$ and $W_{\rm g}$
are generally dominated by the near-surface layers, it follows
that $\omegai E$ depends on the behaviour of the eigenfunction near the
surface which is largely a function of frequency.
The same is true of $\CP_{\rm f}$.
Thus, approximately, we can write \Eq{eq:stochamp} as 
\begin{equation}
\langle V^2 \rangle \simeq  {\CF(\omega) \over E} \; .
\label{eq:stochamp1}
\end{equation}
It follows that, at given frequency, the power in the mode scales
approximately as $E^{-1}$;
equivalently, given \Eq{eq:ekin}, the mode energy is predominantly
a function of frequency.

Based on theoretical estimates of the damping rates and energy input it
is evidently possible to predict the expected oscillation amplitudes
as functions of stellar parameters \citep{Christ1983}.
\cite{Kjelds1995b} showed that the results could be fitted by the
simple relation $v_{\rm osc} \propto L/M$, where $v_{\rm osc}$ is the surface
velocity.
Using Gough's and Balmforth's treatment of convection 
\cite{Houdek1999} confirmed 
that the predicted amplitudes generally increased
with increasing luminosity along the main sequence.
However, the estimated damping rates and energy input are obviously
rather uncertain.
In fact it appears that the observed amplitudes of stars hotter than
the Sun, including Procyon, are substantially lower than these simple
predictions \citep[e.g.,][]{Gillil1993, Martic2004}.
\cite{Houdek2002a} found that the predicted amplitudes for $\alpha$~Cen A and
$\beta$~Hyi (both of roughly solar effective temperature) approximately
agreed with the observed values, while confirming that the predicted surface
velocity for
Procyon, which is substantially hotter than the Sun, exceeded the observed
value by roughly a factor of two.
He speculated that the dominant problem in the calculations might be the
computed damping rate, which arises from a combination of several terms of
different sign;
in particular, he pointed to the neglect of incoherent scattering as a potential
deficiency of the damping-rate computation.
Hydrodynamical simulations of convection can constrain the parameters
of simpler formulations \citep{Samadi2003a, Samadi2003b}.
\cite{Stein2004} made an extensive calculation of the energy input from
detailed simulations of a number of stars, concluding, as in the case of
the simpler calculations, that the excitation increases 
with increasing effective temperature and decreasing gravity.
An overview of the excitation of solar-like oscillations
in pre-main-sequence and main-sequence models was provided by
\cite{Samadi2005}.

For stars other than the Sun the duration or signal-to-noise ratio
of the observations have not allowed resolution of the Lorentzian profile.
However, mode lifetimes can be estimated from the statistics of the 
observed power spectra or inferred frequencies.
From such analyses \cite{Kjelds2005} found that the mode lifetimes in
$\alpha$~Cen A and B were around 3\,d at the maximum in power,
similar to the Sun;
a similar result was obtained by \citet{Fletch2006}.
Analysis of amplitude and frequency scatter 
for the metal-poor subgiant $\nu$~Ind by
\citet{Carrie2007} indicated a lifetime of at least 9\,d, substantially longer 
than for the Sun.
On the other hand, \cite{Stello2006} found that the solar-like oscillations
in the red giant $\xi$ Hya \citep{Frands2002} had a lifetime of only
a few days, far shorter than the predictions by \cite{Houdek2002b}.
As noted by Stello {\etal}, 
this could severely compromise the possibilities
of using frequencies of red giants for asteroseismology.
\citet{Leccia2007} recently obtained a life time of around 2 days for
modes in Procyon, largely consistent with unpublished estimates
based on the calculations by \citet{Houdek2002b}.
This would indicate that the excessive predicted amplitude for Procyon
results from problems with the energy input.
Interestingly, \citet{Houdek2006a} noted, based in part on hydrodynamical
simulations by \citet{Stein2004}, that there may be partial cancellation
between contributions to the forcing from the fluctuating turbulent 
pressure and the turbulent gas pressure.
Further calculations, as well as more precise observations of a broader range
of stars, are needed to check whether such effects may account for the
observed amplitudes.

\begin{figure}
\centering
  \includegraphics[width=0.75\textwidth]{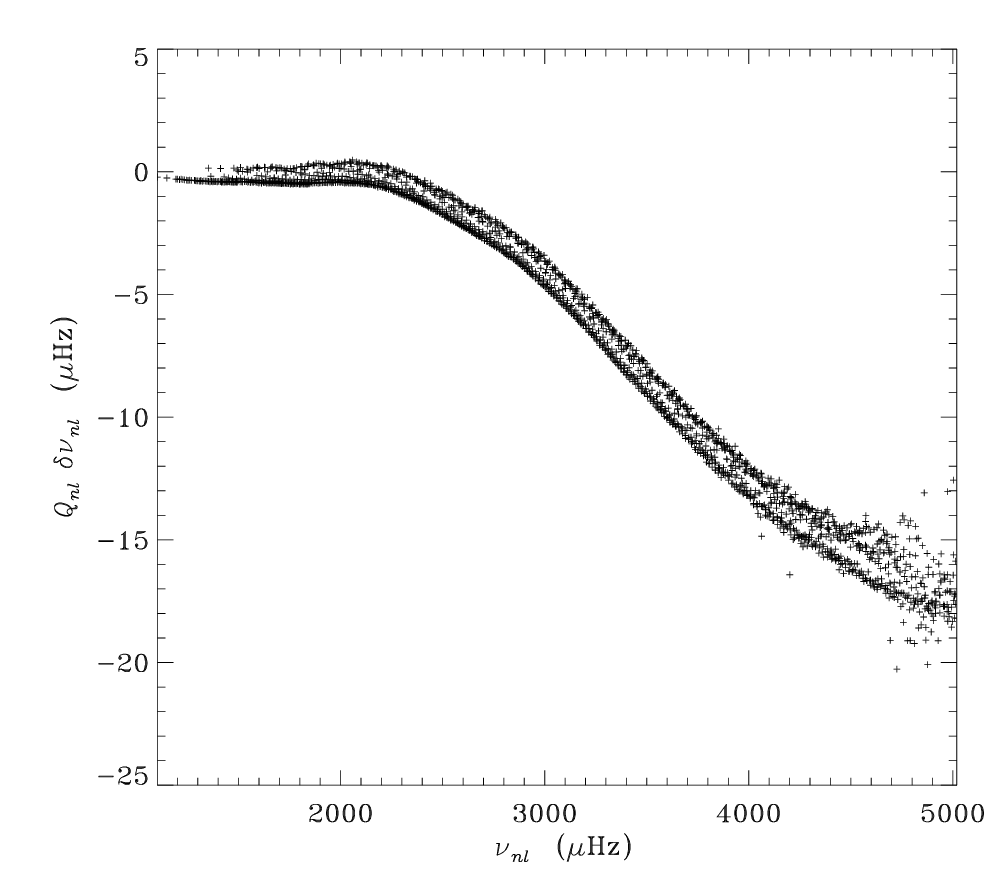}
\caption{Frequency differences, in the sense (Sun) -- (Model),
between observed frequencies from the MDI instrument on the SOHO spacecraft
\citep{Kosovi1997} and Model S of \cite{Christ1996}.
Modes with $l \le 100$ have been included.
To compensate for the variation with degree of the mode inertia,
the frequencies have been scaled by $Q_{nl} = E_{nl}/\bar E_0(\omega_{nl})$
where $E_{nl}$ is the mode inertia (cf.\ \Eq{eq:inertia})
and $\bar E_0(\omega_{nl})$ is the inertia of a radial mode, interpolated to
the frequency $\omega_{nl}$ of the mode.
}
\label{fig:soldif}       
\end{figure}

The effects of near-surface convection also contribute uncertainties in
the calculation of stellar oscillation frequencies.
In most cases frequency calculations assume adiabatic oscillations and
neglect the dynamical effects of convection, in the form of turbulent
pressure, both in the equilibrium model and the pulsations.
A full nonadiabatic calculation would require taking into account the
effects of the perturbation to the convection flux;
this is further complicated by the fact that the convective timescale
in the region of substantial nonadiabaticity is generally similar to
the oscillation period in the case of solar-like oscillations.
Such near-surface effects are characterized by being predominantly 
functions of frequency,
apart from an essentially trivial dependence on the mode inertia $E$,
and by being small at relatively low frequency \citep{Christ1997};
modes in the low range of the frequencies of stochastically excited modes
are evanescent near the surface 
and hence are less affected by the near-surface errors in the computations.
As illustrated in Fig.~\ref{fig:soldif} these near-surface 
errors generally dominate the differences between computed and observed
frequencies of the Sun \citep[see also][]{Christ1988}.
\cite{Rosent1999} found that including an averaged hydrodynamical simulation,
instead of the normal mixing-length treatment, in models of the solar envelope
substantially reduced this near-surface frequency error.
Similar results were obtained by \cite{Robins2003} by including results
from simulations in the physics used in the computation of solar models.
Earlier, \cite{Patern1993} found that the differences between the
solar and model frequencies were decreased if the \cite{Canuto1991}
convection treatment was used instead of the \cite{Bohm1958} mixing-length
treatment used for the model illustrated in Fig.~\ref{fig:soldif} \citep[see also][]{mt96};
\cite{Kosovi1995} noted that including turbulent pressure in the
calculation of the equilibrium model similarly reduced the differences.

\subsection{Seismic inference\label{inference}}
The level of information that can be obtained from asteroseismic data
evidently depends crucially on the amount and quality of the data.
With only a few modes, or the basic structure of the spectrum of
solar-like oscillations, we can determine global parameters of
the star such as the mass and radius.
At the opposite extreme, we may hope to obtain a sufficient number
of well-identified modes to be able to carry out inverse analyses
to infer details of the structure of at least parts of the star.
Regardless of the asteroseismic potential for a given star, 
it is very important to supplement the asteroseismic data with
other types of data, as a test of the consistency of the results
or to extend the usefulness of the oscillation frequencies.

The development and testing of procedures for asteroseismic inference
are far from trivial.
A commonly used, and very effective, procedure is the so-called
`hare-and-hounds' exercise.
Here artificial data, based, say, on the frequencies of some model,
are created by the `hare';
the `hounds' analyse these data with only the information about
the underlying model that would be available from real observations.
Subsequently the inferences of the hounds are compared with the 
truth about the original model.
Evidently, the generation of the data should be as realistic as possible,
for example taking into account the stochastic nature of the excitation
of solar-like oscillations. An example of an application to seismic inferences
for solar-type stars has been given by \citet{mcdt02}.
Application to `classical' pulsating stars, within the context of
the CoRoT mission, was considered by \cite{Thoul2003},
whereas \cite{Appour2006} gave an exhaustive survey of the 
expected performance of the CoRoT mission based on an extensive
series of exercises; \cite{Mazumd2005} carried out a detailed
analysis of asteroseismic inferences for one of these cases.
An interesting recent example of this type of exercise, applied to
time-series analysis of simulated solar data, is provided by
\cite{Chapli2006}.

\subsubsection{\label{basics}Basic asymptotic signatures}

In many cases observed modes are of high radial order
and valuable information can be extracted from the asymptotic
properties of the frequency spectrum.
This is true both for acoustic modes, particularly as observed in
solar-like pulsators, and for the long-period g modes observed in,
for example, slowly pulsating B stars, $\gamma$~Doradus stars and white dwarfs.

\begin{figure}
\centering
  \includegraphics[width=0.75\textwidth]{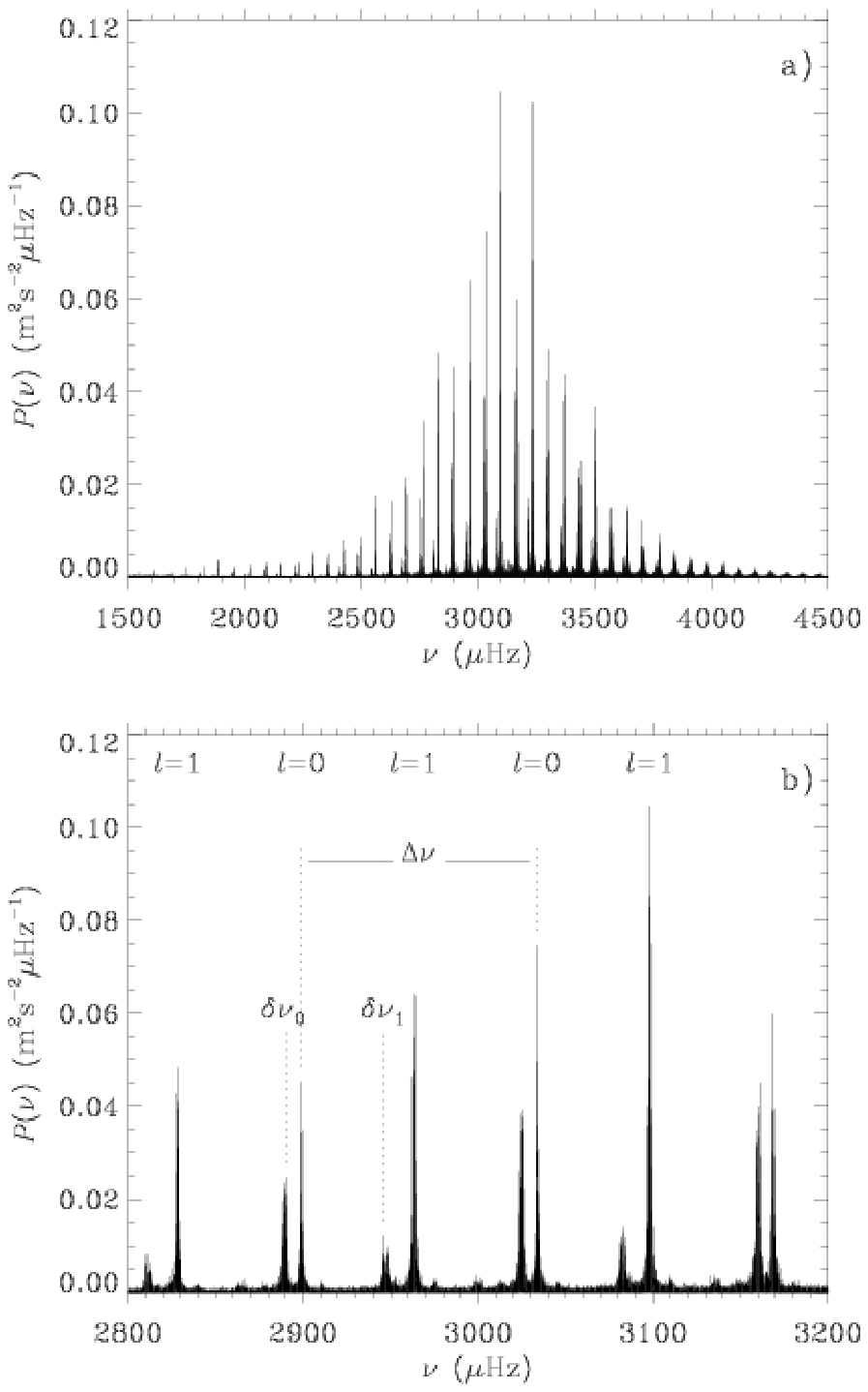}
\caption{Power spectrum of solar oscillations,
obtained with the six-station BiSON network, using Doppler observations
in light integrated over the disk of the Sun;
the data span the full 15-year period over which the complete BiSON network
has been operating.
The ordinate is normalized to show velocity power per frequency bin.
Panel (b) provides an expanded view of the central part of the
frequency range.
Here some modes have been labelled by their degree $l$,
and the large and small frequency separations $\Delta \nu_{nl}$ and
$\delta \nu_{nl}$ [cf.\ Eqs (\ref{eq:pasymp}) and (\ref{eq:smallsep})]
have been indicated. (Note that for simplicity only the relevant indexes were used in the figure.)
\citep[Data kindly provided by the BiSON group. See][]{Chapli2007}.
}
\label{fig:bisspect}       
\end{figure}

The basic asymptotic properties of low-degree high-order acoustic modes
were elucidated by \cite{Vandak1967} and \cite{Tassou1980, Tassou1990};
they can also be obtained from JWKB analysis of the asymptotic 
\Eq{eq:gougheq} \citep[e.g.,][]{Gough1986, Gough1993}.
The result is that the cyclic frequencies $\nu_{nl}$ for acoustic modes of radial
order $n$ and degree $l$ are given, in this asymptotic limit, by the
expression 
\begin{equation}
\nu_{nl}  \simeq
\left( n + {l \over 2 }+ {1 \over 4 }+ \alpha \right) \Delta \nu_0
- [A l(l+1) - \delta ] {\Delta \nu_0^2  \over \nu_{nl}} \; ,
\label{eq:pasymp}
\end{equation}
where
\begin{equation}
\Delta \nu_0 = \left(2 \int_0^R {\dd r \over c} \right)^{-1}
\label{eq:lsep}
\end{equation}
is the inverse sound travel time across a stellar diameter, and
\begin{equation}
A = {1 \over 4 \pi^2 \Delta \nu_0}
\left[ {c(R) \over R }- \int_0^R {\dd c  \over \dd r} {\dd r \over r }\right]
\; ;
\label{eq:pcorr}
\end{equation}
also, $\alpha$ (which in general is a function of frequency)
is determined by the reflection properties near the surface
and $\delta$ is a small correction term predominantly related to the
near-surface region.
To leading order, neglecting the last term, \Eq{eq:pasymp} predicts 
a uniform spacing
of modes of the same degree.
This difference in frequency of modes of the same degree and consecutive order -- 
($\Delta \nu_{n\,l} = \nu_{n+1\,l}-\nu_{n\,l}$) -- 
is known as the {\it large frequency separation} and is,
to leading order, approximately equal to $\Delta \nu_0$.
Also to leading order, \Eq{eq:pasymp} predicts a degeneracy 
between frequencies of modes with degree of the same parity,
\begin{equation}
\nu_{n\,l} \simeq \nu_{n-1 \, l+2} \; .
\label{eq:asdeg}
\end{equation}
Finally, modes of odd degree fall halfway between modes of even degree. 
This pattern is clearly observed in solar data 
(see Fig.~\ref{fig:bisspect}a)
and is one of the clearest indicators for the detection
of solar-like oscillations.

The departure from the degeneracy in \Eq{eq:asdeg} is reflected in
the {\it small frequency separation\/}
\begin{equation}
\delta \nu_{nl} = 
\nu_{n\,l} - \nu_{n-1 \, l+2} \simeq
- ( 4 l + 6 ) {\Delta \nu_0 \over 4 \pi^2 \nu_{nl}}
\int_0^R {\dd c \over \dd r }{\dd r \over r } \; ,
\label{eq:smallsep}
\end{equation}
where we neglected the small term in $c(R)$ in \Eq{eq:pcorr}.
This frequency structure is also clearly visible in solar data,
as shown in Fig.~\ref{fig:bisspect}b.
It is evident that the separation depends on the degree;
this can be used to identify the degrees of the observed modes,
as was done also in the early phases of helioseismology
\citep[e.g.,][]{Christ1980a}.

\begin{figure}
\centering
  \includegraphics[width=0.75\textwidth]{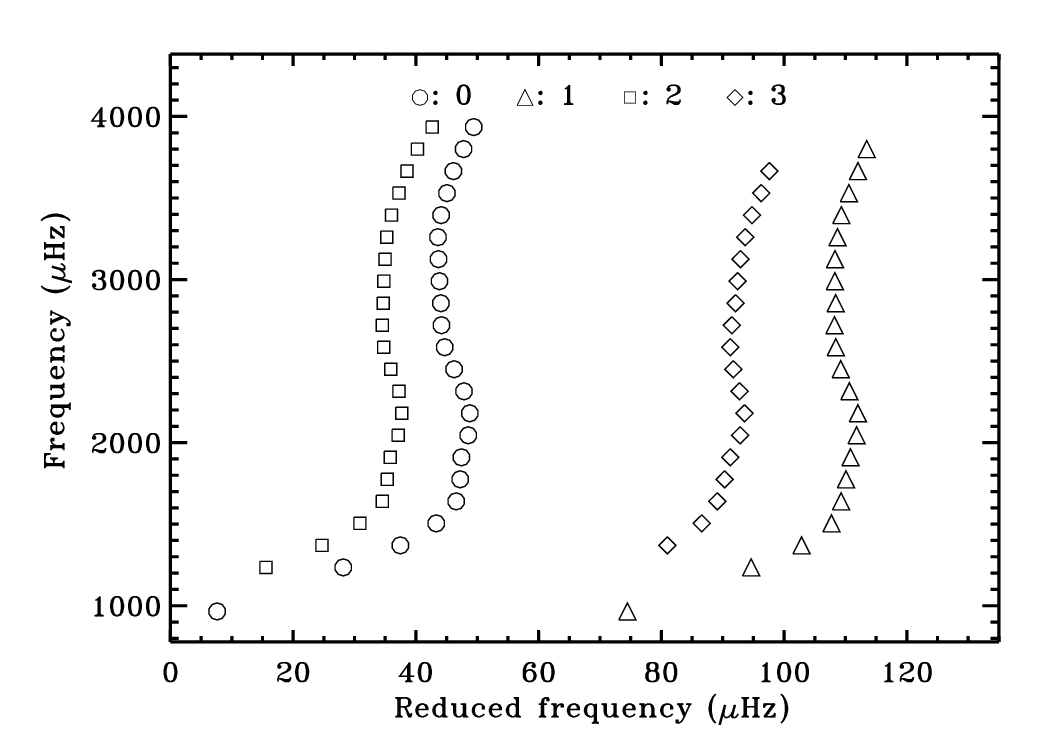}
\caption{Echelle diagram for observed solar frequencies
obtained with the BiSON network \citep{Chapli2002a},
plotted with $\nu_0 = 830 \muHz$ and
$\langle \Delta \nu \rangle = 135 \muHz$ (cf.\ \Eq{eq:echelle}).
Circles, triangles, squares and diamonds are used for modes of degree
$l = 0, 1, 2$ and $3$, respectively.}
\label{fig:bisechelle}       
\end{figure}

A convenient way to illustrate the details of the frequency spectrum
is in terms of an {\it echelle\/} diagram.
Graphically this corresponds to dividing the spectrum in segments of
length $\Delta \nu_0$ and stacking the segments.
Thus we express the frequency as
\begin{equation}
\nu_{nl} = \nu_0 + k \langle \Delta \nu \rangle + \tilde \nu_{nl} \; ,
\label{eq:echelle}
\end{equation}
where $\nu_0$ is a arbitrary reference frequency, $k$ is an integer,
$\langle \Delta \nu \rangle$ is a suitable average of $\Delta \nu_{n\,l}$
and the {\it reduced frequency\/} $\tilde \nu_{nl}$ 
is between $0$ and $\langle \Delta \nu \rangle$.
An example, for observed solar frequencies, is illustrated in
Fig.~\ref{fig:bisechelle}.
Had the asymptotic relation in \Eq{eq:pasymp} been exact, the result
would be largely vertical sets of points, separated by the small separation.
As shown there are significant departures from this behaviour;
in particular, the curvature arises largely from the term in 
$\alpha = \alpha(\omega)$ in the asymptotic relation;
as discussed in Section \ref{sharpfeatures}
this behaviour carries information about the helium content in
the stellar envelope.

Owing to the factor $r^{-1}$ in the integral in \Eq{eq:smallsep} the
small separation is very sensitive to the sound-speed structure of the
stellar core.
{}From a physical point of view this is the result of the propagation
regions of the modes.
The $l = 2$ modes are reflected at the inner turning point $r_{\rm t}$
(cf. \Eq{eq:rt} and Fig.~\ref{fig:raypath}) whereas the radial modes 
penetrate essentially to the centre.
In other parts of the star the modes are very similar; hence the 
dominant difference in their frequencies arises from the core.
This property makes acoustic modes of low degree
(so far the only degrees that are observable in distant stars
with solar-like oscillations)
particularly valuable as diagnostics of stellar cores.

Other frequency combinations which suppress the dominant large-scale
structure of the frequency spectrum carry similar information, at
least to the extent that the asymptotic description is valid.
An example is
\begin{equation}
\delta^{(1)}\nu_{n\, l} = \nu_{n\, l} - {1 \over 2} ( \nu_{n-1 \, l+1} + \nu_{n \, l+1})
\simeq - ( 2 l + 2 ) {\Delta \nu_0  \over 4 \pi^2 \nu_{nl}}
\int_0^R {\dd c  \over \dd r }{\dd r \over r } \; .
\label{eq:smallsep1}
\end{equation}
The combination of $\delta \nu_{n\, l}$ and $\delta^{(1)}\nu_{n\, l}$ 
is particularly important in the case of intensity observations where
generally only modes of degree $l = 0, 1$ and 2 are detectable.

\begin{figure}
\centering
  \includegraphics[width=0.75\textwidth]{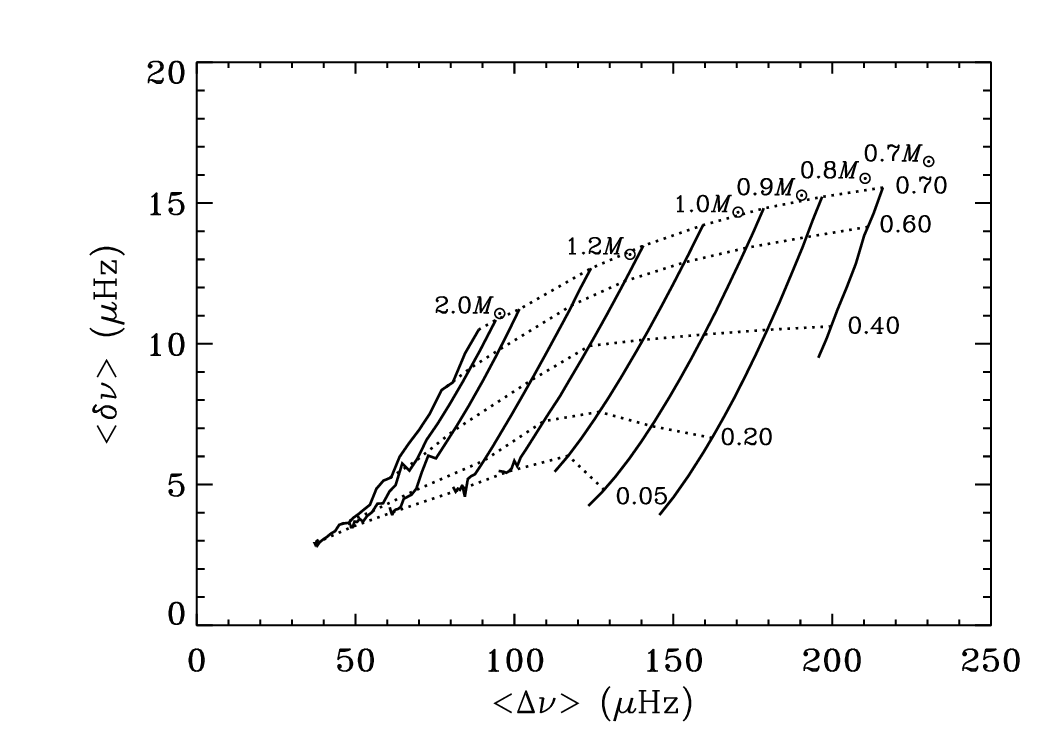}
\caption{Stellar evolution in terms of the average large separation
$\langle \Delta \nu \rangle$ and small 
separation $\langle \delta \nu_{n0} \rangle$
(cf.\ Eqs (\ref{eq:pasymp}) and (\ref{eq:smallsep}));
the averages were evaluated through least-squares fitting of the asymptotic
behaviour (cf.\ \Eq{eq:pasymp}) to computed frequencies, in the manner
of \cite{Scherr1983}.
Solid lines show evolution tracks at the masses indicated and dashed
lines show models at constant central hydrogen abundance, also
indicated in the figure.
}
\label{fig:oschr}       
\end{figure}

The usefulness of the large and small frequency separations as asteroseismic
diagnostics lies in the fact that they can often be determined with 
reasonable precision, even when the observed data do not allow reliable
determination of individual frequencies.
A recent interesting example of this type of analysis was provided by
\cite{Roxbur2006} who showed that a quantity equivalent to the small
separation could be determined from the autocorrelation function of
the observed time series, even in cases where the noise level was too high
to allow reliable frequency identification.

\begin{figure}
\centering
  \includegraphics[width=0.75\textwidth]{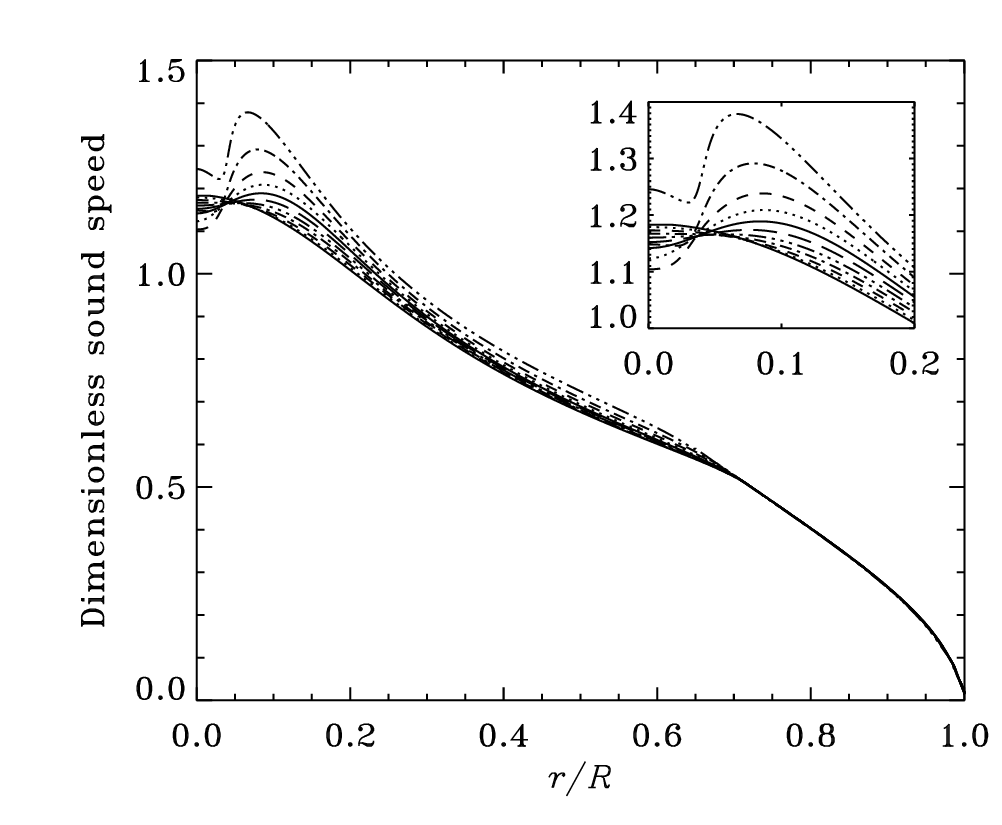}
\caption{Dimensionless sound speed, in units of $(G M/R)^{1/2}$,
in a 1 M$_\odot$ sequence, corresponding to Model S of \cite{Christ1996}.
Results are shown for models of age 0, 1, 2, 3, 4, 5, 6, 7, 8, 9, and 10 Gyr,
in the order of increasing sound speed at $r = 0.1\, R$.
The insert provides a magnified view of the core.
The increase in the scaled central sound speed between the last two models
arises from the increase in the radius (relevant for the scaling) and the
increasing importance of electron degeneracy.
}
\label{fig:cevol}       
\end{figure}

The dependence of the sound speed on the chemical composition
(cf.\ \Eq{eq:csqideal})
makes the small frequency separations sensitive measures of the
evolutionary state of main-sequence stars.
This can be illustrated by presenting the effects of stellar evolution
in a $(\langle \Delta \nu \rangle, \langle \delta \nu \rangle)$ diagram
\citep[e.g.,][]{Christ1984, Christ1988, Ulrich1986, Gough1987},
where $\langle \Delta \nu \rangle$ and $\langle \delta \nu \rangle$ are
suitable averages of $\Delta \nu_{n\, l}$ and $\delta \nu_{n\, l}$, respectively. 
An example is shown in Fig.~\ref{fig:oschr}.
The large separation varies with stellar properties essentially
as $t_{\rm dyn}^{-1}$, whereas the small separation directly reflects
the change in the sound speed in the core resulting from
the changing compositional structure as the star evolves.
The effect of evolution on the sound speed 
is illustrated in Fig.~\ref{fig:cevol}.
This is dominated by the increasing mean molecular weight as
hydrogen is used in the core, leading to a sound-speed inversion.
The effect of the positive sound-speed gradient in the core is evidently
to reduce the integral in \Eq{eq:smallsep} and hence the small separation,
as shown in Fig.~\ref{fig:oschr}.

\begin{figure}
\centering
  \includegraphics[width=0.75\textwidth]{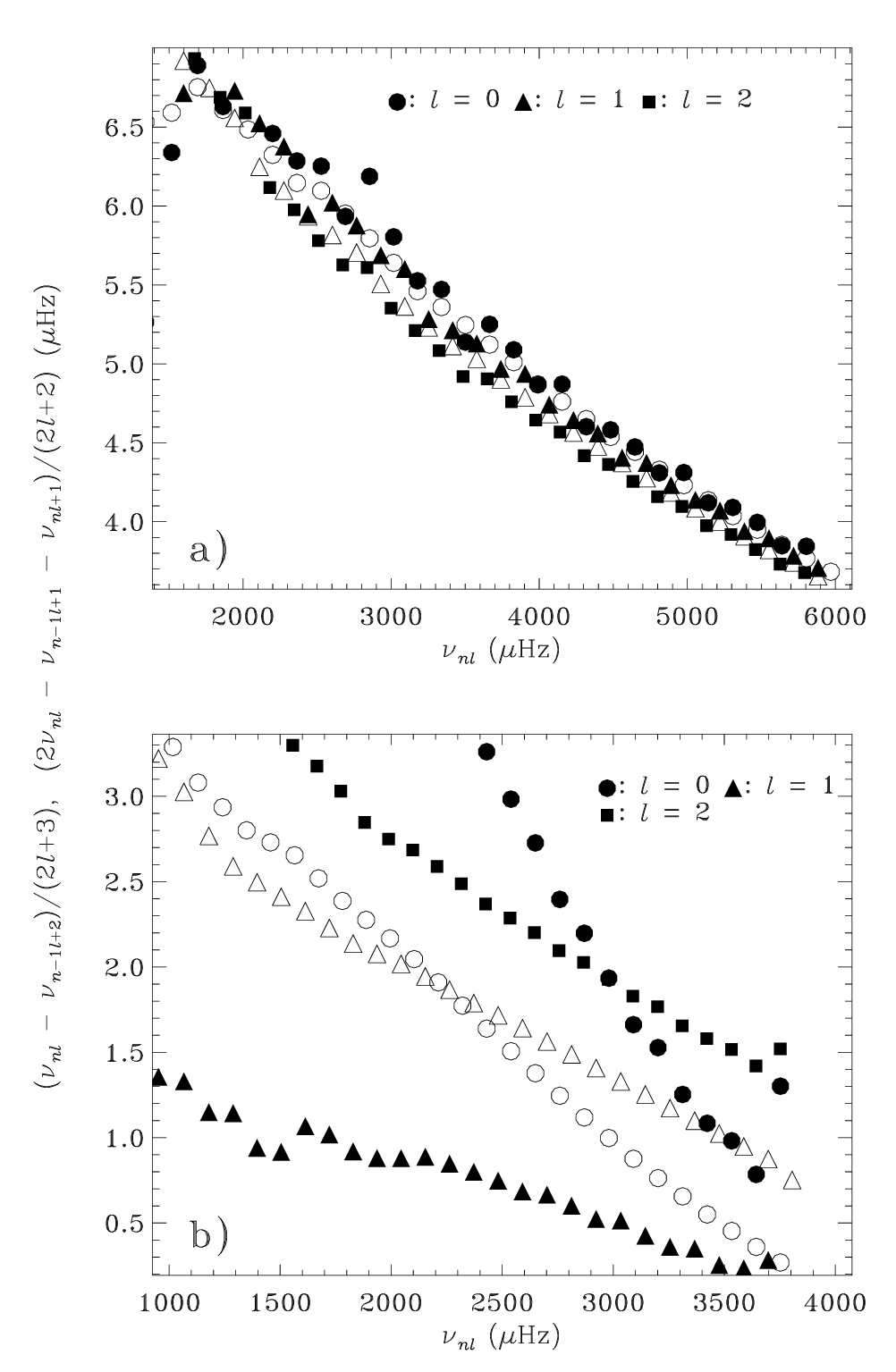}
\caption{Scaled small separations for 1 M$_\odot$ models,
in a sequence corresponding to Model S of \cite{Christ1996}:
a) ZAMS; b) age 8 Gyr.
Open symbols show $\delta \nu_{n\, l}/(2l+3)$ and closed symbols show $\delta^{(1)} \nu_{n\, l}/(l+1)$.
Note that according to Eqs (\ref{eq:smallsep}) and (\ref{eq:smallsep1})
these scaled values should be independent of $l$.
}
\label{fig:smallsep}       
\end{figure}

It should be mentioned that the simple asymptotic expressions
have some limitations.
The analysis is based on the {\it Cowling approximation\/}, neglecting
the perturbation to the gravitational potential which in fact has
a significant effect on the frequencies of the lowest-degree modes
\citep[e.g.,][]{Robe1968, Christ1991}.
Also, the rapid variation of sound speed in the core
of evolved models makes the validity of JWKB analysis questionable.
The latter effect is illustrated in Fig.~\ref{fig:smallsep},
showing $\delta \nu_{n\, l}/(2l+3)$ and $\delta^{(1)} \nu_{n\, l}/(l+1)$ in 
1 M$_\odot$ models on the ZAMS and at an age of 8 Gyr.
According to Eqs (\ref{eq:smallsep}) and (\ref{eq:smallsep1})
these values should fall on a single function of frequency;
while this is approximately the case for the unevolved model,
$\delta^{(1)} \nu_{n\, l}/(l+1)$ is far from this behaviour in the evolved model.
This illustrates the encouraging fact that the frequencies contain more
information than suggested by the simple asymptotic relation.
As discussed in Section \ref{sec:invers}
this underlies the possibilities for carrying out inverse analyses to
infer the structure of stellar cores, even on the basis of just low-degree
modes.
Asymptotic descriptions that provide a better treatment of the
central regions have been developed by \cite{Roxbur1994, Roxbur2000}.

As discussed above (cf.\ Fig.~\ref{fig:soldif}),
near-surface problems
have a substantial effect on computed frequencies, particularly at
high frequency.
Since these effects are frequency dependent they also influence
the large, and to a lesser extent the small, frequency separations.
It was demonstrated by \cite{Roxbur2003} that the effect on the small
frequency separations could be greatly reduced by considering
ratios between small and large frequency separations, such as
\begin{equation}
r_{02}(n) \equiv
{\nu_{n\,0} - \nu_{n-1\,2} \over \nu_{n\,1} - \nu_{n-1\,1}} \; .
\label{eq:seprat}
\end{equation}
The analysis was extended to a broader range of stars by
\cite{Oti2005} and \cite{Roxbur2005},
demonstrating that the separation ratios provide a sensitive measure
of the properties of stellar cores, even given substantial uncertainties
in the treatment of the surface layers.

\begin{figure}
\centering
  \includegraphics[width=0.75\textwidth]{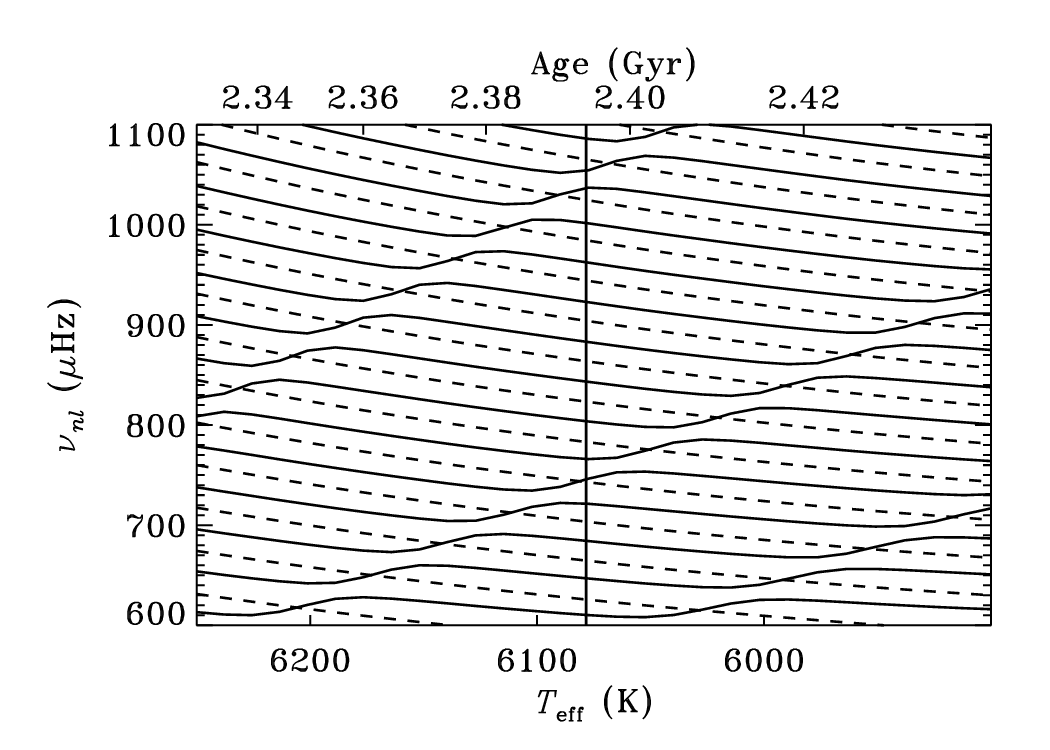}
\caption{
Evolution of adiabatic frequencies for a model of $\eta$~Boo.
The lower abscissa shows the effective temperature $T_{\rm eff}$,
the upper abscissa the age of the model in Gyr.
The dashed lines correspond
to modes of degree $l = 0$, and the solid lines to $l = 1$.
The vertical solid line indicates the location of the
model whose frequencies are illustrated in Fig.~\ref{fig:eboechelle}.
\cite[Adapted from][]{Christ1995}.}
\label{fig:ebofreq}       
\end{figure}


For high-order, low-degree g modes the {\it periods}
$\Pi_{nl} = 1/\nu_{nl}$ are uniformly spaced;
to lowest asymptotic order they satisfy
\begin{equation}
\Pi_{nl} \simeq {\Pi_0 \over \sqrt{l(l+1)}} 
\left(n + {l \over 2} + \alpha_{\rm g} \right) \; ,
\label{eq:gasymp}
\end{equation}
where 
\begin{equation}
\Pi_0 = 2 \pi^2 \left( \int_0^{r_1} N {\dd r \over r} \right)^{-1} \; ,
\label{eq:gspacing}
\end{equation}
and $\alpha_{\rm g}$ is a phase constant \citep[e.g.,][]{Vandak1967, Smeyer1968,
Tassou1980}.  Thus the period spacing depends on the degree, leading to a more
complex structure of the spectrum than in the case of the acoustic modes.
Equations (\ref{eq:gasymp}) and (\ref{eq:gspacing}) assume a star with a
radiative core, and with an outer turning point (where $\omega = N$) at $r =
r_1$; in the cases illustrated in Fig.~\ref{fig:charfreq} the turning point is
typically close to the lower boundary of the convective envelope.  If the model
has a convective core, the term in $l/2$ is not included in \Eq{eq:gasymp} and
the integration limits in \Eq{eq:gspacing} must be suitably modified.  Such
a situation occurs in the case of the observed high-order g modes in the SPB stars
\cite[e.g.,][]{wael91,decat02,Aerts2006b} and $\gamma\,$Doradus stars
\cite[e.g.,][]{aerts04,moya05,dupret05b}. Unfortunately, it turned out too
difficult so far to disentangle the period spacings due to the rotational
splitting effects, because of lack of identification of $l$ and $n$. It has thus
not yet been possible to exploit the period spacings seismically for these
main-sequence stars, as was done for the white dwarfs
\citep{Winget1991, Winget1994}.  

Higher-order corrections to the expression  given by \Eq{eq:gasymp} were discussed by,
for example, \cite{Tassou1980}, \cite{Ellis1984, Ellis1986}
and \cite{Provos1986}.


\subsubsection{Signatures of mixed modes}
\label{mixmodes}

As mentioned in Section \ref{modeprop} the clean separation between 
acoustic and gravity modes disappears in evolved stars where the buoyancy
frequency in the deep interior reaches values corresponding even
to frequencies characteristic of solar-like pulsations.
A typical example is the case of $\eta$~Boo, where solar-like
oscillations were first detected by \cite{Kjelds1995a}.
Models of this star \citep[e.g.,][]{Christ1995, Guenth1996, DiMaur2004}
show that it is likely in the subgiant phase, 
with energy generation in a hydrogen-burning shell. 
This is also reflected in the large values of $N$ 
(cf. Fig.~\ref{fig:charfreq}),
arising from the compact core and resulting in very high values of $g$
(cf. \Eq{eq:buoyideal}).
Figure~\ref{fig:ebofreq} shows the evolution with age
of selected frequencies in such a model.
The radial modes are purely acoustic, with frequencies decreasing as
the radius of the star increases, largely in accordance with \Eq{eq:dyntime}.
The $l = 1$ modes generally follow the same behaviour; also,
as expected from \Eq{eq:pasymp}, their frequencies are roughly halfway
between those of the $l = 0$ modes.
However, there is evidently a second class of modes, with frequencies
{\it increasing\/} with age.
These are g modes predominantly trapped under the peak in the buoyancy
frequency in the deep interior of the star;
with increasing age, the buoyancy frequency increases and so, therefore,
do the g-mode frequencies.
Where modes of the two classes approach each other,
the frequencies undergo an {\it avoided crossing\/}, never quite meeting
\citep[e.g.,][]{Osaki1975, Aizenm1977, Christ1980b};
indeed, this behaviour is characteristic of
eigenvalue problems with several classes of solutions
\citep[e.g.,][]{vonNeu1929}.
At the point where the modes are closest they have a mixed character,
with substantial amplitudes both in the g-mode and the p-mode propagation
regions (see Fig.~\ref{fig:charfreq}).
If such modes can be observed, they would provide information about
the core of the star.

For stochastically excited modes it follows from \Eq{eq:stochamp1} that the
amplitudes are inversely related to the mode inertias 
defined in \Eq{eq:inertia}, normalized by the surface amplitude.
This is typically large for modes partly trapped
under the buoyancy frequency in the deep interior.  However, as is
evident from Fig.~\ref{fig:charfreq} the evanescent region between the
g-mode and p-mode trapping regions is quite thin for $l = 1$ modes in
the $\eta$~Boo model.  Consequently, the mode inertia of the mixed
modes only exceeds that of the purely acoustic modes by less than a
factor 4 and the modes are likely to be excited stochastically to
observable amplitudes.  For $l > 1$, on the other hand, the evanescent
region is broader; consequently the interaction in the avoided
crossing takes place in a smaller interval of age and frequency and
the amplitudes of the mixed modes are therefore likely smaller.

\begin{figure}
\centering
  \includegraphics[width=0.75\textwidth]{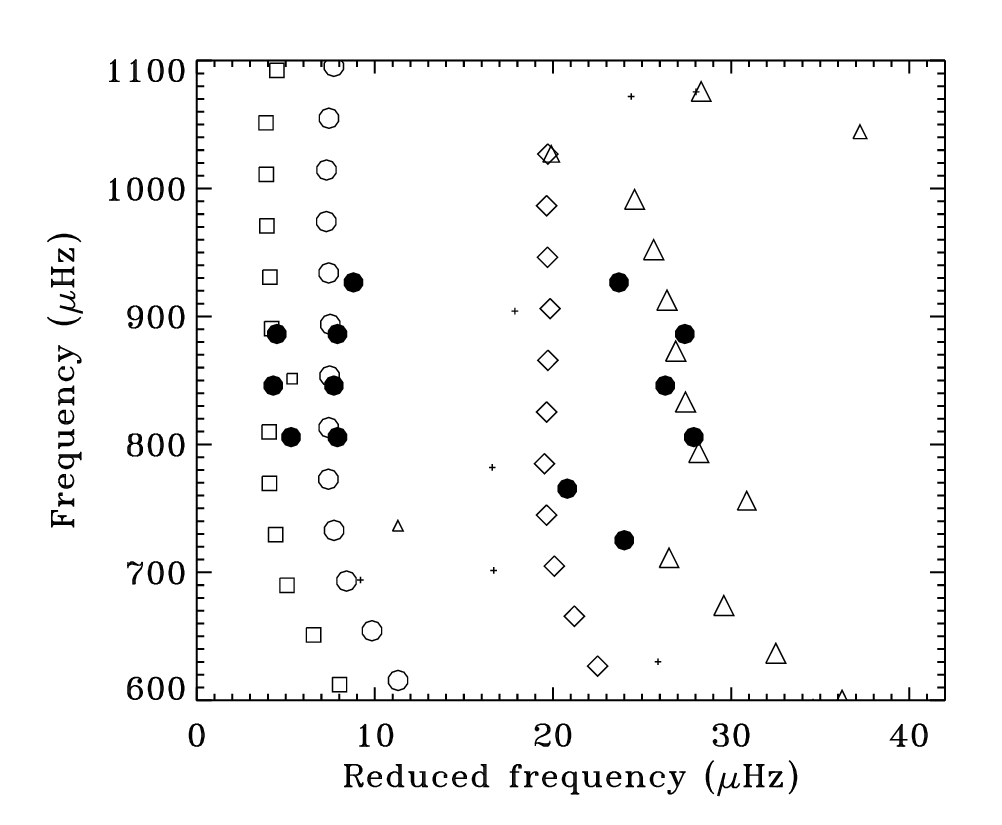}
\caption{
Echelle diagram for $\eta$~Boo,
with a frequency separation of $\langle \Delta \nu \rangle = 40.3\muHz$
and a reference frequency of $\nu_0 = 846 \muHz$.
The filled circles show observed frequencies from \cite{Kjelds1995a}.
The open symbols show computed frequencies for a model of 
the star \citep{Christ1995};
to compensate for the near-surface effects the computed frequencies
have been decreased by $10 \muHz$, to achieve a reasonable average match to
the observations.
Circles are used for modes with
$l = 0$, triangles for $l = 1$, squares for $l = 2$ and
diamonds for $l = 3$.
The size of the symbols indicates the expected amplitude of the modes,
relative to a radial mode of the same frequency, and assuming that all
modes are excited to the same energy (cf.\ Eqs~(\ref{eq:ekin})
and (\ref{eq:stochamp1}));
symbols that would otherwise be too small have been replaced by crosses.
}
\label{fig:eboechelle}       
\end{figure}

It is evident that frequencies of modes undergoing avoided crossing
do not satisfy the p-mode asymptotic relation, \Eq{eq:pasymp}.
Indeed, as shown in Fig.~\ref{fig:ebofreq} the modes with g-mode
behaviour effectively add extra modes between the regular pattern
of acoustic mode.
This is very visible in the echelle diagram shown in Fig.~\ref{fig:eboechelle}
\citep{Christ1995}.
For modes of degree $l = 2$ and 3 the presence of avoided crossings
barely affects the behaviour of the frequencies;
for $l = 1$, on the other hand, there are obvious departures from the 
regular behaviour.
It is interesting that the observed frequencies of \cite{Kjelds1995a}
show a qualitatively similar behaviour,
indicating that the effect of the g-mode propagation region can be
observed in this star.
This has largely been confirmed by subsequent observations and analyses
\citep{Guenth1996, Kjelds2003, Guenth2004b, Carrie2005}.
It was noted by \cite{DiMaur2004} that an alternative would be to model 
$\eta$~Boo as being in the phase of central hydrogen burning, but with
some convective-core overshoot.
Such a model does not show effects of mode mixing in the relevant
frequency band and could therefore
be excluded if the presence of avoided crossings could be definitely
established in the observed frequencies.

Less evolved stars show mode mixing at lower frequencies.
Thus such effects are likely in stars pulsating due to the heat-engine
mechanism, where the frequency range of the unstable modes typically
covers the low-order acoustic and gravity modes.
An example may be the $\beta$~Cephei star $\nu$~Eri,
where modelling by 
\citet{Pamyat2004}, based on frequencies and mode identification obtained by 
\citet{Handle2004} and \cite{DeRidd2004},
indicated an avoided crossing between the lowest-order dipolar p and g modes.
Also, the implied effect of an avoided crossing allowed \citet{Mazumd2006}
to place strong constraints on models of the $\beta$~Cephei star
$\beta$~CMa, based on only frequencies of two identified modes.

With further evolution the buoyancy frequency in the deep interior of
the star increases dramatically and the density of g modes becomes
extremely large, in the observationally relevant frequency interval.
An example of such a star is the G7 giant $\xi$ Hya,
where \cite{Frands2002} found evidence for solar-like oscillations.
The oscillations of a model of this star were discussed in some
detail by \cite{Christ2004}.
For the nonradial modes there are still cases where the frequency
resonates with the outer, acoustic, propagation region, resulting
in modes of predominantly acoustic nature and hence with a mode inertia
similar to that of a radial mode of similar frequency.
These modes might in principle be excited to an observable level
through stochastic excitation.
However, it was noted by \cite{Dziemb2001} that the very high order of 
the modes in the g-mode propagating region would lead to strong radiative
damping, despite the rather small amplitudes of the modes in this region,
possibly to an extent that the non-radial modes would be undetectable.
In fact, 
\Eq{eq:workint} shows that the integrand in the work integral depends
on the divergence of the flux perturbation;
assuming radiative transfer the flux perturbation contains the
gradient of the temperature perturbation (cf.\ \Eq{eq:radlum}),
which, relating it to the density perturbation through the approximate
adiabatic relation and using the equation of continuity,
is itself given by a derivative of the displacement.
Thus the integrand of the work integral contains a third derivative of
the displacement which, for high-order modes, is obviously potentially
large, even for a relatively small displacement in the g-mode propagating
region.
The interpretation of the observations by \cite{Frands2002} is
consistent with only radial modes being detected.
Nonetheless \citet{Hekker2006}
found strong evidence that the dominant modes in $\epsilon$~Oph
had $l = 2$.
This apparent conflict between theory and observations has still
to be resolved.

\subsubsection{Signatures of sharp features}
\label{sharpfeatures}

Abrupt transitions in a star's internal structure cause small departures from the regular p-mode asymptotic spacing discussed earlier.  
Abrupt changes in stratification (on scales smaller than or comparable with the local wavelength) 
-- e.g., the edge of a convective region or an ionization zone -- 
add to the frequencies (considered say as a function of $n$) a periodic component
\begin{equation}
\widetilde{\delta\nu} \sim A \cos \left[2(2\pi\nu \tau_d + \phi)\right]\;,
\end{equation}
where $\tau_d$ is the location of the feature in acoustic depth,
$\phi$ is a surface phase, and $A$ is an amplitude.
The contribution of the signature from the Sun's second helium ionization zone to the large separation, for instance, 
is large enough that it should be explicitly taken into account when calibrating the value of the large separation
\citep{mcdt02}.

A more-studied signature of a sharp feature is that from the base of a convection envelope such as the Sun's.
\cite{mcdt02} explored the behaviour of this signature as a function of mass and age of the star, using frequencies
of stellar models. In particular, they computed the amplitude $A$ of the signal (at a fiducial frequency) and 
inferred the acoustic depth $\tau_d$ of the
base of convection zone, as functions of stellar mass and central hydrogen abundance. 
If the mass and age are already known, departures from this behaviour can be indicative of, 
e.g., convective overshooting \citep[see, e.g.,][]{mt00}.

\begin{figure}
\centering
  \includegraphics[width=0.75\textwidth]{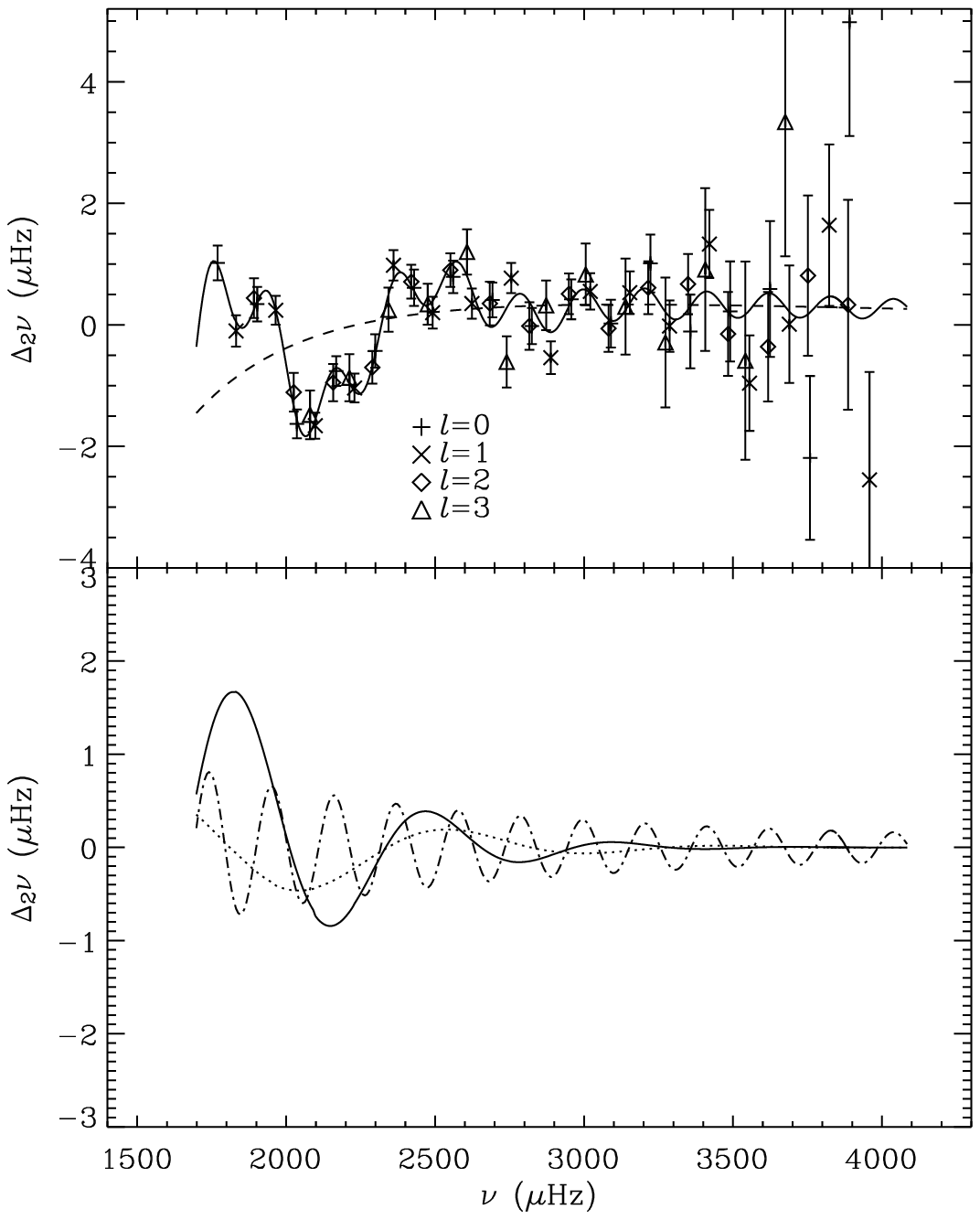}
\caption{
In the upper panel the symbols show second differences
$\Delta_2\nu_{nl} = \nu_{n+1 \, l} - 2 \nu_{n\,l} + \nu_{n-1 \, l}$
for the values of $l$ indicated in the figure, 
in simulated data for a 1\,M$_\odot$ model of age 5.54\,Gyr.
The simulation was based on two periods of 4-months observations,
separated by one year,
with the SONG network \citep[see][]{Grunda2007}.
The solid curve shows a fit to $\Delta_2\nu$ (where the subscripts $n\, l$ have been dropped for simplicity) based on the analysis
by \citet{Houdek2007}.
The dashed curve is a smooth contribution,
modelled as a third-order polynomial in $\nu^{-1}$, which
represents possible near-surface effects (cf.\ Fig.~\ref{fig:soldif})
as well as likely departures at low frequency from the asymptotic expressions
underlying the fit.
The lower panel identifies the remaining individual contributions to the fit.
The dotted and solid curves show contributions from the first and
second helium ionization zones and the dot-dashed curve is the
contribution from the base of the convective envelope.
}
\label{fig:hgsong}       
\end{figure}

Various authors have looked at the use of the signature of the second
helium ionization zone on stellar oscillation frequencies
\citep{mt98, mt05, Perez1998, Miglio2003, basuetal04, Houdek2004}.
Basu et al.\ found that the fractional abundance of helium by mass, $Y$,
can be determined to an accuracy of $0.01$ to $0.02$ 
in most cases with realistic noise estimates. 
For masses greater than 1.4\,M$_\odot$, however, they found 
that the signal fit becomes unreliable.
\citet{Houdek2007} carried out a careful combined analysis of the effect of the helium I and helium II
ionization zones on the asteroseismic signatures and obtained a substantially
improved fit to the computed behaviour 
\citep[see also][]{Houdek2006b}.
As an example of the application of their technique, 
Fig.~\ref{fig:hgsong} shows the resulting fit to simulated data for 
a solar-like star.

\cite{Mazumd2005} proposed a new model calibration tool, 
using the large separation and the inferred acoustic depth of (say) the convective envelope. 
Using this with classical constraints, he obtained impressive accuracy 
in identifying the model properties in a 
hare-and-hounds exercise aimed at testing data-analysis techniques for
the CoRoT mission \citep[see][]{Appour2006}.

The effect of sharp transitions on stellar oscillations has also been studied in the context of roAp stars and white dwarfs. In the former case, \citet{vauclair04} have analysed frequency data for the roAp star HD~60435, and have found evidence for a signature of a steep helium gradient, as expected from results based on models of these stars including helium settling in the stellar envelope. 
Concerning white dwarfs, the layering resulting from settling in the
strong gravitational field has substantial influence on the g-mode spectrum
and hence on the asteroseismic diagnostics of white-dwarf properties
\citep[e.g.,][]{Bradle1993, Bradle1994}.
An interesting analysis of the effects of sharp features,
with application to white dwarfs, was presented by \citet{Montgo2003}.

More recently, the effect on the oscillation frequencies of sharp
transitions associated with the edge of convective cores has also been
considered theoretically for stars significantly more massive than the
Sun \citep{roxburgh01}, and for main-sequence solar-like pulsators
\citep{cunha07}. Moreover, attempts to identify the signatures of convective cores in solar-like
pulsators from simulated data have  been carried out
by \citet{Mazumd2006}. In a main-sequence solar-like pulsator the edge of
the convective core is located close to the inner turning point of low
degree acoustic modes. Consequently, unlike all other cases discussed
above, in this case the signal added by the sharp transition
associated with the edge of the core is not periodic.  Nevertheless,
based on a theoretical analysis of the signal and on simulated data, \citet{cunha07} have shown that such signal
should be detected when the data for solar-like pulsators reaches the
level of precision expected from space-based dedicated instruments,
such as CoRoT, and that the detection of the expected signal in real
data of solar-like pulsators will provide unprecedented information
about the cores of these stars.

\subsubsection{Direct Fitting}
\label{dirfit}

Clearly, even the most basic seismic signatures, such as the large and small
separations discussed above, provide additional constraints to
traditional stellar modelling.  The first approach to asteroseismic inference
is by direct fitting, or forward modelling. Such an approach
consists of fitting a set of measured observables for a given star by
running forward in time a stellar structure and evolution code which
takes as input parameters the (unknown) basic stellar parameters we
are trying to determine.  Then at each point in time (age) the code
produces as output the model observables, which can be compared with
the measured ones. Furthermore, the code also produces detailed
numerical information on the internal structure of the star.

For a single star, the ``classical'', non-asteroseismic observables are the
effective temperature $T_{\rm eff}$, the logarithm of surface gravity $\log g$,
and 
the surface metallicity $[{\rm M/H}]_{\rm surf}$, which in PopI stars can be identified with the logarithmic surface abundance of iron relative to that in the Sun $[{\rm
Fe/H}]_{\rm surf}$
\footnote{More precisely defined as $\log(N_{\rm Fe}/N_{\rm H}) - \log(N_{\rm
Fe}/N_{\rm H})_\odot$, where $N_{\rm Fe}$ and $N_{\rm H}$ are the number
densities of iron and hydrogen in the stellar atmosphere.}.
Moreover, whenever the distance is available, also the luminosity $L$ can be derived directly from observations and, so far for only a limited
number of stars, also the interferometrically determined radius, $R$. (Notice
that $T_{\rm eff}$, $L$, and $R$ are interdependent, even if they are three
different observables.) The issue of how to include the observed oscillation
frequencies as observables into the fits is still under study. Before the advent
of spectrographs such as HARPS and UVES, the quantity that could most
confidently be determined was the large frequency separation, and perhaps also
the small frequency separation. As a result, most of the modelling work done to
date used those quantities as the asteroseismic observables for the
fits. However, it is now possible to resolve individual oscillation modes, which
contain additional information that is lost if only the separations are included
in the fits, as pointed out by \citet[][and references therein]{Metcal2005} from
his experience in modelling pulsating white dwarfs.  A possible way to include
the information from the individual modes, suggested by \cite{Beddin2004}, is to
carry out fits to the frequency ridges for modes with the same degree, $l$. From
the results of such fits one may then define the following set of asteroseismic
observables:
$\{\nu^{\rm (ref)}_0, \langle \Delta\nu \rangle, r^{\rm (ref)}_{02}\}$,
where the superscript `${\rm ref}$' indicates suitable reference values, e.g., evaluated
at peak power.
These three (or other similar) parameters may then be
used as the asteroseismic observables which, together with the non-asteroseismic
observables, must be successfully matched by a stellar model.

In order to obtain a model for the internal structure of a star at a
given age, a stellar structure and evolution code will take as input
parameters the stellar mass, $M$, and the initial abundances of
hydrogen and helium for the star, $X_0$ and $Y_0$, respectively. It
starts by calculating the zero-age model for the internal structure of
that star, and then evolves it in time in steps until the required
age. In addition to those three parameters, such codes also allow some
choices regarding details of the physics in the stellar interior. As
mentioned earlier, these are mostly related to the treatment
of convection, diffusion and settling of heavy elements, the equation
of state, and opacities, among others. In particular, 
a parameter is introduced in the formulation of convection.
The mixing-length treatment (MLT)
describes the characteristic length of turbulence as scaling directly with the local
pressure scale height: the scaling factor is known as the mixing
length parameter, $\alpha_{\rm ML}$,
which is essentially unconstrained and
left as a free parameter in the models. When convection is dealt with
in the approach of Canuto and Mazzitelli \citep{Canutoetal1996} a free
parameter is also introduced. The latter formulation has produced only
rather small improvements in the results with respect to the MLT, as
shown by \cite{Miglio05}. The extent of mixing at the
borders of convectively unstable regions is also not fully understood.
Overshoot is dealt with in stellar codes by introducing another free
parameter describing its extent by scaling with the local pressure
scale height. Since the observational data available do not yet
allow for testing this parameter 
in solar-like pulsators, we will not consider it in what
follows. That will hopefully change in the near future, when better
data, particularly of more evolved solar-like pulsators become
available.

The problem of direct fitting is thus to fit a set of observables,
$\{y_i\} = \{T_{\rm eff}, L, [{\rm Fe/H}]_{\rm surf}, R,
\nu^{\rm (ref)}_0, \langle \Delta\nu \rangle, r^{\rm (ref)}_{02}\}$,
\footnote{Here we have assumed that the parallax, and hence the luminosity,
have been determined.
If this is not the case $\log g$ can be included as observable instead.}
with associated uncertainties $\{\sigma_i\}$, with a model having as
input parameters $\{a_j\} = \{M, X_0, Y_0, \alpha_{\rm ML}, age\}$,
where the input
parameters are what we are trying to determine;
additional parameters may be needed to model the effects of the
near-surface problems, particularly in the case of solar-like
oscillations (cf.\ Fig.~\ref{fig:soldif}). 
The best way to
approach this problem is to use an objective, automated fitting
procedure that is free to search for the best solution within the
parameter space defined by physical constraints to the input
parameters (for example, the age has to be positive and smaller than
the age of the Universe). One such fitting procedure is the
Levenberg-Marquardt method \citep[e.g.,][]{BR2003}, which
has been successfully used in nonlinear multi-parameter fitting
problems in different  
contexts but with similar needs \citep[e.g.,][]{teixeira1998,Miglio05}. 
This is an iterative least-squares minimization
method which locates a $\chi^2$ minimum by combining gradient search
when far from the minimum, with expansion of the $\chi^2$ surface near
the minimum. The fit is carried out by minimizing 
\begin{equation}
\chi^2 = \sum\limits_{i}
[(y_i^{\rm (obs)}-y_i^{\rm (mod)})/\sigma_i]^2
\end{equation}
with respect to parameters $\{a_j\}$,
where $y_i^{\rm (obs)}$ are the observed values of the observables and
$y_i^{\rm (mod)}$ are the corresponding computed model values for the given
set of parameters.   
This procedure produces fast convergence, and is well suited to make
use of distributed computing for the 
calculation of the derivatives for the gradient search. Unfortunately
and like most methods, it cannot guarantee convergence to the global
minimum, and it may thus be advisable to make several searches with
different starting values with respect to the input parameters. 
This can be carried out in a systematic way by utilizing various forms of
non-linear optimization techniques, such as genetic algorithms
\citep[e.g.,][]{Metcal2000, Metcal2003} or Markov Chain Monte Carlo 
(MCMC) techniques.%
\footnote{An application of MCMC to frequency determination for
solar-like oscillations was discussed by \citet{Brewer2007}.}

For a binary system the fitting problem is similar to the
one described above for a single star, only it involves a larger
number of observables and input parameters and some extra
constraints. 
In the case of an ``ideal'' binary system 
\citep[cf. $\alpha$~Cen AB,][]{Miglio05},
the masses of the individual components (A and B) are well determined, which
provides us with two more observables. The
non-asteroseismic observables involved in the fit are then $\{M_{\rm A},
T_{\rm eff,A}, L_{\rm A}, [{\rm Fe/H}]_{\rm A}, R_{\rm A}, M_{\rm B},
T_{\rm eff,B}, L_{\rm B}, [{\rm Fe/H}]_{\rm B}, R_{\rm B}\}$. If both
components have reliably determined individual oscillation
frequencies, it is possible to define asteroseismic observables for
each component, as explained above for a single star:
$\{\nu^{\rm (ref)}_{0,{\rm A}}, \langle \Delta\nu \rangle_{\rm A}, r^{\rm (ref)}_{02, {\rm A}}, 
\nu^{\rm (ref)}_{0,{\rm B}}, \langle \Delta\nu \rangle_{\rm B}, r^{\rm (ref)}_{02, {\rm B}}\}$.
One can then
include all these observables simultaneously in the fit, or only a
subset of them. As for the model, it is assumed that both components
of the binary have the same age and initial composition, while the
mixing lengths are generally allowed to differ. With these constraints,
the input model
parameters are then $\{M_{\rm A,mod}, M_{\rm B,mod}, X_0, Y_0,
\alpha_{\rm ML,A}, \alpha_{\rm ML,B}, age\}$, 
where $M_{\rm A,mod}$ and
$M_{\rm B,mod}$ are variable model masses and must be distinguished
from the observed values, $M_{\rm A}$ and $M_{\rm B}$. Finally, the same
fitting procedure as described above performs the least-squares
minimization and produces models simultaneously for both components of
the binary.

Rather than making a combined fit to the `classical' observables and the
oscillation frequencies, \cite{Guenth2004a} and \cite{Metcal2005} recommended
fits to the frequencies alone, possibly supplemented by a separate fit to
the classical variables $\{T_{\rm eff}, L, [{\rm Fe/H}]_{\rm surf}\}$.
Any discrepancy between the results of such separate fits would obviously
indicate inconsistencies in the modelling (or the data).
This technique was applied by \cite{Guenth2004b} to the analysis of
observations of $\eta$~Boo.
It should be pointed out, however, that inconsistencies would presumably
also be apparent in the residuals from a combined fit, as discussed above.
Thus an evaluation of the relative merits of these two different approaches
probably requires further investigations, including hare-and-hounds exercises.

Direct fitting is a relatively simple way to find a good first solution to
the problem of stellar model fitting.  Applications of this method to stars for which both asteroseismic and interferometric data are available are discussed in Section~\ref{sec3}. Direct fitting is also a crucial step
towards providing a close enough starting point for the inverse
methods, which will give us access to the actual details of the
stellar interior.

\subsubsection{Inverse approach}
\label{sec:invers}

Solving the inverse problem is predicated on solving, at least in some approximation, 
the forward problem. Here the forward problem is to compute the global resonant frequencies 
given a stellar model, since the observables to be inverted are global resonant frequencies.
The simplest approach to the inverse problem would be to compute frequencies for different
models and find which match the observables most closely, in some sense. This is the 
direct fitting method described in Section~\ref{dirfit}. At its crudest, this could involve a 
straightforward match to the frequencies themselves, but this would not distinguish between
the myriad possible contributions to any frequency mismatch. Better, as described above, would
be a match to certain signatures computable from the frequencies, such as the large and 
small frequency separations.
   
More sophisticated inverse methods allow one to ``step outside'' a given space of models. Most of the 
inverse techniques are linear, though the forward problem is inherently nonlinear. Thus it
is certainly advantageous, and probably necessary, to have a starting model that is ``near''
the truth. Herein lies one difference between asteroseismology and helioseismology. In
helioseismology, we have models of the Sun that are already very close to the truth; whereas
the inherent uncertainties in a star's global parameters may mean that even finding a 
good starting model for a linear inversion becomes part of the inverse process. For some 
general considerations, see \cite{gough85b, mjtjcd02}.

Linear inversion methods used in helioseismology -- and just starting to 
be used in asteroseismology -- fall mainly into two categories: 
least-squares fitting of the observational data, or optimally localized
averages methods. For the sake of illustration, we suppose we have
data $d_i$ ($i=1,\ldots,M$) which are related linearly to some 
unknown property $\Omega ({\bf r})$ of the interior of the star, where ${\bf r}$ is
in general a 3D position vector, according to 
\begin{equation}
d_i\ =\ \int_V K_i({\bf r})\Omega({\bf r})\,{\rm d}{\bf r}\ +\ \epsilon_i \; ,
\label{eq:mjt1}
\end{equation}
where the integral is over the whole interior volume $V$ of the 
star.
The {\it kernels} $K_i({\bf r})$ are presumed to be known functions 
describing the sensitivity of the data to the unknown function
$\Omega$. The $\epsilon_i$ are data noise, whose statistical 
properties are assumed to be known: for simplicity, we suppose
they are independent, normally distributed with zero mean and 
standard deviation $\sigma_i$.

The method of Optimally Localized Averages (OLA),
or the Backus-Gilbert method,
is based on the work of \cite{backus68,backus70}.
To estimate $\Omega$ at some location ${\bf r}_0$, one seeks coefficients 
$c_i({\bf r}_0)$ so as to construct an {\it averaging kernel}
\begin{equation}
\CK ({\bf r}_0,{\bf r}) \ = \ \sum_{i=1}^M c_{i}({\bf r}_0) K_i ({\bf r})
\end{equation}
that is peaked around ${\bf r} = {\bf r}_0$ and is small elsewhere, and
has unit integral
\begin{equation}
\int_V \CK({\bf r}_0,{\bf r}){\rm d}{\bf r}\ =\ 1\;.
\label{eq:mjt3}
\end{equation}
Taking the same linear combination both sides of Eq.~(\ref{eq:mjt1}) gives
\begin{equation}
\sum_{i} c_{i}({\bf r}_0 ) d_i \ = \
\int_V \CK ({\bf r}_0,{\bf r}) \Omega ({\bf r}) {\rm d}{\bf r} \;+\;\sum_i c_i({\bf r}_0)\epsilon_i\;.
\label{eq:mjt4}
\end{equation}

If one succeeds in localizing $\CK ({\bf r}_0,{\bf r})$ about ${\bf r}={\bf r}_0$, then one can
regard the left-hand side of Eq.~(\ref{eq:mjt4}) as an estimate 
${\bar\Omega}({\bf r}_0)$ 
of the unknown function $\Omega$ at
${\bf r}={\bf r}_0$. For this estimate to be useful, the choice of coefficients must be
moderated so that the error in the estimate propagated from the errors in
the data is not too large: thus there is a trade-off between localizing the
averaging kernel and reducing the effect of data errors on the solution. 

One way of determining the coefficients $c_i({\bf r}_0)$
\citep{gough85b} is to choose them to minimize
\begin{equation}
\int_V J({\bf r}_0,{\bf r}) \CK({\bf r}_0,{\bf r})^2 {\rm d}{\bf r}\;+\;
\mu \sum_{i} \sigma_i^2 c_i({\bf r}_0)^2 \;.
\end{equation}
Here 
$\mu_0$ is a parameter whose value is chosen
to balance between the conflicting aims of minimizing the first term, 
and hence localizing the averaging kernel, and minimizing the
second term, and hence keeping the effect of data errors on the solution
small. The function $J$ is chosen to penalize the averaging kernel for 
being non-zero far from the target position ${\bf r} = {\bf r}_0$, and should generally 
increase as $|{\bf r}-{\bf r}_0|$ increases.
The normalization (\ref{eq:mjt3}) is imposed as an
exact constraint.

An alternative formulation of OLA is
Subtractive OLA (SOLA) \citep{pijper92, pijper94}. In the SOLA method
the coefficients $c_i({\bf r}_0)$ are chosen to minimize
\begin{equation}
\int_V \bigl({\cal T}({\bf r}_0,{\bf r})\;-\;\CK({\bf r}_0,{\bf r})\bigr)^2 {\rm d}{\bf r}\;+\;
\mu \sum_{i,j} E_{ij}c_i({\bf r}_0) c_j({\bf r}_0)\;,
\end{equation}
where ${\cal T}({\bf r}_0,{\bf r})$ is a chosen target form that the averaging kernel is 
to resemble  and $ E_{ij}$ is the error variance-covariance matrix of the observed frequencies. A reasonable choice of target function is a Gaussian in the 
radial and -- if appropriate -- the latitudinal directions.
In addition to being computationally more efficient, 
it has been found in a number of helioseismic applications that 
it is easier to control the localization of the averaging kernel 
with the SOLA formulation of OLA. 

Another widely used class of inversion methods is called 
Regularized Least Squares (RLS),
also known as least squares with Tikhonov regularization 
\citep{tikhon77}.  
The unknown function $\Omega({\bf r})$ is first approximated by a linear
combination of chosen base functions $\phi_j$ ($j = 1,\cdots,N$):
\begin{equation}
\Omega({\bf r})\ =\ \sum_j x_j\phi_j\;.
\end{equation}
Often, each $\phi_j$ is localized to a small region in ${\bf r}$. 

It follows that 
Eq.~(\ref{eq:mjt1}) becomes a matrix equation,
\begin{equation}
{\bf A}{\bf x}\ =\ {\bf b}\;,
\end{equation}
where the elements of matrix ${\bf A}$ are given by
\begin{equation}
A_{ij}\ =\ \int_V K_i({\bf r})\phi_j({\bf r})\,{\rm d}{\bf r}\;,
\end{equation}
${\bf x}$ is the vector of the $N$ expansion coefficients $x_i$, and 
${\bf b}$ is the vector of the $M$ data $d_i$. 
In the RLS method, the expansion
coefficients are determined by minimizing
\begin{equation}
\| {\bf A}{\bf x} - {\bf b}\|_2^2\;+\;\lambda^2\|{\bf L}{\bf x}\|_2^2\;,
\end{equation}
where $\|\cdots\|_2$ denotes the 2-norm of a vector.  
Typically ${\bf L}$ is an approximation to some derivative 
operator, and $\lambda$ is a parameter whose value is chosen to determine
the relative importance of minimizing the first term (hence fitting the
data) and minimizing the second term (and hence keeping the solution
``smooth'' in some sense). Most commonly, ${\bf L}$ has been chosen to 
be a second-derivative operator in helioseismic applications.


As discussed in Section \ref{nonsphere}
rotation affects the star's frequencies and hence is amenable
to seismic inference.
Inversion for the Sun's internal rotation is well established: see for example 
\cite{schou98} and \cite{thomps03}. 
Inversion for stellar rotation is, in principle, also possible. 
The Sun is a slow
rotator and hence rotation can be studied with 
just first-order perturbation theory (cf.\ \Eq{eq:rotsplit}).
In this case, the relation between the observable rotational splittings and
the unknown angular velocity is essentially of the form assumed in
\Eq{eq:mjt1} and the techniques discussed above can be directly applied.
For more rapidly
rotating stars, higher-order perturbation theory, or nonperturbative methods,
are necessary (see Section~\ref{nonsphere}).

Inversions of artificial data, with observational characteristics as expected for CoRoT targets,
indicate that it may be possible successfully to invert for rotation in parts of the star 
\citep{lochard05}. These authors applied a SOLA inversion technique to artificial data
computed for a stellar model of 1.55 solar masses, sufficiently evolved to possess some 
mixed modes in the frequency range expected to be stochastically excited. Their mode
set contained 50 $l=1,2$ modes including 3 mixed modes.
With this set, and for realistic
noise, they were able to obtain sufficient radial resolution to determine the rotational 
profile in the core with reasonable accuracy between fractional radii 0.1 and 0.3, 
as well as a 
rather broad average of the rotation in the outer 40 per cent of the star by radius. Thus
for suitable stars one may obtain some indication of rotation both in the core and in the envelope. 
Suitable stars are those where some mixed modes can be observed as well as p modes, and where
the rotation is not so slow that the fractional error on the rotational splitting measurements
is too big for the inversion results to be significant. 

\begin{figure}
\centering
  \includegraphics[width=0.90\textwidth]{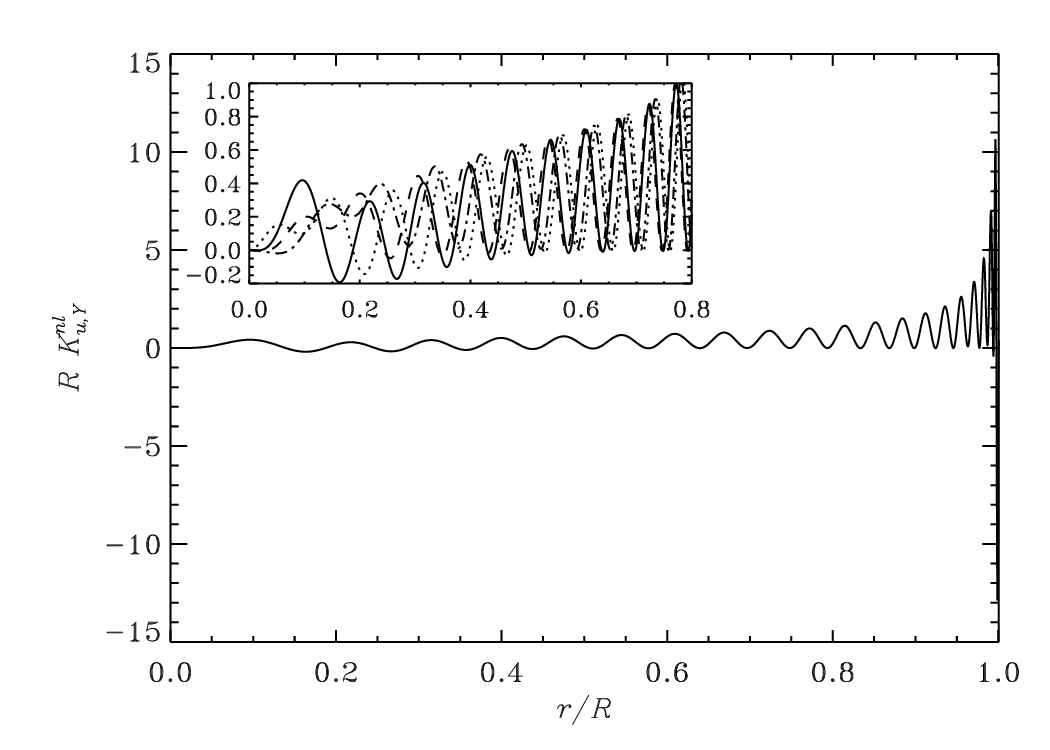}
\caption{
Structure kernels $K_{u,Y}^{nl}$ relating the change in $u = p/\rho$ to
frequency changes, at fixed helium abundance $Y$, for modes in a model
of the present Sun with frequencies near $3000 \muHz$;
the abscissa is fractional distance to the centre.
The main panel shows a radial mode.
The insert shows the behaviour in the bulk of the interior for modes
with $l = 0$ (solid curve), $l = 1$ (dotted curve),
$l = 2$ (dashed curve), and $l = 3$ (dot-dashed curve).
}
\label{fig:strucker}       
\end{figure}


The above methods can be extended to where there is more than one
unknown function.
In the case of inferring the radial hydrostatic structure an additional
complication is the strongly nonlinear dependence of the observed oscillation
frequencies on structure.
As proposed by \citet{Gough1978}
this is dealt with by linearization around a known reference model,
e.g., characterized by sound speed $c_0(x)$ and density $\rho_0(x)$,
where $x$ is distance to the centre, in units of the surface radius,
and using the variational property of linear adiabatic oscillations
\citep[e.g.,][]{Chandr1964}.
Assuming that the differences $\delta c/c = (c(x) - c_0(x))/c_0(x)$
and $\delta \rho/\rho = (\rho(x) - \rho_0(x))/\rho_0(x)$ are 
sufficiently small, the resulting frequency differences
$\delta \omega_{nl}/\omega_{nl} = 
(\omega_{nl} - \omega_{nl, 0})/\omega_{nl,0}$,
$\omega_{nl,0}$ being the frequencies of the reference model,
can be approximated by
\begin{equation}
{\delta \omega_{nl} \over \omega_{nl,0}} =
\int_0^R \left( K^{nl}_{c, \rho} {\delta c \over c_0} 
+ K^{nl}_{\rho, c} {\delta \rho \over \rho_0} \right) \dd x \; .
\label{eq:strucinv}
\end{equation}
(To this expression must be added a term describing the effects of 
the near-surface errors in the model -- cf.\ the discussion in connection
with Fig.~\ref{fig:soldif}.)
Thus the inverse problem is reduced to a slight generalization of the problem
discussed above and can be treated with similar techniques.
Instead of $(c, \rho)$ other equivalent pairs of structure variables,
involving the adiabatic compressibility $\Gamma_1$, can be used;
transformations between these different sets of variables are carried out
using the constraint of hydrostatic equilibrium.
Further transformations are possible if other aspects of the physics of
stellar interiors, such as the equation of state, are assumed to be known
\citep[e.g.,][]{Gough1990b}.
A detailed discussion of inversion techniques for this problem,
with special emphasis on the helioseismic problem, was provided by
\citet{Rabell1999}.

\begin{figure}
\centering
  \includegraphics[width=0.75\textwidth]{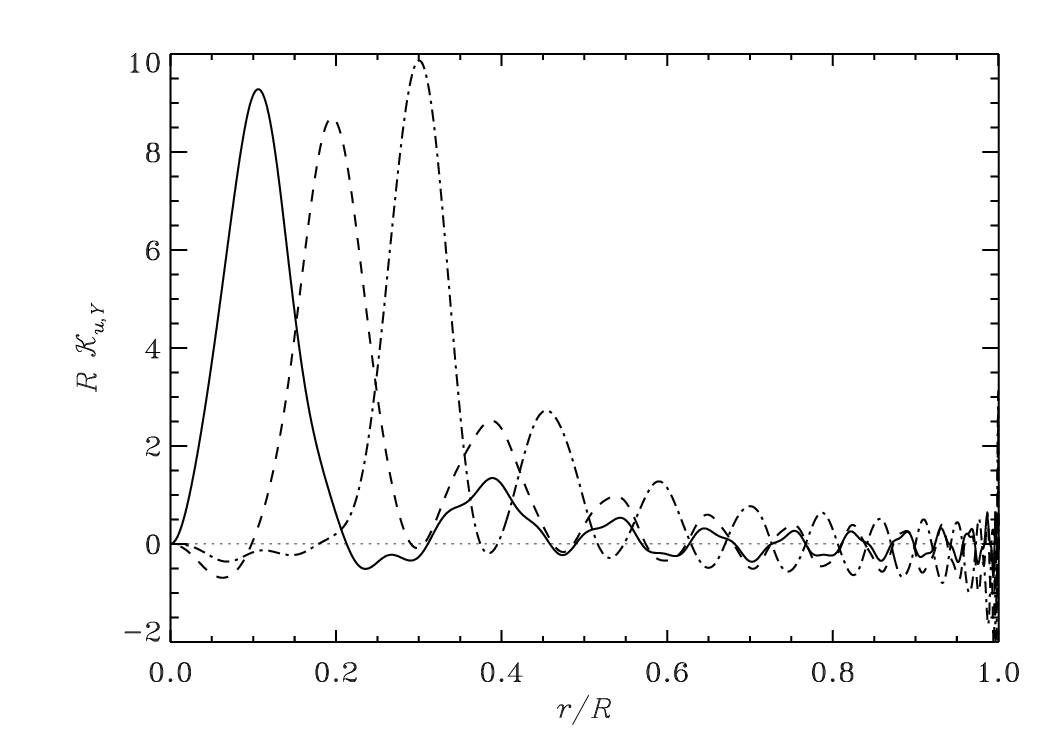}
\caption{
Averaging kernels for inversion to infer differences in $u$,
defined such that the inferred $\delta u /u$ at $r = r_0$ is
approximately given by $\int \CK_{u,Y}(r_0, r) (\delta u/u)(r) \dd r$.
The inversion is
based on using a model of the present Sun as reference, and including
representative modes with degree $l$ from 0 to 3.
The inversions have target locations $r_0 = 0.1 R$ (solid curve),
$0.2 R$ (dashed curve) and $0.3 R$ (dot-dashed curve).
}
\label{fig:strucavker}       
\end{figure}

Inversions for the radial hydrostatic structure are well established 
in helioseismology
\citep[see for example][]{gough96, basu97}.
Such inversions 
have been demonstrated with artificial data for other stars.
For example, \cite{basuetal02}
considered the quantity $u=p/\rho$,  which is
a proxy for the (squared) adiabatic sound speed, 
in combination with the helium abundance $Y$ and assuming the equation of
state to be known, so that $\Gamma_1$ is known in terms of 
$(p, \rho, Y)$.
Examples of the corresponding kernels $K^{nl}_{u,Y}$, for acoustic
modes characteristic of solar-like oscillations, are shown 
in Fig.~\ref{fig:strucker}.
In the bulk of the model the kernels have a similar behaviour, 
essentially reflecting the asymptotic behaviour of the acoustic modes.
However, in the core the properties of the kernels depend on degree,
in part reflecting the variation of the position of the lower turning point
(cf.\ \Eq{eq:rt}).
This richness of behaviour allows inferences to be made beyond the simple
asymptotic properties,
as was also indicated by the non-asymptotic behaviour of the small
frequency separations in Fig.~\ref{fig:smallsep}.
Thus, Basu {\etal}\ demonstrated that corrections to $u$
can be well determined in the core of 
a star like the Sun -- though not elsewhere -- with only low-degree modes.
This is illustrated by the relatively well localized averaging kernels
shown in Fig.~\ref{fig:strucavker},
indicating that some resolution is possible of the correction to $u$.
Further experiments along these lines would evidently be interesting.

\subsection{Current limitations to asteroseismic studies\label{limitations}}

There are a number of observational aspects that may limit the success of asteroseismic studies. While some of these limitations result from intrinsic properties of the pulsations - and would exist even if the data were perfect - others may hopefully be overcome in the future, with the improvement of the seismic and complementary data available for pulsating stars, including data provided by interferometric observations.

\subsubsection{Frequency determination}

Clearly, our ability to extract information from the frequency spectra depends
directly on whether or not the time series from which the spectra are derived
allow us to resolve the individual frequencies and to determine each frequency
with enough precision. As explained in Sections \ref{spectra} and
\ref{groundbased}, to accomplish that we need data sets that are sufficiently
long and which have an appropriate coverage, to avoid complicated window
functions that may prevent us from identifying which frequencies are real. These
can often be obtained for some classical pulsators, traditionally observed
photometrically from small or medium-size telescopes, but not for solar-like
pulsators, which require either observations from space, or with high-resolution
spectrographs available only on very few telescopes.

While the development of means to acquire lengthy and continuous data of
pulsating stars is a technical challenge to which the research community can -
and is already - responding, there are limitations to the precision with which
the individual frequencies may be known in some pulsators, that are related to
the excitation mechanism and cannot be avoided. In classical pulsators the modes
are typically phase coherent over periods of years or decades, allowing
extremely accurate determinations of the frequencies.
{In the case of white uncorrelated noise with average zero and constant
variance $\sigma_N^2$ in time, and when no interference occurs between the
different true frequencies and the noise peaks, the following error
estimates are found, respectively, for the amplitudes, $A$, frequencies, $\nu$, and phases, $\delta$, of modes with infinite lifetime,
\begin{equation}
\displaystyle{\sigma_A=\sqrt{\frac{2}{N}}\ \sigma_{\rm N},\ \ 
  \sigma_\nu=\frac{\sqrt{6}\ \sigma_N}{\pi\sqrt{N}\ A\ T},\ \ 
  \sigma_\delta=\frac{\sigma_N}{\pi\ \sqrt{2N}\ A} },
\label{fout}
\end{equation}
where, as before, $T$ is the total time span of the data and $N$ is the number of data points
\citep{Bloom76, montg99}.  
}
In contrast, for solar-like oscillations the frequency error scales
as $T^{-1/2}$, assuming that the observing time is much longer than
the mode lifetime, which typically is of the order of a few days
for the largest-amplitude modes \citep[e.g.,][]{libbrecht92}.
As an example, the observations of the star FG~Vir
\citep{Breger2005}, intermittently covering the period 1992 -- 2004, have
a relative frequency error as low as around $10^{-8}$
as estimated from Monte-Carlo simulations (M. Breger, private communication);
for comparison, the frequency errors of BiSON full-disk solar data obtained
with nearly continuous observations over a similar period \citep{Chapli2002b},
while still impressively small, exceed $10^{-6}$.

\subsubsection{Mode identification}
\label{modeid}
The observational potential for asteroseismology depends also greatly
on our ability to identify the modes of oscillation in the frequency
spectra - {\it i.e.} the wavenumbers $(l,m,n)$. This is often not an
easy task, particularly in classical pulsators.  As a result of the
mechanism that limits the amplitudes of heat-engine-driven pulsations,
in classical pulsators the distribution of
mode amplitudes is highly irregular (cf.\ Section~\ref{heatengine}).
Moreover, the mechanism
responsible for exciting the modes does not assure that all modes in a
given range of frequency are excited and in most cases the order of the modes excited is such that the  asymptotic regime is not applicable and the frequencies in the spectra are not approximately equally spaced. Consequently, it is often
difficult or impossible to identify the modes on the basis of the
frequency distribution alone; such identification of the observed
modes with those of stellar models is evidently required to make use
of the information contained in the oscillation frequencies.

By reducing the level of observational noise, more modes may become
detectable, improving the chances of identifying the pattern of
frequencies; a striking example is the observations of the $\delta$
Scuti star FG~Vir
\citep[e.g.,][]{Breger2005}.  However, there is a risk that the
spectrum becomes impossibly confused by the high density of
rotationally split modes of moderate degree, possibly excited to
substantial amplitudes and hence visible even in stellar observations
with sufficiently high sensitivity
\citep{Daszyn2005a}.
In fact, for both $\delta$~Scuti stars and $\beta$~Cephei stars the
predicted mode spectra are so dense that the splitting due to
rotation implies merging of the multiplets whenever the rotation
velocity is a significant fraction of the critical velocity (say $>
20\%$), a situation quite often encountered in practice for such 
stars.  This explains why the biggest challenge in seismic
applications to these stars is the identification of the modes of oscillation.
On the other hand, the relatively high amplitudes of the dominant modes make it
realistic to observe these stars with long-term multicolour photometry and
high-resolution spectroscopy, and hence use amplitude ratios and phase
differences as an aid to the mode identification \citep[e.g.,][]{Viskum1998,
Garrid2000, Daszyn2005b, Handle2006}. Additional details on mode identification techniques particularly important in the context of classical pulsators will be presented in Section~\ref{sec3}.

For the $\gamma\,$Doradus stars and the slowly pulsating B stars, the
situation is even worse. Besides the difficulties in detecting a large
number of modes (cf.\ Section~\ref{mspulsators}), brought about by
the range of frequencies in which the oscillations are excited, the
predicted frequency spectra are so dense \citep{Pamyat1999,Dupret2005a} that we have not yet reached the stage of seismic
modelling of individual targets within these classes, although
promising attempts were recently made by \citet{moya05} and
\citet{dupret05b} for $\gamma\,$Doradus stars and by \citet{Aerts2006b} for an SPB discovered with MOST.

In solar-like pulsators the oscillation power,
given by Eqs (\ref{eq:stochamp}) or (\ref{eq:stochamp1}),
varies slowly with
frequency and generally little with degree at a given frequency
(as discussed in Section~\ref{mixmodes} stars with mixed modes are an exception).
For relatively short observing runs the stochastic nature of the
excitation may cause a few modes to be excited to exceptionally high
or exceptionally low amplitudes; for longer runs, or in average power
spectra, one would expect to detect almost all modes in the frequency
range where the modes are excited to substantial amplitudes, with the
cut-off in degree mentioned in Section~\ref{origin}, determined
by the spatial averaging over the stellar disk ($l\le 2$ 
for observations in broadband intensity or
$l\le 3$ for observations in Doppler velocity
\cite[e.g.,][]{Dziemb1977, ChristGough1982}. Also, since generally the modes are acoustic
modes of fairly high order, they approximately satisfy the simple
asymptotic relation given in \Eq{eq:pasymp}.
This frequency distribution may be used to identify the degrees of the
modes, by identifying the small frequency separation
as was indeed done in early analyses of helioseismic data
\citep[e.g.,][]{Christ1980a}.
Also, since most modes in a given frequency range at low degree are
detected, the number of frequencies available to asteroseismic analysis
tends to be large for solar-like pulsators.

\subsubsection{Fundamental parameters of pulsating stars}
Last, but not least, the success of asteroseismic studies depends
greatly on the availability of accurate complementary data for the
pulsating stars.

An excellent review of the status of seismic modelling of classical
pulsators is available in \citet{Michel2006}. In this paper, the author
made it clear that, besides hitherto ignored physical effects and lack
of mode identification, the unavailability of high-precision basic
stellar parameters such as the effective temperature, the gravity and
the metallicity, is a serious obstacle to make progress in the
modelling.  On one hand, the abundances may be available from
spectroscopy, but that represents only the surface value and we have
rather limited knowledge on how the abundances behave as a function of
depth.  On the other hand, typical uncertainties for the effective
temperatures and gravities of OB field dwarfs are 1000\,K and 0.2 dex,
mainly due to the lack of accurate hot calibrators. These numbers
decrease to 200\,K and 0.1 dex for AF-type main sequence stars, but
for those rapid rotation is usually involved adding additional
uncertainty.  When rotation is slow, these stars
become chemically peculiar, and, thus, the determination of their
fundamental parameters is subject to large systematic errors.
The uncertainty in the fundamental parameters is significantly larger
for giant and supergiant stars of OBA spectral type.

The lack of precise fundamental parameters implies that we cannot eliminate a
sufficient number of stellar models when we match the sparse number of
unambiguously identified oscillation modes. Moreover, it also implies that we
might not be able to find a model of the star which is sufficiently close to the
truth to guarantee the success of the inverse procedure.  The delivery of an
accurate independent estimate of either the mass or the radius of the star would
therefore imply a major step forward in seismic modelling of classical
pulsators, and this is where interferometry will hopefully help a great deal in
the near future.

\section{Interferometry\label{sec2}}

\subsection{Principles of astronomical interferometry\label{principles}}
In the current section, we provide a basic introduction into optical and
infrared interferometry.  We limit ourselves to the basic principles  
only and
will focus on the quantities that are of relevance for asteroseismic
applications. More general basic reviews on interferometry are  
available in
\cite{paresce97}, \cite{lawson00}, 
\cite{quirrenbach01a}, \cite{bergeron02}, and \cite{monnier03}.

\subsubsection{Imaging}

The theoretical framework for astronomical imaging is the theory of
diffraction. For an incoherent object, i.e.\ an object for which  
individual
constituents emit lightwaves that are uncorrelated to each other, the  
image
${\cal I}$
produced at the focus of a telescope is the convolution of the {\it  
Object
Intensity Distribution\/} OID by the telescope  
{\it Point
Source Function\/} PSF,
\begin{equation}
{\cal I} (\alpha,\beta)=\left ({\rm OID} \star {\rm PSF} \right)  
(\alpha,\beta),
\end{equation}
where $\alpha$ and $\beta$ are angular coordinates. 
 In the presence of the atmosphere, the PSF encompasses the degradation of image quality by turbulence.
Transformed to  
spatial
frequency space, this relation becomes,
\begin{equation}
\tilde{{\cal I}}(u,v)=\tilde{{\rm OID}}(u,v) \times  \tilde{{\rm  
PSF}} (u,v),
\label{eq:otf}
\end{equation}
where $u$ and $v$ are spatial frequency coordinates connected to the  
angular
coordinates through the Fourier transform, 
\begin{eqnarray}
{\cal I} (\alpha,\beta)\ \  & \mapsto^{\hspace{-0.6cm}\rm\small  
^{Fourier}}\ \  &
\tilde{{\cal I}} (u,v)=\int\!\!\!\!\int
{\cal I} (\alpha,\beta)\; e^{-2\eye\pi(\alpha u + \beta v)}\; \dd \alpha \dd \beta.
\end{eqnarray}
Here, $u$ and $v$ are linear telescope pupil coordinates normalized  
by the
wavelength. Thus, the larger the pupil, the higher the largest  
reachable spatial
frequency and the larger the resolution in the image. The Fourier  
transform of
the ${\rm PSF}$ is called the {\it optical transfer function\/} (OTF)  
whose
value drops from one at zero spatial frequency to zero at the cut-off  
frequency
of the telescope. The latter is exactly ${\cal D}/\lambda$
in the absence of atmospheric turbulence, with ${\cal D}$ the diameter of the  
telescope and $\lambda$ the wavelength at which one is observing.  The
inverse of the cut-off frequency is termed the {\it angular  
resolution}.  The
OTF can also be considered as the autocorrelation of the telescope pupil
function. The process of imaging of an object is therefore equivalent to
performing a low-pass filtering of the object's spatial spectrum.

\subsubsection{Spatial coherence}

An interferometer is nothing but a particular case of an imager. For  
the sake of
clarity, let us choose a very simple interferometer with only two  
telescopes and
let us assume the telescope diameter to be zero.  Formally, this  
defines a
telescope with a particular pupil function made of two subpupils, the
autocorrelation of which has three peaks in spatial frequency space: one
centred on zero (the OTF of a single telescope) and two symmetric peaks
centred on $\pm {\cal B}/\lambda$, with ${\cal B}$ the distance  
between the two
telescopes, due to the correlation of the two different pupils. The
interferometer performs a band-pass filtering of the object's spatial  
spectrum
at ${\cal B}/\lambda$, the reciprocal of which sets the spatial  
resolution. The
PSF associated to the OTF is a fringe pattern with 100\% fringe  
contrast.

The $\pm {\cal B}/\lambda$ peaks in the OTF express that the  
interferometer
performs the correlation of the light waves collected by telescopes 1  
and 2,
denoted as $E_{1}\left(\vec{x},t\right)$ and $E_{2}\left(\vec{x},t 
\right)$, and
measures the spatial coherence of light at the two telescope locations,
\begin{equation}
\gamma_{12}(\vec{{\cal B}}/\lambda)= {\cal V}(\vec{{\cal B}}/\lambda)=\frac 
{\left <
E_{1}\left(\vec{x},t\right).E_{2}^{\star}\left(\vec{x},t\right)
\right>}{\sqrt{\left < |E_{1}\left(\vec{x},t\right)|^2 \right>\left <
|E_{1}\left(\vec{x},t\right)|^2 \right> }},
\end{equation}
with the time average $\left < . \right >$ taken over ${\Delta t\gg  
\lambda/c}$
\citep{goodman85}. This correlation can be performed in different ways  
as beams can
be mixed either with a beam splitter or with a focusing optics whose  
operations
are equivalent in the zero telescope diameter case (i.e.\ when the  
baseline is
large with respect to each individual pupil). The theory of the spatial
coherence of light states that the correlation factor, i.e.\ the  
mutual degree
of coherence or complex visibility, is the Fourier transform of the  
source
normalized spatial intensity distribution,
\begin{equation}
{\cal V}
(\vec{{\cal B}}/\lambda)=\frac{\int\!\!\int {\rm OID}
(\alpha,\beta)\; e^{-2\eye\pi(\alpha
{\cal B}_\alpha/\lambda + \beta {\cal B}_\beta/\lambda)}\;
\dd \alpha \dd \beta}{\int\!\!\int
{\rm OID} (\alpha,\beta)\; \dd \alpha \dd \beta},
\label{eq:zvc}
\end{equation}
with $ {\cal B}_\alpha$ and ${\cal B}_\beta$ the components of the
interferometer baseline $\vec{{\cal B}}$ projected on the sky plane  
in the
direction of the source. This result is known as the {\it Zernike -  
Van Cittert
theorem}. A multi-telescope interferometer therefore measures the  
source spatial
frequency spectrum at the spatial frequencies defined by its baselines.

\subsubsection{The modulus and phase of the visibility function}

The interferometric observable is the complex visibility, which is  
the spatial
spectrum of the source. Using Eq.(\ref{eq:otf}) one can understand  
that the
measured fringe pattern can be approximated by 
the product of the visibility with the  
fringe pattern
obtained on a point source, which, by definition or using Eq.(\ref 
{eq:zvc}), has
${\cal V}=1$. Information can therefore be derived from the fringe
contrast, i.e.\ from the modulus of ${\cal V}$ (denoted as $|{\cal V}| 
$) and from
the phase of ${\cal V}$. The latter is termed the fringe pattern  
position
$\delta$. It is usually given a value relative to the zero optical path
difference translated in phase: $2\pi\delta/\lambda$.

The modulus of the visibility is related to the size of the source.  
Indeed, the source characteristic angular size is proportional to the reciprocal of the visibility  function
width. Thus, the larger the source, the smaller the fringe
contrast. The phase of the visibility is related to the amount of  
asymmetry in
the object. The visibility of a perfectly point-symmetric source is  
real and
therefore has a phase of $0 (\pi)$ whereas the phase of the  
visibility of an
asymmetric source may take any value in the range $[-\pi,+\pi]$.

The simplest and most famous visibility function is that of a uniform  
disk of
angular diameter $\theta_{\rm UD}$ whose modulus is, 
\begin{equation}
\left|{\cal V}({\cal B},\lambda,\theta_{\rm UD})\right| = \left|
\frac{2 J_1(\pi {\cal B}\,\theta_{\rm UD}/\lambda)}
{\pi {\cal B}\,\theta_{\rm UD}/\lambda}
\right|,
\label{historical}
\end{equation}
with a first null at $\lambda/{\cal B} = 1.22\ \theta_{\rm UD}$, and where $J_1$ is the first-order Bessel function of the first kind. This  
result
provided the historical method to measure stellar diameters.

It is now evident from Eq.\,(\ref{historical}) that long baseline  
interferometry
can, in principle, provide very precise measurements of angular size 
$ \theta$.
Combined with a precise measurement of the trigonometric parallax, $\pi$, the
linear diameter $D$ of stars in the solar neighbourhood can be  
derived from
$\theta$ in terms of the solar value $D_\odot$ through the simple  
relation
\begin{equation}
D[D_\odot] = 107.47\ \theta[{\rm mas}] / \pi[{\rm mas}].
\label{diameter}
\end{equation}

\subsubsection{Aperture synthesis}
\label{synthesis}

Traditionally, visibilities are fitted with models that best  
represent the
source spatial intensity distribution to measure parameters that  
characterize
it, such as its diameter, limb-darkening coefficients or star  
separations and
position angles in the case of multiple systems, etc.  The ultimate  
goal of
interferometers is to perform imaging with a spatial resolution much  
higher than
that of classical telescopes. An excellent example in the radio  
domain is the
VLA. The primary data provided by the interferometer, the complex  
visibilities,
need to be assembled and inverted through Eq.~(\ref{eq:zvc}) to  
reconstruct an
image. In order to do so, and ideally, the OTF of a telescope with an  
equivalent
size needs to be sampled with the interferometer baselines.  Doing  
so, the
interferometer synthesizes a monolithic telescope OTF (or, more  
exactly, the
support of the OTF) and this process is known as aperture synthesis. The
difficulty is to have enough telescopes to achieve this. The Earth  
rotation
helps as the baselines projected in the direction of the source  
rotate at the
pace of the Earth rotation. In practice the sampling is not as  
perfect as if it
were a regular grid and the image cannot be obtained with a simple  
inverse
Fourier transform. Instead, an image reconstruction algorithm that  
makes up for
missing or irregularly sampled visibilities needs to be used. More  
and more
images are currently produced by optical interferometers \citep[e.g.,][]{kraus07}.

\subsubsection{Interferometers in practice}

The specificity of optical radiation is that only intensity can be  
detected
efficiently. As a consequence, lightwaves cannot be directly measured  
at each
telescope and need to be propagated down to the recombination point.  
This
requires a large number of mirrors in practice -- at the cost of  
photometric
efficiency -- to guide the beams. In addition, atmospheric turbulence  
has a very
short coherence time during which interference fringes have to be  
measured. This
forces individual exposure times to be short. The combination of  
these two
effects leads to a smaller sensitivity than for classical instruments  
and
interferometers need to be operated in wide photometric bands which  
are in
practice limited by the transmission of the atmosphere. Besides spatial
coherence of light, interferometers are limited by the temporal  
coherence of
light. As a matter of fact, the number of fringes in the fringe  
pattern is
proportional to the reciprocal of the photometric bandwidth. The  
fringe pattern
is therefore very localized and delay lines are required to keep the  
optical
path difference very close to zero to find fringes.

To summarize, an optical interferometer comprises telescopes, an  
optical train
to guide the beams to a delay line system, and subsequently to the  
beam combiner
where lightwaves are mixed to produce fringes. These are the basic  
sub-systems
to which adaptive optics can be added to allow the use of larger  
apertures. In
this case, spatial coherence needs to be restored and a fringe  
tracking system
must be used that allows to stabilize the zero optical path  
difference, despite
the jitter due to turbulence and vibrations.

\subsection{Stellar physics with interferometers}
\label{stphys}

This section summarizes astrophysical applications of optical and infrared
interferometry, with emphasis on those topics that are relevant for the synergy
with asteroseismology. A few attempts have already been made to combine
information from both techniques for the modelling of stellar properties
\citep{PTG03,kervella03b,kervella04c,th05}.
Advances in the observing capabilities will increase the potential for combined
interferometric and asteroseismic studies in the future.

\subsubsection{Stellar diameters}

The most straightforward observation with an interferometer is the measurement
of a stellar angular diameter: by measuring ${\cal V}$ or ${\cal V}^2$ with a
range of baselines of different lengths, one can determine $\theta_{\rm UD}$
from Eq.~(\ref{historical}) through a straightforward $\chi^2$ minimization
process. Note that $\theta_{\rm UD}$ is somewhat smaller than the physical
diameter of the star, since Eq.~(\ref{historical}) assumes that the star is a
uniform disk and thus neglects the effect of limb darkening. A correction
factor can be computed from a grid of stellar atmospheres, and applied to
$\theta_{\rm UD}$ to calculate the limb-darkened diameter $\theta_{\rm LD}$
\cite[e.g.,][and references therein]{quirrenbach01a}.
This procedure is
frequently better than using a limb-darkened model from the beginning, because
interferometric data obtained on baselines ${\cal B} \lessim \lambda /
\theta_{\rm UD}$, which resolve the star only partially, cannot distinguish
well between uniform and limb darkened disks, or different limb darkening
models. It is therefore advantageous to publish model-independent observational
results (represented by $\theta_{\rm UD}$) and the adopted model-dependent
quantities (represented by the limb darkening correction factor) separately. In
general, limb darkening corrections are smaller in the near-infrared than at
visible wavelengths; this favours $H$ band or $K$ band observations for precise
determinations of stellar diameters, provided that sufficiently long baselines
are available.

If the parallax is known, it can be combined with $\theta_{\rm LD}$ to compute
the linear radius of the star. At present, i.e., after {\it Hipparcos\/} and
before {\it Gaia\/}, this is most interesting for low-mass main-sequence stars;
these are sufficiently close so that good parallaxes are available, and they can
be resolved with present interferometers. We discuss such potential targets
for asteroseismology in Section~\ref{gp} (see also Table~\ref{best_stars}).
It seems that
in the mass range 0.5\,M$_\odot$ to $0.8$\,M$_\odot$ observed radii are
systematically larger than predicted by theoretical mass-radius relations 
\citep{lane01,seg03,berger06}.

One of the most fundamental applications of interferometry to stellar
astrophysics is the calibration of the stellar temperature scale. The effective
temperature of a star is defined by
\begin{equation}
T_{\rm eff} \equiv \left(\frac{L}{4 \pi \sigma R^2} \right)^{1/4} =
\left(\frac{4 f_{\rm bol}}{\sigma \theta^2_{\rm LD}} \right)^{1/4}~~,
\end{equation}
where $f_{\rm bol}$ is the
bolometric flux and $\sigma$ the Stefan-Boltzmann constant. The most direct
and model-independent way of measuring effective temperatures is thus the
combination of bolometric fluxes with angular diameters. More indirect methods
such as the infrared flux method described by 
\citet{black77}, which uses
the ratio of the total integrated flux to the flux in the $K$ band
as temperature indicator,
can be validated by comparison with directly determined effective temperatures
\cite[e.g.,][]{han74a,belle99,morz03,kervella04b}.

Diameter measurements of Cepheid variables are of particular interest, as they
can contribute to the calibration of their period-luminosity relation, and thus
to the distance scale in the local Universe. Several methods have been employed
that use different combinations of angular diameters, radial-velocity curves,
light curves and trigonometric parallaxes. Most useful, however, are
observations that measure not only the average diameter of the Cepheid, but
also the diameter variations due to the pulsations 
\citep{lane00,arm01,kervella04a, merand05}.
Such
data can be combined with radial-velocity curves to determine geometric
distances, provided that the {\it projection factor\/} $\cal G$ 
is known, which is
needed to convert observed radial velocities to the radial motion of the
stellar atmosphere. The factor $\cal G$ 
is usually obtained from theoretical model atmospheres,
but it is also possible to use interferometric observations together with the
radial-velocity curve and the trigonometric parallax to calibrate $\cal G$.

\subsubsection{Limb darkening and wavelength-dependent diameters}
\label{limbd}

The visibility function of a single star is the Fourier transform of the
center-to-limb surface brightness profile $I(\alpha)$:
\begin{eqnarray}
\label{vis} {\cal V} (k{\cal B}) 
& = & \int_0^{\theta/2} \int_0^{2\pi} \cos(k{\cal B}\alpha\cos\phi) 
\,I(\alpha) \,\alpha
\,\rm{d}\alpha \,\rm{d}\phi \nonumber \\
       & = & 2 \pi \int_0^{\theta/2} J_0(k{\cal B}\alpha) \,I(\alpha) \,
\alpha \,\rm{d}\alpha~~,
\end{eqnarray}
where $\alpha$ is an angular distance coordinate, $k\equiv 2\pi/\lambda$, 
and $J_0$ the Bessel
function of zeroth order. For stars, the intensity distribution is customarily
given as a function of $\mu = \sqrt{1-(2\alpha/\theta)^2}$, 
rather than $\alpha$ itself.
For polynomials
\begin{equation}
\label{polylimb} I(\mu) = \sum_i a_i \mu^i~~,
\end{equation}
Eq.~(\ref{vis}) leads to 
\begin{equation}
\label{polyvis} {\cal V} (k{\cal B}) = \frac{1}{C} \sum_i a_i
2^{^i\!\!/\!_2} \Gamma\left(\frac{i}{2}+1\right)
\frac{J_{^i\!\!/\!_2+1}(k{\cal B}\theta/2)}{(k{\cal
    B}\theta/2)^{^i\!\!/\!_2+1}}, 
\end{equation}
with
\begin{equation}
C = \sum_i \frac{a_i}{i+2}
\end{equation}
\citep{Quirre1996}.
For a disk of uniform brightness $I \equiv 1$, this formula reduces to the
familiar Airy pattern  as in Eq.~(\ref{historical}).

Since the correct treatment of limb darkening is crucial for precise
measurements of effective temperatures, it is important to perform observational
checks of theoretical limb darkening curves. This is a fairly difficult task,
because data are required around and beyond the first null of the visibility
function (as in Eq.~(\ref{historical})), where the signal-to-noise ratio is
low. The first such measurement was carried out with the Narrabri Intensity
Interferometer; data from a 203-hour(!) integration on Sirius showed that the
height of the second maximum of the visibility function was consistent with the
prediction from a model atmosphere \citep{han74b}.  Similar observations have
been carried out with modern interferometers for Arcturus \citep{Quirre1996},
for $\alpha$\,Ari and $\alpha$\,Cas \citep{haj98}, for HR\,5299, HR\,7635, and
HR\,8621 \citep{Wittko2001}, for $\psi$\,Phe \citep{Wittko2004}, for
$\alpha$\,Ori and $\alpha$\,Her \citep{perrin04}, for $\gamma$\,Sge
\citep{Wittko2006}, and for $\alpha$\,Cen\,B \citep{Bigot06}.  These data are
complementary to those obtained with other methods (e.g., transits of extrasolar
planets), as many of the targets are giant stars, and all of them are bright and
nearby, so that a wealth of additional information is available, including
high-resolution spectra.

The wavelength dependence of the limb darkening correction obviously leads to a
variation of $\theta_{\rm UD}$ with $\lambda$; the ratio $\theta_{\rm
UD}(\lambda_1) / \theta_{\rm UD}(\lambda_2)$ can therefore be used as a
diagnostic tool for the properties of the stellar atmosphere \citep{morz91}. An
even stronger dependence of the diameter on wavelength can be observed in cool
evolved stars. At each wavelength, an interferometer measures essentially the
diameter of the surface at optical depth value 1.  In cool stars with low
gravity, the location of this surface can vary by a significant fraction of the
stellar radius between wavelengths at which the opacity is high (i.e., at
wavelengths corresponding to molecular absorption bands), and those at which it
is low (i.e., in the continuum between absorption lines). Strongly
wavelength-dependent diameters are therefore measured for Mira stars
\cite[e.g.,][]{eis07} and semi-regular variables \cite[e.g.,][]{perrin05}, but
this effect has also been observed in ``normal'' giants of luminosity class III
\citep{Quirre1993}.

\subsubsection{Rapidly rotating stars}
\label{rotstars}

\begin{figure*}
\centering
\includegraphics[width=0.9\textwidth]{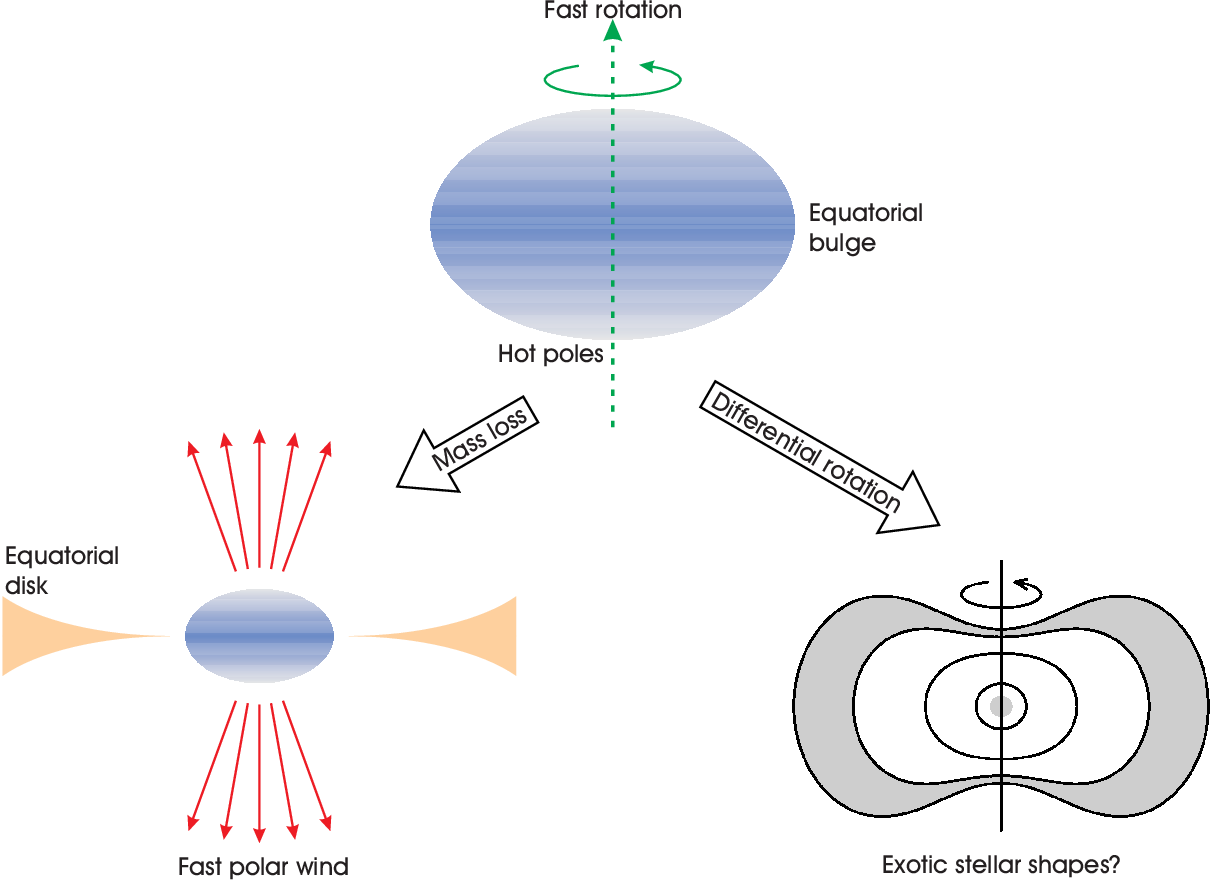}
\caption{Rapidly rotating stars develop a pronounced equatorial bulge, which is
substantially cooler than the polar regions (top). Radiation pressure drives a
fast wind from the polar regions if they are sufficiently hot, whereas disks
are frequently formed in the equatorial plane (bottom left). Stars in which the
rotation rate decreases strongly with distance from the axis may assume
non-convex shapes, shown here in cross-section with convective regions shaded
(bottom right, adopted from MacGregor et al.\ 2007). Future interferometric
observations will be needed to determine whether stars with such exotic shapes
actually exist.}
\label{fig:rotation}       
\end{figure*}

Rapidly rotating stars deviate from spherical symmetry in two ways: they are
noticeably oblate (i.e., flattened), and their surface temperature increases
from the equator to the poles (see Fig.~\ref{fig:rotation}, top). 
The latter effect is known as the von Zeipel effect or {\it gravity darkening\/}
\citep{zeip}.  The combination of both effects leads to an asymmetric brightness
distribution on the sky and thus to visibilities that depend not only on the
length, but also on the orientation of the interferometer baseline, and to
non-zero closure phases. This has been observed for the bright stars Altair
\citep{belle01,ohishi04,Domici2005,peterson06b,Monnier2007}, Achernar
\citep{Domici2003}, Regulus \citep{McAlister2005}, and Alderamin
\citep{belle06}.  Interferometric observations also lead to the conclusion that
Vega is a rapid rotator seen nearly pole-on \citep{Auf06,peterson06a}.
In general, reasonable agreement is found between the observations and
simple models based on the assumptions of rigid rotation, mass concentration
towards the center of the star, and a fully radiative atmosphere. There are
some small discrepancies, which may indicate that one or more of the underlying
assumptions need to be relaxed; however, at present the data are not sufficient
yet to provide strong constraints on more sophisticated models.

Rapid stellar rotation has two important consequences for determination of
fundamental properties of stars, and for the comparison of observations with
theoretical models: first, the stellar structure and evolution is affected by
rotation; second, the star radiates non-isotropically, and thus its observed
properties (spectral energy distribution, colours, spectral type, 
etc.) depend on
the inclination of the rotation axis with respect to the line of sight -- much
more dramatically than one might naively expect, as demonstrated by the example
of Vega. Strong differential rotation might further complicate the situation,
as stars with different masses and internal structure can appear identical to
the observer, while  
in extreme cases, stars might even have
non-convex shapes \citep{MacGre2007} (see Fig.~\ref{fig:rotation}, bottom right). 
Direct
interferometric observations of the shapes and surface brightness distributions
of rapid rotators are thus an indispensable tool for the correct modelling of
these stars.

A further consequence of rapid stellar rotation in hot stars is the enhanced
radiation pressure near the poles due to the von Zeipel effect, which can drive
a polar wind (see Fig.~\ref{fig:rotation}, bottom left). 
This is the most
likely explanation of the elongation of $\eta$\,Car \citep{boekel03}
and Achernar \citep{ker06}
observed
interferometrically.

\subsubsection{Binary stars}

In addition to measurements of stellar diameters, observations of close binary
stars is one of the classical applications of optical interferometry. 
The squared visibility\footnote{Single-baseline interferometers normally
measure ${\cal V}^2$, not ${\cal V}$, see \citet{shao92}.}
of a binary
is given by
\begin{equation}
\label{binvis} {\cal V}^2 = \frac{1}{(1+{\cal R})^2} \left[ {\cal V}_1
+ {\cal R}^2 {\cal V}_2 
+ 2 {\cal R} {\cal V}_1 {\cal V}_2 \cos \left( \frac{2 \pi {\cal S} {\cal B} 
\cos \psi}{\lambda} \right) \right],
\end{equation}
with ${\cal R} \leq 1$ 
the brightness ratio of the two stars, ${\cal V}_1$ and ${\cal
V}_2$ their individual visibilities, $\cal S$ their
separation, and $\psi$ the angle in the $(u,v)$ plane between the interferometer
baseline and the line joining the two stars \cite[e.g.,][]{lawson00}.
To determine ${\cal S}$, $\psi$, and
the stellar parameters uniquely from measurements of ${\cal V}^2$, data have to
be obtained for many points in the $(u,v)$ plane. This can be accomplished by
observations with multiple baselines, by Earth-rotation synthesis, by taking
data at multiple wavelengths (since $u$ and $v$ scale with the wavenumber), or
by a combination of these techniques. Repeated observations with good coverage
of the orbital phases can then be used to determine a visual orbit
\citep{arm92a}.
A better method was developed by \citet{hu93}
and applied in most subsequent analyses. Here, the seven orbital elements
and the stellar parameters are fitted directly to the observed visibilities.
This global approach has the advantage that the fast orbital motion of
short-period binaries is taken into account properly.

\begin{table}[t]
\caption{\label{SB2tab}Interferometrically determined orbits and component
masses for double-lined spectroscopic binaries. More details are available
in the references listed below the table. The star 
$\kappa$\,Peg is a triple system; in this case the ``wide''
(A -- Ba/Bb) and ``narrow'' (Ba -- Bb) orbits are listed separately.}
{\scriptsize
\tabcolsep=3pt
\begin{tabular}{lr@{+}lr@{.}lr@{.}l@{$\pm$}lr@{.}l@{$\pm$}lllll}
System & \multicolumn{2}{c}{Spectral} & \multicolumn{2}{c}{$a''$} &
\multicolumn{3}{c}{$M_1$ [$M_\odot$]} & \multicolumn{3}{c}{$M_2$ [$M_\odot$]} &
Instr. & Ref.\\
& \multicolumn{2}{c}{Types} & \multicolumn{2}{c}{[mas]} &
\multicolumn{3}{c}{} & \multicolumn{3}{c}{} &
& \\
\hline \vspace{-2mm}\\
HD\,27483     & F6V     & F6V     &   1 & 3 &  1 & 38 & 0.13  & 1&39  & 0.13  &
PTI     & K04 \\
$\alpha$ Vir  & B1III-IV& B3V:   &    1 & 5 & 10 & 9  & 0.9   & 6&8   & 0.7   &
Narrabri & H71 \\
$\kappa$ Peg B& F5IV    & K0V:    &   2 & 5 &  1 &662 & 0.064 & 0&814 & 0.046 &
PTI     & M06 \\
V773 Tau A    & K2      & ?       &   2 & 8 &  1 &54  & 0.14  & 1&332 & 0.097 &
KI      & B07 \\
$\theta$ Aql  & B9.5III & B9.5III &   3 & 2 &  3 & 6  & 0.8   & 2&9   & 0.6   &
Mk\,III & H95 \\
$\beta$ Aur   & A2V     & A2V     &   3 & 3 &  2 &41  & 0.03  & 2&32  & 0.03  &
Mk\,III & H95 \\
12 Boo        & F9IV    & F9IV    &   3 & 4 &  1 &435 & 0.023 & 1&408 & 0.020 &
PTI     & B00 \\
              & \multicolumn{2}{c}{}& 3 & 5 &  1 &416 & 0.005 & 1&374 & 0.005 &
combined & B05b\\
$\sigma$ Sco  & B1III   & B1V     &   3 & 6 & 18 &4   & 5.4   &11&9   & 3.1   &
SUSI    & N07b \\
$\gamma^2$ Vel& O7.5II  & WC8     &   3 & 6 & 28 &5   & 1.1   & 9&0   & 0.6   &
SUSI    & N07c \\
BY Dra        & K4V     & K7.5V   &   4 & 4 &  0 &59  & 0.14  & 0&52  & 0.13  &
PTI     & B01 \\
$o$ Leo       & F9      & A5m     &   4 & 5 &  2 &12  & 0.01  & 1&87  & 0.01  &
combined & H01 \\
HD\,9939      & K1IV    & K0V     &   4 & 9 &  1 &072 & 0.014 & 0&838 & 0.008 &
PTI     & B06 \\
$\sigma$ Psc  & B9.5V   & B9.5V   &   5 & 6 &  2 &65  & 0.27  & 2&36  & 0.24  &
PTI     & K04 \\
64 Psc        & F8V     & F8V     &   6 & 5 &  1 &223 & 0.021 & 1&170 & 0.018 &
PTI     & B99b \\
93 Leo        & G5III   & A7V     &   7 & 5 &  2&25   & 0.29  & 1&97  & 0.15  &
Mk\,III & H95 \\
$\zeta^1$ UMa & A2V     & A2V     &   9 & 6 &  2&51   & 0.08  & 2&55  & 0.07  &
Mk\,III & H95 \\
              & \multicolumn{2}{c}{}& 9 & 8 &  2&43   & 0.07  & 2&50  & 0.07  &
NPOI    & H98 \\
$\iota$ Peg   & F5V     & G8V     &  10 & 3 &  1&326  & 0.016 & 0&819 & 0.009 &
PTI     & B99a \\
$\eta$ And    & G8III   & G8III   &  10 & 4 &  2&59   & 0.30  & 2&34  & 0.22  &
Mk\,III & H93 \\
$\alpha$ Equ  & G2III   & A5V     & ~12 & 0 &  2&13   & 0.29  & 1&86  & 0.21  &
Mk\,III & A92b \\
27\,Tau       & B8III   & ?       &  13 & 1 &  4&74   & 0.25  & 3&42  & 0.25  &
combined & Z04 \\
HD\,195987    & G9V     & ?       &  15 & 4 &  0&844  & 0.018 & 0&665 & 0.008 &
PTI     & T02 \\
$\zeta$ Aur   & K4Ib    & B5V     &  16 & 2 &  5&8    & 0.2   & 4&8   & 0.2   &
Mk\,III & B96 \\
$\theta^2$ Tau & A7III  & A:      &  18 & 6 &  2&1    & 0.3   & 1&6   & 0.2   &
Mk\,III & T95 \\
              & \multicolumn{2}{c}{}&18 & 8 &  2&15   & 0.12  & 1&87  & 0.11  &
combined & A06 \\
$\lambda$ Vir & Am      & Am      &  19 & 8 &  1&897  & 0.016 & 1&721 & 0.023 &
IOTA & Z07 \\
HD 98800\,B   & K5V     & ?       &  23 & 3 &  0&699  & 0.064 & 0&582 & 0.051 &
KI & B05a \\
$\phi$ Cyg    & K0III   & K0III   &  23 & 7 &  2&536  & 0.086 & 2&437 & 0.082 &
Mk\,III & A92a \\
$\alpha$ And  & B8IV    & A:      &  25 & 2 &  5& \multicolumn{2}{@{}l}{5:} &
2 & \multicolumn{2}{@{}l}{3:} &
Mk\,III & P92,T95 \\
$\beta$ Cen   & B1III   & B1III   &  25 & 3  & 11&2   & 0.7   & 9&8  & 0.7    &
SUSI    & D05,Au06 \\
$\beta$ Ari   & A5V     & G0V:    &  36 & 1  &  2&34  & 0.10  & 1&34  & 0.07  &
Mk\,III & P90 \\
$\lambda$ Sco & B1.5IV  & B2V     &  49 & 3  & 10&4   & 1.3   & 8&1   & 1.0   &
SUSI    & T06 \\
12 Per        & F8V     & G2V     &  53 & 2  &  1&382 & 0.019 & 1&240 & 0.017 &
CHARA   & Ba06 \\
$\alpha$ Aur  & G1III   & G8III   &  55 & 7  &  2&56  & 0.04  & 2&69  & 0.06  &
Mk\,III & H94 \\
$\delta$ Equ  & F7V     & F7V     & 231 & 9  &  1&193 & 0.012 & 1&188 & 0.012 &
PTI     & M05 \\
$\kappa$ Peg  & F5IV    & F5IV    & 235 & 0  &  1&549 & 0.050 &
\multicolumn{3}{c}{composite} &
PTI     & M06 \\
\hline
\end{tabular}\\
}
{\small
References: A92a: \citet{arm92a};
A92b: \citet{arm92b};
A06: \citet{arm06};
Au06: \citet{Aus06};
Ba06: \citet{bag06};
B96: \citet{benn96};
B99a: \citet{Boden99a};
B99b: \citet{Boden99b};
B00: \citet{Boden00};
B01: \citet{Boden01};
B05a: \citet{Boden05a};
B05b: \citet{Boden05b};
B06: \citet{Boden06};
B07: \citet{Boden07};
D05: \citet{davis05};
H71: \citet{herb71};
H93: \citet{hu93};
H94: \citet{hu94};
H95: \citet{hu95};
H98: \citet{hu98};
H01: \citet{hu01};
K04: \citet{kon04};
M05: \citet{munt05};
M06: \citet{munt06};
N07b: \citet{north07b};
N07c: \citet{north07c};
P90: \citet{pan90};
P92: \citet{pan92};
T95: \citet{Tom95};
T02: \citet{torres02};
T06: \citet{Tan06};
Z04: \citet{Zwahlen-2004:a};
Z07: \citet{zhao07}
}
\end{table}

The goal of binary star observations is normally the determination of
fundamental parameters such as the masses, radii, and luminosities of the two
components. It is usually necessary to combine two or more techniques to achieve
this goal, since not all orbital elements can be determined with any single
method alone. A particularly useful case are double-lined spectroscopic
binaries, for which the visual orbit can also be obtained. However, there is
little overlap between the two classes, since spectroscopic binaries tend to
have small orbits that are difficult to resolve. Even orbits of double-lined
spectroscopic binaries obtained with adaptive optics or speckle techniques are
rarely precise enough to give masses better than $\sim 10\%$. This is one of the
reasons why interferometric observations of double-lined spectroscopic binaries
are of great importance for the determination of fundamental stellar parameters.

Adding the orbital inclination 
from the interferometric orbit to the spectroscopic
elements allows computation of the component masses, and combining the angular
diameter of the orbit with the physical scale set by the spectroscopy yields
the distance, or {\it orbital parallax}. 
Because of the fundamental importance of
these data, extensive observations of double-lined spectroscopic binaries 
have been carried out with almost
every interferometer in existence. They are summarized in Table~\ref{SB2tab},
which has been updated from \citet{quirrenbach01b}.
The orbital solutions and
error estimates are taken from the references cited, and are therefore not
uniform. The error bars refer formally to 1\,$\sigma$, but some authors may be
more conservative than others in assessing systematics in the data or in
dealing with discrepancies between different subsets of the data (e.g.,
different eccentricities from the spectroscopic and interferometric orbits). It
should also be pointed out that determining the scale of the orbit (in angular
units), and the subsequent computation of the orbital parallax, requires
knowledge of the effective central wavelength of the interferometric
observations, which depends on the stellar colour \citep{hu94}.
Systematic errors in this quantity may easily go unnoticed since they do not
affect the $\chi^2$ of the orbital fit. In many cases, however, the precision of
the mass determination is limited by the spectroscopic, not by the
interferometric orbit.

It is instructive to compare Table~\ref{SB2tab} with the masses of eclipsing
binaries compiled by \citet{An91}.
Only a handful of the interferometrically
determined masses meet Andersen's accuracy criterion for being useful for
critical tests of main-sequence stellar models, which he set at 2\%.
Furthermore, the baselines used in the observations compiled in the table are
generally too short to give good stellar radii (with the exception of Capella).
On the other hand, the agreement for the component masses of $\beta$ Aur, a
system in common between the two samples, is encouraging. Furthermore, analyses
of pairs with evolved components such as Capella, $\phi$ Cyg, and $\alpha$ Equ
provide useful tests of post-main-sequence evolutionary models 
\cite[e.g.,][]{arm92a}.
The availability of orbital parallaxes giving good luminosities
is a clear advantage in this respect.

Comparing Table~\ref{SB2tab} with the previous version in \citet{quirrenbach01b}
also shows that quite some progress has been made over the past few years, and
that many interferometers have started to cover particularly interesting types
of stars, such as $\beta$\,Cephei stars ($\beta$\,Cen and $\lambda$\,Sco) and
pre-main-sequence objects (V773\,Tau and HD\,98800\,B). The key to further
progress will be observations of stars with well-determined spectroscopic
elements and state-of-the-art determination of the metal abundance.
Comprehensive tests of stellar models require covering all regions of the HR
diagram. Many of the eclipsing systems in \citet{An91} are also accessible
to the new instruments, which could provide improved distances and better
luminosity ratios for partially eclipsing systems. The good instantaneous
coverage of the $(u,v)$ plane afforded by the multiple baselines and wavelength
channels of the new arrays will allow determination of orbits from snapshot
observations, making them very efficient instruments for binary programmes.

\subsubsection{Circumstellar material}

Optical/infrared interferometry has been used extensively to characterize the
geometry and physical properties of circumstellar material. Typical examples
are measurements of the dust distribution around late-type stars \citep{dan94},
the proof that Be stars are surrounded by geometrically thin disks
\citep{Quirre1997},
and the spatially resolved detection and
compositional analysis of silicates in the innermost two astronomical units of
protoplanetary disks \citep{boekel04}.
As already mentioned in
Section~\ref{rotstars}, the hot polar regions of rapidly rotating early-type
stars may drive bipolar winds; these winds may coexist with equatorial disks
\cite[e.g.,][]{mal07}.

In the present context, the presence of circumstellar material is relevant
mainly because it affects the stellar parameters inferred from photometric,
spectroscopic, or interferometric observations. It is therefore advisable to
perform interferometric observations with a range of baseline lengths and
orientations, and to check whether the observed visibilities follow
Eq.~(\ref{polyvis}) or, in the case that only short baselines are used,
Eq.~(\ref{historical}). The presence of circumstellar material will normally
manifest itself in a visibility function that does not converge to ${\cal V}
\rightarrow 1$ for ${\cal B} \rightarrow 0$. It is then possible to either
correct the derived stellar parameters for the effects of the circumstellar
material, or to reject the affected stars from a sample for which
high-precision data are sought.

\subsection{Present instruments and capabilities}

\label{presentinf}

\subsubsection{A short overview of current interferometers}

As is clear from the discussion in Section~\ref{principles}, an interferometer (and in particular
an optical interferometer) is very different from a focal instrument at a
telescope (e.g.,\ a spectrograph).  While a few parameters are sufficient to
qualify a spectrograph (e.g.,\ spectral resolution and operating wavelength),
both the telescopes and recombiners have to be taken into account for an
interferometric instrument.  Most degrees of freedom are related to
the interferometer and not to the focal instrumentation.

\begin{figure*}
\centering
\includegraphics[width=0.9\textwidth]{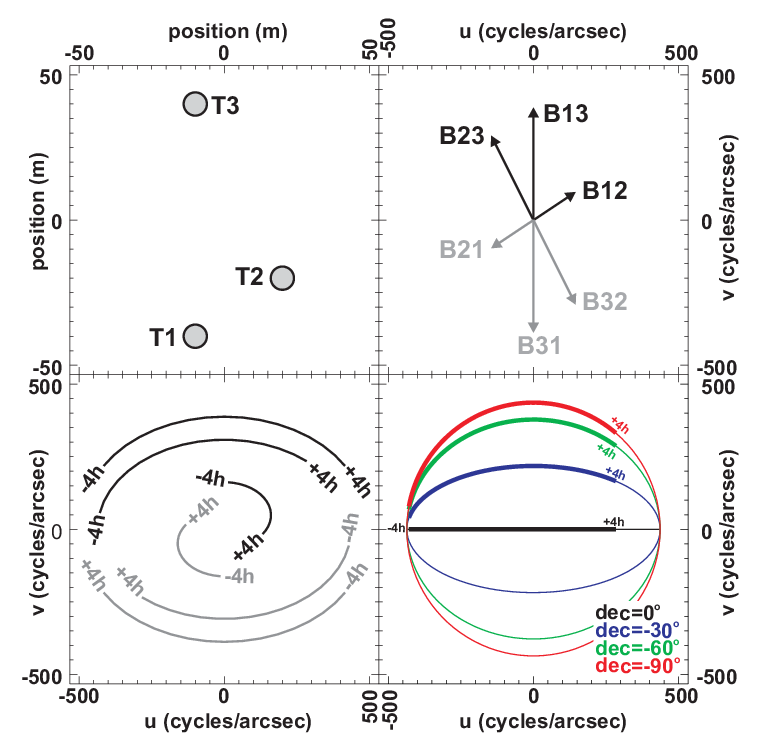}
\caption{\label{fig:uvcoverage} The $(u,v)$ coverage of an interferometer. Top
    left: projection onto the $(E,N)$ plane of the physical positions of the
    telescopes at the observatory.  Top right: instantaneous $(u,v)$ coverage at
    a wavelength of 1~$\mu$m, assuming the source is at a declination of
    $-45^\circ$ and passing through the meridian. The telescopes are at a
    latitude of $-45^\circ$. Although there are two centrosymmetric baselines
    for each telescope combination, they yield no extra information (the object
    brightness distribution being a real function, its Fourier transform is
    centrosymmetric).  Bottom left: dependence of the $(u,v)$ coverage on hour
    angle (-4h to +4h), for the same parameters as the top right figure.  Bottom
    right: dependence of the $(u,v)$ coverage on the source declination for the
    B23 baseline. These curves are ellipse arcs, ranging from a circle arc for a
    source at a celestial pole, and collapsing to a line segment for a source at
    the equator.}
\end{figure*}

Since interferometers measure visibilities and visibilities encode information
on the surface brightness distribution of the source at a given angular
frequency, the first step in the qualification of an interferometer is to define
the angular frequencies made available to constrain the astronomical object
brightness distribution.  These angular frequencies are the $(u,v)$ space
coverage and they depend on 
\begin{enumerate}
\item the baselines (the  physical positions of the telescopes) and
  on the operating wavelength;
\item the declination  of the source, the latitude  of the observatory
  and the hour angle;
\item the number of telescopes;
\end{enumerate}
\citep[e.g.,][]{Dyck2000}.
Figure~\ref{fig:uvcoverage} illustrates some of these dependencies.
More specifically, the baselines and operating wavelength determine the maximum
achievable angular frequency ($\mathcal{B}/\lambda$).  It limits the minimum
angular diameters of the objects that can be resolved to $\theta \sim
\lambda/\mathcal{B}$.  Small angular sizes can be reached either with large
baselines or small operating wavelengths (e.g.,\ 
observing in the $V$ band, in
contrast with the $K$ band, is equivalent in $(u,v)$ space to increase the
baselines by 
a factor of 4).  The source declination, observatory latitude and hour
angle determine how the baseline vector projects onto the sky as the Earth
rotates. Thanks to the Earth rotation, different points in the $(u,v)$
space become accessible.

Finally, the number of telescopes defines the instantaneous number of available
baselines and, therefore, the instantaneous number of points where the
visibility can be measured. Earth rotation can in some circumstances overcome
the limitation of a small number of telescopes. However, the maximum increase in
number of $(u,v)$ points due to rotation is a factor of 10.  In contrast, the
increase of number of telescopes implies that the number of $(u,v)$ points can
increase factorially.  In order to use all the available telescopes, a focal
instrument must be available to combine them.

Once the telescopes positions are defined, an instrument (recombiner) will
combine the light and detect the fringes.  Two classes of recombiners are
available.  One class disperses the fringes and detects a visibility spectrum
$\mathcal{V}(\lambda)$ at each $(u,v)$ point. This is, in a sense, analog to a
spectrograph. The other class detects the visibility in a narrow wavelength
range (with one or very few wavelength ``pixels'') and is, in a sense, analog to
a filter based system.

The performance or limiting magnitude of an interferometer is usually defined
for an unresolved source ($\mathcal{V}=1$).  At the detector the fringe contrast
is the visibility.  
The observation of an object with a larger angular diameter implies a decrease
of the visibility amplitude and thus also of the fringe contrast at the
detector. 
The limiting magnitude decreases by
$-2.5\log(|\mathcal{V}|)$. In practice, several aspects determine the limiting
magnitude of an interferometer, the most important one for asteroseismology
being fringe tracking\footnote{The ability to freeze the fringes on the detector
and integrate for larger times $\lesssim 100$~s} and the availability of
adaptive optics at the telescopes.\footnote{The intensity loss is proportional
to the strehl ratio at the telescope. Therefore, the capacity to reach lower
wavelengths (and therefore higher angular frequencies in $(u,v)$ space) is
increased.}

The scientific results discussed in Section~\ref{stphys} provide an overview
of the capabilities of today's interferometers -- and, by inference, of their
limitations. For the precise determination of fundamental stellar parameters a
combination of good spatial and spectral resolution, high sensitivity, broad
wavelength coverage (including access to the visible wavelength range), and
high operational efficiency (to facilitate the calibration) is vitally
important. Unfortunately, none of the present interferometers provides all of
these characteristics simultaneously.

At the time of writing this article (mid-2007), seven optical and infrared
interferometers are in operation; the main properties of these facilities are
summarized in Table~\ref{insttab}. The results presented in Section~\ref{stphys}
have been obtained with these instruments and a few additional ones that have
already been closed, notably the Mk\,III Interferometer on Mt.\ Wilson in
California and
the IOTA on Mt.\ Hopkins in Arizona.
We refer the 
reader to the interferometer web
pages for up-to-date information\footnote{Interferometers in the WWW:
  \begin{itemize}
    \item VLTI: {\tt http://www.eso.org/projects/vlti/instru/}
    \item CHARA: {\tt http://www.chara.gsu.edu/CHARA/}
    \item KI: {\tt http://planetquest.jpl.nasa.gov/Keck/keck\_index.cfm}
    \item NPOI: {\tt http://www.nofs.navy.mil/projects/npoi/}
    \item SUSI: {\tt http://www.physics.usyd.edu.au/astron/susi/}
  \end{itemize}
}.  
Table~\ref{insttab} clearly shows that the limiting magnitude of most
interferometers does not depend on the aperture diameter and is currently at
$m_{K}=6-7$.  Visibilities as small as ${\mathcal V}^2=2\times10^{-4}$ have been
measured \citep{Monnier2007}.  
It is also evident from Table~\ref{insttab} that
there are large differences in the capabilities of the various facilities. In
addition to the parameters listed, one has to consider the available spectral
resolution, stability of the fringe servo system, speed of acquisition and
observing efficiency, site characteristics, calibration procedures, and other
factors for assessing the suitability of a specific facility for any given
scientific question.  Particularly interesting for asteroseismology is the
combination of a large number of telescopes.

\begin{sidewaystable}
\caption{\label{insttab}  Properties of  major operational
  interferometers   and  associated   instrumentation,   adapted  from
  \cite{McAlister2000}. The  limiting magnitudes quoted  are those for
  the  most  sensitive  instrument  configuration, for  an  unresolved
  object (published data  or  available documentation  were used  to
  determine the values presented). The limiting magnitudes 
should be considered with caution,
  since experimental facilities such as the KI do not have the same stringent
  requirements as general user facilities, such as the VLTI. For interferometers
  with hybrid telescope sizes curved brackets are used.  }
  \centering
\tabcolsep=2pt
\begin{tabular}{lccccccc}
\hline\noalign{\smallskip}
Interferometer & number of  & telescope    & maximum
&operating
                   &limiting  &spectral & Ref.\\[3pt]
\& Instrument    & telescopes & diameter (m) & baseline (m)
&wavelength ($\mu$m)&magnitude &resolution & \\\hline
\tableheadseprule\noalign{\smallskip}
VLTI (Paranal, Chile)     &4+(4)&8 (1.8)&130 (200)&0.45-20  &n.a.&n.a. & G00\\
VLTI/AMBER&    3&n.a.   &130 (130)&1.0-2.5  &$m_{\rm K}=7 (5)$  &30-12000 & \\
VLTI/MIDI &    2&n.a.   &130 (130)&   8-13  &$m_{\rm N}=4 (0.7)$&30-230 & \\
SUSI (Narrabri, AUS)  &   11&   0.14&      640&0.4-0.9  &n.a.     &n.a. & D99\\
SUSI/red  &    2&   n.a.&       80&0.7      &$m_{\rm V}$=4
&$\Delta\lambda=80$~nm & \\
SUSI/blue &    2&   n.a.&       50&0.44     &$m_{\rm B}$=3
&$\Delta\lambda=1.5$~nm & \\
KI/K (Mauna Kea, HI)     &    2&     10&         85&2.2    &$m_{\rm K}$=8
&$\Delta\lambda=1.5$~$\mu$m & C04\\
PTI (Mt. Palomar, CA)      &    2&    0.4&        110&1.6-2.2&$m_{\rm K}$=6
&$\Delta\lambda=0.1$~$\mu$m & C99 \\
NPOI  (Flagstaff, AZ)    &    6&    0.12&       64&0.55-0.85&$m_{\rm V}$=6
&$\Delta\lambda=19$~nm & A98\\
CHARA (Mt. Wilson, CA)    &    6&    1.0&       331&0.45-2.4 &n.a&n.a.& T05\\
CHARA/Classic/FLUOR&    2&n.a.&       331&1.6-2.4&$m_{\rm K}$=6
&$\Delta\lambda=350$~nm & \\
CHARA/MIRC&     4&  n.a.&       331&1.5-2.5&$m_{\rm K}$=6&20-300 & \\
\noalign{\smallskip}\hline
\end{tabular}\\
{\small
References: 
A98: \citet{arm98};
C99: \citet{col99};
C04: \citet{col04};
D99: \citet{davis99};
G00: \citet{glindemann00};
H04: \citet{hale04};
T05: \citet{tenBrummelaar}
}
\end{sidewaystable}

Most technical information on optical/infrared interferometers gets published in
the proceedings of the conferences on this topic that are part of the bi-annual
SPIE conferences on astronomical instrumentation. The most recent volumes in
this series are \citet{lena00,traub03,shao03,traub04} and \citet{monnier06}.
The reader is referred to these proceedings for more details on existing,
planned, and historical interferometers. Summary overviews can also be found in
the reviews by \citet{quirrenbach01a} and \citet{monnier03}.

\subsubsection{Interferometric images in the optical domain}

Most of the astrophysical results summarized in Section~\ref{stphys} have been
obtained by fitting model parameters to visibilities observed with
interferometers comprised of a single baseline or at most a few baselines. This
approach works very well if a physical model of the target can be constructed a
priori (e.g., a uniform or limb-darkened disk, a von Zeipel model of a rapid
rotator, or a binary), but it becomes problematic if the brightness distribution
is not known beforehand. For example, the thin atmospheres and low gravities of
Mira stars result in strong deviations from spherical symmetry, which can be
detected with very limited data sets \cite[e.g.,][]{Quirre1992, rag06}.  However,
for a detailed analysis and an understanding of the causes of these asymmetries
one would have to obtain true images that show the shapes and brightness
distribution of these ``jellyfish-like'' objects.

As pointed out in Section~\ref{synthesis}, interferometric imaging is possible
only with arrays that provide good coverage of the $(u,v)$ plane, and
sophisticated data analysis algorithms are needed to reconstruct images from
visibility data \cite[e.g.,][and references therein]{bald02,quirrenbach01a}.
While these theoretical foundations of optical aperture
synthesis imaging are well-understood, this technique is still in its infancy
because of the limited number of telescopes available in present-day arrays
(see Table~\ref{insttab}), and because of the difficulties inherent in building
and operating complex many-telescope beam combiners.

The first images reconstructed from optical interferometers only
showed very simple
objects such as the binary Capella \citep{bald96}; the capabilities of the
six-telescope beam combiner of NPOI was first demonstrated by imaging the
$\eta$\,Vir triple system \citep{hu03}.  More recently, the surface of Altair
has been imaged with the CHARA Array, directly showing the rotational flattening
and von Zeipel effect \citep{Monnier2007}.  It is certainly encouraging to see
interferometric images that carry truly useful astrophysical information, and
with the emergence of more instruments capable of combining four or more
telescopes simultaneously, the impact of such images is expected to increase in
the near future.

The combination of data taken in multiple spectral channels can further help to
increase the $(u,v)$ plane coverage, since the spatial frequency is inversely
proportional to the wavelength. This technique obviously works best if the
source structure is independent of wavelength, as for example in binary systems
with two components of equal spectral type \citep{bens97}.  A generalization is
possible if the wavelength dependence can be easily parameterized, such as a
field of unresolved stars or a spotted stellar surface, which can be described
by brightness and colour for each point.

\subsection{Limitations of interferometers for the measurement of stellar
diameters\label{limitinter}}

We focus now on the fundamental limits of the accuracy of visibility  
modulus
measurements in the context of stellar diameter measurements.  
Visibility phases
are not discussed here as they are not used to constrain diameters.  
Whenever we
speak of an {\it accurate\/} visibility modulus measurement, we mean  
that this
measurement is {\it precise\/} and {\it unbiased}. It is important to  
make this
distinction, because an estimate may be based on measurements with a  
very small
statistical dispersion but with a large bias. In this case, it cannot be
regarded as accurate. Reaching high accuracies therefore requires to  
be able to
reduce biases to magnitudes smaller than the final error bar.

\subsubsection{Sources of visibility errors and biases}

Interferometers do not provide a direct measurement of the spatial  
coherence of
a source, because causes of coherence loss can be numerous,
the major ones being
atmospheric turbulence, polarizations, longitudinal dispersion and
vibrations. All these sources of loss will imply that
visibilities are observed with a reduced value compared with the case  
where no
losses occur.
As a consequence, it is possible to calibrate the losses by normalizing
visibilities with the help of the
fringe contrast observed for calibrator stars, as long as
the coherence losses for the target and for the calibrator are  
sufficiently
similar, 
\begin{equation}
{\cal V}_{\rm obj}=\frac{C_{\rm obj}}{T} =\frac{C_{\rm obj}}{C_{\rm cal}} 
\times {\cal V}_{\rm
   cal}^{\rm exp},
\label{eq:cal}
\end{equation}
where $T$ is the transfer function or the point source response of the
interferometer defined in Eq.(2) of \citet{perrin03} 
while $C_{\rm obj}$, $C_{\rm cal}$ and ${\cal V}_{\rm cal}^ 
{\rm exp}$
are the fringe contrast measured on the source, the fringe contrast  
measured on
the calibrator, and the expected visibility of the calibrator,  
respectively.
For a point-like calibrator, we have $T=C_{\rm cal}$.  
The two fringe  
contrast
measurements are subject to various noise sources (detector noise,  
photon noise
and background noise) and biases which are discussed below. The expected
visibility of the calibrator has an intrinsic error and is estimated  
from a
model with an uncertainty and a potential bias. To reduce sources of  
errors due
to calibrators and instabilities of the interferometer response, the  
object
observations are often bracketed by observations of several calibrators.

\begin{figure*}
\centering
  \includegraphics[width=0.75\textwidth]{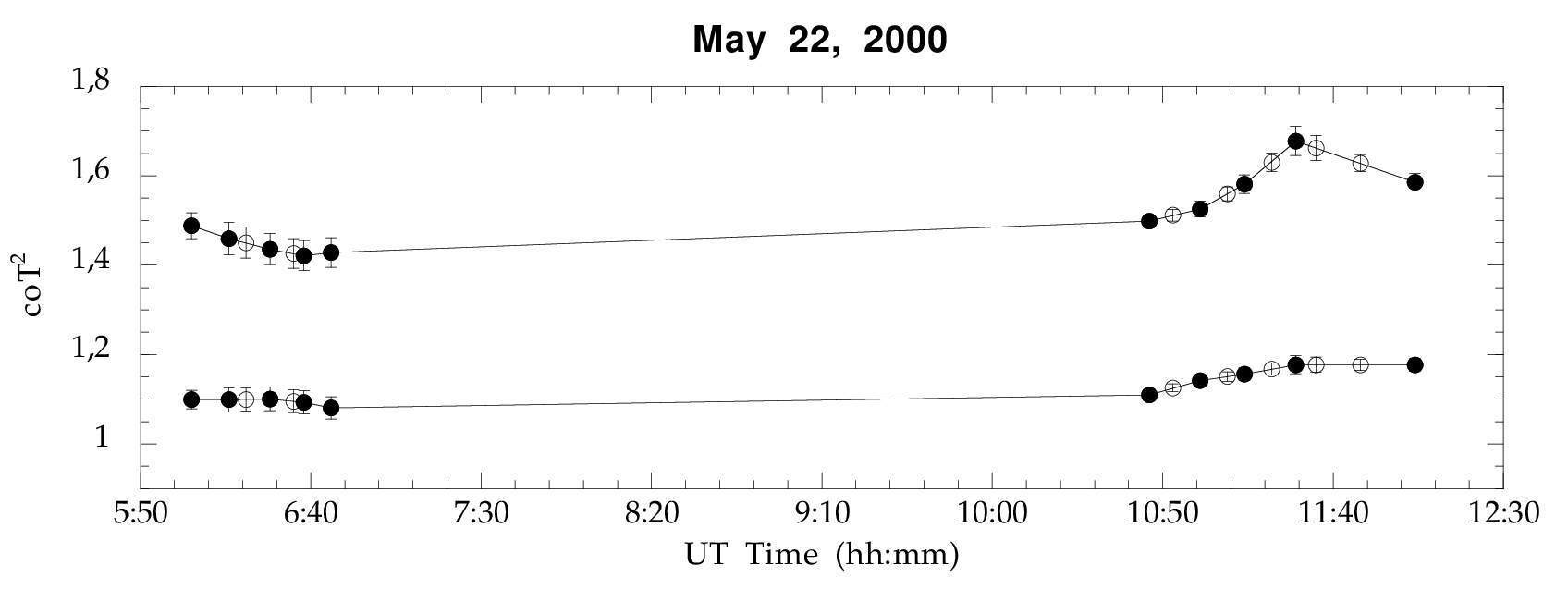}
  \includegraphics[width=0.5\textwidth]{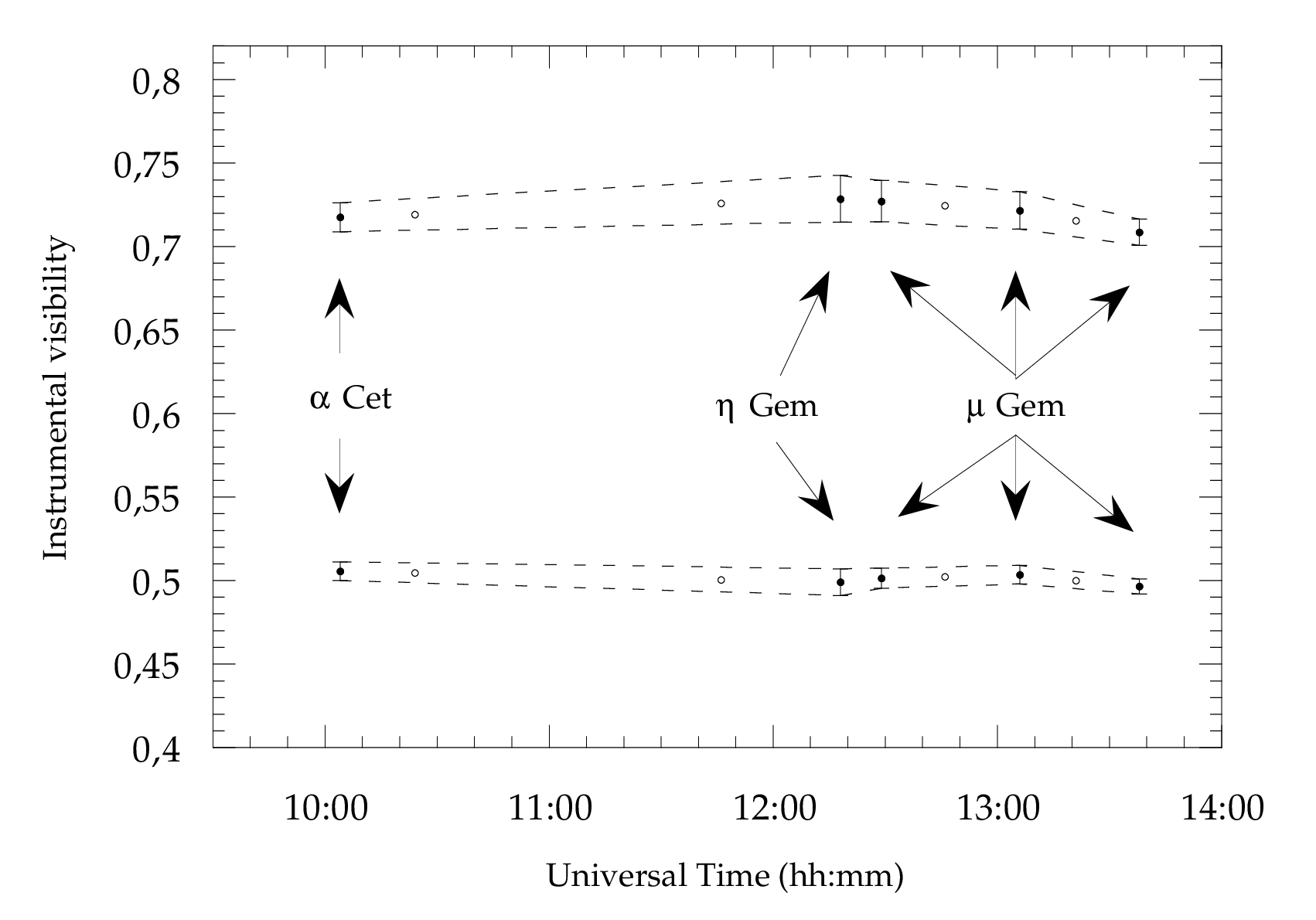}
\caption{Examples of transfer functions of the FLUOR instrument at IOTA. Top
panel: squared co-transfer function for the two channels of the interferometer.
The calibrator measurements are indicated by filled circles; the open circles
represent interpolated values at the time the scientific source was observed
\citep[from][with permission]{perrin03}. Bottom panel: the instrumental visibility, which is
the visibility measured for three different calibrators. The dashed lines are
the $1\sigma$ upper and lower limits. The corresponding transfer function is
stable to within $1\sigma$ on a time scale of 2 hours
\citep[from][with permission]{perrin04}. }
\label{fig:T}       
\end{figure*}

\subsubsection{Atmospheric turbulence and spatial filtering}

The paramount source of error and degradation of the accuracy on the  
visibility
measured by interferometers is atmospheric turbulence. In regular  
conditions on
the ground, only a fraction of the incident photons are coherent.  
This fraction is given by the coherent energy, which depends on the
turbulence strength, the zenith angle and the considered wavelength \cite[see e.g.,][for a definition and discussion]{quirrenbach00}. The coherent energy
is typically of order 30\% but can be increased somewhat by using
adaptive optics.
The coherent energy is a fluctuating quantity because the seeing is fluctuating. It is,
therefore, difficult to calibrate and this implies a strong  
limitation on the
quality of visibility measurements. Turbulence at the scale of the
interferometer, i.e.\ the average phase over each pupil, also plays a
role. These phase differences are called (differential) piston  
errors. They
induce jitter of the fringe pattern, which degrades visibilities.  
Vibrations
have effects similar to turbulence on the stability of  
interferometers, which is
why they are usually merged with turbulence, as we will assume in the  
following.

The interferometer's response can be written as the product of a  
slowly varying
response $T_{\rm inst}$ with an unstationary atmospheric response.  
The latter
comprises the coherent energy $T_{\rm atm}=E_c$ and the piston effect  
$T_{\rm
piston}$. One way to help solve the issue of fluctuating coherent  
energy is to
benefit from adaptive optics or to reduce the telescope diameter  
below $r_0$. In
either case, residual phase fluctuations remain. The best procedure  
consists in
spatially filtering almost flat wavefronts with a pinhole, or, even  
better, with
a single-mode fiber. A single-mode fiber flattens the wavefronts or more 
exactly selects the fraction of incident lightwaves whose wavefronts are flat.  
This operation
thus trades phase fluctuations against intensity fluctuations that  
can easily be
monitored. The new interferometer response becomes:
\begin{equation}
T=T_{\rm inst}\times T_{\rm piston}\times\frac{2\sqrt{P_AP_B}}{P_A+P_B},
\end{equation}
with $P_A$ and $P_B$ the intensities coupled in the fibers at the  
focus of
telescopes $A$ and $B$, respectively. As these intensities are  
measured in real
time, the response of a single-mode interferometer is $T=T_{\rm inst} 
\times
T_{\rm piston}$. This has led to great improvements in the  
measurements of
diameters.The piston term can be further reduced by servoing the  
fringe position
with a fringe tracker or by freezing piston fluctuations with an  
adequately
short exposure time. Under these circumstances and after calibration,  
the piston
effect is more a source of noise than a bias.

\subsubsection{Longer time scale sources of errors}

The instrumental parts of the coherence losses, apart from vibrations,  
have long
time scales of variation. They are mostly due to polarization and, to  
a lesser
extent, to longitudinal dispersion.

Polarizations with different orientations will not perfectly  
interfere nor will
they necessarily produce fringes if they are crossed. Differential  
birefringence will  
cause
interferograms in two orthogonal polarization axes to be shifted and  
to add up
incoherently, the worst case being a shift of half a fringe. Although  
the
effects can be dramatic, recipes are well known to solve the issue,  
e.g.,\ the
use of identical or symmetric optical trains are enough to guarantee  
very high
fringe contrasts.

The nature of longitudinal dispersion is to make the zero optical path
difference wavelength dependent. This happens when the index of  
refraction is
different in the two arms of the interferometer or if the two arms have
different lengths and if the index of refraction is chromatic. The  
former case
may happen when single-mode fibers are used to propagate the beams.  
In this
case, dispersions need to be matched by adjusting the fiber lengths  
and using
homogeneous fibers. The latter case may happen in the bluest part of the
spectrum when propagating beams in unevacuated and unequal  
interferometer
arms. In both cases, solutions are well-known and these degradations  
can be
avoided.

As long as the calibrator(s) and the science target are subject to  
the same
coherence losses (which assumes similar reflections on mirrors, e.g.,  
and,
therefore, the sources to be nearby and observed with a short time  
difference if
the interferometer properties are temperature dependent) $T_{\rm inst} 
$ is quite
easy to measure. Fig.~\ref{fig:T} shows examples of measurements with  
the
single-mode interferometer FLUOR at IOTA.

\begin{figure*}
\centering
\includegraphics[width=0.7\textwidth, bb=60 58 565 775 , angle = -90] 
{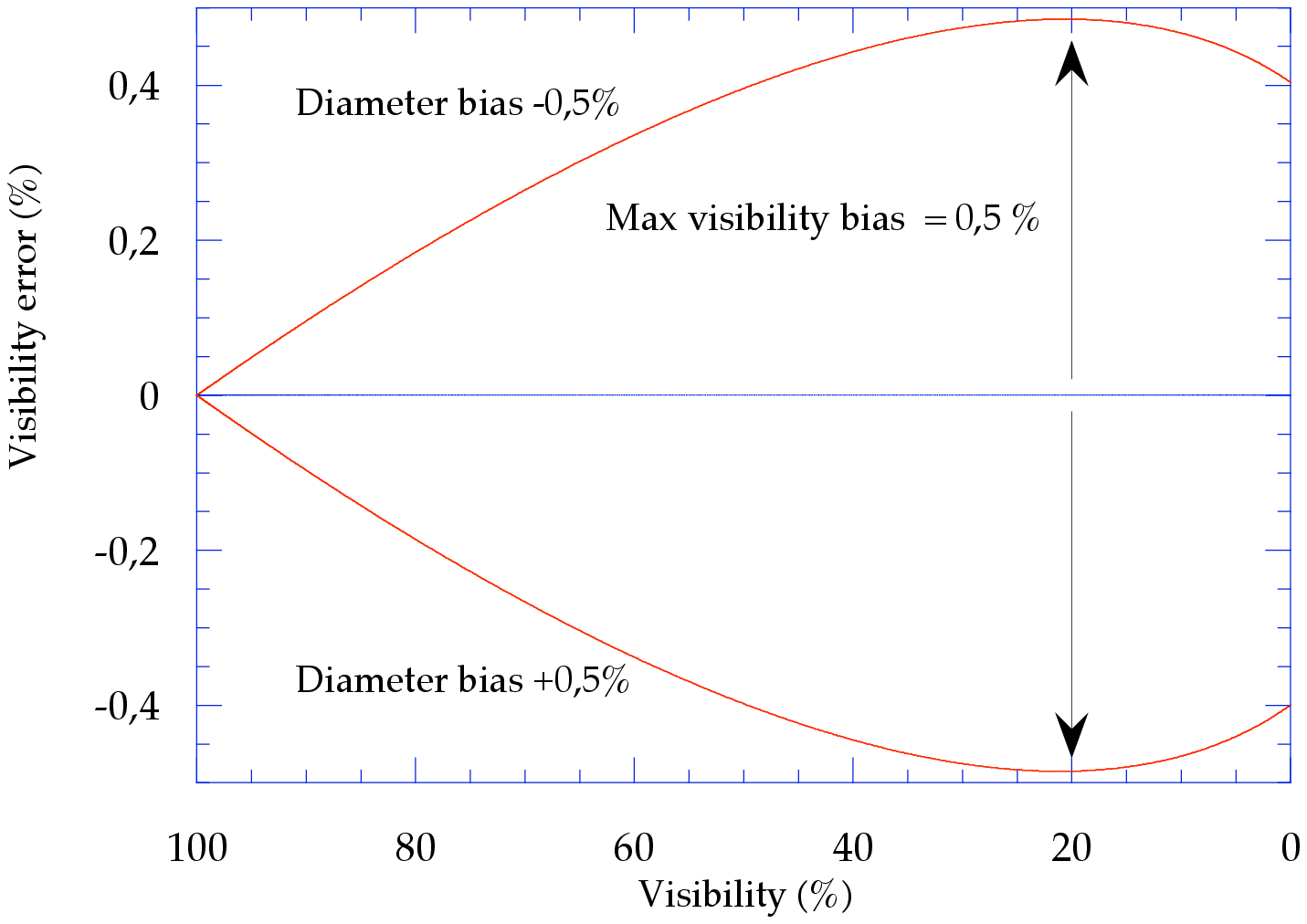}
\caption{Error on the predicted visibility of a uniform disk as a  
function of
visibility for a relative error on the diameter of 0.5\%. The error  
remains
smaller than 0.1\% as long as the visibility value is above 90\%.}
\label{fig:error}       
\end{figure*}

\subsubsection{Calibration and calibrators}

As we have discussed, atmospheric errors can be eliminated by going to space, or
thanks to single-mode optics. Vibrations can be solved by servoing optical
paths. Polarization and dispersion can be tackled by a good design of the
interferometer. One remaining difficulty is the computation of the expected
visibility of calibrators.  As Eq.~(\ref{eq:cal}) shows, the ultimate error on
the visibility will come from this estimate, even if other sources of errors
have been perfectly calibrated. An ideal calibrator is a point-like source as
its visibility is exactly 1. However, such a source emits no photons.  Realistic
sources have a finite diameter. Errors on the expected visibility may come from
uncertainties on the source angular diameter but also on its geometry. It is
necessary to assume spherical symmetry for the source and it is mandatory that
this assumption is correct at the resolution of the interferometer, i.e., one
must make use of a simple featureless star with a well known diameter and a very
compact atmosphere. Assuming such a star, one difficulty to compute expected
visibilities arises from limb-darkening as it changes the apparent angular
diameter from one wavelength to another and a correction is
required. Theoretical models exist \citep[e.g.,][]{cl02} and are tested against
interferometric data \citep[e.g.,][]{Wittko2004}. The effect is lower at infrared
than at visual wavelengths because the difference between a uniform and a
limb-darkened disk is only of order 2\% at red wavelengths.

Angular diameters, and therefore visibilities, can be predicted with two
different methods. The first one is a direct method based on a direct
measurement of the source angular diameter and the visibility is extrapolated
using a uniform or a limb-darkened disk model. The best demonstrated accuracy on
angular diameter measurements is 0.5\% in the $K$ band for $\alpha$~Cen A and B
\citep{kervella03b}. Extrapolating this to the longest baseline of VLTI (200~m)
and to the $J$ band leads to the measurements of angular diameters with an
accuracy of 0.5\% for all stars larger than 1~mas. This provides a list of
primary calibrators but is not enough to calibrate all VLTI observations for
example at the highest accuracy of 0.1\%, as the required calibrators then are
rather 0.1~milliarcseconds (mas) in diameter (see below).

Secondary calibrators whose angular diameters are computed with indirect methods
are required.  They aim at predicting a zero-magnitude stellar diameter
$\theta_{\rm zm}$ as a function of a color index or a spectral type. Stellar
diameters for any magnitude then follow from $\theta_{m}= \theta_{\rm
zm}\times10^{-m/5}$. The typical error is 5\% if all stars are taken into
account. The prediction error can be reduced to 1.2\% for carefully selected A0
to M0 giants using accurate photosphere and atmosphere modelling
\citep[][]{borde02}. \citet[][]{kervella04c} have shown that this can still be
improved using surface brightness relationships for selected dwarfs. These
relationships are calibrated with interferometric measurements. The best
correlation is found for dereddened ($B-L$) colours: the residual prediction
error is better than 1\% and can be smaller than 0.5\%. Fortunately, metallicity
does not seem to play a role, which eliminates one potential source of bias.

As a conclusion, one may consider that the state-of-the-art accuracy for the
prediction of diameters for single stars is 0.5\%. The accuracy on the derived
visibility will then depend on the resolution of the
interferometer. Fig.~\ref{fig:error} shows the error on the predicted visibility
assuming a 0.5\% error on the diameter as a function of the calibrator
visibility or, equivalently, to the spatial resolution. The error is always
smaller than 0.5\% and is smaller than 0.1\% as long as the visibility is larger
than 90\%. This sets the minimum level of visibility for calibrators to keep the
calibration error of the same magnitude or smaller than other errors.

\subsubsection{Fundamental limits on accuracies}

Sources of errors are many and only major ones have been discussed  
here. More
thorough discussions would be beyond the scope of this paper.  
However, the
reader has to be made aware that other errors may arise when modelling  
the
visibilities. For example, one must be careful and compute models  
with the same
estimator as the visibility estimator to derive the correct diameter.  
This is
all the more important as the band is wide. Such effects are  
discussed in \citet[][]{perrin05}. Also, as discussed in this same paper, the  
sensitivity to an
effect such as piston or chromaticity depends on the choice of the  
visibility
estimator. From Eq.~(\ref{eq:cal}) one can establish the following  
error
budget,
\begin{equation}
\frac{\sigma_{{\cal V}_{\rm obj}}}{{\cal V}_{\rm obj}}=\sqrt{\left[ \frac{\sigma_{C_ 
{\rm
	  obj}}}{C_{\rm obj}}
\right]^2 + \left[ \frac{\sigma_{C_{\rm cal}}}{C_{\rm cal}} \right]^2  
+ \left[
\frac{\sigma_{{\cal V}_{\rm cal}^{\rm exp}}}{{\cal V}_{\rm cal}^{\rm exp}} \right]^2}
\end{equation}
The first two errors can be very small for the brightest sources (photon and
detector noise) and are limited by piston noise. At this stage, vibrations can
be assimilated along with piston errors. Piston errors can be further reduced
with a fringe tracker. The third error term is the dominant one for the
brightest sources. In practice it currently is of the order of 0.1\%.  It is,
therefore, the fundamental limitation on visibility accuracies. One way to
narrow down or clear its influence and push the limits of accuracy further, is
to perform a relative calibration as described in Section~\ref{calib}. With
relative calibration one avoids to rely on the exact value of the diameter of
the calibrator.

\subsubsection{\label{calib}Relative calibration}

Relative calibration is an efficient way to search for relative variations of
parameters in sources such as those that can be the goal in asteroseismology,
where one could attempt to measure the stellar diameter variations due to
oscillations.  These diameter variations may be small and even below the
visibility accuracy imposed by calibrators. The way to get around this
difficulty is to use a single stable calibrator. Different visibilities recorded
at different baselines and/or different times will share the same calibration
fluctuations because the calibration errors are correlated from one measurement
to another.

The contributions of correlated and uncorrelated noise for  
interferometric
measurements were already described in \citet[][]{perrin03}. When fitting a  
model of
the source, absolute parameters (such as the average diameter) will  
have an
accuracy limited by the accuracy on the calibrator diameter, since  
correlated
errors will not average out. Relative parameters, such as those due  
to the
relative amplitude of the oscillations, will, however, only be  
limited by
photon, detector and piston noise. These noise sources can be reduced by
adopting longer exposure times. This can be easily understood with  
the following
simple example. Let us assume that the diameter of a pulsating star  
with a
relative pulsation amplitude of $10^{-3}$ in diameter is compared with
that of
quiet stars whose diameters are known to within an accuracy of 10\%.  
If the
diameter of the pulsating star is compared with a different calibrator  
at each
observation, then the small pulsation will not be detected because  
the scatter
in diameter measurements will be at least 10\% divided by the square  
root of the
number of calibrators. If a single calibrator with poorly known  
diameter is
chosen for each observing period, then the absolute effect of the  
pulsation on
the diameter will still not be detected. However, the oscillation,  
measured in
units of the calibrator diameter, will be detected if the noise  
permits it.
This technique may turn out to be powerful for this kind of application.

\subsection{Future projects\label{futureinterf}}

Most of the facilities listed in Section~\ref{presentinf} are planning upgrades
to their respective infrastructure, and new beam combining instruments. A new
imaging interferometer is currently under construction at the Magdalena Ridge
Observatory \citep{Busher2006}.  In the following sections we describe possible
future developments that could open new vistas for interferometric studies of
the structure and surfaces of stars.

\subsubsection{The future of the VLTI}

The VLTI is unique among all interferometers worldwide in that a large number of
institutions are contributing their expertise to its instrumentation programme,
and in its ambition to provide an interferometric facility that is accessible to
a large number of users and well-suited for diverse observing programs. At
present, two instruments are operational at the VLTI: AMBER \citep{petrov07} and
MIDI \citep{lein03}.  AMBER is a near-infrared instrument working in the $J$,
$H$, and $K$ bands with a spectral resolution up to 10,000. AMBER can combine
the beams from three telescopes simultaneously and is thus capable of measuring
closure phases. MIDI is a single-baseline instrument that measures correlated
spectra with a resolution up to 230 in the $N$ band ($8$ to $13\,\mu$m).

The PRIMA (Phase-Referenced Imaging and Microarcsecond Astrometry) facility
\citep{Quirre1998,delpl06} will substantially enhance the capabilities of the
VLTI by providing real-time fringe tracking and co-phasing.  The astrometric
mode of PRIMA will be used for an extensive extrasolar planet survey aimed at
measuring their masses and orbits \citep{Quirre2004}.
The phase-referenced imaging mode of PRIMA will push the sensitivity limit of
the VLTI by several magnitudes, because once the interferometer is co-phased,
its point-source sensitivity (within the isoplanatic patch) is close to that of
a single telescope with equal collecting area. In this observing mode, PRIMA
can also serve as a fringe tracker for AMBER and MIDI.

At present, concepts for three second-generation VLTI instruments are being
studied by international consortia: GRAVITY \citep{gil06}, MATISSE
\citep{lop06}, and VSI \citep{mal06}.  GRAVITY is a dual star instrument
optimized for precision narrow-angle astrometry and interferometric phase
referenced imaging of faint objects within a $2''$ field. Its science case is
centered on measuring motions in dense stellar clusters, in particular in the
cluster surrounding the Galactic Center. MATISSE will be a successor of MIDI,
capable of combining 
four telescopes simultaneously, and expanding the wavelength
range shortward to 2.7\,$\mu$m, thus including the $L$ and $M$ bands in addition
to $N$. The most important scientific drivers for MATISSE are studies of the
formation of stars and planetary systems, and of dust shells around late-type
stars. 
By combining up to six telescopes, VSI, the successor of AMBER, will bring
true interferometric imaging and spectro-imaging in the near-infrared to the
VLTI, with applications to stellar surface imaging as well as detailed
investigations of circumstellar matter around young and evolved stars.

\subsubsection{Interferometric high-resolution spectroscopy}

\begin{figure}[t]
\begin{center}
\begin{tabular}{c}

\includegraphics[height=7.4cm]{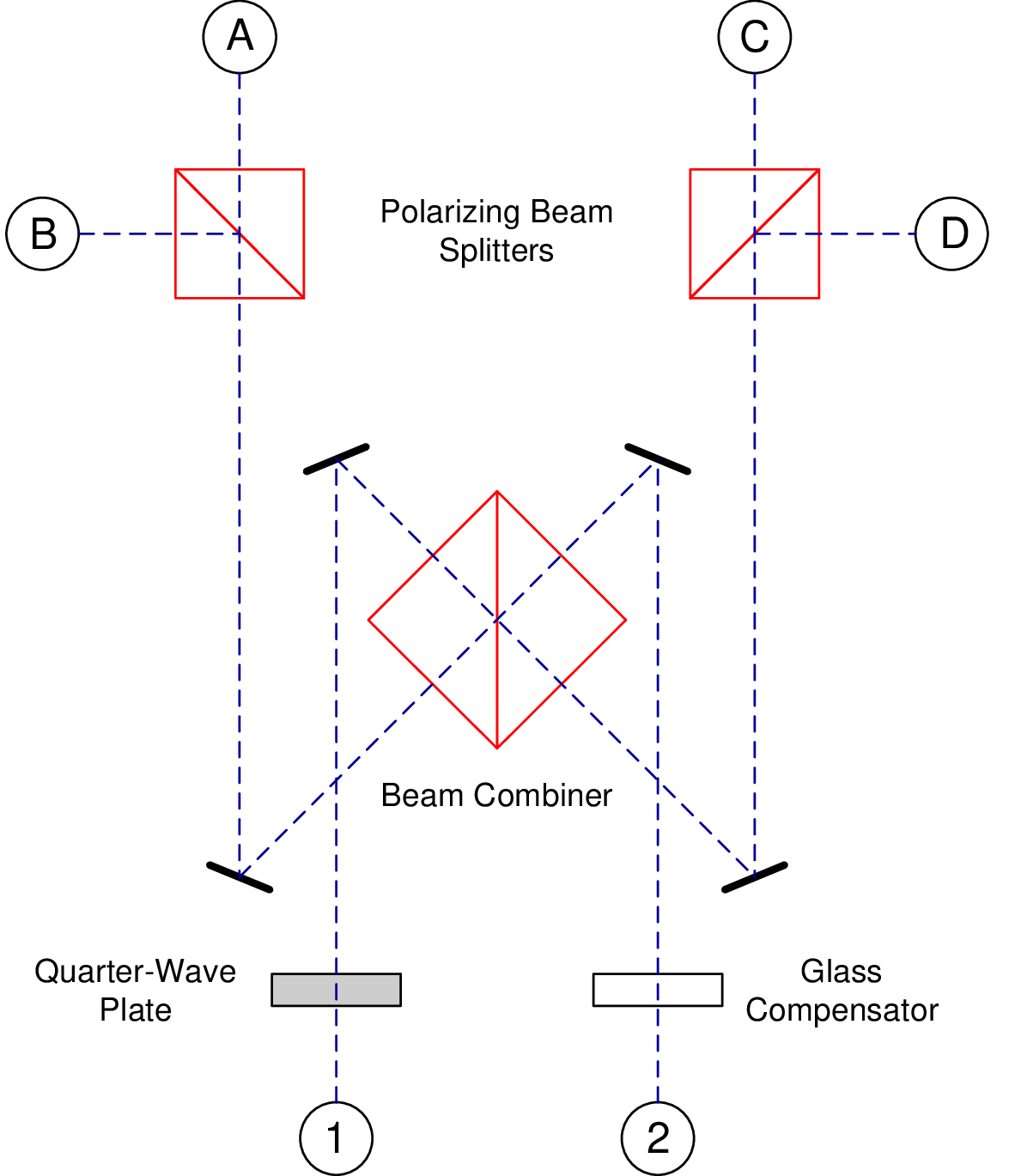}
\end{tabular}
\end{center}
\caption
{\label{beamcomb} Sketch of a four-output beam combiner. An external fringe
tracker stabilizes the pathlength difference between the beams from two
telescopes, which are relayed to the inputs at the bottom of the figure. The
quarter-wave plate produces an achromatic $\pi / 2$ phase shift between the two
polarization states in one beam. The beams from both telescopes are combined at
the main beam combiner; light emanating from the right-hand side of the beam
combiner is shifted by $\pi$ with respect to the left side. The two
polarizations in each arm are separated by polarizing beam splitters, and the
four outputs coupled into fibers for relay to detectors or spectrograph inputs.
The phases in the four outputs $A$, $B$, $C$, and $D$ are shifted by 0,
$\pi/2$, $\pi$, and $3\pi/2$, respectively.}
\end{figure}

Frequently interferometry is regarded as a method that can provide
high-resolution {\it images}; the possibility of obtaining both high {\it
spatial} and {\it spectral} resolution simultaneously with the same observation
is often overlooked. Such observations enable many completely new approaches to
fundamental questions in stellar astrophysics. Examples of astrophysical
problems that can be addressed by interferometric spectroscopy include:
\begin{itemize}
\item{{\it Late-Type Giant Stars:} Measuring the variation of the stellar
diameter with wavelength, or even better wavelength-dependent limb darkening
profiles, provides a sensitive probe for the structure of strongly extended
atmospheres of cool giant stars. Such data can be directly compared with
predictions of theoretical models, and provide qualitatively new tests of
stellar model atmospheres \citep{Quirre1993,Quirretal04}.}
\item{{\it Mira stars:} Mira variables possess exceedingly complicated
atmospheres, with global asymmetries that change on time scales as short as a
few weeks \citep{Quirre1992}. Tracing the propagation of shocks in
these atmospheres gives new information on their density structure, and on
pulsation mechanisms.}
\item{{\it Cepheids:} It is possible to perform direct measurements of the
projection factors of Cepheid pulsations, which relate the true velocity of the
pulsation to the observed radial velocity. Uncertainties in the projection
factor 
${\cal G}$,
which presently has to be computed from theoretical models, are a serious
limiting factor in current estimates of Cepheid distances with the
Baade-Wesselink method \citep{sabb95,mar02}.}
\item{{\it Generalized Doppler Imaging:} The chemical and magnetic properties of
stellar photospheres can be mapped with classical Doppler imaging
\cite[e.g.,][]{rice02,Kochukhov04,Kochukhovetal04}.  However, the reconstruction of stellar
surface features from line profile variations alone is plagued with
ambiguities. These can largely be resolved by the additional phase information
contained in interferometric data \citep{Jankov01}.  For stars with
inhomogeneous surface properties, interferometry enables studies of individual
surface regions.}
\item{{\it Rotational Axes:} Determining the rotation axis orientation of the
components in binary systems can provide new test for binary formation
theories. Stellar rotation induces a phase difference between the red wings and
the blue wings of stellar absorption lines. Measuring the direction of the
phase gradient on the sky allows determination of the orientation of the
stellar axis \citep{petrov89,chelli95}.
More detailed modelling of
the interferometric signal can also provide the inclination of the stellar
rotation axis with respect to the line of sight \citep{Domici2004}.
One can thus check whether the rotation axes of binaries are aligned
with each other, and with the orbital angular momentum.}
\item{{\it Differential Rotation:} Along with the oscillation spectrum, surface
differential rotation is a powerful diagnostic of the interior structure of a
star. Unfortunately, it is difficult to observe surface differential rotation
with classical spectroscopy because of degeneracies between inclination, limb
darkening, and rotation \cite[e.g.,][]{gray77}.  These degeneracies can be
resolved by the additional information from interferometric spectroscopy
\citep{Domici2004}.  High spectral resolution is essential for this
application.}
\item{{\it Circumstellar disks:} Velocity-resolved interferometric observations
in emission lines can be used to determine the geometry and velocity field of
outflows from pre-main-sequence stars and of disks around young stars and Be
stars.}
\end{itemize}
The required spectral resolution depends on the width of the stellar lines under
consideration, of course. For cool stars with narrow lines, the information
content increases dramatically with spectral resolution, calling for an
interferometric instrument providing a resolution $ \gtrsim 50,000$,
substantially more than what is available today.

Scanning through the fringe packet can in principle provide the desired
spectroscopic information; the envelope of the fringe packet is the Fourier
transform of the correlated source spectrum (multiplied by the bandpass of the
instrument). This technique is known as double-Fourier interferometry 
\citep{mar88,mek90}; it is appealing because it does not
require any special hardware in addition to the scanning delay line. This
method should work quite well for observations of strong isolated emission
lines, but it is unsuitable for measurements of stellar absorption line
spectra. In this case, the spectral resolution and the integrity of the
observed spectrum are most certainly limited by calibration problems, because
the shape of the whole fringe envelope has to be determined very precisely to
allow the extraction of spectral information by Fourier transforming.

It thus appears that standard dispersive spectroscopy is much better suited for
most of the applications delineated above. The spectrograph can easily be linked
to the interferometer with optical fibers \citep{Quirre04a}.  Full information
on the complex visibility ${\cal V}$ is contained in the four fringe
quadratures, i.e., one has to count the photons at phases 0, $\pi/2$, $\pi$, and
$3\pi/2$ \cite[e.g.,][]{shao92}.  A convenient way of generating these four bin
counts $A$, $B$, $C$, and $D$, is shifting one polarization by $\pi/2$ with
respect to the other, as shown schematically in Fig.~\ref{beamcomb}. An
alternative would be a two-stage beam combiner, which sends 50\% of the light
from each interferometer arm to a ``standard'' beam combiner, giving the 0 and
$\pi$ outputs, and 50\% through an achromatic $\pi/2$ phase shifter to a second
beam combiner, which thus produces the $\pi/2$ and $3\pi/2$ outputs. The
achromatic phase shifter can be realized with adjustable dispersive elements
\citep{mier00}.  In either case, for any wavelength $\lambda$, the full
interferometric information is thus contained in the intensities $A(\lambda)$,
$B(\lambda)$, $C(\lambda)$, and $D(\lambda)$ carried in the four output beams of
the beam combiner. These can be routed to a spectrograph with optical
fibers.\footnote{Note that the pathlength of the optical fiber is irrelevant,
because it is located after the beam combiner. Consequently, there is no need to
use a single-mode fiber; multi-mode fibers will do just as well.}

In the case of the VLTI, one could take advantage of the instrument 
UVES/FLAMES,
which already has a fiber feed for multi-object spectroscopy \citep{pas03}.
This instrument provides a resolution of $ \approx 60,000$ when a fiber with
core diameter of 70\,$\mu$m is chosen. The required fiber length for coupling
UVES to the VLTI is $\sim 150$\,m; fibers of this length and of the type used
for the multi-object FLAMES link have a transmission $\ge 80$\,\% over the
wavelength range from 0.6\,$\mu$m to 1\,$\mu$m. The interface between the fibers
and the spectrograph could be very similar to that of FLAMES. A similar setup
would be possible for the infrared high-resolution spectrograph CRIRES in the
$J$ and $H$ bands, or for other interferometers that are co-located with an
observatory at which high-resolution optical or near-infrared spectroscopy is
performed.

\subsubsection{Kilometric Optical Interferometer}

The present generation of interferometers provide baselines up to $\sim 400$\,m,
and thus an angular resolution of a few mas in the near-IR; consisting of two to
six telescopes, they provide only limited coverage of the $(u,v)$ plane (see
Section~\ref{presentinf}).  A number of authors
\cite[e.g.,][]{ridg,arnold02,Quirre04b,lar07} have discussed concepts for a
potential future interferometer array with much better sensitivity, higher
angular resolution, and much-improved imaging capabilities. A preliminary
science case and an assessment of the key enabling technologies for such a
Kilometric Optical Interferometer (KOI) have been compiled by
\citet{sur05,sur06}.

\begin{table}[t]
\caption{ \label{parameters} Summary of strawman parameters and
characteristics for an interferometer with kilometric baselines. For more
details, see \citet{Quirre04b}.}
\begin{center}
\begin{tabular}{lcl}
Parameter & Value & Comment\\
\noalign{\smallskip} \hline \hline \noalign{\smallskip}
Number of  telescopes & 27 & Needed for snapshot imaging \\
Telescope phasing & Autonomous & Adaptive optics, laser guide stars \\
Array co-phasing & External & Dual-star operation \\
Sky coverage & $\gtrsim 10\%$ & At $R$ band, near Galactic pole \\
Telescope diameter & 8-10\,m & Needed to get sky coverage \\
Efficiency & 25\% & To limit telescope size \\
Wavelength range & $0.5 \ldots 20\,\mu$m & Could be reduced to $0.5 \ldots
2.2\,\mu$m \\
\end{tabular}
\end{center}
\vspace{-0.5cm}
\end{table}

With baselines up to ${\cal B} = 10$\,km 
in length and operating at wavelengths down
to $\lambda = 0.5$\,$\mu$m, KOI would deliver images with 10\,$\mu$as
resolution, two orders of magnitude better than any other telescope
contemplated at the moment. Combined with a sensitivity (for compact objects)
that equals or surpasses present-day large monolithic telescopes, this
spectacular angular resolution enables a wealth of completely new observing
programmes, including:
\begin{itemize}
\item{Imaging of Jupiter-size objects (e.g., brown dwarfs) at a distance of
$\sim\,10$\,pc;}
\item{High-quality imaging of stellar surfaces;}
\item{Asteroseismology based on modes of higher degree;}
\item{Detailed imaging of young stellar objects and pre-main-sequence disks;}
\item{Studies of many types of binary systems;}
\item{Baade-Wesselink distances of pulsating stars, novae, and supernovae;}
\item{Dynamics of dense clusters;}
\item{Orbits of stars orbiting the black holes 
in the centers of nearby galaxies;}
\item{Detailed imaging and imaging spectroscopy of broad-line regions in active
  galactic nuclei;}
\item{Geometric distances of quasars through reverberation mapping;}
\item{Resolving the afterglows of gamma-ray bursts.}
\end{itemize}
In the present context it is particularly interesting that observations 
of solar-like oscillations with degree as high as 60 would allow inversion
to infer the structure and rotation of the radiative interior and lower part
of the convection zone in a star like the Sun, including the 
rotational transition layer at the base of the convection zone which is
likely crucial for the operation of the solar dynamo
\citep[e.g.,][]{thomps03}.
The above are just a few examples of what an interferometer with kilometric
baselines could do; many other topics of current interest could be addressed as
well.

A specific example of the concepts mentioned is an interferometric array with 27
telescopes and baselines of up to 10\,km, as presented by \citet{Quirre04b} and
summarized in Table~\ref{parameters}. Such an array could probably be built
today with existing technologies, but the cost would be prohibitively high.
However, advances in telescope building as needed for the construction of
extremely large telescopes with $30$ to $42$\,m diameter, and improvements in
optical fiber technology, could make a next-generation interferometric facility
affordable within the next decade.


\section{The Interferometry-Asteroseismology connection}
\label{sec3}

Some of the major limitations currently faced in asteroseismic studies have been
highlighted at the end of Section~\ref{sec1}. These include the large
uncertainties associated with the determination of global parameters of
pulsating stars and the difficulties in identifying the modes of oscillation,
particularly in classical pulsators. In the present section we address these
issues further, emphasising the unique role that interferometry can play in
finding ways to overcome the aforementioned limitations. We review the first
attempts ever made to combine asteroseismic and interferometric data in studies
of particular pulsators. Moreover, we briefly look at current plans to study
additional pulsating stars for which seismic and interferometric data are
expected to be available in the near future.

\subsection{\label{gp}Improving the determination of global parameters of pulsating stars}
\label{addcontr}

There are different ways in which interferometry may help in reducing the
uncertainties associated with the determination of global parameters of
pulsating stars, hence in helping constrain the models that are to be used in
asteroseismic studies. Besides the direct measurement of the angular diameter of
bright stars in the solar neighbourhood, for which accurate parallaxes are also
available, interferometry can put constraints on the mass and distance of
pulsating stars that are members of binary systems. Moreover, the determination
of radii of stars in the solar neighbourhood, through the combination of
interferometric and astrometric data, will allow us to test for the presence of
systematic uncertainties in the determination of global parameters through other
methods, such as high-resolution spectroscopy or multicolour photometry.  In
this way, the interferometric estimates can provide us with new calibrations
that may subsequently be used to reduce the uncertainties in the values of
global parameters of pulsating stars that are not within reach of current
interferometric instruments.

\subsubsection{Impact of having an interferometric measurement for the radius\label{impact}}

The radius of a star, derived from the combination of its angular diameter and parallax, provides an additional observational constraint that can be used in the fitting procedure described in Section~\ref{dirfit}. However, it remains to be seen whether adding this interferometric constraint will improve in any significant way the results of the fitting.  The question of whether the uncertainties in the fitted model parameters are expected to be reduced by the addition of the interferometric constraint, and, in particular, of which model parameters will benefit most from this addition, was recently addressed by \citet{creevey07}. 

\begin{figure}
\centering
\includegraphics[height = 5cm]{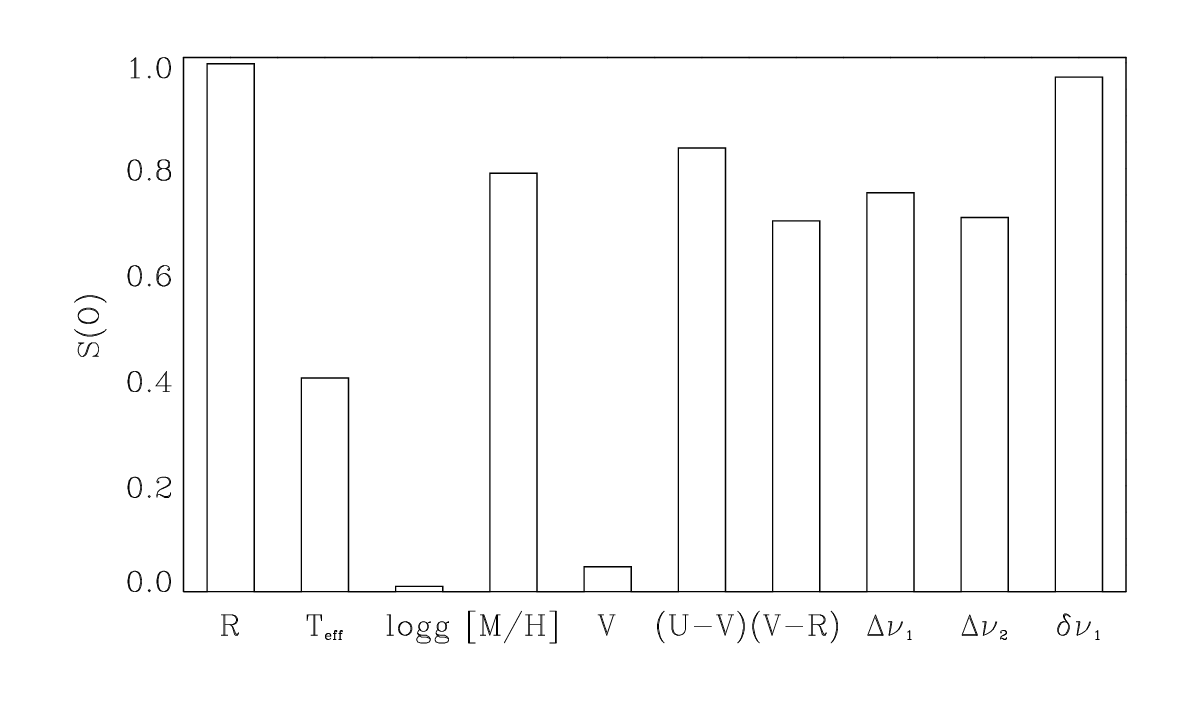}
\caption{Relative significance, $S$, of each of the individual observables, 
$O$, for the determination of the model parameters \citep[see][for the mathematical definition of $S$]{brownetal1994}.
A larger value of $S$ implies that the observable has an overall stronger impact on the determination of the parameter solution. The observables are, from left to right, the radius, the effective temperature, the logarithm of surface gravity, the surface metallicity, the visual magnitude, the magnitude difference in the $U$ and $V$ bands, and in the $V$ and $R$ bands, the large separations for two pairs of modes, both with degree $l = 0$,
and one small separation between modes of $l=0$ and $2$.
The analysis assumed errors of $0.01 {\rm R}_\odot$ in $R$, 
50\,K in $T_{\rm eff}$, 0.3 in $\log g$, 0.05 in [M/H] and $V$,
0.005 in $(U-V)$ and $(V-R)$ and $1.3 \muHz$ in the oscillation frequencies.
From \citet{creevey07} with permission. }
\label{significance}
\end{figure}

Let us suppose that we have a set of classical and seismic
observables, with associated measurement errors, and that we want to
determine, through direct fitting, a set of model input parameters with
corresponding uncertainties.  The impact of including the observable
{\it radius} on the determination of each of the model parameters
can be studied through the {\it Singular Value Decomposition} of the
derivative matrix, i.e, the matrix containing the derivatives of the
observables with respect to the model input parameters, divided by
the measurement errors \citep{pre86}. 
As discussed by \citet{brownetal1994}, and emphasised by
\citet{creevey07}, the role that a given observable plays in the
determination of the parameter solution depends crucially on the
combination of observables used in the fitting and on the
corresponding measurement errors.
The {\it significance\/} $S_i$ of an observable $O_i$ can be
quantified from this analysis as a measure of the extent to which
a 1-$\sigma$ change in $O_i$ shifts the inferred parameters towards
the 1-$\sigma$ error ellipsoid in the parameter space
 \cite[see][for details]{brownetal1994}.
We consider the case where the set of observables includes
spectroscopic, photometric and 
various combinations of seismic measurements for a solar-like star.
Figure~\ref{significance} shows the
significance of each observable for the determination of the parameter
solution in this case. 
Clearly the radius measurement plays a significant role in the
parameter determination, and is partially responsible for the low
impact of adding simultaneously, the observables {\it visual
magnitude\/}, $V$, and {\it surface gravity\/}, $g$. 

According to the results of \citet{creevey07}, the radius and the large separation play complementary roles in the determination of the model parameters. When the error in the radius increases, the consequent decrease of its significance is accompanied by an increase of the significance of the large separation. Likewise, an increase of the error in the large separation is accompanied by a decrease of its significance, and an increase of the significance of the radius.

\begin{figure}
\hskip -0.5cm
\centering
\hskip 0.5cm \includegraphics[width = 11cm]{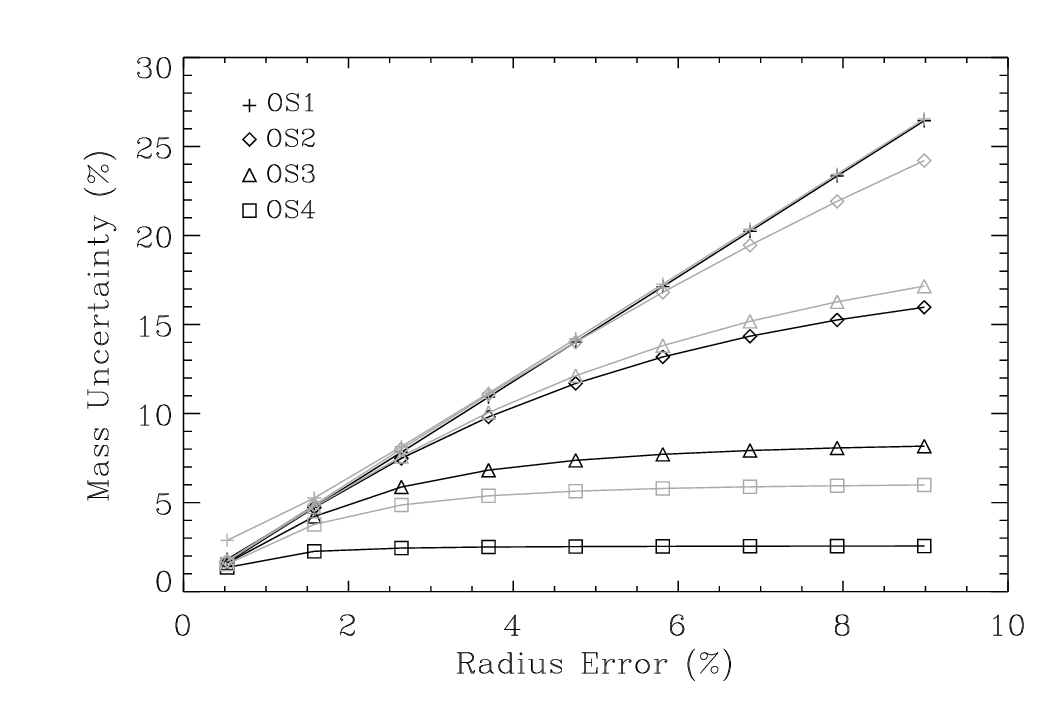}\\
\includegraphics[width = 11cm]{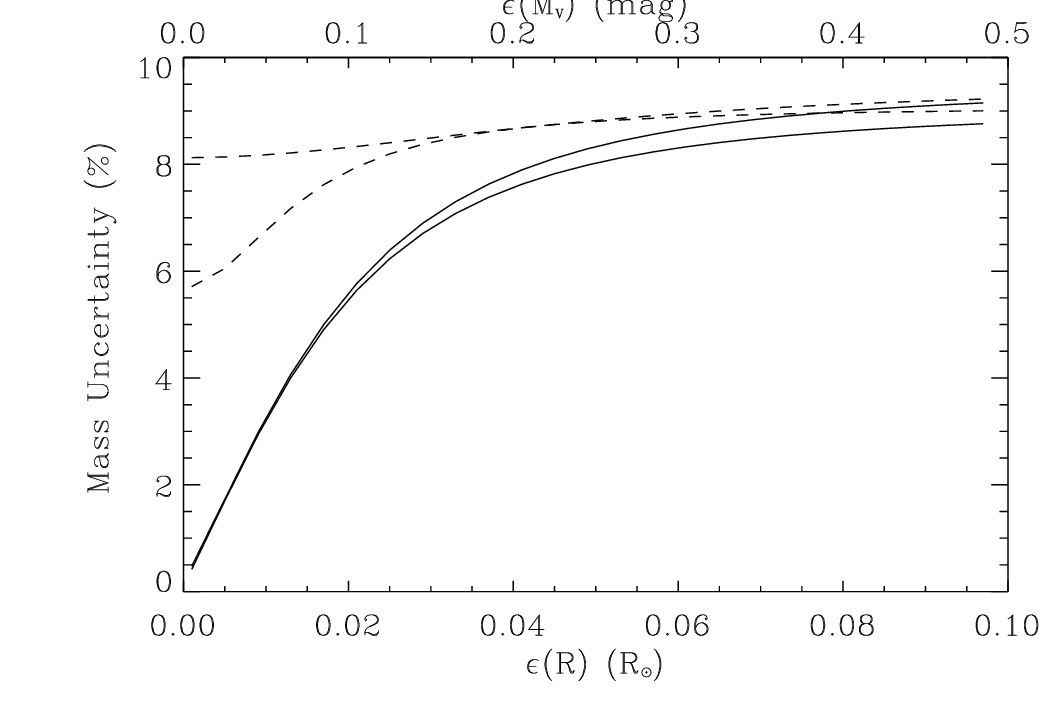}
\caption{Upper panel: Theoretical uncertainty in the mass as a function of the
error in the radius, for different combinations of observables. All
sets of observables include measurements of the radius, effective
temperature and metallicity;
apart from radius and frequencies the errors in the observables are as in 
Fig.~\ref{significance}.
Moreover, the set of observables includes:
$\langle\Delta\nu\rangle$ and $\langle\delta\nu\rangle$ (crosses), a range of 
$\Delta\nu_{nl}$ (diamonds), an additional five small frequency separations 
(triangles) and individual frequencies (squares).
Dark lines show the results when
the error in the frequencies is $\epsilon(\nu) = 0.5 \mu$Hz, while the grey are for $\epsilon(\nu) = 1.3 \mu$Hz. 
Lower panel: Theoretical uncertainty in the mass as a function of the error in the radius (full lines, lower abscissa) as compared with the mass uncertainties as function of the error in the absolute magnitude (dashed lines, upper abscissa),
when the latter is used as an observable instead of the radius. From \citet{creevey07} with permission.
}
\label{fig-massuncertainties}
\end{figure}

Even though Fig.~\ref{significance} shows that a precise measurement of the stellar radius can have a significant impact on the determination of the model input parameters, it does not tell us which parameters this measurement benefits the most. In this respect, the study mentioned above indicates that a measurement of the radius will have a particularly strong impact on the determination of the model mass.
The upper panel of Fig.~\ref{fig-massuncertainties} illustrates how the uncertainty in the mass determination depends on the radius error, for different sets of observables.
When the error in the radius drops below 3\%, all curves,
except for the one with the squares, converge.
Since these curves show the results when different sets of seismic observables are included (different symbols),
and for different errors in the seismic data (different tones),
the fact that they converge when the error in the radius gets sufficiently small
indicates that in this case it is mostly the error in the radius that defines the uncertainty in the model mass.

The curve with squares,
obtained by including individual frequencies in the set of observables,
shows the only case in which the seismic data can impact substantially on the mass determination, even when the error in the radius is below 3\%.
Unfortunately, as mentioned in Section~\ref{sec1},
the poor modelling of the outer layers of solar-like stars results in significant systematic errors in the computation of model oscillation frequencies.
Thus, using combinations of frequencies that are less subject to the systematic errors may be a better option;
full use of the last case considered in Fig.~\ref{fig-massuncertainties} requires procedures for correcting for the near-surface errors.
Once the mass is determined with sufficient precision,
the individual frequencies can provide a powerful tool to distinguish between different possible solutions and,
eventually, provide detailed diagnostics of the properties of stellar interiors.

The lower panel of Fig.~\ref{fig-massuncertainties} illustrates the importance of using a direct measurement of the radius,
rather than an observable such as the 
absolute magnitude, for the mass determination.
If the radius is not included in the set of observables,
then, even when the errors in the visual magnitude and in the effective temperature are both small,
the uncertainty in the mass determination remains large.

To test the results for the theoretical uncertainties discussed above, \citet{creevey07} also performed simulations of observations and
attempted to recover the true input parameters in an automatic
fashion. All the model parameters were recovered to within a 1-$\sigma$
standard deviation.  According to their results, reaching 0.1 -- 3.0\% error in the radius, allows for the 
determination of the mass to within 1.0 -- 4.0\%.  The simulation results
show similar qualitative trends to the theoretical trends, but for
errors in the radius larger than 3\% all of the results were quantitatively
better than those shown in Fig.~\ref{fig-massuncertainties}.

\subsubsection{Asteroseismic targets within reach of interferometric 
measurements}

As discussed in Section~\ref{sec2}, long baseline interferometry can provide
very precise measurements of the angular size $\theta$ of stars in the solar
neighbourhood.  Moreover, the combination of the latter with precise
measurements of the trigonometric parallaxes $\pi$, through
Eq.~(\ref{diameter}), provides the linear diameters $D$ of the corresponding
stars, in terms of the solar value $D_\odot$.

Naturally, the overall impact of interferometric determinations of stellar radii
on our understanding of stellar structure and evolution depends on the number of
stars for which a high-precision estimate of $D$ can be achieved.  To assess
that impact, in the following paragraphs we estimate the number of stars for
which both precise interferometric angular size {\it and\/} trigonometric
parallax can be measured.

Following the classical example of requiring a $3\%$ relative accuracy on the
radius of a star for it to be useful in constraining stellar structure models
(cf.\ Section~\ref{impact} \citep[see also, e.g.,][]{An91}, we set a $3\%$
relative uncertainty requirement on the linear diameter. This translates into a
relative uncertainty of about 2\% for both the angular diameter and the
parallax, given that they play a symmetrical role in Eq.\,(\ref{diameter}). For
peculiar stars with poorly understood physics it may be useful to relax this
requirement to several more per cent, but our estimate of the impact is focused
on normal stars.

To adopt a realistic approach, we consider the {\it proven\/} capabilities of
the infrared single-mode interferometric instruments FLUOR and VINCI
\citep{kervella03a}. They represent the state-of-the-art in terms of measurement
precision, with a typical relative uncertainty on fringe visibility 
${\cal V}_{\rm obj}$ of 
$\sigma_{{\cal V}_{\rm obj}^2} =
\delta {\cal V}_{\rm obj}^2/{\cal V}_{\rm obj}^2 \simeq 1$\% 
for less than 10 minutes of 
observing time.  To
relate this uncertainty with the one for the angular diameter, $\sigma_{\theta}=
\delta \theta/\theta$, we define the amplification factor $A_f$
\begin{equation}
A_f = \left[ \delta {\cal V}_{\rm obj}^2/{\cal V}_{\rm obj}^2\right] / 
\left[ \delta \theta/ \theta\right],
\end{equation}
from which we find
\begin{equation}
\sigma_\theta = \sigma_{{\cal V}_{\rm obj}^2} / |A_f|.
\end{equation}
We recall that the visibility ${\cal V}_{\rm obj}$ is a function of the 
interferometric baseline
length, ${\cal B}$, and of the wavelength, $\lambda$, at which the 
observations are
carried out. 

In order to determine the smallest angular diameter that can still be measured
with low enough relative uncertainty, we consider the longest possible baseline of the
VLTI \citep{glindemann04}, that is, ${\cal B} = 202$\,m. For the wavelength, we
consider the $H$ band ($\langle\lambda \rangle = 1.64\,\mu$m) which is accessible by using
the AMBER \citep{petrov03} instrument on the VLTI. This setup (${\cal
B}=202\,m$, $\lambda=1.64\,\mu$m) is almost equivalent to the usage of the
maximum baseline of ${\cal B}=330$\,m at which the CHARA Array can operate
equipped with the FLUOR instrument working in the $K$ band at
$\lambda=2.2\,\mu$m \citep{tenbrummelaar03}.  For the VLTI-AMBER $H$ band
configuration and the 1\% visibility relative uncertainty adopted here, the amplification
factor criterion $|A_f| > 0.5$ gives a minimum requirement on the angular
diameter of $\theta > 0.52$\,mas.

For the parallax, we have considered the values listed in the {\it Hipparcos\/}
catalogue \citep{perryman97}. While this catalogue generally gives the best
estimate for nearby moderately cool to moderately hot stars, this is not always
the case for the very late-type red dwarfs beyond M2V and for the hottest OB
stars.  However, we choose not to consider them in the present
discussion. Dwarfs beyond M2V are in most cases too small to be within reach of
accurate angular diameter measurements while there are hardly any massive OB
stars within the close solar vicinity.  Typically, the {\it Hipparcos\/}
parallaxes of the considered stellar sample, excluding the coolest and hottest
dwarfs, have errors $\simeq 0.9$\,mas. This sets a limit of $\pi > 45$\,mas,
i.e.\ $d<22$\,pc to fulfill the uncertainty requirements. We selected all the
stars in the {\it Hipparcos\/} catalogue fulfilling this criterion and found a 
total of 1196.

To determine the number of stars that satisfy both criteria derived above, i.e.\
$\theta > 0.52$\,mas and $ \pi > 45$\,mas, we have to estimate the angular
diameter of all {\it Hipparcos\/} stars within a distance of 22\,pc. To
accomplish this, we used broadband photometry and the surface brightness
relations calibrated by \cite{kervella04c}. In particular, we used their
relation 
\begin{equation}
\log \theta_{\rm LD} = 0.2752\, (V-K) + 0.5178 - 0.2 V,
\end{equation}
where $\theta_{\rm LD}$ is the limb-darkened angular diameter (in mas),
$V$ is the visual magnitude ($V$ band) and $K$ the infrared one ($K$ band).  
We extracted the visual and infrared photometry from the
{\it Hipparcos\/} and 2MASS \citep{cutri03} catalogues, respectively.
Interstellar reddening can be neglected for such nearby stars, so the result of
our computation is a list of angular diameters for all 1196 stars in our sample.
These angular diameters are then used to compute the relative uncertainties 
$\sigma_\theta$ using the amplification factor $A_f$ defined above. 
The relative uncertainty of the {\it Hipparcos\/} parallax $\sigma_\pi$ is subsequently
used to estimate the total relative uncertainty on the linear radius through the relation  $\sigma_R \approx \sqrt{\sigma_\theta^2 + \sigma_\pi^2}$.  In some cases, the
interferometer can fully resolve the star and measure its limb darkening. In
those cases, the angular diameters can in principle be measured with very high
accuracy and we set $\sigma_\theta = 0$.

\begin{figure}[t]
\centering
\includegraphics[bb=0 0 360 288, width=9cm]{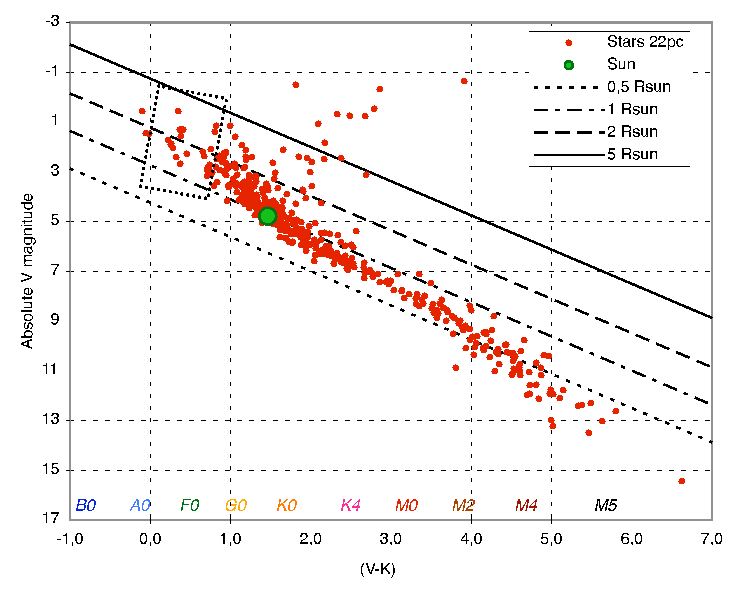}
\caption{Positions in the HR diagram of the stars within 22\,pc accessible to a
linear radius measurement with a total relative uncertainty below 3\% through a
combination of interferometric angular diameter and {\it Hipparcos\/}
parallax. The diagonal lines are radius isocontours, and the shaded box on the
left is the lower part of the classical instability strip.}
\label{stars_hr}
\end{figure}

The result of this selection is that the radii of $\simeq 450$ stars can be
derived with a relative uncertainty smaller than 3\%. Their positions in the
Hertzsprung-Russell $(M_V,V-K)$ diagram are presented in
Fig.~\ref{stars_hr}. Decreasing the radius relative uncertainty to 1\% results in a
sample of 77 stars, well spread over the main sequence. The 19 stars whose
radius can in principle be obtained within 0.5\% relative uncertainty are listed in
Table~\ref{best_stars}. As a remark, the best radius estimates to date ($\simeq
0.2\%$) have been obtained on $\alpha$\,Cen A and B by \cite{kervella03b} and
by \cite{Bigot06} on $\alpha$\,Cen B.

Overall, it appears that the linear radii of a broad variety of dwarfs can be
derived with good to excellent precision using a combination of interferometry
and trigonometric parallaxes. These stars cover the main sequence from A0V down
to M4V (with the addition of {\it Proxima Centauri}, M5.5V). A few tens of giants and
subgiants are also within reach, mostly in the G-K spectral range. Among the
measurable hot dwarfs, many are known to be fast rotators (e.g.,\ Altair and
Vega), and the geometrical deformation of their photospheres can be measured
using different baseline orientations and/or closure phase measurements
\citep{Domici2003}.

Currently, the main factor preventing a higher number of stars from being
resolved by interferometry is the interferometric baseline length. However, even
with an infinitely long baseline, only $\simeq 800$ stellar radii would be
measurable with a relative uncertainty smaller than $3\%$, due to the
uncertainty in the {\it Hipparcos\/} parallaxes. In many cases, the parallax is
by far the limiting factor in the radius estimation. In particular, the CoRoT
targets are currently outside the $3\%$ accessible domain as defined above, due
to large uncertainties associated with their parallax determinations. Measuring
accurate parallaxes of bright and nearby stars thus appears to be the key to
increasing the precision of stellar radius estimates.  One of the central goals
of {\it Gaia\/}, 
an ESA mission with expected launch in 2011, is the determination of
accurate positions of a large number of stars in our galaxy. As a consequence,
major improvements are expected in this context in a not too distant future.

\begin{table}
\caption[]{List of the nearby stars whose linear radius is measurable with a
relative uncertainty $\sigma_R$ below 0.5\%. ``LD" in the $\sigma_\theta$
(relative error on the angular diameter) column designates the stars that are
fully resolved by the AMBER-VLTI instrument in the $H$ band, and that are
therefore accessible to limb darkening measurements.
 \label{best_stars}}
\centering
\begin{tabular}{llccc}
\noalign{\bigskip}
\hline
Name & Sp. Type & $\sigma_\pi$ & $\sigma_\theta$ & $\sigma_R$ \\
\hline
$\alpha$ Cen A & G2V & 0.19\% & LD & 0.19\% \\
$\alpha$ Cen B & K1V & 0.19\% & LD & 0.19\% \\
$\epsilon$ Ind & K5V & 0.25\% & LD & 0.25\% \\
61 Cyg B & K7V & 0.25\% & 0.08\% & 0.27\% \\
GJ\,411 & M2V & 0.23\% & 0.13\% & 0.26\% \\
$\epsilon$ Eri & K2V & 0.27\% & LD & 0.27\% \\
$\tau$ Cet & G8V & 0.29\% & LD & 0.29\% \\
Procyon A & F5IV-V & 0.31\% & LD & 0.31\% \\
GJ\,887 & M2/M3V & 0.29\% & 0.18\% & 0.34\% \\
$\eta$ Cas A & G0V SB & 0.37\% & LD & 0.37\% \\
$\delta$ Pav & G5IV-Vvar & 0.40\% & LD & 0.40\% \\
82 Eri & G8V & 0.33\% & 0.23\% & 0.40\% \\
GJ\,380 & K5V & 0.39\% & 0.11\% & 0.41\% \\
$\sigma$ Dra & K0V & 0.27\% & 0.30\% & 0.40\% \\
$\chi$ Dra A & F7Vvar & 0.39\% & 0.14\% & 0.41\% \\
Sirius & A0m & 0.42\% & LD & 0.42\% \\
Vega & A0Vvar & 0.43\% & LD & 0.43\% \\
AX Mic & M1/M2V & 0.45\% & 0.15\% & 0.47\% \\
Altair & A7IV-V & 0.48\% & LD & 0.48\% \\
\hline
\end{tabular}
\end{table}

\subsubsection{Interferometry of binary stars}

Even though a number of stars in the classical instability strip are
found to be within reach of interferometric measurements, hardly any
of the well-established classical pulsators with precise
parallax determinations satisfy that condition. Naturally, new bright
classical pulsators might be discovered from space photometry
assembled by missions such as WIRE and MOST.  A nice such example is
the bright star Altair which was found to be a $\delta\,$Scuti star from
the analysis of WIRE white-light photometry \citep{Buz05}. Unfortunately, despite the small relative uncertainty in its parallax, which is below 0.5\%, rapid
rotation, between 70\% and 90\% of its critical velocity, and the lack
of mode identification,  limit considerably the seismic
modelling of this star \citep{Su05}.

Given the small number of known classical pulsators within reach of interferometric measurements,
the major contribution of interferometry to asteroseismic studies of this type of pulsators will most likely come from the observation of binaries with a pulsating component.
As mentioned in Section~\ref{sec1} the
binarity can be an asset for asteroseismology because it in principle
allows for a reduction of the observational error box associated with the
fundamental parameters of the components. Quite a
number of bright classical pulsators are found among spectroscopic binaries
 \citep[see, e.g.,][for a recent review]{Pig06}.  This is particularly the
case for the pulsating B stars \citep[for overviews, see][]{dc00,
adc03}, but also for $\delta\,$Scuti stars, as, e.g., $\theta^2\,$Tau
\citep{Ken96} and $\gamma\,$Doradus stars \citep{DeCat06}.

Historically, {\it visual binaries\/} were first recognised, due to their
proximity on the sky and their relative orbital motion.  Even though Kepler's
third law ties the size and period of the orbit with the sum of the masses of
the two components, the angular nature of the observations prevents that
sum from being accessible unless the distance to the system is known.

As a consequence of the improvement of the observation techniques, the
two stars are no longer required to be seen for an orbit to be
derived.  For instance, astrometric binaries are double stars where
one component is often too faint to be detected but its mass disturbs
the space trajectory of the primary, causing a periodic astrometric
wobble around the trajectory of a genuine single star.  Although the
distance is known in such cases, only the absolute size of the primary
orbit is accessible, thus limiting again the use of Kepler's third law.

Switching to interferometry does not really help for the cases of visual or
astrometric binaries.  While an interferometer greatly improves the resolving
power, the relative angular positions suffer the same limitations as the old
micrometric visual data. They yield the angular size of the relative orbit which
can only be changed into the mass sum if the distance to the system is known.

Spectroscopy, especially radial velocities, offers a valuable alternative to
positional astronomy.  Unlike visual binaries, {\it spectroscopic\/} ones are
not distance limited in the sense that the binary signature does not deteriorate
with the distance to the system.  Another advantage is that, in case of
double-lined systems, the mass ratio is directly available.  Unfortunately, the
size of the orbit is entangled with the orbital inclination which prevents
derivation of the sum of masses and, therefore, the individual masses.

Even though an adjective -- visual or spectroscopic -- often describes how a
binary was detected, the observation methods often overlap after the discovery.
Thanks to the improvement of both spectroscopy towards small radial velocity
amplitudes and interferometry towards close binaries, the overlap has expanded
substantially over the past 20 years.  Whereas none of the above types of
binaries makes the distance or individual masses available, resolving
spectroscopically and visually the components of a binary yields both the
individual masses and the distance to the system, the latter through the
computation of the so-called {\it dynamical parallax}.  This leads to
hypothesis-free stellar masses and distances and provides the desired
constraints on the error box.  A catalogue of spectroscopic binary orbits is
available in, e.g., \citet{Pourbaix-2004:b} and provides a list of candidates
for further interferometric observations \citep{McAlister-1976}.

In order for the binarity to be of help in asteroseismology, we need
to be able to determine the stellar masses with a relative uncertainty
of at most a few percent.  In order to achieve that, interferometric
data must be accompanied by accurate spectroscopy.  Many meanwhile
interferometrically resolved binaries still have only 50+ years old
spectroscopic orbits.  
It is also worth realizing that recent precise radial velocities do not {\it per se} lead to accurate spectroscopic orbit determination.
As shown for $\alpha$~Cen
\citep{Dravins-1990:a,Dravins-1990:b,Pourbaix-2002:a}, the systematic
shift in radial velocities caused by, for instance, the gravitational
redshift and the convective blueshift, can largely exceed the quoted
precision of the measurements.  Such systematic effects can only be
quantified well from a full coverage of the orbit.  It is therefore
wise to select binaries whose period is sufficiently short for the
orbit to be covered at least once with one and the same instrument.
Although this condition on the period decreases the likelihood of
being able to resolve the binary interferometrically, it does not make
the task impossible, as seen from the studies of, e.g., Capella \citep{Bagnuolo-1989} and Atlas
\citep{Zwahlen-2004:a}.  However, it does require long-term monitoring.

For massive binaries with classical pulsators, the requirement of a few percent
precision on the masses is very stringent. Improvements in the component masses,
through the determination of the dynamical parallax from the combination of
high-precision spectroscopy covering the orbit and interferometric data, were
achieved for the $\delta\,$Scuti star $\theta^2\,$Tau \citep{Tom95}, and
for the $\beta\,$Cephei stars $\beta\,$Cen \citep{Aus06} and
$\lambda\,$Sco \citep{Tan06}. The hypothesis-free masses of these stars
are, however, not yet precise enough to be of sufficient help in the seismic
modelling (see also Section\,\ref{future}).

\subsubsection{\label{Ii}Indirect impact on global parameter determination}

In the absence of a direct measurement of the stellar radius, a combination of
the star's luminosity and effective temperature is often used to derive a value
for the radius. However, classical methods used to derive the effective
temperature, such as multi-band photometry and spectroscopy, either rely on
calibrations or on atmospheric models. Consequently, the effective temperatures
commonly used to place the stars in the HR diagram, and thus to constrain their
seismic analysis, are often subject to systematic errors which can only be
tested for, and hopefully corrected, when independent measurements of the
same quantities are available.

Interferometry is expected to play an important role in this context. The
determination of the angular radii allows new determinations of the effective temperature of these
stars that can be used to test for the presence of systematic errors in the
values derived when using other techniques. Even though to some degree the new
effective temperature determinations may also depend on atmospheric models,
since a bolometric correction has to be applied to determine the bolometric flux of the star, one might expect that the extra information gained by
the direct measurement of the angular diameter of the star will make the two measurements largely independent of each other,
hence allowing us to test for systematic errors.

The new, interferometric based, determinations of the effective temperatures of
nearby stars will also allow us to test and improve the calibrations that are
commonly used in the photometric determination of effective temperatures of
stars for which no spectroscopic data is available. The improvement of these
calibrations is of particular importance when studying pulsators among peculiar
stars and low-metalicity stars.  The uncertainty in the effective temperature
determinations for these stars can easily exceed 200~K. However, their study is
tremendously important, as they provide different environments in which models
of stellar interiors and of stellar atmospheres can be tested.
 
In roAp stars, for instance, the emission is so anomalous, with heavy line
blanketing in the ultraviolet and consequent redistribution of the flux over
larger wavelengths \citep{leckrone73}, that the photometric calibrations used
for normal stars cannot be relied upon when studying them. Often, the values for
the effective temperature of these stars derived from different methods do not
agree within their formal errors, which is a clear indication that some form of
systematic or calibration error is present.  A convincing indication that systematic errors are
likely to be present in the values available for the effective temperatures of
roAp stars is given by \cite{matthews99}. In their work, the authors found that
the parallaxes predicted from asteroseismic data of roAp stars are
systematically larger than the parallaxes determined by {\it Hipparcos} for the same
stars. This result was corroborated by the work of \cite{cunha03} in which the
authors have tried, without success, to bring the seismic and astrometric data
for one of the roAp stars in the sample into agreement, by exploring a grid of
models with different physics and chemical composition. Unfortunately, the
number of potential roAp stars that may be resolved by current interferometric
instruments is very small. Nevertheless, new values for the effective
temperature of only a few of these stars, derived from their angular diameters,
could have a strong impact on the seismic study of the whole class.

The potential of interferometry to improve indirectly global parameter
determination of stellar pulsators goes even beyond the detection of systematic
errors in classical methods or the possibility of deriving new, more adequate
calibrations. In fact, the stellar limb darkening measured by the modern
interferometric instruments is sufficiently accurate to be used as
diagnostics on the stellar outer layers, in particular on the temperature
profiles. These, in turn, put direct constraints on atmospheric models. To take
advantage of the instrumental performance of the interferometers, a similar
effort in numerical modelling of stellar atmospheres has to be done. In
particular, the standard 1D hydrostatic models of atmospheres such as ATLAS
\citep{Kurucz1992} or Phoenix \citep{ha99} have to be improved. Indeed, these models
are based on very limiting assumptions. Even if the temperature structure in the
1D codes can always be calibrated by a judicious choice in the free parameters
(e.g.,\ mixing length), it can never reproduce the temperature profile in the
transition (surface) region since the coupling between radiation and gas
dynamics is essential there. On the contrary, 3D radiative hydrodynamical 
models initiated two decades ago by {\AA}. Nordlund, R. Stein and coworkers to
model the solar convective surface \citep{no82,Stein1989,sn98} 
do not suffer from
these unrealistic physical hypotheses. They solve the complete set of
hydrodynamical equations coupled with radiative transfer with realistic
equations-of-state and opacities to obtain 3D, time-dependent, inhomogeneous
models of stellar atmospheres. The realism of these simulations is now such that
they can reproduce almost perfectly solar and stellar line profiles \citep[for both
shifts and asymmetries, e.g.,][for FeI lines]{Asplund00}
and provide reasonable modelling of aspects of solar envelope structure and
acoustic oscillations \citep[e.g.,][]{Rosent1999}
and granulation \citep[e.g.,][]{Stein1989, Nordlu1991, 
Trampedach98, Svenss2005}.

The use of these 3D radiative hydrodynamical codes in the context of interferometry has been proposed
recently by \cite{be76} for Procyon, and by \cite{Bigot06} for the K-dwarf
$\alpha$~Cen B for which new visibility measurements were obtained with
VINCI/VLTI.  There are significant differences between the structures of the
atmosphere obtained by 1D and 3D radiative hydrodynamical approaches, especially for stars with
not-so-late spectral type (such as F stars) and for metal poor stars. The amplitude
and position of the second and higher order lobes of
visibility are very sensitive to the limb darkening and therefore to the
temperature structure of the atmosphere, which provides then strong constraints
on the models.
The importance of such accurate limb darkening measurements forces us to
prepare a grid of stellar 3D limb darkening model descriptions, similar to the
limb darkening laws produced by \cite{cl02} using Kurucz's grid of 1D atmospheres, covering
the whole HR diagram with solar and low metallicities. Such a grid is necessary
to prepare and exploit future interferometric angular diameter
measurements. Moreover, these 3D radiative hydrodynamical approaches have to be used for deriving
other fundamental parameters of the stars as well, such as the surface gravity and
the abundances combined with a non-local thermodynamic equilibrium treatment of the line transfer \citep{ti99}.
An interesting, if controversial, aspect of the 3D modelling is its application
to the determination of solar abundances, yielding abundances of carbon, 
nitrogen and particularly oxygen well below the previously assumed values
\citep[for a review, see][]{Asplun2005}.
Models computed with these abundances are in stark disagreement with the
results of helioseismology
\citep[e.g.,][]{Turck2004, Antia2005, Bahcal2006, Basu2007},
and no obvious solution to this difficulty has yet been found
\citep{Guzik2006}.

A promising recent development has been the modelling of rather larger
parts of the solar convection zone \citep{Stein2006},
covering supergranular scales and a somewhat deeper region and hence 
potentially more realistic models of solar oscillations.
Similar results can be expected for other stars in the foreseeable future.
When combined with asteroseismic diagnostics from mode excitation and 
frequency shifts (cf.\  Section~\ref{stochastic})
we can finally expect a reliable treatment of the near-surface regions.

\subsection{\label{mi}Mode identification}
 
As mentioned earlier, for asteroseismology to reach its full potential, it is
not only necessary to measure the oscillation frequencies. It is also necessary
to identify the modes in terms of the number of radial nodes $n$, and the number
and position of nodal lines over the surface, which is characterised by the spherical
wavenumbers $(l,m)$. 

With sufficiently accurate measurements of the stellar effective
temperature, luminosity and/or radius, and a mass inferred from
fitting an evolutionary track, it is possible, in principle, to deduce
expected oscillation frequency patterns.  The large frequency
separation is a special case of such a pattern for p-mode oscillations
in the asymptotic frequency regime of slowly rotating stars.
Similarly, period spacings can be used for g-mode oscillations in the
asymptotic regime.  Direct observational constraints on the
separations may also become available whenever numerous oscillation
modes are observed. According to Eq.~(\ref{eq:pasymp}), the ratio
of a p-mode oscillation frequency, corrected for surface effects and
for rotational splitting, and half of the large separation, should be
an integer number approximately equal to $2n + l$. Hence, knowledge of
the theoretically expected {\it or\/} 
observationally determined large separation reduces the likely
range of appropriate values of $l$ and $n$, for any measured
oscillation frequency, to the combination $2n+l$. However, this still
leaves considerable ambiguity in the mode identification and it
assumes that rotation does not prohibit the selection of the central
frequency peak of the rotationally split multiplets. This comfortable
situation is seldom encountered, except perhaps in solar-like stars
(see Section\,\ref{cs} below for examples).

When only a limited number of modes is excited to observable amplitude, or when
the modes do not follow particular frequency separations, or whenever a very
dense frequency spectrum is predicted, the frequency values alone are
insufficient to derive the wavenumbers.  In this case, one could in principle
proceed with seismic modelling considering {\it all\/} values for $(l,m,n)$ for
any of the detected frequencies. In order to limit the computation time of such
forward modelling, the values of the degree $l$ are usually limited from the
geometric arguments mentioned in section \ref{origin}.
It is then customary to consider modes with $l\leq 3$ and to assume $m=0$, when no obvious evidence for rotational splitting is found in the Fourier transform of the time series.

This procedure is not very satisfactory, though, because rotation can easily
result in non-equidistant splitting and imply merging of frequency multiplets in
such a way that they cannot be unravelled. Moreover, quite a number of classical
pulsators show evidence for modes with degree $l\geq 4$, while not showing evidence for modes of
lower degree, which presumably are not excited to observable amplitudes. In these cases, the assumption
of $m=0$ is unjustified.  The quest for {\it empirical mode identification\/}
has therefore become an extended topic by itself in asteroseismology.  With this
term we mean the assignment of $l$ and $m$ values to
each of the observed frequencies from the data themselves, without relying on the
(unknown) details of the model properties of the star. It  is usually
impossible to obtain a correct mode
identification for each detected oscillation frequency. However, even only one correct $(l,m)$ identification, e.g.,\ the one
for the dominant mode, can imply a serious reduction of the free parameter space
in the modelling and is therefore worthwhile to attempt. 

Empirical mode identification is a tedious and sophisticated task. It
requires a detailed confrontation between oscillation theory applied to the
outer stellar atmosphere and observational characteristics, such  
the observed amplitudes and phases of lightcurves, of radial-velocity
curves or of line-profile variations. 
Essentially two types of diagnostics are in use to identify the
modes. One of them is based on time series of multicolour photometry and the
other one relies on time series of line-profile variations detectable from
high-resolution spectroscopy.

\subsubsection{Mode identification from multicolour photometry}

The amplitude and phase behavior of an oscillation mode can be markedly
different when measured in different filters of a photometric
system. These differences depend on the degree $l$ of the mode and form the basis of a photometric mode identification method. 
Only the oscillation frequencies observed in all the filters are considered for mode identification.
The idea is that the degree $l$ of a given oscillation mode, whose
frequency is detected in all the filters of the photometric system,
may be derived from the comparison of the amplitude and phase values
measured with different filters.

Different versions of this mode identification method are present in the
literature. It was originally proposed by \citet{Stamfo1981}, relying on the work
by \citet{Dziemb1977} and \citet{BalSt79}. Subsequently, \citet{Watson1988} improved the
method by bringing it into applicable form, while \citet{Hey94}
included the perturbation of the limb darkening into it. For an extensive review of the
method, we refer to \citet{Garrid2000}.  All these versions are based on adiabatic
oscillation theory, and treat the non-adiabaticity of the oscillatory behaviour
in the outer atmosphere by means of an unknown ad-hoc parameter $R_{\rm nad}$
(usually denoted as $R\in [0,1]$, although this is very confusing as it is the
standard symbol for the stellar radius, hence our notation of $R_{\rm
nad}$). The theoretical expressions for the amplitude and phase of the light
curve in the different filters (i.e.\ as a function of wavelength) depend,
among other things, on the geometrical configuration of the nodal lines with
respect to the observer, i.e.\ on the values of $(l,m,i)$, where $i$ is the
inclination angle between the axis of symmetry of the oscillation and the
line-of-sight. The axis of symmetry of
the oscillation is usually taken to be the rotation axis, except for the roAp
stars where the magnetic axis is a more natural and hence better
choice.

It was already realized by \citet{Stamfo1981} that the functional
dependence of the amplitude and phase on the mode geometry allows one to group
the $m$ and $i$ parameters into one single factor which is independent of
wavelength. It is therefore possible to eliminate this factor, and with it the very
disturbing and unknown inclination angle, by considering amplitude ratios and
phase differences among the different filters. This is the procedure that is
always adopted nowadays.  The disadvantage is that the information on
the $m$-value is lost and, consequently, only the degree $l$ of the mode can be identified.  The theoretical expressions for amplitude ratios and
phase differences require the computation of the perturbed version of the
adopted limb darkening law and of the perturbed stellar flux as a function of
effective temperature and gravity. This brings us back to the need of good
atmospheric models (see Set.\,\ref{Ii}). In particular, it turns out that this
identification method is rather sensitive to the adopted treatment of convection
for stars with outer convection zones, such as $\delta\,$Scuti and $\gamma\,$Doradus
stars  \citep{Garrid2000}. 

A big step forward was achieved with the new versions of the method developed by
\citet{Bra95} for white dwarfs, by \citet{Bal99} for $\delta\,$Scuti stars, and by
\citet{Dup03} for all main-sequence oscillators. In these works, a non-adiabatic
treatment of the oscillations in the outer atmosphere was included, with
different levels of sophistication, through which the unknown ad-hoc factor
$R_{\rm nad}$ was eliminated.  \citet{Dup03} illustrated their method for
$\beta\,$Cephei stars, slowly pulsating B stars, $\delta\,$Scuti stars, and
$\gamma\,$Doradus stars. The dependence on the adopted theory of convection remains
for the latter two classes of pulsators.  A  non-adiabatic treatment similar to that of \citet{Dup03} was presented by \citet{Ran05b} in the
context of pulsating subdwarf B stars. However, the latter does not contain an equally
detailed treatment of the oscillations in the outer atmosphere.

The theoretical amplitude ratios and phase differences are dependent on the
stellar fluxes, which are determined by the radius, effective temperature and
mass (or equivalently, the gravity) of the star. As discussed above, these
parameters are often not known with high precision. Their uncertainties must be
propagated into the final selection of the best value for $l$ from the observed
amplitude ratios. This was often ignored in the past, but is taken care of in
modern applications of this method following \citet{Bal99}.  Examples of such
applications are available in \citet{Han03, Han05, Handle2006}, \citet{DeRidd2004} and
\citet{Sho06} for $\beta\,$Cephei stars, in \citet{DeCat2005, DeCat07} for slowly
pulsating B stars, in \citet{dupret05b,Dup05c} for $\delta\,$Scuti and $\gamma\,$Doradus
stars, and, finally, in \citet{Jef04, Jef05} and \citet{Tre06}
for subdwarf B stars. We refer the reader to these papers for more detailed
information.

\subsubsection{Mode identification from line-profile variations}

The introduction of high-resolution spectrographs with sensitive detectors in
the 1980s, had a large impact on the field of empirical mode identification.
Spectroscopic data indeed offer a very detailed picture of the pulsational
velocity field.  It remains a challenge to obtain spectra covering the overall
beat period of the multiperiodic oscillations, with a high resolving power
(typically above 40\,000) and with a high signal-to-noise ratio (typically above
200 and preferably much higher than that), for a good temporal resolution
(typically below a few percent) in the sense of the ratio of the integration
time to the oscillation period. The latter condition is necessary in order to
avoid smearing out of the oscillations during the cycle.

Methodology to derive the full details of the pulsational velocity field (at
least six unknowns -- see below) tends to be complex. For this reason,
multicolour photometric observations, which can only lead to an estimate of the
wavenumber $l$ but which can be obtained from small telescopes, are still of
utmost importance for mode identification.  Such data are especially
suitable for the study of long-period pulsations because small telescopes are
available on longer time scales.

Each spectral line is subject to different broadening mechanisms, among which
atomic broadening which is usually negligible, pressure broadening giving rise
to a Lorentz profile, and thermal broadening characterised by a Gaussian
profile. These three mechanisms act on a microscopic scale and lead to a global
time-independent broadening of the spectral line. This is why they are usually
treated together by means of the so-called {\it intrinsic line profile}, whose
shape is the convolution of two Lorentz profiles and a Gaussian profile. This
convolution results in a {\it Voigt profile}. In practice one tries to avoid
having to use spectral lines that are sensitive to pressure broadening (such as
the hydrogen lines) for mode identification because the oscillatory signal is
much less visible in their wings compared with lines that are mainly thermally
broadened. In the latter case, it is justified to assume a Gaussian for the
intrinsic line broadening function. This is particularly justified for hot stars
with low to moderate density.

{\it Rotational broadening\/} of the spectral lines is observed as being
time-independent as long as no surface inhomogeneities occur.  Stellar
oscillations, on the other hand, give rise to periodic broadening of the line
profiles.  The shape of the line profile is completely determined by the
parameters occurring in the 
expression for the pulsation velocity which can be obtained from
Eq.(\,\ref{eq:gougheq}).
In particular it is
dependent on the wavenumbers $(l,m)$ of all the oscillation modes.

The {\it Doppler effect\/} determines how much a spectral line centred at
the wavelength $\lambda_0$ is broadened, according to:
\begin{equation}
\label{doppler}
\displaystyle{\frac{\Delta\lambda}{\lambda _0}=\frac{v}{\tilde c},}
\end{equation}
in which $v$ is the component of the total velocity field $\vec{v}$ at the
stellar surface in the line of sight and $\tilde c$ is the speed of
light.
The velocity
vector $\vec{v}$ consists of a component due to the surface rotation and a
component due to the oscillations. In the case of a constant rotation over the
stellar surface, 
the rotational component $\vec{v}_{\rm rot}$
is easy to derive and can be written in terms of the rotational frequency
$\Omega$, giving rise to the {\it equatorial rotational velocity\/} $\Omega R$,
and the inclination angle.

In the case of spheroidal modes in the approximation of a non-rotating star, the
pulsation velocity expressed in a system of spherical coordinates
$(r,\theta,\varphi)$ centred at the centre of the star and with polar axis
along the axis of symmetry of pulsation, is
given by~:
\begin{equation}
\label{nrp}
\displaystyle{\vec{v}_{\rm puls}=
\left(v_r,v_{\theta},v_{\varphi}\right)=N_{l}^mv_{\rm p}
\left(1,K\frac{\partial}{\partial\theta},\frac{K}{\sin\theta}
\frac{\partial}{\partial\varphi}\right)Y_{l}^m(\theta,\varphi)
\exp{\left(-{\rm i}\omega t\right).}}
\end{equation}
In this expression, $N_{l}^m$ is a normalization factor for the
$Y_{l}^m(\theta,\varphi)$, $v_{\rm p}$ is proportional to the pulsation
amplitude, and $K$ is the ratio of the
horizontal to the vertical velocity amplitude. The latter can be found from the
boundary conditions and turns out to be $K \simeq GM/(\omega^2R^3)$ in the
approximation of a non-rotating star \citep[e.g.,][]{Unno1989}.

In order to compute the line profile shape, we have to determine the normalized
flux at a particular wavelength (or velocity).  Consider a system of spherical
coordinates $(r',\theta',\varphi')$ with the polar axis coinciding with the
direction to the observer.  In order to compute the theoretical line profile,
one subdivides the visible stellar surface into infinitesimal elements
$({\theta'}_i,{\varphi'}_j)$, $i=1,\ldots,N; j=1,\ldots,M$ where $N$ and $M$ are
typically 180 and 360, respectively, thanks to the present-day computational
power.  The velocity field due to the rotation and the pulsation leads to a
Doppler shift at a point $(R,\theta',\varphi')$ on the visible equilibrium
surface of the star.  The local contribution of a point $(R,\theta',\varphi')$
to the line profile is proportional to the flux at that point.  We assume that
the intensity $I_{\lambda}(\theta',\varphi')$ is the same for all points of the
considered surface element.  The flux through the surface element surrounding
the point $(R,\theta',\varphi')$ thus is the product of the intensity
$I_{\lambda}(\theta',\varphi')$ and the projection on the line of sight of the
surface element around the considered point~:
\begin{equation}
I_{\lambda}(\theta',\varphi')\ R^2\ \sin\theta'\ \cos\theta'\ \dd \theta'\ 
\dd \varphi'.
\end{equation}
An important effect that changes the flux over the visible surface is the limb darkening.
For line-profile variation calculations, the linear approximation of
the limb darkening law largely suffices, because the profile variations are
dominated by the Doppler shifts due to the surface velocity.  The flux of a
surface element centred around the point $P(R,\theta',\varphi')$ of the
equilibrium surface with size $R^2\sin\theta'\ d\theta'\ d\varphi'$ then is
\begin{equation}
\label{energie}
I_0\ h_{\lambda}(\theta')\ R^2\ \sin\theta'\ \cos\theta'\ 
\dd \theta'\ \dd \varphi',
\end{equation}
where $h_{\lambda}$ is the adopted limb darkening law and $I_0$ the intensity at $\theta=0$.
As discussed earlier, interferometry may play a role in improving our knowledge of
$h_{\lambda}$ and thus in the interpretation of line-profile variations.

Perturbations of the intensity and of the surface due to the oscillations change
the line profile. Usually, however, these effects are far less important than
the velocity effect for slowly rotating classical pulsators without surface
inhomogeneities, and one therefore often assumes \mbox{$\delta
I_{\lambda}(\theta',\varphi')=0$} during the oscillation cycle.  However, one
can easily generalize any line profile generation code to include the
non-adiabatic perturbation of the intensity, $\delta [I_0\
h_{\lambda}(\theta')]$, as well as the perturbed surface size of the elements
due to the oscillation.  

In order to take into account intrinsic broadening
effects, the local line profile is convolved with an intrinsic profile, which,
in the simplest approximation, is taken to be Gaussian with variance
$\sigma_{\rm th}^2$, where $\sigma_{\rm th}^2$ depends on the spectral line
considered. Generalizations to an intrinsic Voigt profile or a profile derived
from a stellar atmosphere model are easily performed, but imply much longer
computation times.  

The time dependence of the temperature eigenfunction may be important for the
computation of the intrinsic line profile when the spectral line is sensitive to
small temperature variations. For this reason, one carefully selects the best
spectral line for mode identification.  It is advantageous to use an unblended,
deep, thermally broadened line which is insensitive to small temperature changes
in the line-forming region in the atmosphere \citep[e.g.,][]{DeRi02}. This choice
thus depends on the spectral type of the star. For $\beta\,$Cephei stars, e.g., the
best line is the Si\,III\,4560\AA\ triplet \citep{adc03} while for slowly
pulsating B stars the Si\,II\,4130\AA\ doublet is ideally suited
\citep{Ae99}. For very fast B-type rotators, these multiplet lines are
unfortunately blended and one has little choice but to consider the isolated
He\,I\,6678\AA\ line \citep[e.g.,][]{Bal97} or other helium lines
\citep[e.g.,][]{Riv03}.  Temperature effects on line profile variations of
$\delta\,$Scuti and $\gamma\,$Doradus stars have not been studied carefully.

We represent by $p(\lambda)$ the line profile and by ${\lambda}_{ij}$ the
Doppler-corrected wavelength for a point on the star with coordinates
$(\theta'_i,\varphi'_j)$, i.e.,
\begin{equation}
\label{dopplerc}
\displaystyle{\frac{{\lambda}_{ij}-{\lambda}_0}{{\lambda}_0}=
\frac{\lambda (\theta'_i,\varphi'_j)-\lambda_0}{{\lambda}_0}=
\frac{\Delta \lambda(\theta'_i,\varphi'_j)}{{\lambda}_0}=
\frac{v(\theta'_i,\varphi'_j)}{\tilde c}} \; ,
\end{equation}
with $v(\theta'_i,\varphi'_j)$ the component of the sum of the pulsation and
rotation velocity of the considered point in the line of sight. An explicit
expression for $v(\theta'_i,\varphi'_j)$ can be found in, e.g.,
\citet{Ae92}.   The line profile is then given by
\begin{equation}
\label{conv}
\displaystyle{p(\lambda)=\sum_{i,j}
\frac{I_0h_{\lambda}(\theta'_i)}
{\sqrt{2\pi}\sigma}
\exp{
\left(
{-
\frac{(\lambda_{ij}-\lambda)^2}{2\sigma^2}
}
\right)
} R^2\sin\theta'_i\cos\theta'_i\Delta\theta'_i\Delta\varphi'_j} \; ,
\end{equation}
where the sum is taken over the visible stellar surface
($\theta'\in[0^{\circ},90^{\circ}], \varphi'\in[0^{\circ},360^{\circ}[$) and
where we have assumed a constant Gaussian intrinsic profile and a non-variable
surface normal for simplicity.  As discussed above, this formula for the
computation of line profile variations should be generalized in order to take
into account the following additional time-dependent effects: a variable surface
normal, a variable flux through non-adiabatic temperature and gravity
variations, a time-dependent intrinsic profile, Coriolis and centrifugal
correction terms to the pulsation velocity expression. The most up-to-date
line-profile generation codes take into account most of these effects, except
for those due to the centrifugal force 
\citep[e.g.,][]{Tow97,Schrij97,DeRi02,Zima06}.

The principle of line-profile fitting as a mode identification method is
obvious: one generates theoretical line profiles $(\lambda, p(\lambda))$ over
the oscillation cycle from Eqs\,(\ref{dopplerc}) and (\ref{conv}), or their more
sophisticated version including temperature and Coriolis effects, and one
compares them with the observed ones to select the set of parameters that
results in the closest resemblance to the data.  In order to do this
objectively, one must construct an atlas of theoretical profiles for different
values of $(l,m)$ and for realistic ranges of the the other line-profile
parameters (velocity amplitude of each of the modes, rotation velocity,
inclination angle, and intrinsic profile width).  This method is relatively easy
and straightforward to apply to a monoperiodic oscillator.  Whenever more than
one mode is present, however, the method becomes unrealistic in computation time
because one cannot scan a large enough parameter space. The latter has 6
dimensions for one mode and increases with 3 for any additional mode, in the
approximation where one neglects temperature and Coriolis effects.

To overcome the computational obstacle of line-profile fitting, and to make the
identification more objective, quantitative mode identification methods have
been developed since the late 1980s. With each of these, one replaces the
observed line-profile variations with carefully studied diagnostics derived from
the data.  One such method is based on the moment variations of the spectral
lines and was first introduced by \citet{Bal86a, Bal86b, Bal87} and further
developed by \citet{Ae92}, by \citet{Ae96} and by \citet{Bri03}. 
This method relies on the mathematical property that the line profile is
fully characterised by its first three velocity moments.  The latest version of
the moment method has been applied to many different types of classical pulsators.
 This method is
very powerful for low-degree modes ($l\leq 4$) in slow rotators ($v\sin i\leq
50$\,km\,s$^{-1}$).

\begin{figure}
\centering
\includegraphics[width=11cm]{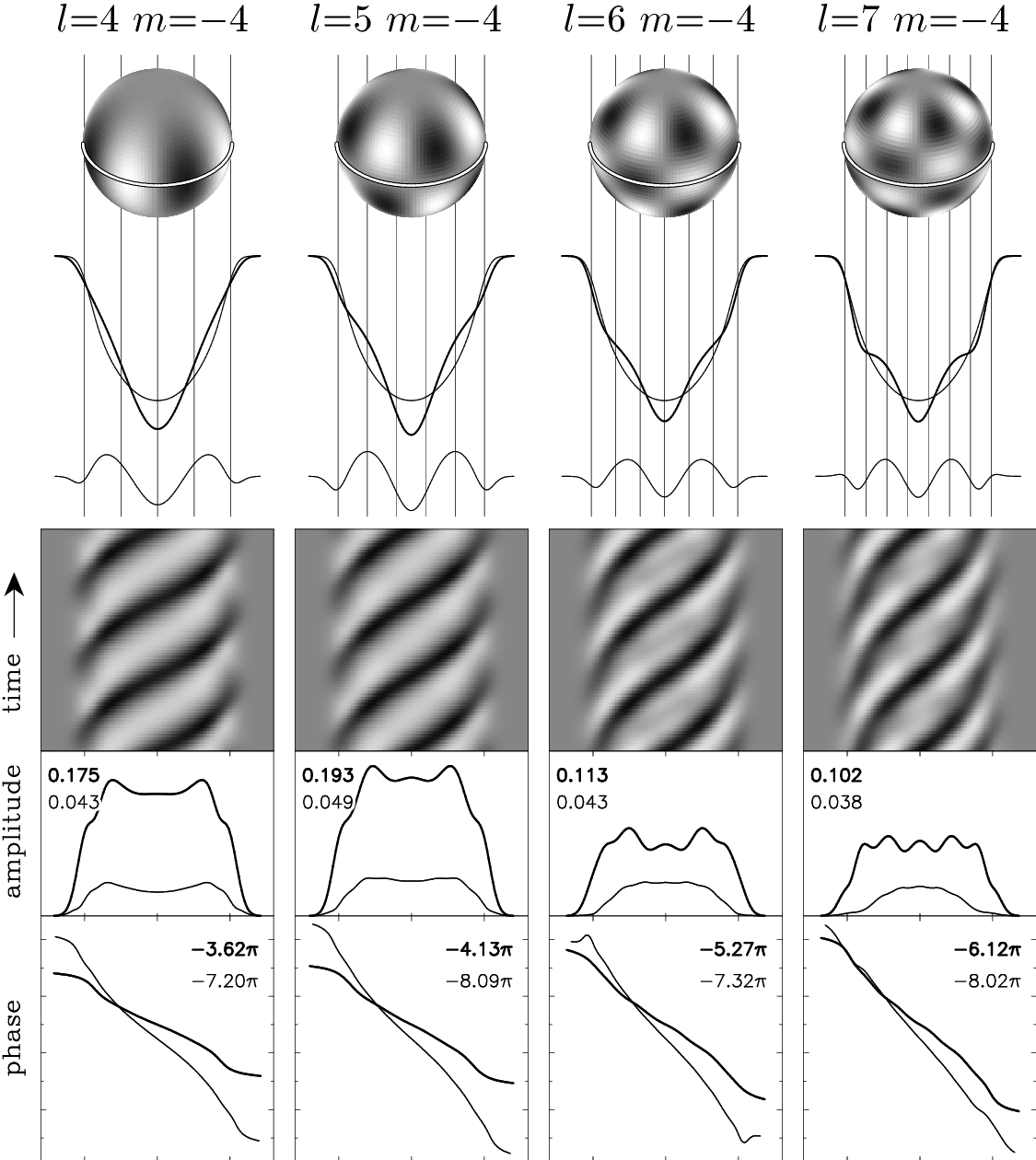}
\caption{Simulated line-profile variations due to non-radial oscillations of
different $(l,m)$. From top to bottom we show: a representation of the radial
part of the eigenfunction, the line profile due to the mode at a particular
phase in the cycle in comparison with the profile without an oscillation, the
difference between the two profiles, a grey-scale representation of the profiles
with respect to the mean during three cycles, the distribution of the amplitude
across the pulsation-induced line-profile variations (thick line) and its first
harmonic (thin line) with the maximum values indicated, the distribution of the
phase across the pulsation-induced line-profile variations
(thick line) and its first harmonic (thin line) in
units of $\pi$ radians with the blue-to-red phase differences $\Delta\psi_0$ and
$\Delta\psi_1$ used in Eqs\,(\protect\ref{ts1}) and (\protect\ref{ts2})
indicated. The projected equatorial rotation velocity is indicated by the outer
vertical lines in the top panel. Figure
reproduced from \citet{TelSchr97} with permission.}
\label{john}
\end{figure}

A second method, which will be shown below to be more relevant in terms of the
interplay with interferometry, was first introduced by \citet{Gies88}, and
further developed by \citet{KeWa96}, \citet{Schrij97}, \citet{Man00} and
\citet{Zima06}. Its use is illustrated and explained in Fig.\,\ref{john}. It is
based on the properties of the amplitude and phase distribution across the line
profiles for each oscillation frequency and its harmonic.  These properties are
linked to the $(l,m)$-value of the mode, and to the inclination angle, as can be
seen from Fig.\,\ref{john}.

The computation of the amplitude and phase behaviour across the profile is
particularly suited to analyse line-profile variations due to high-degree
($l\geq 4$) modes in rapid rotators ($v\sin i\geq 50$\,km\,s$^{-1}$), because we
need a high resolving power within the lines to interpret small moving
subfeatures.  The method can also be applied to slow rotators with low-degree
modes, however, when combined with the moment method \citep[e.g.,][]{Tel97}.  
In contrast to the moment method, no rigorous mathematical derivation
for the amplitude and phase variations across the profile as a function of $l$
and $m$ is available, despite efforts \citep[e.g.,][]{Hao98}. For this reason, 
\citet{TS98} performed an extensive Monte Carlo simulation study to
exploit the method visualized in Fig.\,\ref{john} in terms of mode
identification for p modes, including effects of the Coriolis force in the
velocity eigenfunction. From these simulations, they reached the following
conclusions:
\begin{itemize}
\item
There exists a strong correlation between the phase difference $\Delta\psi_0$ at
the blue and red edge of the profile for the oscillation frequency $\omega$ and
the degree of the mode. A good estimate of $l$ is
\begin{equation}
l\ \approx\ (0.10\ +\ 1.09\ |\Delta\psi_0|/\pi)\ \pm\ 1.
\label{ts1}
\end{equation}
\item
There exists a clear but less strong correlation between the phase difference
$\Delta\psi_1$ from blue to red for the first harmonic of the oscillation
frequency $2\omega$ and the azimuthal number of the mode. A good estimate of $m$
is
\begin{equation}
m\ \approx\ (-1.33\ +\ 0.54\ |\Delta\psi_1|/\pi)\ \pm\ 2.
\label{ts2}
\end{equation}
\end{itemize}
The simulations of \citet{TS98} clearly showed that the original
suggestion by \citet{Gies88} to associate the phase differences of
the oscillation frequency 
with a measure of its $m$-value, 
is not appropriate. A similar simulation study for g
modes is not available.

\begin{figure}
\centering
\includegraphics[width=11cm]{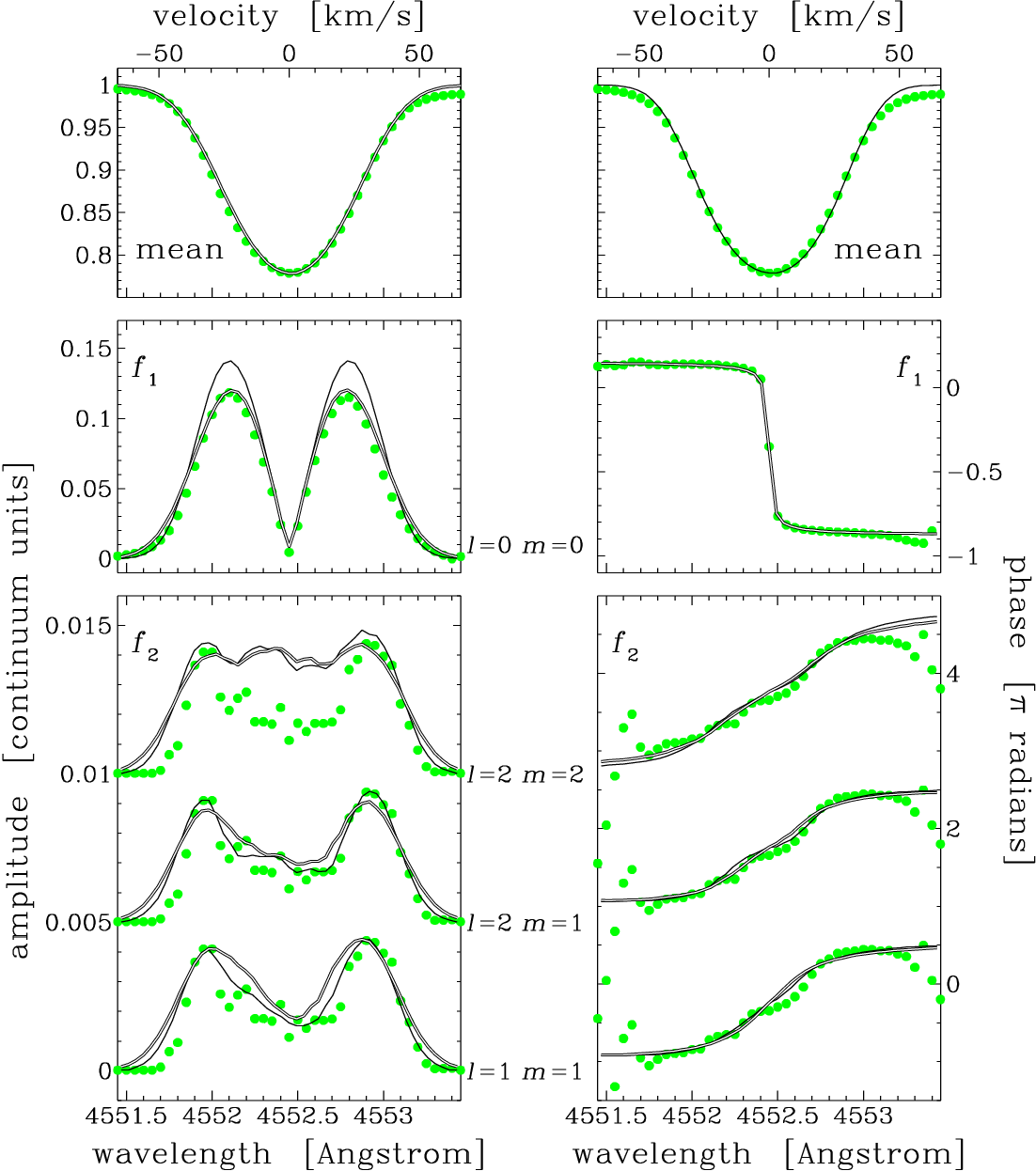}
\caption{Line-profile computations (full lines) are compared with data (dots)
for the star $\beta\,$Cephei. The top panel shows the average profile, the
middle panels the amplitude and phase across the profile for the dominant radial
mode and the lower panels the amplitude and phase of the three most likely
identifications of the small-amplitude non-radial mode ($f_2$).  Figure
reproduced from \citet{Tel97} with permission.}
\label{bcepjohn}
\end{figure}

The fitting formulae (\ref{ts1}) and (\ref{ts2}) are easy to apply once the
oscillation frequencies are determined. However, they provide only a crude
estimate of the degree and azimuthal order with a large uncertainty,
particularly for low-degree modes. It is therefore necessary to model the
amplitude and phase across the profile in full detail to achieve a reliable
identification. In order to do that, one computes theoretical line profile
variations from Eqs\,(\ref{doppler}) and (\ref{conv}), derives their amplitude
and phase across the profile as in Fig.\,\ref{john} and compares them with those
derived from the observations.  The earliest such application was made for the
star $\beta\,$Cephei by \citet{Tel97} and is depicted in Fig.\,\ref{bcepjohn} for
the dominant radial mode and for the three best solutions found for the
identification of the second, low-amplitude mode.

\citet{Zima06} introduced a statistical significance test into the \citet{TS98}
method. In this way, he was able to discriminate more easily
between different mode identification solutions. He tuned and applied his
method, which he termed {\it pixel-by-pixel method\/}, to observed
line-profile variations of the $\delta\,$Scuti star FG\,Vir \citep{Zimaetal06}.

\subsubsection{Improvements through interferometric data}

The pixel-by-pixel method described above exploits the amplitude and
phase across the profile as mode identification diagnostic by relying,
through Eqs\,(\ref{doppler}) and (\ref{conv}), on the expression for
the pulsational velocity in terms of $l$ and $m$.  The Doppler effect
is considered to be the dominant source of information in interpreting
the variations in Eq.\,(\ref{conv}) in terms of $l$ and $m$.  A new
interesting idea was put forward by \citet{Ber03a}.  They inverted a
time series of line-profile variations, turning in this way the data
into a stellar surface brightness distribution. This comes down to an
image reconstruction method, also termed Doppler Imaging in the
context of spotted stars. They applied this inversion without assuming
any pre-knowledge of the physical cause of the variations of the line
profiles. After having performed the inversion, the authors assumed
that the most important cause of the line-profile variations in
Eq.\,(\ref{conv}) are surface brightness variations superposed onto a
time-independent broadened Doppler profile.  Rather than focusing on
$v({\theta'}_i,{\varphi'}_j)$ in the interpretation of
Eq.\,(\ref{conv}), they thus considered $\delta
[I_{\lambda}(\theta',\varphi')]$ to be the dominant information for
the mode identification. Such a situation may occur for rapidly
rotating stars, whose velocity perturbation due to the oscillations is
very small compared with its rotational broadening. In such a case, the
pulsation-induced intensity perturbations gain importance with respect
to the velocity perturbations.  \citet{Ber03b} applied their method to
the rapidly rotating $\beta\,$Cephei star $\omega^1\,$Sco and found it to
be capable of recovering $l$ and $m$ of the oscillation, which had
been derived before from the Pixel-by-pixel method by \citet{TS98}.
This brings us to the capability of combining surface brightness
distributions and variations derived from interferometry, with surface
velocity variations derived from high-resolution spectroscopy. We
first describe how such brightness distributions can be measured
interferometrically.

Long before the availability of appropriate instrumentation, \citet{Vak92} 
suggested to study surface variations due to non-radial oscillations
of rapidly rotating stars from
long-baseline interferometry. When the baseline is long enough,
images of stellar surfaces
can be obtained without {\it a priori\/} information on the targets. This is
possible with the CHARA array \citep{tenBrummelaar} and an instrument 
operating in the visible, or
with the OHANA project \citep{perrin06}. As an example, the installation of the
instrument GI2T/REGAIN \citep{Mourard} at the combined focus of CHARA could give
access to the location of abundance spots in chemically peculiar stars via
direct visibility measurements in the second and third lobes of the visibility
function.  However, with the baseline lengths and numbers of baselines currently
available from interferometers, there are still quite severe limitations to the
number of resolution elements over stellar surfaces. To overcome these
limitations, a combination of interferometric information coming from exis\-ting
instruments with lower angular resolution (such as the VLTI \citep{Rantakyro})
and classical observables may be used to map submilliarcsecond (sub-mas)
structures in stellar surfaces. This technique, known as differential
interferometry \citep{beckers82,aime84} couples spatial and spectral information
by measuring fringe phases throughout a spectral range (around an
absorption/emission line for instance). The ``fringe phase'' observable being
proportional to the stellar photocenter, photocenter displacements throughout
absorption/emission lines can be measured (at a sub-mas level) and local
features generating such lines can be located. By itself, this technique may be
used, for instance, to map the surface spots in chemically peculiar pulsators,
such as roAp stars, and, thus, provide additional non-seismic constraints about
the atmospheric structure and magnetic fields, that are extremely valuable when
studying these pulsators.

\begin{figure}
\centering
\rotatebox{-90}{\resizebox{9.cm}{!}{\includegraphics{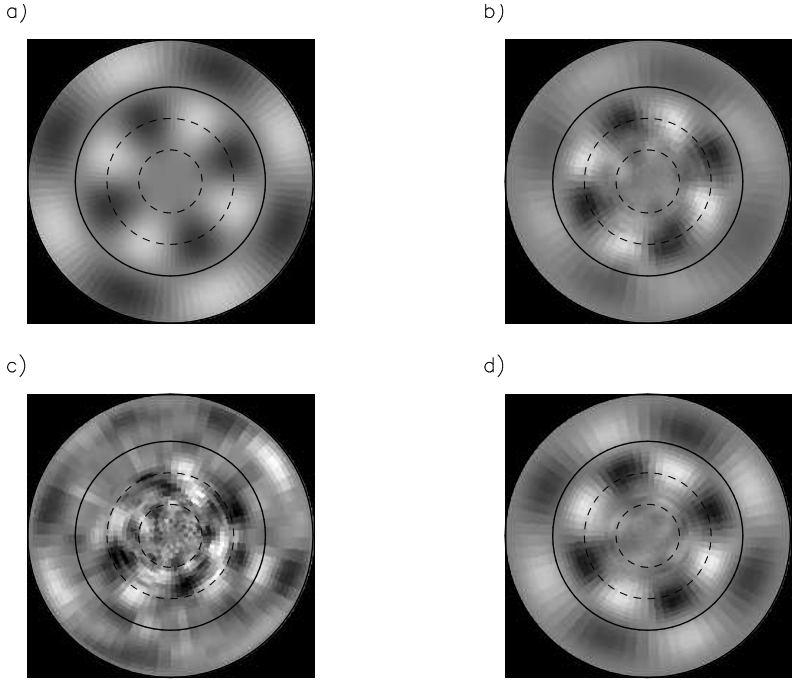}}}
\caption{(a) Simulation of the pole-on projection of the stellar surface
brightness perturbations due to an $l=5, m=4$ mode on a star with an inclination
angle of $i=45^\circ$.  
The equator and the latitudes $30^\circ$ and $60^\circ$ are
represented by full and dashed circles, respectively.
Reconstruction based on (b)
simulated flux spectra, (c) photocenter shifts, and (d) combined flux spectra
and photocenter shifts.  Figure reproduced from \citet{Jankov01b} with
permission. }
\label{jankov}
\end{figure}

From the point of view of mode identification, the combination of the two
techniques mentioned above, namely time-resolved spectroscopy and differential
interferometry, seems to be a promising approach.  As already anticipated by
\cite{Jankov01}, this combined technique can be successful in identifying
oscillation modes with $l>2$ in rapid rotators, providing information on the
modes that can perhaps not be obtained from each of the two methods separately.
The flux variations due to the non-radial modes introduce a complex pattern in
the $uv$ plane. This can be disentangled by comparing the associated
photocenter displacements due to the oscillations with predicted monochromatic
intensity maps of a non-pulsating star. In practice, one simulates photocenter
displacements as a function of $(l,m,i)$. Such a simulation defines a particular
``spatial filter'' for each $(l,m,i)$. Applying one-by-one all these spatial
filters to the data allows one to identify the true nature of the mode. This is
illustrated in Fig.\,\ref{jankov}, in which the original signal in panel (a) is
compared with a map (b) recovered from spectra alone with a method similar to
the one of \citet{Ber03a}, as well as to the map based on the
photocenter shifts alone (panel (c)), or a combination of both (panel (d)).  The
limitations of panels (b) and (c) are particularly apparent in the
reconstruction of the features below the equator, where a loss of contrast
occurs.  A significant improvement with respect to these separate
reconstructions is obtained using both spectra and photocenter shifts
simultaneously as in panel (d).

\citet{Souza02, Domici2003} and \citet{Jan04} showed that
measurements of the displacement of the photocenter across the stellar disk
using closure phase allows mapping of the surface brightness, but 
requires a minimum of three telescopes in an interferometric
array in such a way that 
fringes are collected for all three baselines. The simulation
study by \citet{Jan04}
anticipates that the closure phase measurements are sufficiently sensitive
to detect a mode of low $(l,m)$.  In general, however, numerous modes are
simultaneously excited.  In such more realistic cases, the photocenter
displacements are `washed out' by the averaging effect of the many
$(l,m)$-values.  In that case, one can perform {\it stroboscopic
interferometry\/}.  The aim of this is to pursue an identification for a fixed
number of oscillation frequencies which have been derived from time series
analysis of observables of any kind. When carrying out the interferometric
measurements, a selected oscillation frequency is used to phase-lock the data
to this frequency. In this way, all surface structure that is not associated
with this frequency is assumed to be removed. This would greatly improve the
phase closure signal strength.

It is important to note that the phase-locking does not need to be carried out
during the observations, which would create insurmountable problems in its
practical implementation. Rather, the interferometric fringes may be collected
at the normal cadence which needs to be sufficiently high compared with the
oscillation frequencies. The frequency filtering can be done as a
post-processing step by an appropriate weighting procedure. The design of such
weights is a standard time-series analysis/inverse problem. A further advantage
of doing the latter is that it would be possible to design the appropriate
weights for each of the measured oscillation frequencies separately, and use the
same set of interferometric observations to constrain the identification of all
the oscillation modes whose frequencies are known from other diagnostics.

\subsection{\label{cs}Current studies of pulsating stars including
  interferometric constraints}

Sections \ref{gp} and \ref{mi} give a clear indication that results from
observations obtained with current interferometric instruments are likely to
have a very significant impact in asteroseismic studies in the very near
future. 
At present, however, the number of pulsating stars for which
sufficiently accurate interferometric and asteroseismic data are available is
still limited. In this section we review the results of a few case
studies of stellar pulsators involving a combination of interferometric and
asteroseismic constraints.

\subsubsection{$\alpha$ Cen}

As emphasised earlier, the study of binary systems with combined interferometric
and asteroseismic constraints is of particular interest.
The visual binary $\alpha$~Cen is amongst the most promising systems in
this context. Besides the availability of precise luminosities, masses and
chemical composition for both components, interferometric and seismic constraints
have recently been obtained for this system. The linear radii of $\alpha$~Cen A
and B are known with high precision, thanks to the combination of precise
parallaxes \citep{Soderhjelm99} and angular diameters measured with VINCI, at
VLTI \citep{kervella03b, Bigot06}.  Moreover, solar-like oscillations have been
detected, first in $\alpha$~Cen A by \citet{Bouchy02}, and then in component B
by \citet{Carrier03}. The oscillations of both components were meanwhile studied
in more detail by \citet{Beddin2004} and \citet{Kjelds2005}.

The numerous and precise observational constraints available make the binary
system $\alpha$~Cen a suitable target to test stellar structure and
evolution models in conditions that are slightly different from those in the
Sun. This is the reason why $\alpha$~Cen has been the subject of many
theoretical studies, in particular since the detection of solar-like
oscillations in $\alpha$~Cen A \citep[see e.g.,][]{Thevenin02,Thoul03,
Eggenberger04, Miglio05, yildiz07}.

\citet{Miglio05} modelled $\alpha$~Cen by means of the
Levenberg-Marquardt non-linear fitting algorithm 
\citep[e.g.,][]{BR2003}
that performs a simultaneous least-squares
adjustment of 
the model parameters ($M_{\rm A}$,$M_{\rm B}$,$Y_0$,$Z_0$,$\alpha_{\rm A}$,
$\alpha_{\rm B}$,age) to all the observational constraints.
A goal was to study how the results of the fit
depend on the `physics' included in the stellar models, and on the
choice of classical, interferometric, and seismic observables
considered in the fitting procedure. Here, the subscripts A and B refer, respectively, to the components A and B of the system. The addition of a precise linear
radii to the observational constraints resulted in a reduction of the
size of the error box associated with the position of the star in the
HR diagram, and, in turn, in a reduction of the uncertainties in the
parameters derived from the fitting.  This reduction in
the uncertainties associated with the derived parameters is
particularly important when evaluating the significance of adding an
extra free parameter in the modelling, as, for instance, when
considering two distinct mixing-length parameters for components A and
B of the system. However, the authors also found that when using the
linear radii, instead of $T_{\rm eff}$, as the observable, the
agreement between the observed and the model average large separations
becomes worse. As discussed in Fig.\ \ref{fig:soldif}, this is likely
a consequence of the inaccurate modelling of the surface layers of the
star, which directly influences the value of the large separation at
high frequency. In order to avoid this problem, when deriving global
parameters of $\alpha$~Cen through the fitting procedure,
the authors followed the suggestion of \citet{Roxbur2003}, and
considered as seismic observables, combinations
of frequencies that are largely independent of near-surface layers
(cf.\ Eq.~(\ref{eq:seprat})).

\begin{figure}
\centering
\includegraphics[width=0.95\textwidth]{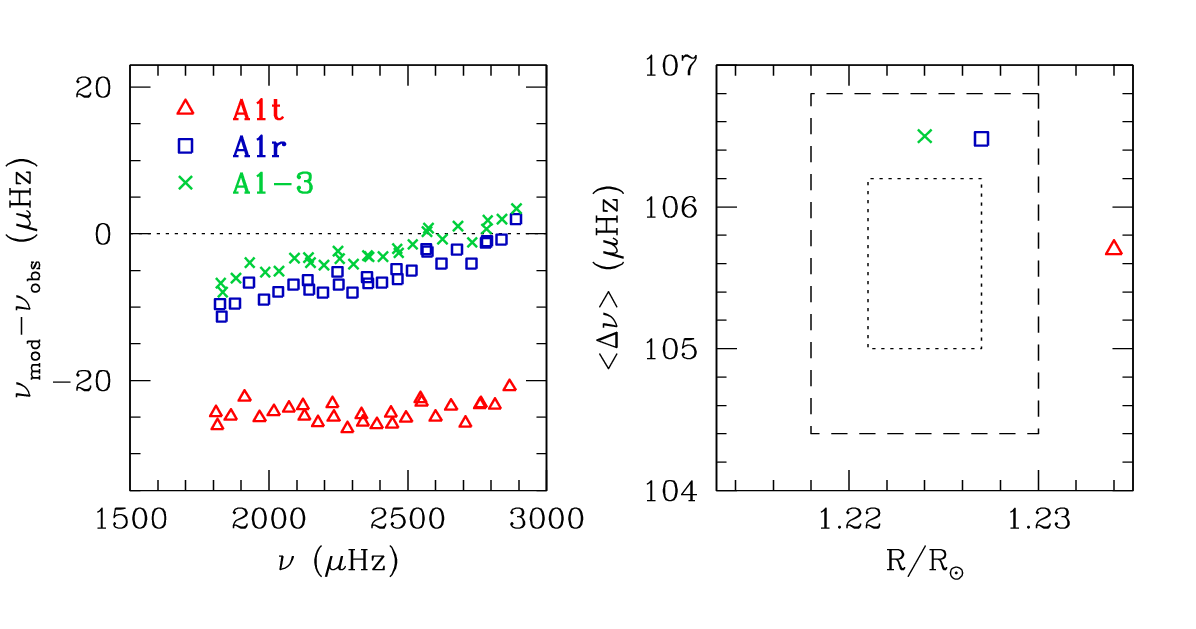}
\caption{Difference between theoretical and observed frequencies for models of $\alpha$~Cen~A ({\it left}), and a radius - $\langle \Delta\nu \rangle$ diagram ({\it right}) comparing the values predicted by the models \citep[see][]{Miglio05} to the 1- and 2-$\sigma$ error boxes in $R$ and $\langle \Delta\nu\rangle$. The models are calibrated considering among the constraints: $T_{\rm eff}$ and $\langle\Delta\nu \rangle$ (A1t), the radius and $\langle\Delta\nu\rangle$ (A1r) and, in the case of model A1-3, including in the fit the radius and the combination of frequencies suggested by \citet{Roxbur2003}. Though the observed $\langle\Delta\nu\rangle$ is well reproduced, model A1t predicts a radius which is inconsistent with the observed value and is responsible for a large shift between observed and theoretical frequencies. When the interferometric radius is included in the fit (A1r, A1-3), the predicted $\langle\Delta\nu\rangle$ is larger than the observed: this discrepancy (which is well known in the solar case) is likely to be related to surface effects.}
\label{miglio}
\end{figure}

 Figure~\ref{miglio} compares the results for $\alpha$~Cen A obtained when different observables are included in the fitting. When $T_{\rm eff}$ and the average observed large separation, $\langle \Delta\nu \rangle$, are used (model A1t) a large discrepancy is found between model and observed frequencies. Moreover, in this case the radius of the model is inconsistent with the interferometric one. On the other hand, when the interferometric radius is used instead of $T_{\rm eff}$ (models A1r and A1-3), the difference between model and observed frequencies is significantly reduced. However, the average large separations derived from the models are found to be larger than the observed one. Model A1-3, obtained when using the interferometric radius and the combination of seismic observables suggested by \citet{Roxbur2003}, shows the best agreement with the observations both in what concerns the predicted frequencies and radius. 

\subsubsection{$\tau$~Cet}
\label{tauCet}

$\tau$~Cet 
is a G8\,V star. It was one of the first
asteroseismic candidates to have its radius measured by VINCI at the VLTI 
\citep{PTG03, DFetal04}.
Having an accurate
measurement of the stellar radius allowed Teixeira et al.\ (in preparation) 
to construct preliminary models for $\tau$~Cet, and to
predict both the average large separation, and the location, in
frequency, of the expected envelope for solar-like
oscillations. As a logical follow-up of that work, an asteroseismic
campaign with the HARPS instrument at the $3.6$-m telescope in Chile
was carried out. Despite adverse
weather and unexpected technical problems at the telescope, 32
oscillation frequencies were measured, corresponding to mode degrees
of $l = 0,1,2,3$. The asteroseismic data provided further independent
constraints for the modelling, in addition to the four ``classical'',
non-asteroseismic observables ($T_{\rm eff}$, $L$, [Fe/H], $R$).

A large range in values of $\tau$~Cet's metallicity and effective temperature
is reported in the recent literature. Estimates of [Fe/H] published over the
last four years range from $-0.57$ to $-0.4$, while those of $T_{\rm eff}$ range
from 5270\,K to 5420\,K. In these (not unusual) circumstances, a precise
determination of the stellar radius, through the combination of interferometric
data and the parallax, is particularly important.

To model $\tau$~Cet, the authors used the Aarhus STellar Evolution code
\citep{Christ1982, Christ2007},
and the Aarhus Adiabatic Pulsation code \citep{CDB91},
coupled to an automated fitting
procedure based on the Levenberg-Marquardt optimization method 
running on a cluster with 8 processors. The procedure
searches the parameter space for the combination of stellar parameters ($M$,
$X_0$, $Y_0$, $\alpha$, age) that best reproduces the non-asteroseismic and
asteroseismic observables for the star. Whilst it is relatively straightforward
to include the ``classical'' observables in the fit, the asteroseismic
observables present a bigger challenge.

\begin{figure}
\centering
\includegraphics[width=0.75\textwidth]{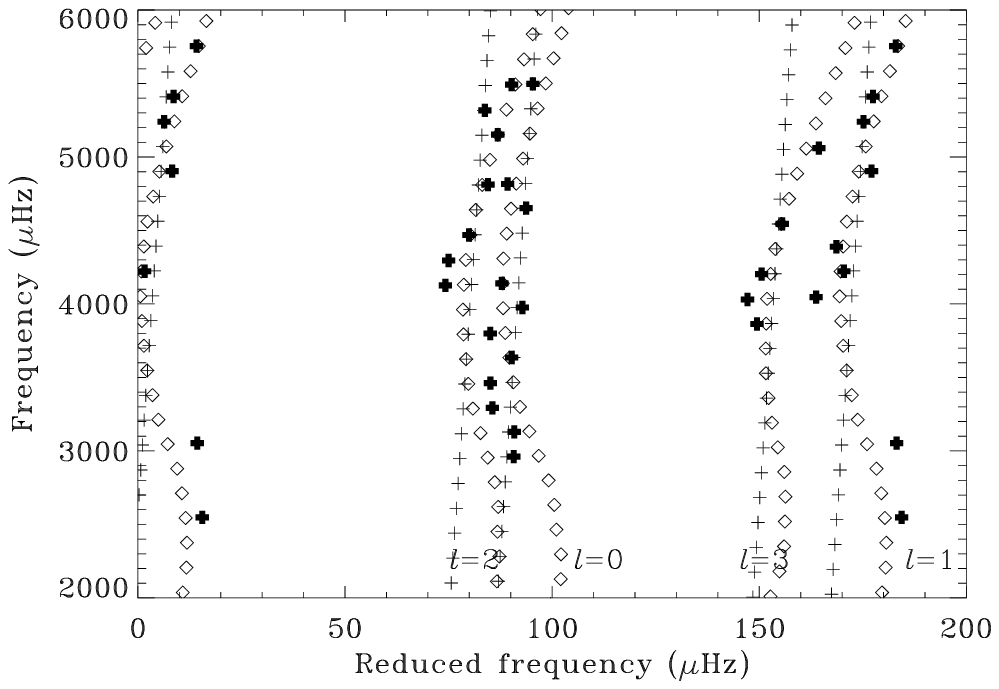}
\caption{Echelle diagram showing the observed frequencies for $\tau$ Cet (thick crosses;
  Teixeira et al., in preparation) and the model frequencies (diamonds), with an average large
  separation $\langle \Delta \nu \rangle = 168.83 \muHz$;
  the thin crosses indicate the asymptotic relation.
  The model and asymptotic relation frequencies
  have been increased by $35\,\mu$Hz to coincide with the observed frequencies.
  }
\label{tct_fig1}
\end{figure}

In the asymptotic regime, the theory of stellar oscillations predicts
the frequencies to exhibit regular spacings with characteristic large
and small separations (cf.\ Eqs~(\ref{eq:pasymp}) and
(\ref{eq:smallsep})). However, for $\tau$~Cet the oscillation
frequencies around peak power for low $l$ values are not yet in the
asymptotic regime. 
None the less, the authors first attempted a fit to the observations based on an asymptotic
relation, to match the large and small separations.
This resulted in finding a best solution with predicted individual
frequencies significantly shifted from the observed ones, as shown
in Fig.~\ref{tct_fig1}.

\begin{figure}
\centering
  \includegraphics[width=0.85\textwidth]{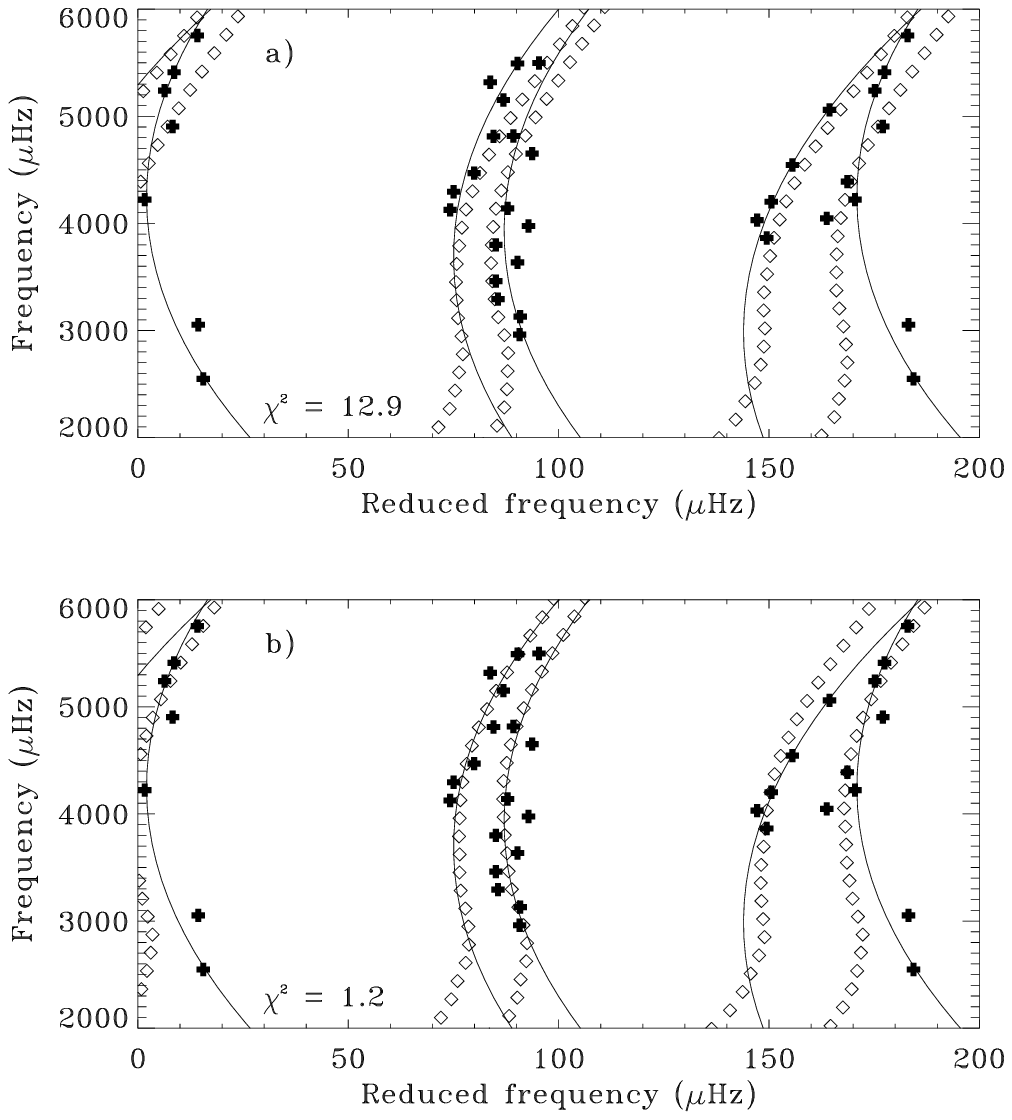}
\caption{Echelle diagrams showing the observed frequencies (thick
  crosses; Teixeira et al., in preparation), the best-matching Bedding et
  al.\ functions (curves), and the model frequencies (diamonds) for two
  cases: {\bf a)} without surface correction, and {\bf b)} with
  surface correction (100\% relative correction).
  An average large separation $\langle \Delta \nu \rangle = 168.83 \muHz$
  was used.
  } 
\label{tct_fig2}
\end{figure}

In order to include the absolute values of the observed frequencies in
the fit, the authors used functions to fit the frequency ridges in the
\'echelle diagram, as suggested by \citet{Beddin2004}.
From those,
an absolute frequency, a large separation, and a small separation were
derived, and these quantities were used 
as asteroseismic observables in the
optimization procedure. The resulting solution was encouraging, as can
be seen in Fig.~\ref{tct_fig2}a. However, the model did not appear to
match particularly well the observations for the higher frequency
modes. Moreover, the mixing-length parameter resulting from the fit of
the seismic data was significantly larger than in the solar case. Once
again, this was interpreted as a consequence of the deficient modelling of the outer layers
of the star (cf.\ discussion in
Fig.\ \ref{fig:soldif}). Consequently the authors
decided to include a surface correction in the fit, by scaling the
empirical solar surface frequency correction to the case of
$\tau$~Cet. That resulted in a very satisfactory fit, as shown in
Fig~\ref{tct_fig2}b, now with a mixing-length parameter close to that
for the Sun.

\subsubsection{$\beta$ Hyi}
$\beta$ Hyi is a bright G2 subgiant. Despite early attempts to detect solar-like
oscillations in this star \citep{frandsen87,edmonds95}, such oscillations were
confirmed only with the works of \cite{bedding01} and \cite{carrier01}, using
UCLES and CORALIE, respectively. Recently, a dual-site campaign, involving HARPS
and UCLES \citep{Beddin2007}, allowed for the identification of 28 oscillation
modes, some of which showing a clear indication of avoided crossings, as
expected from theoretical studies of this star \citep{fernandes03,dimauro03}
(see also Section~\ref{mixmodes}). From these results, the authors derived the
mean density of $\beta$~Hyi to be $\bar\rho= 0.2538 \pm 0.0015$~g\,cm$^{-3}$.

Recently \citep{north07a} have measured the angular diameter of $\beta$~Hyi
using Sydney University Stellar Interferometer, having found a limb-darkened
angular diameter of $2.257\,\pm\,0.019$~mas. Combining the latter with the
Hipparcos parallax for the star, the authors found a linear radius
$R=1.814\,\pm\,0.017$~R$_\odot$, which combined with the mean density derived
from asteroseismic data, allowed them to derive a mass
$M=1.07\,\pm\,0.03$~M$_\odot$. This value of the mass is in agreement with
previous determinations based on model fitting of the observables, but is
significantly better in precision. With these recent asteroseismic and
interferometric constraints, it should be possible in the near future to produce
improved models of $\beta$~Hyi and investigate on physical aspects of its
interior.

\subsubsection{$\theta^2\,$Tau}

As already mentioned above, \citet{Tom95} have combined
spectroscopically and interferometrically measured orbits to deduce the
fundamental parameters of the $\delta\,$Scuti star $\theta^2\,$Tau. They achieved
the following ranges for the masses of the components: $M_A = 2.1 \pm
0.3$\,M$_\odot$, $M_B = 1.6 \pm 0.2\,$M$_\odot$ and a distance of $44.1 \pm
2.2$\,pc. While these are valuable mass determinations for a system including a
classical pulsator, they remain largely above 10\% relative precision.

The star was the subject of a large multisite campaign resulting in multicolour
ground-based photometry \citep{Bre02}, as well as of a WIRE space photometric
campaign \citep{Por02}.  This resulted in 11 established frequencies in the
range 10.8 to 14.6 c\,d$^{-1}$, which corresponds to the region of radial p$_5$
to p$_6$ modes for appropriate stellar models.  The largest stumbling block in
the seismic modelling, is the lack of good mode identification for this star,
despite the accurate measurements of its line-profile variations
\citep{Ken96}. This prevented \citet{Bre02} from reaching a conclusion on the evolutionary
status of the pulsating primary. In particular, it was impossible to deduce if
the $\delta\,$Scuti star is in a post-main-sequence phase (in agreement with
models with or without core overshooting) or is still on the main-sequence
(compatible with computed models including core overshooting).  The main
uncertainties in the modelling, besides the mode identification, are due to the
simple treatment of the convective flux in the hydrogen ionization zone.

\subsection{\label{futurea}Future studies of pulsating stars including 
interferometric constraints}
Besides the studies presented above, a number of additional ones of pulsating
stars combining interferometric and asteroseismic constraints have been proposed
and are awaiting the acquisition of seismic or interferometric data.  Among
these are the binary system {12 Boo}, in which both components are expected to
show solar-like oscillations, and the classical pulsators $\alpha\,$Vir (Spica),
{$\lambda\,$Sco} and $\beta\,$Cen (three $\beta$\,Cephei stars). Moreover, future
interferometric observations with AMBER at VLTI are planned for the $\beta\,$Cephei
star $\kappa\,$Sco and for the slowly pulsating B star HD\,140873.

\subsubsection{12 Boo}

12 Boo is a double-lined spectroscopic binary whose orbit has been resolved by
interferometry: this has allowed 
determination of the masses of
both components ($M_{\rm A}=1.416~{\rm M}_\odot$ and $M_{\rm B}=1.374~{\rm M}_\odot$) 
with a relative precision of about 0.3 \% \citep{Boden05b, Tomkin06}.
Even though the masses of the
components are very similar, their different luminosities, when compared with
theoretical predictions, suggest that the secondary component is still in the
central-hydrogen burning phase, while the primary is more evolved and burning
hydrogen in a shell. As presented by \citet{Boden05b}, however, this conclusion
is highly dependent on the model details and needs further investigation. A
detailed modelling of the system \citep{Miglio07}
shows that
several models of 12\,Boo, based on different theoretical predictions of
extra-mixing processes in the core (e.g., overshooting), are able to reproduce
the available observational constraints.

In order to discriminate among these scenarios, additional and independent
observational constraints are needed. These could be provided by solar-like
oscillations, as these are expected to be excited in both components of the
system.  A precise knowledge of the frequencies of high-order p modes in the
primary component would help determining the evolutionary state of the system
and would give information on the chemical composition gradient built by the
combined action of nuclear burning and mixing in the central regions of the
star.

\subsubsection{$\lambda\,$Sco}

\citet{Tan06} observed the triple system $\lambda\,$Sco, composed of two
early-B stars and a pre-main-sequence star \citep{Uyt04a},
interferometrically with the Sydney University Stellar Interferometer. This
allowed them to determine the elements of the wide orbit with much higher
precision than the previous estimates from spectroscopy alone \citep{Uyt04a}.
The wide orbit turns out to have a low eccentricity, suggesting
that the three stars were formed at the same time, only a few million years ago.

From the brightness ratio and the colour index of the two B stars, \citet{Tan06}
confirmed the previous classification of $\lambda\,$Sco A as B1.5IV and
$\lambda\,$Sco B as B2IV.  They used the mass-luminosity relation for
main-sequence stars to determine the mass ratio of the two B stars and found
$M_B/M_A = 0.76 \pm 0.04$.  The individual masses were subsequently derived from
this mass ratio, the mass function, spectrum synthesis and the requirement that
the age of both components must be the same. This led to $M_A = 10.4\pm 1.3\,$M$_\odot$ and
$M_B = 8.1 \pm 1.0\,$M$_\odot$. 

\citet{Uyt04b} found $\lambda\,$Sco A to be a relatively fast
rotator with a $v\sin i$ of 125\,km\,s$^{-1}$ and to have clear line-profile
variations due to a dominant prograde dipole oscillation mode. Additional
low-amplitude modes were also found and these seem to originate from modes with
degree $l>2$.  Ground-based photometry does not reveal these higher-degree modes
seen in the line-profile variations.  Progress could be achieved from
high-precision space photometry, however. Such data have meanwhile been gathered
with the WIRE satellite (Bruntt et al., in preparation). Hopefully this will
lead to accurate modelling results for this massive B star in the near future.

\subsubsection{$\beta\,$Cen}

\citet{Aus06} have presented dedicated methodology to derive high-precision
estimates of the fundamental parameters of the double-lined spectroscopic binary
$\beta\,$Cen which contains a pulsating $\beta\,$Cephei component in its 357\,d
orbit with an eccentricity of 0.81. Their method is based on high-resolution
spectral time series and interferometric data with a good phase distribution
along the orbit. Such data allowed them to derive the component masses and
dynamical parallax with a relative precision of 6\% and 4\%, respectively, where
the two B stars have almost the same masses and stellar properties.  The
equatorial rotation velocity of the primary is, however, about twice the one of
the secondary, which is quite remarkable and forms a challenge to current
formation scenarios and tidal theories of massive binaries.

Two oscillation frequencies with values expected in the range of $\beta\,$Cephei
stars were derived from the line-profile variations of the broad-lined
primary. Spectroscopic mode identification pointed out that their degree is
definitely higher than 2, but secure values for $l$ could not be
deduced. Ground-based photometry will not help to resolve this ambiguity,
because the binary is photometrically constant. The high-degree modes will
probably be well visible from space photometry, but such data have not yet been
assembled so far (the star is outside the viewing zone of MOST). No evidence of
oscillations in the narrow-lined secondary was found from the high-resolution
spectroscopy. It would definitely be worthwhile to measure this binary from
space with high-precision photometry.

\subsubsection{$\kappa$\,Sco} 

Future interferometric observations of the $\beta\,$Cephei star $\kappa$\,Sco were
approved by the AMBER Science Group in the framework of the Guaranteed Time
Observations in a proposal that included also the slowly pulsating B star
HD\,140873.

The $\beta$\,Cephei star $\kappa$\,Sco has a quite large rotational velocity of
120\,km\,s$^{-1}$.  This target thus provides an opportunity, just as
$\lambda\,$Sco and $\beta\,$Cen, to estimate the influence of rotation on
pulsation, once masses and tidal parameters are determined.  The main pulsation
period has been identified as an $(l=1\,{\rm or}\,2, m=-1)$ mode. Five other
frequencies are present, but in addition to pulsation, rotational modulation is
also invoked to interpret the complex spectral line profile variations
\citep{u05}.  The orbital period of the star is $P=195$\,d, and the eccentricity
is close to 0.5 \citep{u01}.  The mass ratio of this SB2 system is 1.11.  An
estimation of the mass of the main component led to an orbital inclination close
to $80^\circ$, and a maximal visual separation of the order of 10\,mas. The
system can thus be easily resolved with the Auxiliary (ATs) or Unit Telescopes
(UTs) at ESO.

\subsubsection{$\alpha\,$Vir}

The bright $\beta\,$Cephei star $\alpha\,$Vir (Spica) has been measured
interferometrically with the CHARA Array and the FLUOR instrument by
\citet{Auf07}.
Spica is a massive, non-eclipsing, double-lined spectroscopic
binary with a 4-d orbital period and two relatively rapid rotators. The
projected rotational velocity $v\sin i$ is again of order 120\,km\,s$^{-1}$,
just as for $\kappa\,$ and $\lambda\,$Sco and $\beta\,$Cen. Previous
interferometric data with the Sydney University Stellar Interferometer are also
available.  Modelling of the new and archival interferometric and spectroscopic
data is in progress.

On the other hand, Spica was the main target of a 
23-day MOST run from 26 March to 18 April 2007.
This, together with a simultaneous but longer high-resolution
spectroscopic campaign with the CORALIE spectrograph, attached to the 1.2-m
Euler telescope at La Silla and with the spectrograph attached to the 1.22-m 
telescope at the Dominian Astrophysical Observatory, 
should offer a very detailed picture of Spica's
oscillations. This $\beta\,$Cephei star hasn't been revisited since the pioneering
work of \citet{Smith85a, Smith85b} who suggested tidally affected modes for this
star and a tentative mode identification of $l=2$ and 8.

The combination of these state-or-the-art interferometric, spectroscopic and
space photometric data should allow detailed modelling of this $\beta\,$Cephei
binary in the near future.

\subsubsection{HD\,140873} 

The slowly pulsating B star HD\,140873 is member of a B8\,III + A7V binary
system. This double-lined binary has an orbital period of $P=39$\,d and a quite
high eccentricity of 0.73 \citep{dc00}.  Only one pulsation frequency has been
definitively identified as an $(l=1, m=+1)$ mode \citep{DeCat2005} but other
frequencies seem to be present.  The mass ratio of the system amounts to 2. The
estimate of the mass of the primary leads to an inclination angle of about
$50^{\circ}$, implying a maximum separation of 6\,mas and a minimum separation
of 1\,mas. In the latter case, the fringe signal will be poorly sampled with the
UTs.  Among the four known binary slowly pulsating B stars, HD\,140873 is the
one with the longest orbital period, hence allowing the best observational
set-up.

In principle, two interferometric points on the orbit should be enough to derive
the inclination angle if one of these points is on the line of the nodes.
However, due to the uncertainties on the ephemeris, 10 points seem necessary to
derive a proper interferometric orbit, in order to get a precision of a few
percent on the stellar mass.

\section{A look into the future\label{future}}
As has been the case for observations
 of the 5-minute oscillations in the Sun, the most obvious
 way to overcome the fundamental observational limitations to the study of solar-like pulsators, namely observing length and daily aliases, is to establish networks of telescopes and/or observe from  the poles. 

 Several projects are now in preparation or underway to explore these 
 possibilities and include the creation of facilities at DOME-C in 
 Antarctica and global networks of telescopes at different longitudes and
 latitudes. At the time of writing we are not aware of any networks which
 are funded.  A short but incomplete list of ongoing studies includes: {\it SIAMOIS} \citep{Mosser2006}
which 
 aims at using a small telescope with a Fourier transform spectrometer at 
 Dome C, and
 the Stellar Oscillations Network Group (SONG) aiming at a network of 
 several telescopes with spectrographs for high-precision radial velocities
 \citep{Grunda2007}.

 It is to be expected that such efforts as {\it SIAMOIS} or SONG will
 lead to significant progress in the field of asteroseismology. Prior
 to that it would seem fruitful to try to explore the use of existing
 facilities for short, 1--2 weeks, dedicated campaigns of carefully selected
 solar-type targets. For equatorial objects the current location of
 the existing facilities allows for good possibilities of obtaining a
 high duty-cycle if such campaigns can be organized.
An example is the major campaign on Procyon which was organized in January 2007.

The Concordia station at Dome C opens also important new opportunities for
asteroseismic studies of classical pulsators. This station began as a
Franco-Italian scientific base, but has recently expanded under FP6 funding for
the ARENA European Network,%
\footnote{see {\tt http://arena.unice.fr/}} which includes
seven countries (Belgium, France, Germany, Italy, Portugal, Spain, and UK), and
also Australia.  Site testing has been underway for some time, and the first
winter-over and night-time tests were made in 2005.

Dome C is at latitude $-75^{\circ}$ high on the Antarctic plateau (3200\,m
elevation). During the day the median seeing is $0.5^{\prime\prime}$; it is
smaller than $0.3^{\prime\prime}$ for 25\% of the time, and can at times
decrease to only $0.1^{\prime\prime}$ -- nearly space values. A recent
winter-over has shown that at night there is a layer near the ground that
degrades the seeing because of the extreme temperature gradient (night time
temperatures are $\sim -75$\,C), but 30\,m above the ground the exceedingly low
values given above are recovered, so giant telescopes being planned for the site
will be mounted on high platforms.

For asteroseismic photometry the most important characteristic is the low wind
speeds and scintillation. Because the winds of Antarctica are katabatic, they
flow away from Dome C. They are low speed at all altitudes, as tested by balloon
flights -- significantly lower than at Mauna Kea or Paranal, for example. The
result is significantly lower scintillation noise, which is the major noise
source in ground-based high-speed photometry for stars that are not photon-noise
limited. Since the first winter-over indicates photometric conditions occur 80\%
of the time, asteroseismic telescopes at Dome C will be able to obtain data sets
with 80\% duty cycles for over 20 weeks during the long polar night, and with
higher precision than for any other ground-based site. Just as the South Pole
observations dominated helioseismology for many years in the 1980s
\citep{1980Natur.288..541G}, Dome C asteroseismic observations will do so for
stars that cannot be observed with the satellite missions.

Improvements in interferometric facilities, with potential impact to asteroseismology, are also expected over the next decade, as discussed in some detail in Section~\ref{futureinterf}. In particular, the Magdalena Ridge
Observatory interferometer, which is currently  under construction,
will have six 1.4-m telescopes  with baselines up to 340~m, and will operate
in the  red-near infrared  range \citep{Busher2006}.  Being  driven by
interferometric imaging,  this facility  will have a  superior $(u,v)$
plane coverage to the CHARA  array and similar sensitivity. Simultaneously, upgrades  of  existing facilities are expected, with particular emphasis on: combinations of larger numbers of telescopes; extension to shorter wavelengths, in particular pushing facilities currently operating in the near-infrared to visible wavelengths; pushing the limiting magnitude of the interferometers which are still far from the nominal limiting magnitude; and, when the site allows, extension to longer baselines.

Just as in the case of ground-based projects, a number of space projects including a significant asteroseismology
programme may be expected in the future. In North America, the Canadian nanosatellite project BRight Target Explorer (BRITE)
will observe the oscillations of a few bright stars; the NASA Kepler
mission, essentially aimed at detecting planetary transits, will also
include an asteroseismology programme \citep{basri05}; the long
term Stellar Imager project \citep{carpenter05}, presently under
study at NASA, will in principle provide a sufficient resolution at
the surface of nearby stars to study high degree modes up to
$l$=10. On the European side, a new call for
missions was issued by ESA in the Spring of 2007.
A European consortium has prepared and submitted
an answer to this call for an asteroseismology and
exoplanet search mission named PLAnetary Transits and Oscillations of Stars (PLATO) \citep{Catala2005}. The novelty of the
proposed approach will be to search for planetary transits and
analyse stellar oscillations on the same targets, thus providing
strong observational constraints on star and planet evolution,
e.g., by measuring the ages of stars that are hosting planets, and
studying the distribution of planet sizes and orbits as a function of
age. A significant number of these targets will be bright and
sufficiently nearby to envisage high precision interferometric
measurements of their diameters and surface structures. 

More ambitious space projects are being discussed for the longer term
future. One of them, tentatively named AIM (for AsteroInterferometric
Mission), couples interferometric techniques and ultra-high precision
photometry or Doppler velocimetry to resolve stellar disks and
measure high degree modes at the surface of stars. A free flying
fleet of 10 to 100 telescope units of 1-m diameter each, on baselines
of 10 to 40 km, would constitute a powerful interferometer capable of
resolving high degree oscillation modes, up to $l=200$ for a solar-type
star at 10 pc, or $l$=10 for a typical cool star in the Pleiades.

On the interferometry side, it is possible that within a time frame of a decade or so, a small
interferometric  mission  like  PEGASE  \citep{Ollivier2006}  will  be
launched. It will combine two 40-cm siderostats in baselines up to 500
meters  and operate  primarily in  the 1.5-6~$\mu$m  range.  However an
extension to the 0.8-1.5~$\mu$m range would complement existing
ground facilities for stellar diameter measurements. PEGASE
would further allow the following of diameter variations in rapid (few hours to few days)
pulsators that will be uncovered by MOST and CoRoT.

The future of asteroseismology, interferometry and their synergy looks very bright, indeed.

\begin{acknowledgements}
The authors are very grateful to Hans Kjeldsen and G\"unter Houdek for providing
important input to parts of this review. Thanks go also to Michael Bazot, Kara
Burke, Cristina Fernandes, Jo\~ao Fernandes, Michael Gruberbauer, Jo\~ao
P. Marques, Fernando Pinheiro, Jonathan Riley, Marta Santos, Nuno Santos and
S\'ergio Sousa, for useful discussions during the CAUP/OPTICON workshop on
Interferometry and Asteroseismology.  JC-D thanks Michael Kn\"olker for
hospitality at the High Altitude Observatory where much of his contribution was
written.  This work was supported by the OPTICON through funds from the European
Commission.  MC is supported by the EC's FP6, FCT and FEDER (POCI2010) through
the HELAS international collaboration and through the project
POCI~/CTE-AST~/57610~/2004.  
CA is supported by the Research Council of Leuven
University under grant GOA/2003/04.  GP acknowledges support from PHASE, the
space and ground-based high angular resolution partnership between ONERA,
Observatoire de Paris, CNRS and University Denis Diderot Paris 7.
PJVG was
supported in part by the Funda\c{c}\~ao para a Ci\^encia e a Tecnologia through
projects POCI/CTE-AST/55691/2004 and PTDC/CTE-AST/65971/2006 from POCI, with
funds from the European program FEDER.

\end{acknowledgements}


\end{document}